\documentclass{aa}

\usepackage{graphicx}
\usepackage{txfonts}
\usepackage{color}
\usepackage{enumitem}
\usepackage{ulem}
\usepackage{pifont}
\usepackage{color}

\usepackage{gensymb}

\usepackage{subfigure}
\usepackage{float}
\usepackage[colorlinks,linkcolor={blue}, citecolor={blue}, urlcolor={blue}]{hyperref}
\usepackage{etoolbox}
\usepackage{natbib}

\makeatletter

\patchcmd{\NAT@citex}
  {\@citea\NAT@hyper@{%
     \NAT@nmfmt{\NAT@nm}%
     \hyper@natlinkbreak{\NAT@aysep\NAT@spacechar}{\@citeb\@extra@b@citeb}%
     \NAT@date}}
  {\@citea\NAT@nmfmt{\NAT@nm}%
   \NAT@aysep\NAT@spacechar\NAT@hyper@{\NAT@date}}{}{}

\patchcmd{\NAT@citex}
  {\@citea\NAT@hyper@{%
     \NAT@nmfmt{\NAT@nm}%
     \hyper@natlinkbreak{\NAT@spacechar\NAT@@open\if*#1*\else#1\NAT@spacechar\fi}%
       {\@citeb\@extra@b@citeb}%
     \NAT@date}}
  {\@citea\NAT@nmfmt{\NAT@nm}%
   \NAT@spacechar\NAT@@open\if*#1*\else#1\NAT@spacechar\fi\NAT@hyper@{\NAT@date}}
  {}{}

\makeatother

\usepackage{lscape}
\usepackage{adjustbox}
\usepackage{amsfonts}
\usepackage{amssymb}
\usepackage{pifont}
\newcommand{\cmark}{\ding{51}}%
\newcommand{\xmark}{\ding{55}}
\usepackage{multirow}
\usepackage[para]{threeparttable}
\usepackage[labelfont=bf]{caption}
\usepackage{txfonts}
\usepackage[nameinlink, capitalise]{cleveref}
\usepackage[utf8]{inputenc}

\usepackage{placeins}

\makeatletter
\let\@fnsymbol\@arabic
\makeatother

\begin{document} 

   \title{ALMA ACA study of the H$_2$S/OCS ratio in low-mass protostars}
  \authorrunning{T. Kushwahaa et al.}
    \makeatletter
\def\@fnsymbol#1{\ensuremath{\ifcase#1\or *\or \dagger\or \ddagger\or
   \mathsection\or \mathparagraph\or \|\or **\or \dagger\dagger
   \or \ddagger\ddagger \else\@ctrerr\fi}}
    \makeatother

\author{Tanya Kushwahaa\inst{1, 2}\fnmsep\thanks{Email: tkushwahaa.astro@gmail.com} \and Maria N. Drozdovskaya\inst{3} \and \L ukasz Tychoniec\inst{4} \and Beno\^{i}t Tabone\inst{5}}

  \institute{Universit\'e Bourgogne Franche-Comt\'e, 32 Avenue de l'Observatoire, 25000 Besan\c{c}on, France
  \and
  School of Physics and Astronomy, Cardiff University, The Parade, Cardiff CF24 3AA, UK\\
  \email{KushwahaaT@cardiff.ac.uk}
                      \and
    Center for Space and Habitability, Universit\" at Bern, Gesellschaftsstrasse 6, CH-3012 Bern, Switzerland\\
    \email{maria.drozdovskaya@unibe.ch}
                      \and 
    European Southern Observatory, Karl-Schwarzschild-Strasse 2, 85748 Garching bei München, Germany\\
                        \and
    Universit\'e Paris-Saclay, CNRS, Institut d’Astrophysique Spatiale, 91405 Orsay, France\\}

\date{\today}

\abstract {The identification of the main sulfur reservoir on its way from the diffuse interstellar medium to the cold dense star-forming cores and eventually to protostars is a long-standing problem. Despite sulfur's astrochemical relevance, the abundance of S-bearing molecules in dense cores and regions around protostars is still insufficiently constrained.}
{The goal of this investigation is to derive the gas-phase H$_2$S/OCS ratio for several low-mass protostars, which could provide crucial information about the physical and chemical conditions in the birth cloud of Sun-like stars.}
{Using ALMA ACA Band 6 observations, H$_2$S, OCS, and their isotopologs are searched for in 10 Class 0/I protostars with different source properties such as age, mass, and environmental conditions.  An LTE model is used to fit synthetic spectra to the detected lines and to derive the column densities based solely on optically thin lines.}
{The H$_2$S and OCS column densities span four orders of magnitude across the sample. The H$_2$S/OCS ratio is found to be in the range from 0.2 to above 9.7. IRAS 16293-2422 A and Ser-SMM3 have the lowest ratio, while BHR71-IRS1 has the highest. Only the H$_2$S/OCS ratio of BHR71-IRS1 agress within uncertainties with the ratio in comet 67P/Churyumov–Gerasimenko.}
{The determined gas-phase H$_2$S/OCS ratios can be below the upper limits on the solid-state ratios by as much as an order of magnitude. The H$_2$S/OCS ratio depends significantly on the environment of the birth cloud, such as UV-irradiation and heating received prior to the formation of a protostar. The highly isolated birth environment of BHR71-IRS1 is hypothesized to be the reason for its high gaseous H$_2$S/OCS ratio due to lower rates of photoreactions and more efficient hydrogenation reactions under such dark, cold conditions. The gaseous inventory of S-bearing molecules in BHR71-IRS1 appears to be most similar to that of interstellar ices.}

\keywords{Astrochemistry -- Line: identification -- Instrumentation: interferometers -- ISM: molecules -- Stars: protostars}

\maketitle

\section{Introduction}
\label{introduction}

Sulfur (S) is the tenth most abundant element in the Universe (S/H$\sim$1.35$\times$10$^{-5}$, \citealt{Yamamoto2017}). It was first detected as carbon monosulfide (CS) in the interstellar medium \citep{Penzias1971}. S-bearing species have since been detected in different regions including molecular clouds \citep{NavaroAlmaida2020, Spezzano2022}, hot cores (\citealt{Blake1987, Charnley1997, Li2015, Codella2021, Drozdovskaya2018}), comets \citep{Smith1980, Bockelee2000, Biver2021a, Biver2021b}, as well as starburst galaxies (NGC 253; \citealt{Martin2005}). The total abundance of an element in dust, ice, and gas is its cosmic abundance, also called its elemental abundance. The gas-phase abundance of atomic sulfur in diffuse clouds is comparable to the cosmic abundance of sulfur ($\sim$10$^{-5}$; \citealt{Savage1996, Howk2006}). However, the observed abundance of S-bearing species in dense cores and protostellar environments is lower by a factor of $\sim$1000 \citep{Snow1986, Tieftrunk1994, Goicoechea2006, Agundez2018} in comparison to the total S-abundance in diffuse clouds. The forms and mechanisms behind this sulfur depletion in star-forming regions are still unknown. This is often called the "missing sulfur problem".

\begin{table*}
\centering
\caption{Spectral settings of the data sets.}
\label{spectral_setup}
\begin{tabular}{l c c c}
\hline
\multirow{1}{*}{Sky frequency} &  \multicolumn{2}{c}{Channel width} & Number of channels\\
(GHz) & (kHz) & (km s$^{-1}$) & \\
\hline
Project-id: 2017.1.00108.S\\
\hline
214.250-214.500 & 61 & 0.085 & $4~096$\\
216.646-216.771 & 61 & 0.084 & $2~048$\\ 
215.469-215.532 & 61 & 0.085 & $1~024$\\
216.114-216.176 & 61 & 0.084 & $1~024$\\
231.550-231.612 & 61 & 0.079 & $1~024$\\
231.308-231.370 & 61 & 0.079 & $1~024$\\
231.027-231.090 & 61 & 0.079 & $1~024$\\
230.284-230.346 & 61 & 0.080 & $1~024$\\
232.062-233.932 (continuum) & 488 & 0.628 & $4~096$\\
\hline
Project-id: 2017.1.1350.S\\
\hline
216.974-217.222 & 488 & 0.674 & 512\\
216.513-216.762 & 488 & 0.676 & 512\\
217.692-217.941 & 488 & 0.672 & 512\\
215.982-216.230 & 488 & 0.678 & 512\\
219.492-219.616 & 244 & 0.334 & 512\\
219.882-220.006 & 244 & 0.333 & 512\\
218.692-218.817 & 244 & 0.335 & 512\\
220.331-220.449 & 244 & 0.332 & 512\\
230.470-230.707 & 488 & 0.635 & 512\\
231.159-231.408 & 488 & 0.633 & 512\\
230.931-231.180 & 488 & 0.633 & 512\\
232.136-232.384 & 488 & 0.630 & 512\\
233.069-234.937 (continuum) & 977 & 1.260 & $2~048$\\
\hline
\end{tabular}
\end{table*}

Different chemical models have been used to investigate this unknown form of sulfur \citep{Woods2015, Vidal2017, Semenov2018, Vidal2018, Laas2019}. \cite{Vidal2017} have proposed that a notable amount of sulfur is locked up in either HS and H$_2$S ices or gaseous atomic sulfur in cores, depending substantially on the age of the molecular cloud. However, the only solid form of sulfur firmly detected in interstellar ices is OCS \citep{Palumbo1995, Palumbo1997, Aikawa2012, Boogert2015} and potentially also SO$_2$ \citep{Boogert1997, Zasowski2009, Yang2022, McClure2023}. Solid state H$_2$S detection remains tentative to date \citep{Geballe1985, Smith1991}. The initial cloud abundance of S-bearing molecules has been shown to set the subsequent abundances of these molecules in protostellar regions, depending on the free-fall timescales \citep{Vidal2018}. In surface layers of protoplanetary disks, the availability of gaseous S-bearing molecules appears to be strongly linked with the availability of oxygen \citep{Semenov2018}. Observational studies of gas-phase species claim either H$_2$S \citep{Holdship2016} or OCS \citep{Vandertak2003} as the main S-carrier depending on the environment being observed. Other possible reservoirs of sulfur have been proposed in the form of semi-refractory polymers up to S$_8$ \citep{Ahearn1983, Druard2012, 2016MNRAS.462S.253C, Shingledecker2020}, hydrated sulfuric acid \citep{Scappini2003}, atomic sulfur \citep{Anderson2013}, and mineral sulfides, FeS \citep{Keller2002, Kohler2014, Kama2019}. On the other hand, chemical models of the evolution from cloud to dense core with updated chemical networks suggest that sulfur is merely partitioned over a diverse set of simple organo-sulfur ices and no additional form is required \citep{Laas2019}. Matching observed and modeled cloud abundances consistently for the full inventory of gaseous S-bearing molecules to better than a factor of 10 remains challenging \citep{NavaroAlmaida2020}. Laboratory experiments point to the importance of the photodissociation of H$_2$S ice by UV photons leading to the production of OCS ice \citep{Ferrante2008, Garozzo2010, Jimenez2011, Chen2015} and S$_2$ in mixed ices \citep{Grim1987}. \cite{2016MNRAS.462S.253C} claim to have recovered the full sulfur inventory in comets.

Sulfur-bearing species have been proposed to probe the physical and chemical properties of star-forming regions and to even act as chemical clocks \citep{Charnley1997, Hatchell1998, Viti2001, Li2015}. However, it has since been shown that their abundance is sensitive to gas-phase chemistry and the availability of atomic oxygen, which puts their reliability as chemical clocks into question \citep{Wakelam2004b, Wakelam2011}. Studying S-bearing molecules in young Class 0/I protostars is crucial for two reasons. Firstly, their inner hot regions thermally desorb all the volatile ices that are otherwise hidden from gas-phase observations. Consequently, it is more likely to be able to probe the full volatile inventory of S-bearing molecules and investigate the "missing sulfur" reservoir. Secondly, these targets are a window onto the materials available for the assembly of the protoplanetary disk midplane and the cometesimals therein \citep{Aikawa1999, Willacy2007, Willacy2009}. This makes hot inner regions highly suitable targets for comparative studies with comets \citep{Bockelee2000, Drozdovskaya2019}.

The main goal of this paper is to study the physical and chemical conditions in embedded protostars via the H$_2$S/OCS ratio. A sample of 10 Class 0/I low-mass protostars with different physical properties (mass, age, environment) is considered. Such protostars are in their earliest phase of formation after collapse with large envelope masses. In this work, Atacama Large Millimeter/submillimeter Array (ALMA) Atacama Compact Array (ACA) Band 6 observations towards these 10 protostars are utilized. The H$_2$S/OCS ratio is calculated from the column densities of H$_2$S, OCS, and their isotopologs. The details of the observations, model, and model parameters used for synthetic spectral fitting are introduced in \Cref{Method}. The detected lines of major and minor isotopologs of H$_2$S and OCS, their characteristics, and H$_{2}$S/OCS line ratios are presented in \Cref{Results}. The discussion and conclusions are presented in \Cref{Discussions} and \ref{Conclusions}, respectively.

\section{Methods}
\label{Method}

\begin{table*}
\centering
\caption{Observed Class 0/I protostellar systems.}
\label{protostars}
\begin{threeparttable}
    \centering
    \begin{tabular}{l c c c c c c}
    \hline
    Source  & $d$ & $M_{\text{env}}$ & $L_{\text{bol}}$ & $T_{\text{bol}}$ & Class & $v_{\text{LSR}}$\\
    & (pc) & (M$_{\odot}$) & (L$_{\odot}$) & (K) & -- & (km s$^{-1}$) \\
    \hline
    IRAS 16293-2422 A & 141\tablefootmark{a} & 4.0\tablefootmark{d} & $\sim$18\tablefootmark{b} & -- & 0 & +3.2\tablefootmark{c} 
    \\
    IRAS 16293-2422 B & 141\tablefootmark{a} & 4.0\tablefootmark{d} & $\sim$3\tablefootmark{b} &  -- & 0 & +2.7\tablefootmark{c} 
    \\
    NGC 1333-IRAS4A & 299\tablefootmark{e} & 5.6\tablefootmark{f} & 9.1\tablefootmark{f} & 29\tablefootmark{g} & 0 & +7.2\tablefootmark{q} 
    \\
    RCrA IRS7B &  130\tablefootmark{i} & 2.2\tablefootmark{j} & 4.6\tablefootmark{j} & 89\tablefootmark{j} & 0/I & +5.8\tablefootmark{v}
    \\
    Per-B1-c   &  301\tablefootmark{e} & 1.8\tablefootmark{h} & 3.84\tablefootmark{h}  & 48\tablefootmark{k} & 0 & +6.4\tablefootmark{k} 
    \\
    BHR71-IRS1     &  200\tablefootmark{o, p} & 2.7\tablefootmark{q} & 15\tablefootmark{q} & 44\tablefootmark{q} & 0 & -4.4\tablefootmark{q} \\
    Per-emb-25   & 294\tablefootmark{r}  & 0.5\tablefootmark{h} & 1.0\tablefootmark{h} & 68\tablefootmark{h} & 0/I & +5.8\tablefootmark{k} \\
    NGC 1333-IRAS4B & 299\tablefootmark{e} & 3.0\tablefootmark{q} & 4.4\tablefootmark{q} & 28\tablefootmark{g}  & 0  &  +7.4\tablefootmark{q}
    \\
    Ser-SMM3 & 436\tablefootmark{s} & 3.2\tablefootmark{q} & 5.1\tablefootmark{q} & 38\tablefootmark{q} & 0 & +7.6\tablefootmark{q} \\
    TMC1 & 140\tablefootmark{t, u} & 0.2\tablefootmark{q} & 0.9\tablefootmark{q} & 101\tablefootmark{q} & I & +5.2\tablefootmark{q} \\
    \hline
    \end{tabular}
    \begin{tablenotes}\footnotesize
    \tablefoot{The columns represent 1) Source, 2) $d$: distance to the source in pc, 3) $M_{\text{env}}$: mass of the envelope in M$_\odot$, 4) $L_{\text{bol}}$: bolometric luminosity in L$_\odot$, 5) $T_{\text{bol}}$: bolometric temperature in K, 6) Class: stage of the protostar, 7) $v_{\text{LSR}}$: local standard of rest velocity in km s$^{-1}$. References: 
\tablefoottext{a}{\citet{Dzib2018},}
\tablefoottext{b}{\citet{Jacobsen2018},}
\tablefoottext{c}{\citet{Jorgensen2011},}
\tablefoottext{d}{\citet{VanDerWiel2019},}
\tablefoottext{e}{\citet{Zucker2018},}
\tablefoottext{f}{\citet{Taquet2015},}
\tablefoottext{g}{\citet{Tobin2016},}
\tablefoottext{h}{\citet{Enoch2009},}
\tablefoottext{i}{\citet{Neuhauser2008},}
\tablefoottext{j}{\citet{Lindberg2014a},}
\tablefoottext{k}{\citet{Stephens2019},}
\tablefoottext{l}{\citet{Matthews2006},}
\tablefoottext{m}{\citet{Hatchell2007b},}
\tablefoottext{n}{\citet{Hatchell2007c},}
\tablefoottext{o}{\citet{1989A&A...225..192S},}
\tablefoottext{p}{\citet{Straizys1994},}
\tablefoottext{q}{\citet{Kristensen2012},}
\tablefoottext{r}{\citet{Zucker2019},}
\tablefoottext{s}{\citet{Ortiz2018},}
\tablefoottext{t}{\citet{Elias1978},}
\tablefoottext{u}{\citet{Torres2009},}
\tablefoottext{v}{\citet{Lindberg2015}.}
}
    \end{tablenotes}
\end{threeparttable}
\end{table*}

\begin{figure*}
\centering
\begin{minipage}[b]{6.2in}%

    \subfigure{\includegraphics[width=2in]{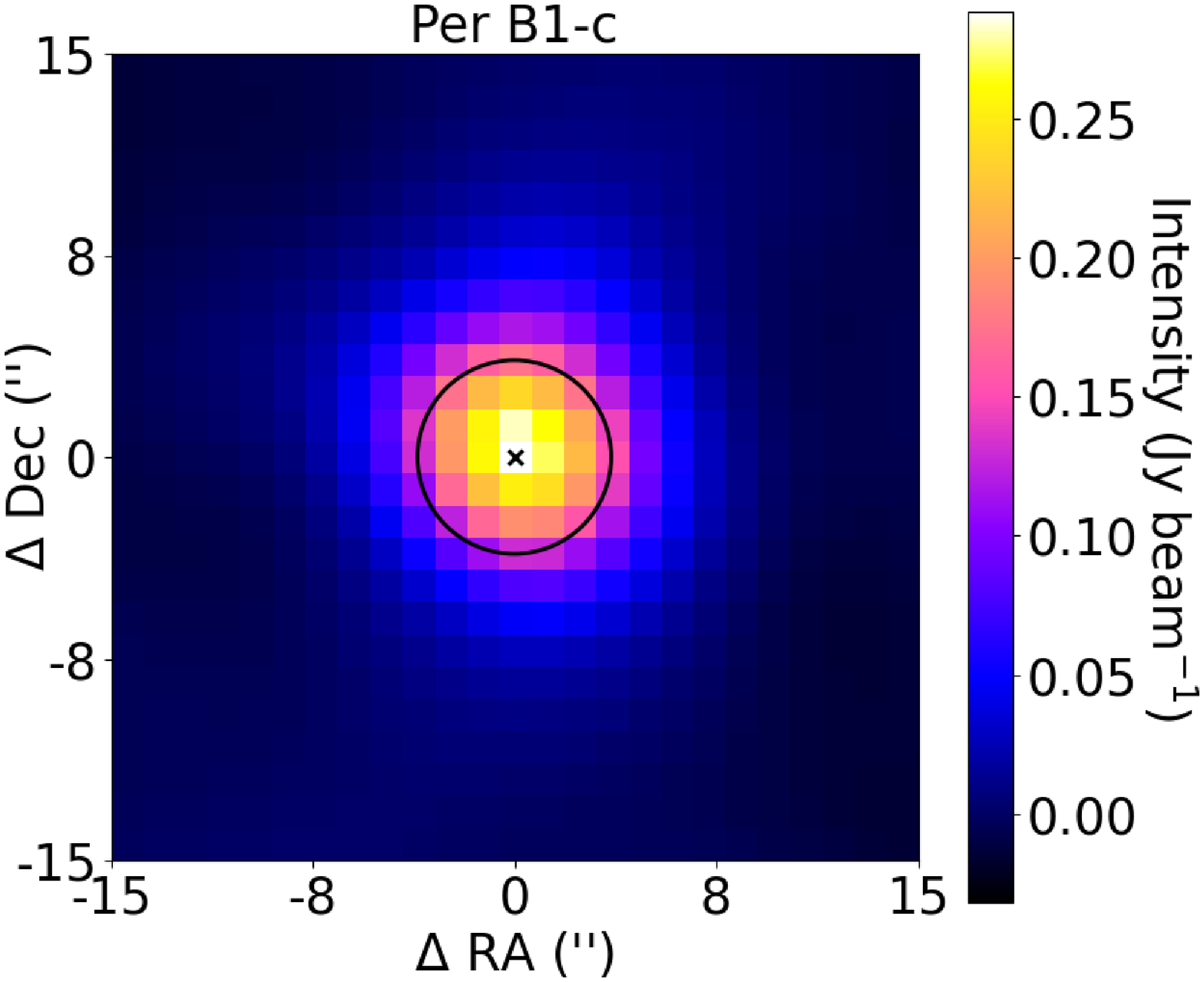}}
    \subfigure{\includegraphics[width=2in]{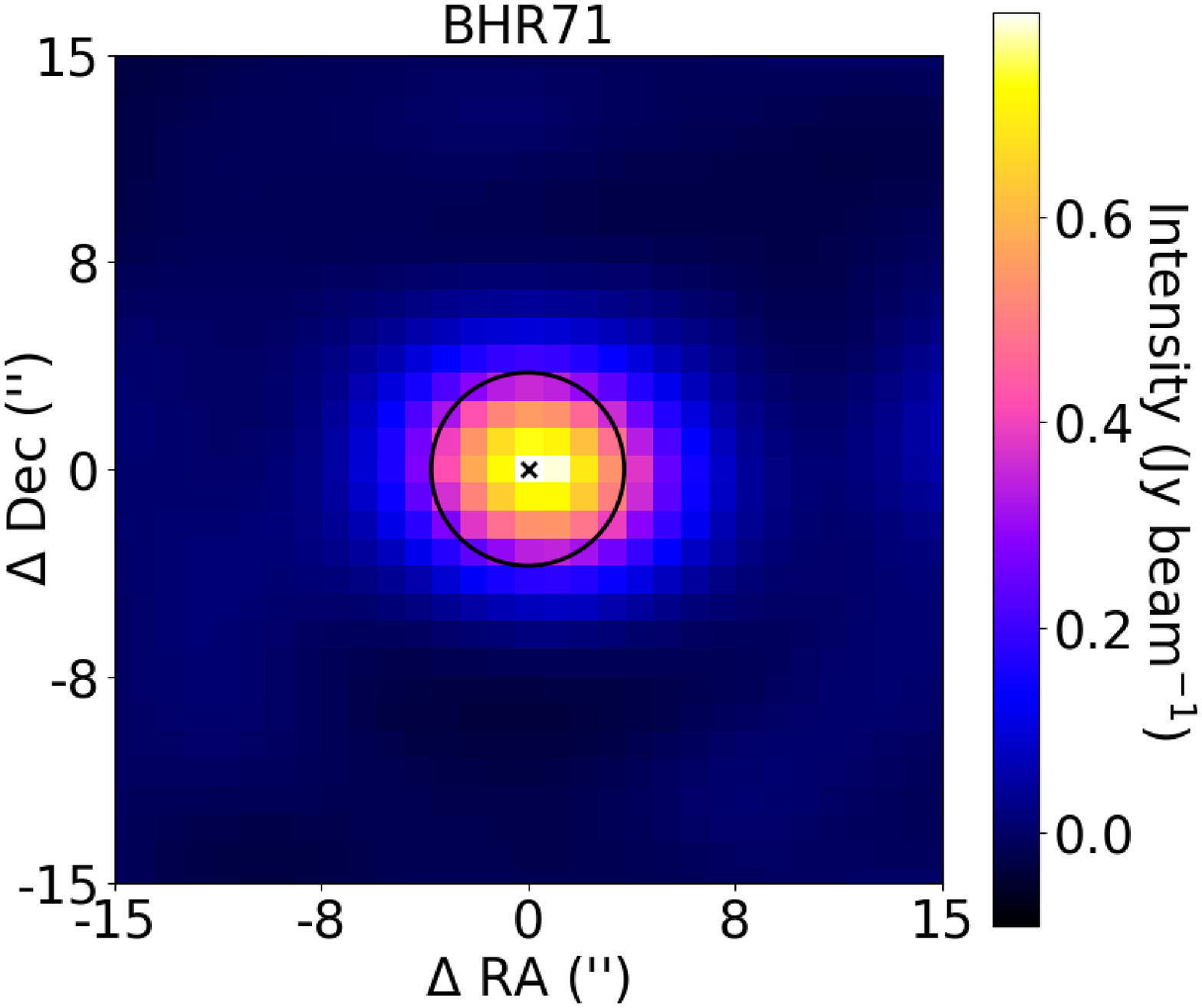}}
    \subfigure{\includegraphics[width=2in]{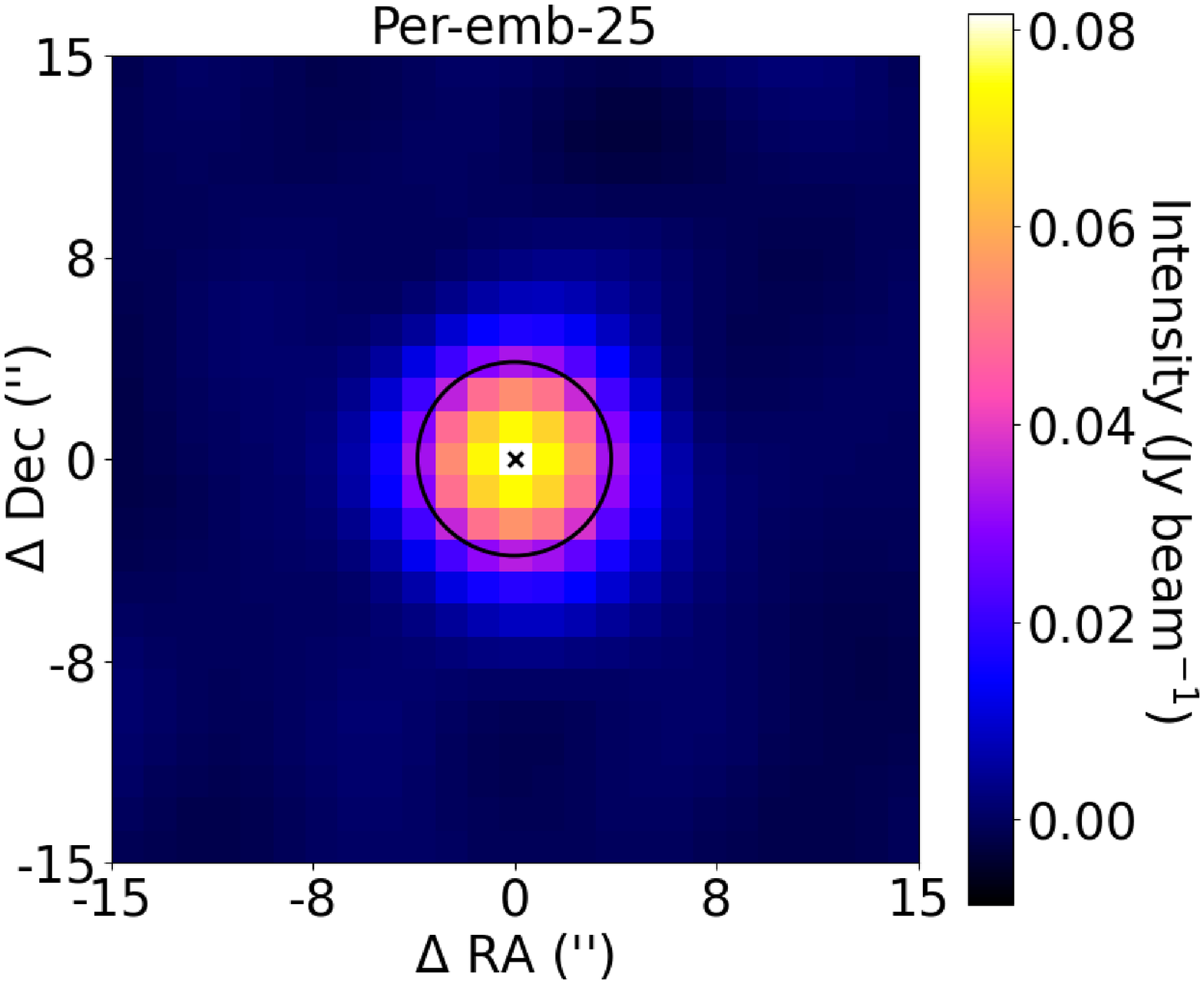}}\\
    \subfigure{\includegraphics[width=2in]{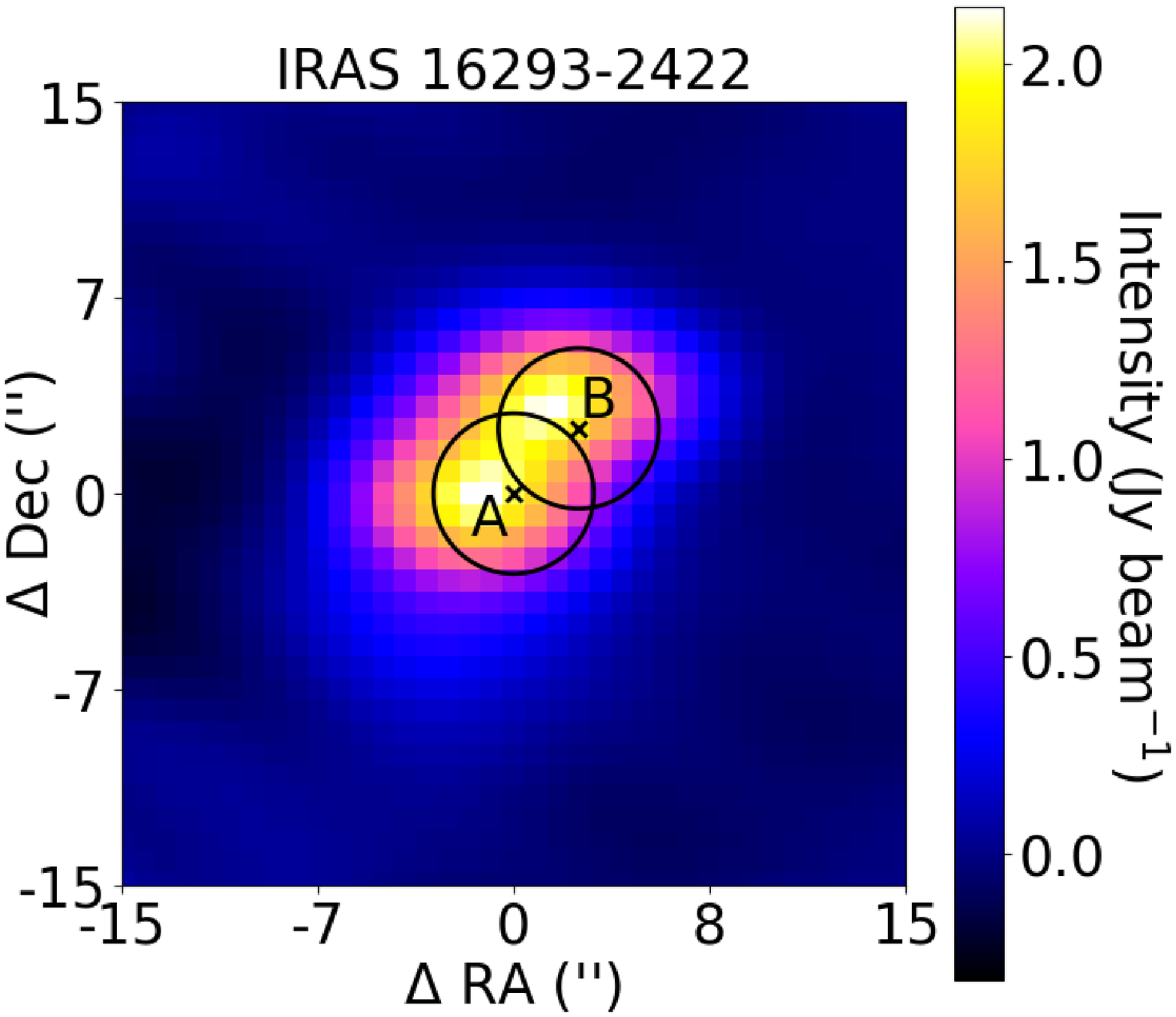}}
    \subfigure{\includegraphics[width=2in]{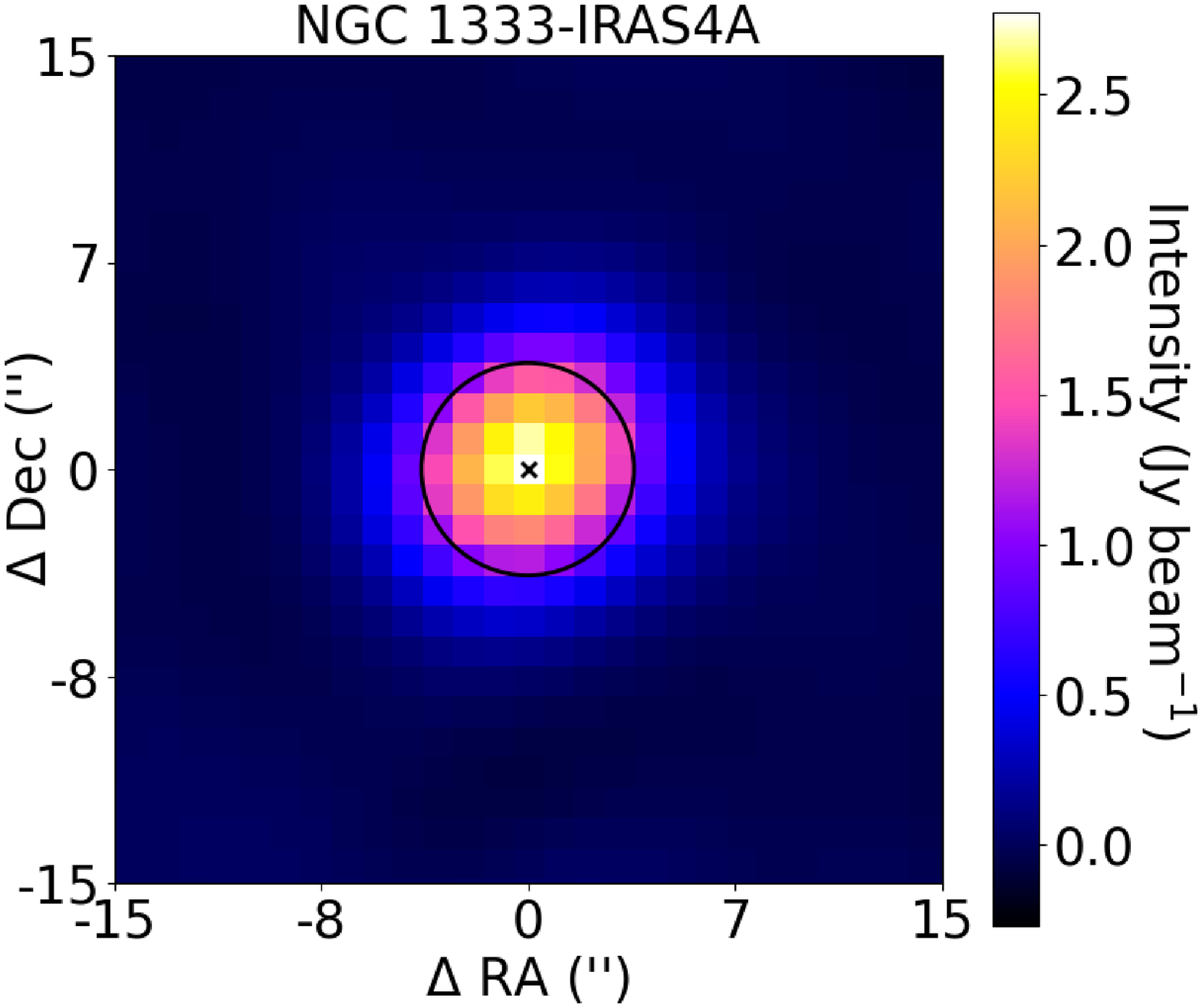}}
    \subfigure{\includegraphics[width=2in]{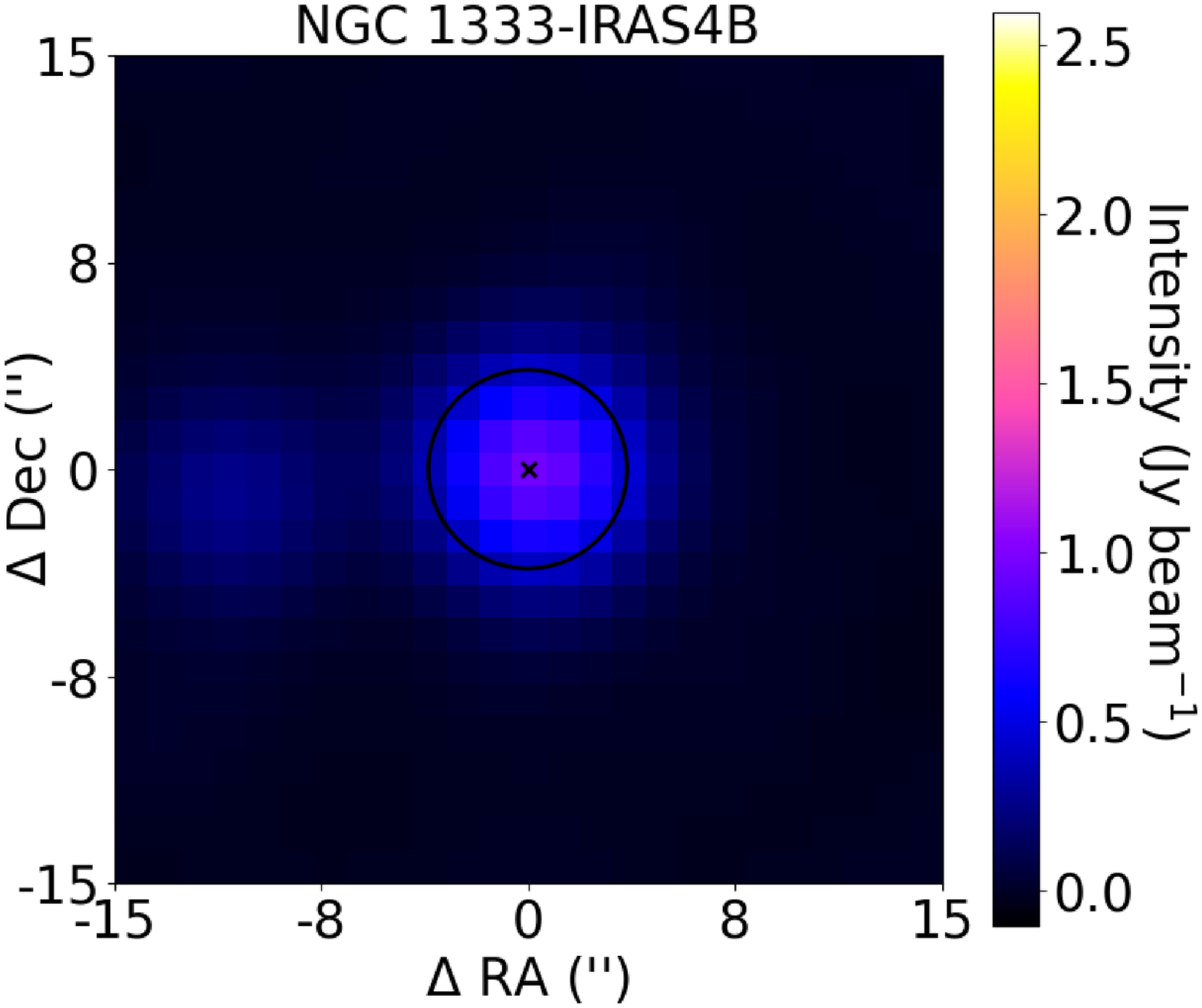}}\\
    \subfigure{\includegraphics[width=2in]{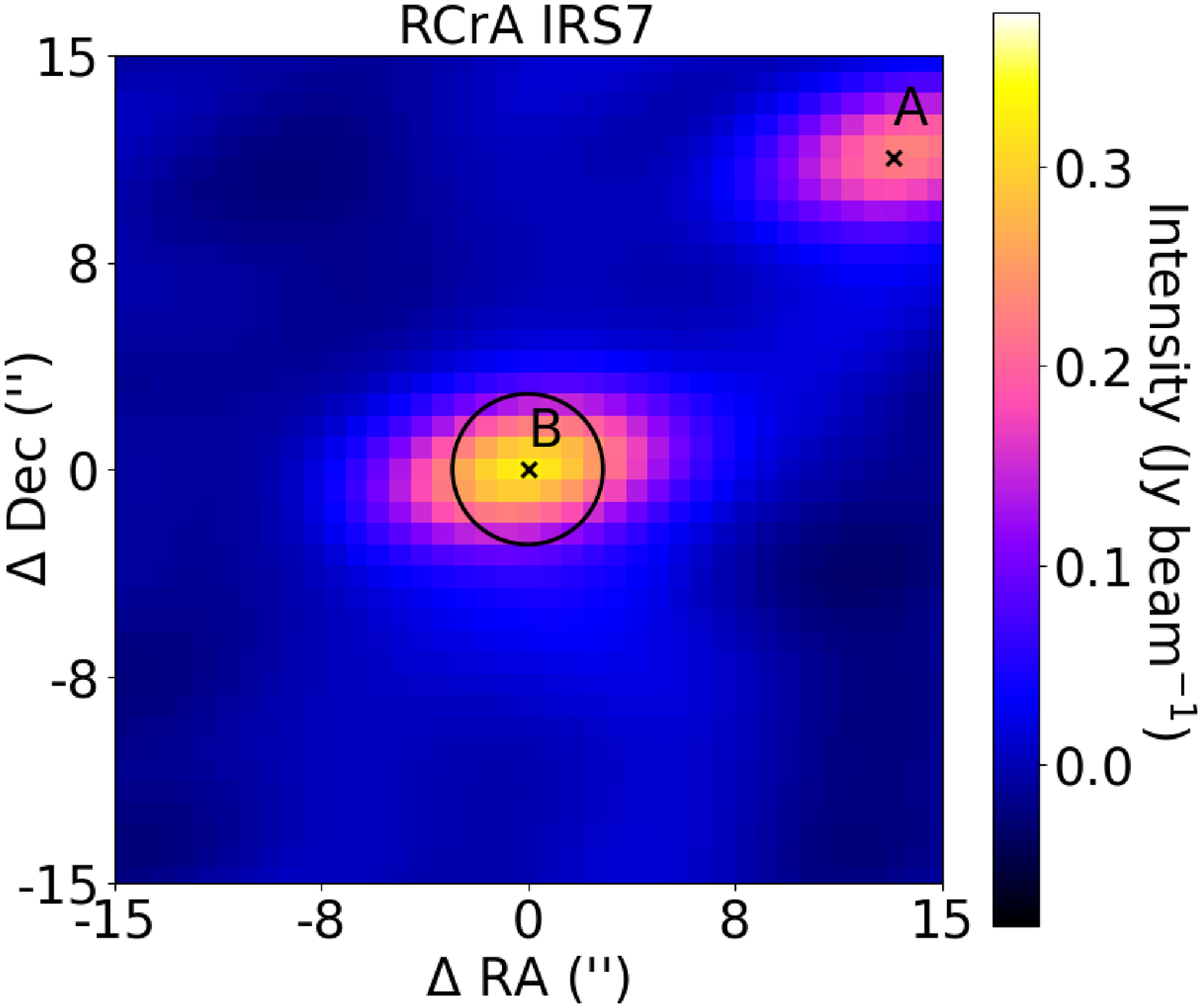}}
    \subfigure{\includegraphics[width=2.1in]{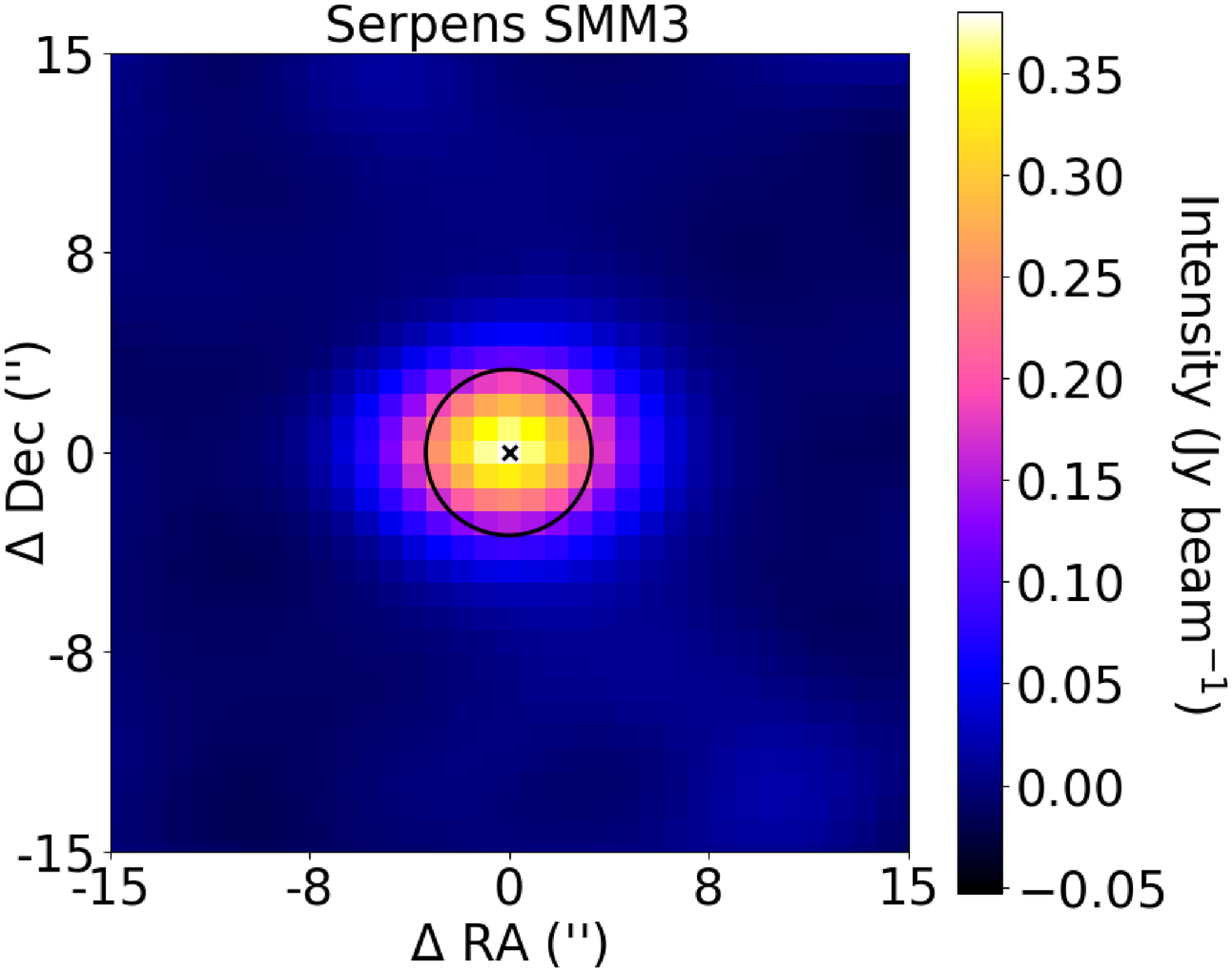}}
    \subfigure{\includegraphics[width=2in]{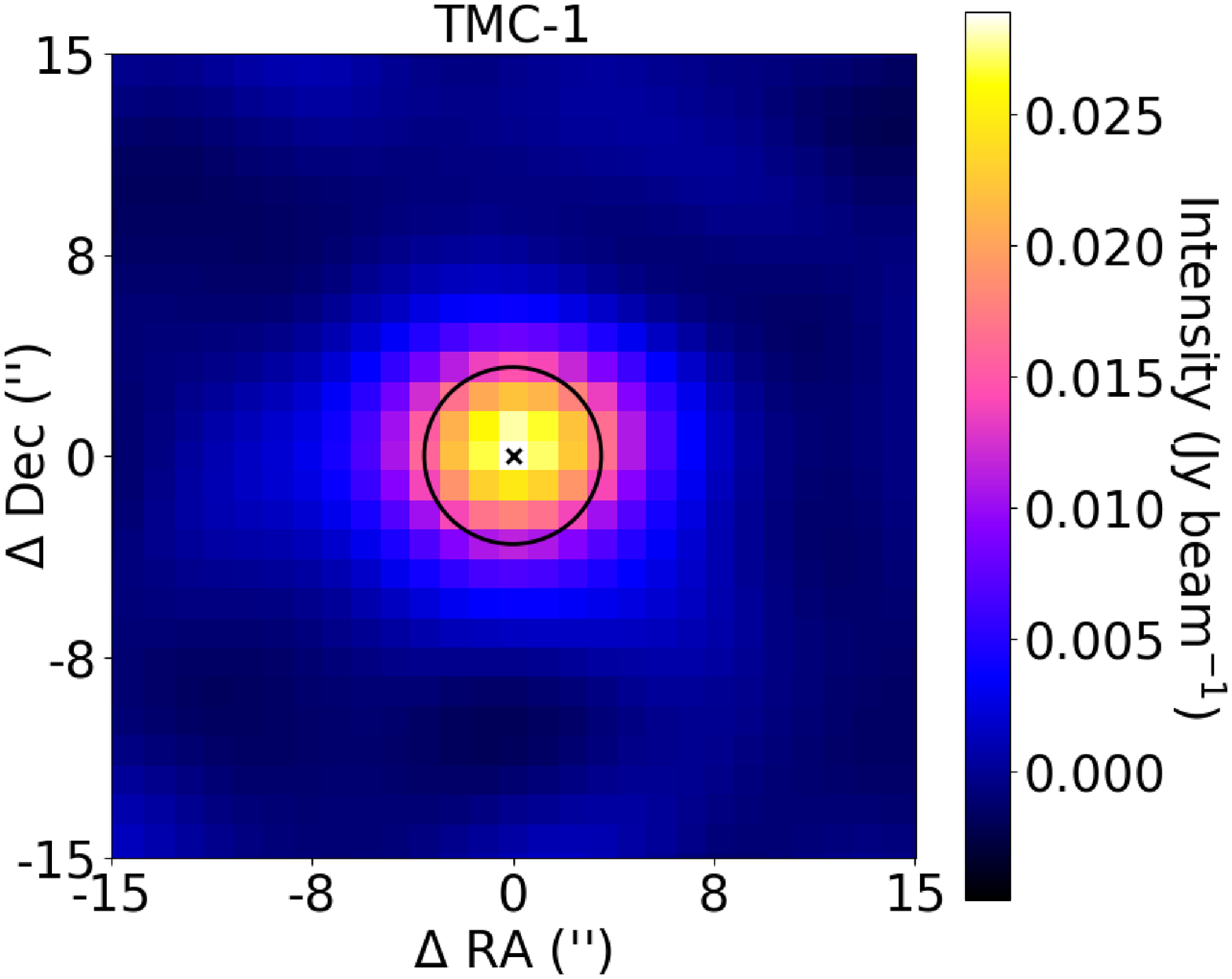}}
\end{minipage}
\caption{ALMA pipeline-produced integrated intensity maps (color scale) with the line channels excluded, which are dominated by dust emission, for the studied sample of sources. On-source spectra are extracted by averaging the flux from the pixels within the circular area centered on `X'.}
\label{maps}
\end{figure*}

\subsection{Observations}

Two sets of observations are jointly analyzed in this paper. The first data set (project-id: 2017.1.00108.S; PI: M.~N. Drozdovskaya) targeted IRAS 16293-2422, NGC 1333-IRAS4A, and RCrA IRS7B. The observations were carried out in Band 6 (211-275 GHz) with the ALMA ACA of 7m dishes. The data set has a spectral resolution of 0.079-0.085 km~s$^{-1}$ (61 kHz), and a spatial resolution of (6.5-9.0)$\times$(4.0-6.3)$''$.  The second data set (project-id: 2017.1.1350.S; PI: \L. Tychoniec) targeted Per-B1-c, BHR71, Per-emb-25, NGC 1333-IRAS4B, Ser-SMM3, and TMC1 with the ALMA ACA 7m dishes, also in Band 6. The data have a similar spatial resolution, (6.1-7.4)$\times$(4.5-6.4)$''$, but a lower spectral resolution of 0.333-0.678 km s$^{-1}$ (244-488 kHz).  The observed frequency ranges of the data are given in \Cref{spectral_setup}. Data cubes were processed through the standard pipeline calibration with CASA 5.4.0-68. For each source, the noise level has been calculated by taking the standard deviation of the flux in the frequency ranges where no emission lines were detected, i.e., regions with pure noise, in the spectral window containing the H$_2$S, 2$_{2,0}$-2$_{1,1}$ line. The noise level of the first data set is 21-32 mJy beam$^{-1}$ channel$^{-1}$, and of the second data set is 7-13 mJy beam$^{-1}$ channel$^{-1}$ (\autoref{extraction_area}). Both data sets have a flux uncertainty of 10\%. The largest resolvable scale of the first and the second data sets are 26.2-29.2$''$ and 24.6-29.0$''$, respectively.

\subsection{Sources}
\label{sources}

The properties of the sources explored in this work are tabulated in \autoref{protostars}. IRAS 16293-2422 (hereafter, IRAS 16293) is a triple protostellar source, consisting of protostars A and B, separated by 5.3$''$ (747 au; \citealt{VanDerWiel2019}) and disk-like structures around the two sources, located in the Rho Ophiuchi star-forming region at a distance of 141 pc \citep{Dzib2018}. This source was studied thoroughly using ALMA under the Protostellar Interferometric Line Survey (PILS; \citealt{Jorgensen2016}) and many preceding observational campaigns (e.g., \citealt{VanDishoeck1995, Caux2011}). Both hot corinos around A and B are rich in a diverse set of complex organic molecules (\citealt{Jorgensen2018, Manigand2020}). The source IRAS 16293 A is itself a binary composed of sources A1 and A2 with a separation of 0.38$''$ (54 au; \citealt{Maureira2020}). IRAS4A is also a binary system, comprised of IRAS4A1 and IRAS4A2, separated by 1.8$''$ (540 au; \citealt{Sahu2019}) in the Perseus molecular cloud, located at a distance of 299 pc \citep{Zucker2018} in the south-eastern edge of the complex NGC 1333 \citep{Looney2000}. IRAS4A1 has a much higher dust density in its envelope than IRAS4A2, but both contain complex organic molecules (\citealt{Sahu2019, DeSimone2020}). IRS7B is a low-mass source, with a separation of  14$''$ ($2~000$ au) from IRS7A \citep{1987ApJ...322L..31B}, and $\sim$8$''$ ($1~000$ au) from CXO 34 \citep{Lindberg2014a}. It is situated in the Corona Australis dark cloud at a distance of 130 pc \citep{Neuhauser2008}. IRS7B has been shown to contain lower complex organic abundances as a result of being located in a strongly irradiated environment (\citealt{Lindberg2015}). 

\begin{table*}
\centering
\caption{Location of the center position (in RA and Dec) and radius (in au and arcsecond) of each circular region, from which the spectra are extracted for the studied protostars. The pixel size (in arcsecond) for each source is also given. The noise levels in mJy beam$^{-1}$ channel$^{-1}$ and mJy beam$^{-1}$ km s$^{-1}$ of the protostars are deduced according to $\sqrt{\frac{\sum{(\text{flux in line-free channel }i)}^{2}}{\text{number of line-free channels}}}$ and $\text{noise}~(\text{mJy}^{-1}\text{beam}^{-1}\text{channel}^{-1}) \times \sqrt{n} \times \text{spectral resolution}~(\text{km s}^{-1})$, where $n$ is the number of channels computed with $\frac{\text{FWHM}~(\text{km s}^{-1})}{\text{spectral resolution}~(\text{km s}^{-1})}$, using the FWHM of the H$_2$S line at 216.710 GHz, respectively.}
\label{extraction_area}
\begin{adjustbox}{width=1\textwidth}
\begin{tabular}{l c c c  c c c c c}
\hline
\multirow{1}{*}{Source} & RA & Dec & \multirow{1}{*}{Pixel size} &  \multicolumn{2}{c}{Radius} & \multirow{1}{*}{FWHM} & \multicolumn{2}{c}{Noise level}\\
 & (J2000) & (J2000)  & ($''$) & (au) & ($''$) & (km s$^{-1}$) &(mJy beam$^{-1}$ channel$^{-1}$) & (mJy beam$^{-1}$ km s$^{-1}$)\\
\hline
IRAS 16293-2422 A &  16h 32m 22.854s & -24\degree 28$'$ 36.465$''$ & 0.8  & 522   & 3.7 & 4.5 & 28 & 17.0\\
IRAS 16293-2422 B &  16h 32m 22.671s & -24\degree 28$'$ 33.145$''$ & 0.8  &  522  & 3.7 & 1.0 & 32 & 9.2\\
NGC 1333-IRAS4A &   03h 29m 10.509s & +31\degree 13$'$ 30.918$''$ & 1.1 &  822  & 3.5 & 1.8 & 29 & 11.2\\
RCrA IRS7B &  19h 01m 56.402s & -36\degree 57$'$ 28.276$''$  & 0.8 & 730   & 4.3 & 1.0 & 21 & 6.0\\
Per-B1-c & 03h 33m 17.880s & +31\degree 09$'$ 31.795$''$  & 1.2 &  879  & 3.0 & 2.2 & 11 & 13.4\\
BHR71-IRS1 &  12h 01m 36.516s & -65\degree 08$'$ 49.298$''$ & 1.0  & 700   & 3.5 & 2.5 & 7 & 9.2\\
Per-emb-25 &  03h 26m 37.514s & +30\degree 15$'$ 27.792$''$ & 1.1  & 600   & 3.0 & 1.0 & 11 & 9.0\\
NGC 1333-IRAS4B &  03h 29m 12.019s & +31\degree 13$'$ 08.010$''$  & 1.2 & 879   & 3.0 & 2.0 & 11 & 12.9\\
Ser-SMM3 &  18h 29m 59.311s & +01\degree 14$'$ 00.365$''$ & 0.9 & $1~526$   & 3.5 & 2.5 & 10 & 13.5\\
TMC1 &  04h 41m 12.700s & +25\degree 46$'$ 34.800$''$ & 1.1 & 303   & 3.0 & 1.0 & 13 & 10.6\\
\hline
\end{tabular}
\end{adjustbox}
\end{table*}

From the second set of sources, IRAS4B (sometimes labeled BI) has a binary component B$'$ (or BII) that is 11$''$ ($3~300$ au) away \citep{Sakai2012, Anderl2016, Tobin2016}. The separation between IRAS4B and IRAS4A is 31$''$ ($9~300$ au; \citealt{Coutens2013}). IRAS4B displays emission from complex organic molecules \citep{Belloche2020} and powers a high-velocity SiO jet \citep{Podio2021}. B1-c is an isolated deeply embedded protostar in the Barnard 1 clump in the western part of the Perseus molecular cloud at a distance of 301 pc \citep{Zucker2018}. B1-c contains emission from complex organic molecules and shows a high velocity outflow \citep{Jorgensen2006, vanGelder2020}. The next closest source, B1-a, is $\sim$100$''$ ($\sim29~500$ au) away \citep{Jorgensen2006}. BHR71 is a Bok globule in the Southern Coalsack dark nebulae at a distance of $\sim$200 pc \citep{1989A&A...225..192S, Straizys1994}. It hosts the wide binary system of IRS1 and IRS2 with a sepration of 16$''$ ($3~200$ au; \citealt{2001ApJ...554L..91B, Parise2006, 2008ApJ...683..862C, Tobin2019}). IRS1 displays pronounced emission from complex organic molecules \citep{Yang2020}. Emb-25 is a single source located in the Perseus molecular cloud \citep{Enoch2009, Tobin2016}. It does not show emission from complex organic molecules \citep{Yang2021}, but powers low-velocity CO outflows \citep{Stephens2019}. TMC1 is a Class I binary source, located in the Taurus molecular cloud \citep{1995ApJ...445..377C, Brown1999} at a distance of 140 pc \citep{Elias1978, Torres2009}. The separation between the two components, TMC1E and TMC1W, is $\sim$0.6$''$ ($\sim$85 au); and neither of the two display complex organic emission \citep{Hoff2020}. SMM3 is a single, embedded protostar located in the SE region in Serpens region; 436 pc away \citep{Ortiz2018}. The next closest-lying source is SMM6 at a separation of 20$''$ ($\sim8~700$ au; \citealt{Davis1999, Kristensen2010, Mirocha2021}). SMM3 launches a powerful jet \citep{Tychoniec2021}, but does not display complex organic molecule emission, which may be obscured by the enveloping dust \citep{vanGelder2020}.

\subsection{Synthetic spectral fitting}\label{fitting}

For the spectral analysis, on-source spectra were extracted from the data cubes of the sources of the two data sets. Circular regions centered on source positions (in RA and Dec) with the radius of spectrum extraction corresponding to one-beam on-source are given in \autoref{extraction_area}. The number of pixels in radius \textit{r} of the circular region to be used was computed by dividing the radius of the circular region with the size of one pixel in arcsecond. The spectroscopy used for the targeted molecules and their isotopologs stems from the Cologne Database of Molecular Spectroscopy (CDMS; \citealt{Muller2001, Muller2005, Endres2016})\footnote{\url{https://cdms.astro.uni-koeln.de/}} and the Jet Propulsion Laboratory (JPL) catalog \citep{Pickett1998}\footnote{\url{https://spec.jpl.nasa.gov/}}. Line blending in the detected lines was checked with the online database Splatalogue\footnote{\url{https://splatalogue.online/}}.

Synthetic spectral fitting  was performed with custom-made Python scripts based on the assumption of local thermal equilibrium (LTE). The input parameters include the full width half-maximum (FWHM) of the line, column density ($N$), excitation temperature ($T_{\text{ex}}$), source size, beam size, and spectral resolution of the observations. Line profiles are assumed to be Gaussian. Further details are provided in section 2.3 of \cite{Drozdovskaya2022}. For some sources, the number of free parameters (such as source size or $T_{\text{ex}}$) can be reduced based on information from other observing programs. These are detailed on a source-by-source basis in the corresponding Appendices. Typically, merely two free parameters were fitted at a time by means of a visual inspection and an exploration of a grid of possible values. The considered range for $N$ was 10$^{13}$-10$^{19}$ cm$^{-2}$ in steps of $0.1 \times N$. Simultaneously, the FWHM of the synthetic fit was adjusted to match the FWHM of the detected line. For some sources (such as NGC 1333-IRAS4A and Ser-SMM3), excitation temperature could not be constrained. Hence, a grid of excitation temperatures was considered with a range between 50 and 300 K.

The line optical depth ($\tau$) was calculated for the best-fitting combination of parameters to check for optical thickness. If a transition of a certain molecule was found to be optically thick, its column density was computed as the average of the column density of the main species derived from its minor isotopologs given by:

\begin{equation}
\overline{N(\text{X})} = \frac{1}{n}\Sigma_{i=0}^n N(\text{X})_i,
\end{equation}

\noindent
where X is H$_2$S or OCS and $N$(X)$_i$ is the column density of H$_2$S or OCS derived from its minor isotopologs. Adopted isotopic ratios for the derivation of main isotopologs from minor isotopologs are given alongside \autoref{main_results_IRAS16293B}.

\section{Results}
\label{Results}

\begin{figure*}
    \centering
    \subfigure{\includegraphics[scale=1.0]{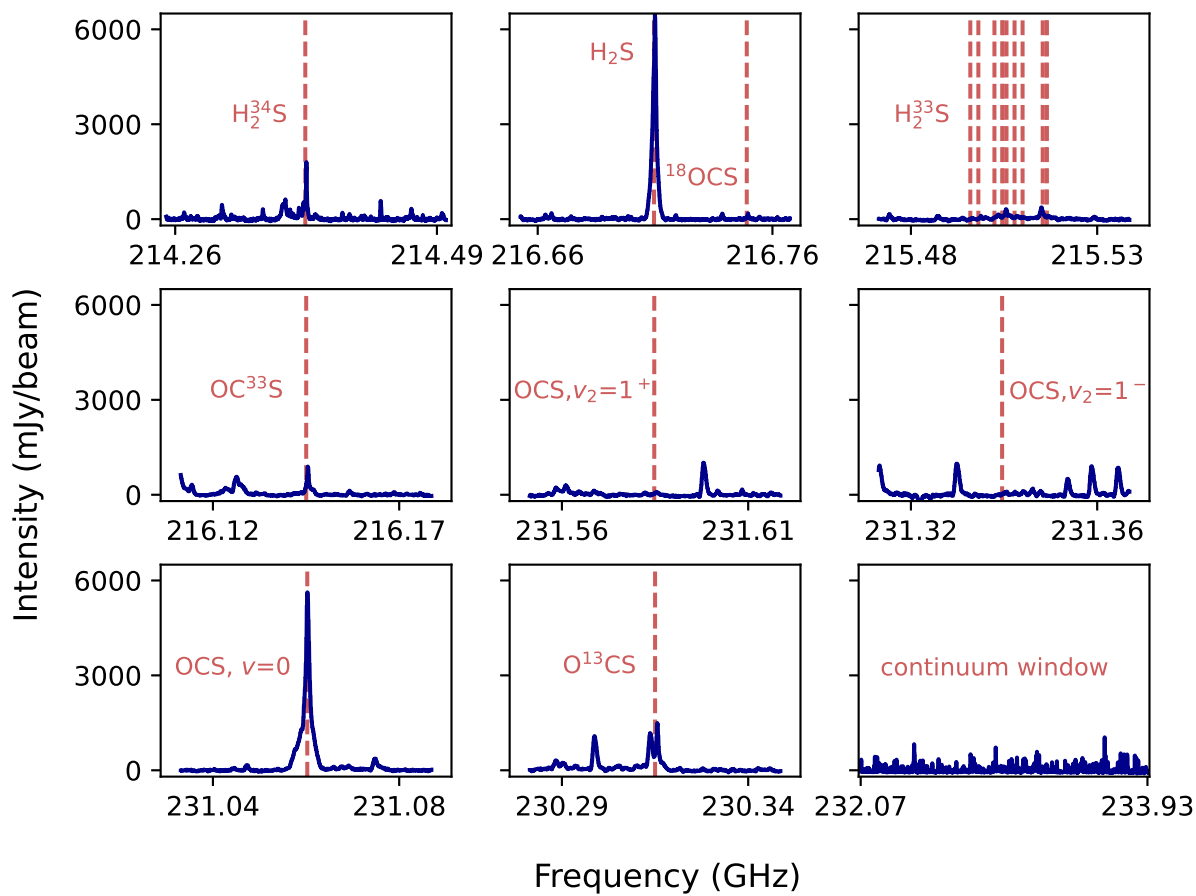}}\\
    \caption{Observed spectral windows of IRAS 16293-2422 B (\autoref{extraction_area}) obtained with ALMA ACA at Band 6 frequencies (\autoref{spectral_setup}). A Doppler shift by v$_{\text{LSR}}$ = 2.7 km s$^{-1}$ has been applied (\autoref{protostars}).}
    \label{Spectra_IRAS16293B}
\end{figure*}

The spectral set up of the first data set allows the targeted sources to be probed for the emission of the main isotopologs of H$_2$S and OCS, $v$=0, their minor isotopologs (HDS, HD$^{34}$S H$_2^{33}$S, H$_2^{34}$S, $^{18}$OCS, O$^{13}$CS, OC$^{33}$S, $^{18}$OC$^{34}$S), and also the vibrationally excited states of OCS ($v_2$=1$^{\pm}$). Consequently, sources IRAS 16293 A, IRAS 16293 B, IRAS4A, and IRS7B were probed for all these species. 

The spectral set up of the second data set allows the other targeted sources (B1-c, BHR71-IRS1, Per-emb-25, IRAS4B, SMM3, and TMC1) to be probed for the main isotopologs of H$_2$S and OCS, $v$=0, their minor isotopologs (HDS, HD$^{34}$S, $^{18}$OCS, OC$^{33}$S, $^{18}$OC$^{34}$S, $^{18}$O$^{13}$CS), and vibrationally excited state of OCS ($v_2$=1$^-$).  All the transitions of the detected molecules have $E_{\text{up}}$ in the range of $84-123$ K and $A_{ij}$ values of $0.69-4.9\times10^{-5}$~s$^{-1}$. The details of the targeted molecular lines are presented in Appendix \ref{A}. Note that the HDS lines probed in the two data sets are not the same - the first data set was probed for HDS, 14$_{2,12}$-13$_{4,9}$ transition at rest frequency 214.325 GHz, while the second data set was probed for HDS, 7$_{3,4}$-7$_{3,5}$ and 12$_{5,7}$-12$_{5,8}$ transitions at 234.046 and 234.528 GHz, respectively. Nevertheless, the $E_{\text{up}}$ is high ($>400$ K) for all three transitions of HDS; and it is not detected in any of these lines in any of the sources. The HD$^{34}$S and OCS, $v_2$=1$^{\pm}$ transitions also have high $E_{\text{up}}$ ($>400$ K). HD$^{34}$S was not detected in any of these lines in any of the sources, but OCS, $v_2$=1$^{\pm}$ was detected in IRAS 16293 A, IRAS 16293 B, and IRAS4A (owing to high OCS column densities and higher sensitivity of the first data set).

All the main and minor isotopologs were detected in IRAS 16293 A, IRAS 16293 B, and IRAS4A except HDS, HD$^{34}$S, $^{18}$OC$^{34}$S, and $^{18}$O$^{13}$CS. In IRS7B, only H$_2$S was detected, the rest of the molecular lines were undetected including OCS, $v$=0. Whereas, only main S-bearing species, H$_2$S and OCS, $v$=0, were detected in B1-c, BHR71-IRS1, and SMM3. IRAS4B showed the rotational transition of OC$^{33}$S ($J=18-17$) in addition to the H$_2$S and OCS, $v$=0 lines. Emb-25 and TMC1 showed no detections of main S-bearing species and their minor isotopologs. Thus, 1-$\sigma$ upper limits on the column densities of H$_2$S and OCS, $v$=0 were derived for Emb-25 and TMC1; and an upper limit on the column density of OCS, $v$=0 was derived for IRS7B. \autoref{line_emissions} provides the CDMS entry, transition quantum numbers, rest frequency, upper energy level, Einstein A coefficient, and the detection/non-detection of each line of all the targeted S-bearing molecules towards all of the sources in the sample.

In \autoref{maps}, the pipeline-produced integrated intensity maps with the line channels excluded are shown for all the sources. These are dominated by the dust emission, but with some degree of contamination by line emission especially for some of the line-rich sources. The circular regions used to extract the spectra of each individual source are also shown. Pixel size of the integrated maps of the sources varies from 0.8 to 1.2$''$. To match the beam size of the observations, the radius of the circular regions was also varied from 3.0 to 4.3$''$. The spatial resolution of the presented ACA observations allowed the binary IRAS 16293 A and B (separated by 5.3$''$) to be resolved as single sources; however, the resolution was not high enough to disentangle the binary components A1 and A2 of IRAS 16293 A. Similarly, the binary components of IRAS4A (with a separation of 1.8$''$), and of TMC1 (with a separation of 0.6$''$) could not be disentangled due to the spatial resolution not being high enough. All other sources are either single sources or binaries separated by large distances; hence, they are spatially resolved as individual sources.

The lower and upper uncertainties on the fitted column densities are derived assuming an error of $\pm$20 K on the assumed excitation temperature and a 1-$\sigma$ noise level. The analysis of the spectra extracted towards IRAS 16293-2422 B is presented in the following \Cref{IRAS16293B} and Appendix \ref{appendix}. Full observed spectral windows towards IRAS 16293 B are shown in \autoref{Spectra_IRAS16293B}. For the other sources, the analysis is presented in Appendices \ref{a_IRAS16293A} through \ref{A_TMC1}.

\subsection{IRAS 16293-2422 B}
\label{IRAS16293B}

\begin{table*}
    \caption{Results from the modeling of synthetic spectra of the detected S-bearing species towards IRAS 16293 B for source sizes of 1$''$ and 2$''$, an excitation temperature of 125 K, and a line width of 1 km s$^{-1}$.}
    \label{main_results_IRAS16293B}
    \centering
    \begin{adjustbox}{width=1\textwidth}
    \begin{tabular}{l c c c c c | c c | c c | c c}
    \hline
    \hline
    \multirow{1}{*}{Species} & \multirow{1}{*}{Transition} & \multirow{1}{*}{Freq.} & \multirow{1}{*}{$E_{\text{up}}$} & \multirow{1}{*}{$A_{ij}$} & \multirow{1}{*}{beam} &
    \multicolumn{2}{c}{$N$} & \multicolumn{2}{c}{Derived $N$} & \multicolumn{2}{c}{$\tau$} \\

    \multirow{1}{*}{} & \multirow{1}{*}{} & \multirow{1}{*}{} & \multirow{1}{*}{} & \multirow{1}{*}{} & \multirow{1}{*}{size} &
    \multicolumn{2}{c}{} & \multicolumn{2}{c}{of isotopologs} & \multicolumn{2}{c}{} \\

    \multirow{1}{*}{} & \multirow{1}{*}{} & \multirow{1}{*}{(GHz)} & \multirow{1}{*}{(K)} & \multirow{1}{*}{(s$^{-1}$)} & \multirow{1}{*}{($''$)} & \multicolumn{2}{c}{(cm$^{-2}$)} & \multicolumn{2}{c}{(cm$^{-2}$)} & \multicolumn{2}{c}{} \\

\hline
& & & & & & 1$''$ & 2$''$ & 1$''$ & 2$''$ & 1$''$ & 2$''$ \\
\hline
&&&&&&&&&&\\

H$_2$S & 2$_{2,0}$-2$_{1,1}$ & 216.710 & 84 & 4.9$\times$10$^{-5}$ & 6.0 & op. thick  & op. thick   & \textbf{$\overline{N(\text{H}_2\text{S})}$=(3.6$\pm$0.6)$\times$10$^{17}$} & \textbf{$\overline{N(\text{H}_2\text{S})}$=(9.2$\pm$1.7)$\times$10$^{16}$} & 30.0 & 8.00\\

&&&&&&&&&&\\

H$_2$ $^{33}$S 

& 2$_{2,0,1}$-2$_{1,1,2}$ & 215.494 & 84& 2.4$\times$10$^{-5}$ &  6.0 &  2.7$^{+0.7}_{-0.4}\times$10$^{15}$  &  7.0$^{+1.7}_{-1.1}\times$10$^{14}$  & $N$(H$_2$S)=3.4$^{+0.8}_{-0.5}\times$10$^{17,\ }$\tablefootmark{c} & $N$(H$_2$S)=8.8$^{+2.1}_{-1.4}\times$10$^{16,\ }$\tablefootmark{c} &  0.02 & 0.004\\
& 2$_{2,0,1}$-2$_{1,1,1}$ & 215.497 &  84 & 2.4$\times$10$^{-5}$   &   &  & &  &  & 0.02 &  0.004\\
& 2$_{2,0,4}$-2$_{1,1,3}$ & 215.501 & 84 & 6.9$\times$10$^{-6}$  &  &   &  &  &  &  0.02 & 0.005  \\
& 2$_{2,0,4}$-2$_{1,1,4}$ & 215.503 & 84 & 4.2$\times$10$^{-5}$  &   &  &  &  &  &  0.10 &  0.030 \\
& 2$_{2,0,2}$-2$_{1,1,3}$ & 215.504 & 84 & 1.7$\times$10$^{-5}$  &  &  &  &  & &  0.02 &  0.010 \\
& 2$_{2,0,2}$-2$_{1,1,2}$ & 215.505 & 84 & 1.9$\times$10$^{-5}$  &  &  &  &  &   & 0.03 &  0.010 \\
& 2$_{2,0,2}$-2$_{1,1,1}$ & 215.508 & 84 & 1.2$\times$10$^{-5}$ &  &   &  &  &    & 0.02 &  0.004 \\
& 2$_{2,0,3}$-2$_{1,1,3}$ & 215.512 & 84 & 2.8$\times$10$^{-5}$ &  &  &  &  &     & 0.06 &  0.020 \\
& 2$_{2,0,3}$-2$_{1,1,2}$ & 215.513 & 84 & 1.1$\times$10$^{-5}$ &  &  &  &  &    & 0.02 & 0.010  \\
& 2$_{2,0,3}$-2$_{1,1,4}$ & 215.513 & 84 & 9.1$\times$10$^{-6}$ &  &  &  &  &    & 0.02 &  0.005 \\

&&&&&&&&&&\\

H$_2$ $^{34}$S & 2$_{2,0}$-2$_{1,1}$ & 214.377 & 84  & 4.7$\times$10$^{-5}$ &  6.0 &   op. thick & $>$1.5$\times$10$^{15}$   &  & &  1.00 & 0.20\\

&&&&&&&&&&\\

OCS, $v$=0 & 19-18 & 231.061 & 111  & 3.6$\times$10$^{-5}$  &  5.6 & op. thick   & op. thick  & \textbf{$\overline{N\text{(OCS)}}$ = (2.7$\pm$0.3)$\times$10$^{17}$} & \textbf{$\overline{N\text{(OCS)}}$ = (7.0$\pm$0.8)$\times$10$^{16}$} &  46.0 & 11.0\\

&&&&&&&&&&\\

OC$^{33}$S & 18-17 & 216.147 & 99 & 2.9$\times$10$^{-5}$  & 6.0  & $>$2.4$\times$10$^{15}$ &  $>$5.6$\times$10$^{14}$ &  &   & 0.40 &  0.10\\

&&&&&&&&&&\\

O$^{13}$CS & 19-18 & 230.318 & 110  & 3.5$\times$10$^{-5}$ & 5.7 &  $>$3.8$\times$10$^{15}$  & $>$8.2$\times$10$^{14}$ &  &  &  0.60 &  0.10\\

&&&&&&&&&\\

$^{18}$OCS & 19-18 & 216.753 & 104  & 3.0$\times$10$^{-5}$ &  6.0  &  4.7$^{+0.6}_{-0.3}\times$10$^{14}$ & 1.2$^{+0.2}_{-0.1}\times$10$^{14}$ & $N$(OCS)=2.6$^{+0.4}_{-0.2}\times$10$^{17,\ }$ \tablefootmark{d} & $N$(OCS)=6.7$^{+1.8}_{-0.6}\times$10$^{16,\ }$ \tablefootmark{d} & 0.07 & 0.02\\

&&&&&&&&&&\\

OCS, $v_2$=1$^-$  & 19-18  & 231.342 & 860 & 3.5$\times$10$^{-5}$ &  5.6 & 8.5$^{+14.5}_{-4.3}\times$10$^{16}$  & 2.5$^{+3.5}_{-1.4}\times$10$^{16}$ & & & 0.03 & 0.01\\
OCS, $v_2$=1$^+$  & 19-18  & 231.584 & 860  &  3.5$\times$10$^{-5}$ &  5.6  &  & & & & 0.03 & 0.01\\

&&&&&&&&&&&\\

\hline
\hline
    \end{tabular}
    \end{adjustbox}
\tablefoot{The columns in the table of line parameters denote the following: (1) detected S-bearing species, (2) transition of the
emission line, (3) frequency of the emission line, (4) upper energy level, (5) Einstein A coefficient, (6) beam size, (7) column density, (8) derived column density of the main isotopologs, and (9) optical depth of the emission line. Directly across from a specific minor isotopolog under ``Derived $N$ of isotopologs’ follows the column density of the main isotopolog upon the assumption of the standard isotopic ratio. In bold in the same column is the average column density of the main isotopolog based on all the available minor isotopologs (only if the minor isotopolog is optically thin and including the uncertainties). The isotopic ratios assumed to derive column densities of the main isotopologs from their minor isotopologs: 
\tablefoottext{a}{$^{12}$C/$^{13}$C = 69 \citep{Wilson1999}, }
\tablefoottext{b}{$^{32}$S/$^{34}$S = 22 \citep{Wilson1999}, }
\tablefoottext{c}{$^{32}$S/$^{33}$S = 125 \citep{2009ARA&A..47..481A}, }
\tablefoottext{d}{$^{16}$O/$^{18}$O = 557 \citep{Wilson1999}.}
}
\end{table*}

\begin{figure}
    \centering
    \includegraphics[scale=0.35]{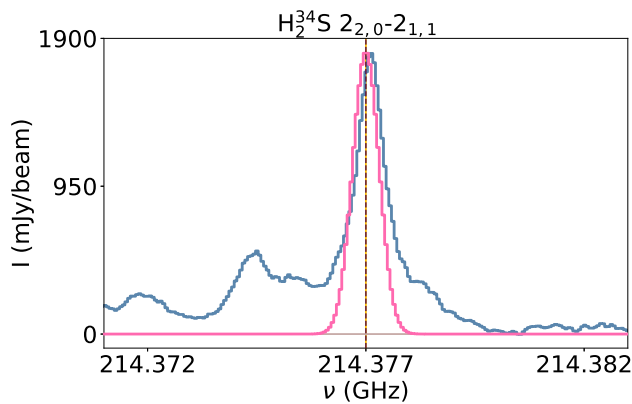}
    \caption{H$_2^{34}$S line detected in IRAS 16293 B. The observed spectrum (in blue), rest frequency of the detected line (brown dashed line), spectroscopic uncertainty on the rest frequency of the detected line (yellow shaded region), and fitted synthetic spectrum (in pink) for source size: 2$''$, excitation temperature: 125 K, and FWHM: 1 km s$^{-1}$.}
    \label{H234S}
\end{figure}

Towards IRAS 16293 B, the main S-bearing species (H$_{2}$S and OCS, $v$=0) and all the targeted minor isotopologs are securely detected, except for HDS due to a very high $E_{\text{up}}$ value ($1~277$ K) and the double isotopologs of HD$^{34}$S, $^{18}$OC$^{34}$S, $^{18}$O$^{13}$CS due to their low abundances (and high $E_{\text{up}}$ for the case of HD$^{34}$S). The detected transitions of H$_{2}$S, 2$_{2,0}$-2$_{1,1}$ and OCS, $v$=0, $J=19-18$ are bright and optically thick ($\tau>>$1). The H$_2^{34}$S line is marginally optically thick ($\tau=0.2$), shown in \autoref{H234S}. The vibrationally excited OCS, $v_2$=1$^\pm$ lines are detected. The lines of the detected molecules do not suffer from blending, except the H$_2^{33}$S, 2$_{2,0,3}$-2$_{1,1,3}$ transition at 215.512 GHz, which is contaminated by the CH$_3$CHO, 11$_{2, 9}$-10$_{2, 8}$ transition. HD$^{34}$S, 7$_{3,4}$-7$_{3,5}$ transition at 232.964 GHz is heavily blended with the CH$_3$CN, $v_8=1$, $J=15-15$, $K=7-5$ transition. Most likely all the emission seen around the rest frequency of HD$^{34}$S comes from CH$_3$CN, because HD$^{34}$S is a minor species ($^{32}$S/$^{34}$S=22, \citealt{Wilson1999}, and D/H $\sim$0.04 incl. the statistical correction by a factor of $2$ to account for the two indistinguishable D atom positions, \citealt{Drozdovskaya2018}) and the $E_{\text{up}}$ of this transition is high (416 K). The spectra of detected and undetected lines are in \autoref{detected_IRAS16293B} and \autoref{undetected_IRAS16293B}, respectively. 

For the analysis of the targeted S-bearing molecules towards IRAS 16293 B, a $T_{\text{ex}}$ of 125 K is assumed. This value has been deduced to be the best-fitting on the basis of ALMA-PILS observations at higher spatial resolution obtained with the 12m array and a full inventory of S-bearing molecules \citep{Drozdovskaya2018}. A FWHM of 1 km s$^{-1}$ is adopted, as it has been shown that this value consistently fits nearly all the molecules investigated towards the hot inner regions of IRAS 16293 B (e.g., \citealt{Jorgensen2018}). For the larger scales probed by the present ALMA ACA observations, a deviation by 2 km s$^{-1}$ from this FWHM can be seen for optically thick lines. This broadening in FWHM is likely due to the opacity broadening effects, which are dominant in optically thick lines, but can be neglected in optically thin lines \citep{Hacar2016}. The synthetic spectral fitting has been carried out for two potential source sizes, 1$''$ and 2$''$ (\autoref{main_results_IRAS16293B}). Column densities depend on the assumed source size and are lower for the larger source size. However, the $N$(H$_2$S)/$N$(OCS) ratio is 1.3$\pm$0.27 and 1.3$\pm$0.28 for source sizes of 1$''$ and 2$''$, respectively. Thus, the ratio is independent of the assumed source size and is robustly determined with the ALMA ACA data.

For a source size of $2\arcsec$, the column density of the vibrationally excited state of OCS, $v_2$=1$^\pm$ derived for IRAS 16293 B (2.5$\times$10$^{16}$ cm$^{-2}$) is an order of magnitude lower than the OCS, $v_{2}=1$ column density (2.0$\times$10$^{17}$ cm$^{-2}$) derived in \cite{Drozdovskaya2018}. For a source size of $1\arcsec$, the here obtained value ($8.5\times10^{16}$~cm$^{-2}$) is in closer agreement with \cite{Drozdovskaya2018}. Likewise, the OCS, $v$=0 column density determined from the minor isotopologs of OCS for a source size of 1$''$ ($2.7\times10^{17}$~cm$^{-2}$) is in a closer agreement with the column density of OCS, $v=0$ (2.8$\times$10$^{17}$ cm$^{-2}$) derived in \cite{Drozdovskaya2018}, also based on minor isotopologs, than for source size of 2$''$ ($7.0\times10^{16}$~cm$^{-2}$). \cite{Drozdovskaya2018} used a smaller source size (0.5$''$) to constrain the column densities of OCS and H$_2$S. These comparisons suggest that the ALMA ACA observations in this work are subject to beam dilution, hence the column densities are likely somewhat underestimated. The column density of H$_2$S could not be constrained to better than a factor of 10 in \cite{Drozdovskaya2018}, namely $1.6\times10^{17}-2.2\times10^{18}$~cm$^{-2}$. This was due to the fact that only deuterated isotopologs of H$_{2}$S were covered by the PILS observations and the D/H ratio of H$_{2}$S is only constrained to within a factor of $10$. Based on values of the H$_2$S column densities for 1$''$ and $2\arcsec$ source sizes obtained in this work, the lower estimate for the H$_2$S column density in \cite{Drozdovskaya2018} seems to be more accurate. In turn, the H$_2$S/OCS ratio obtained in this work ($1.3$) is closer to the lower end of the $0.7-7$ range computed in \cite{Drozdovskaya2018}.

\subsection{Line Profiles}

For the synthetic spectral modeling, Gaussian line profiles are assumed (Section~\ref{fitting}). However, even for optically thin lines, a deviation from Gaussian line profiles is seen in some cases. Two prominent examples are H$_{2}^{34}$S and O$^{13}$CS in IRAS 16293 A (\autoref{detected_IRAS16293A}), where the high spectral resolution of the data set clearly allows multiple peaks to be spectrally resolved in these lines. Likely, the reason for this is that this source is a compact binary \citep{Maureira2020} with multiple components within the ACA beam of these observations. Another prominent example is the OCS, $v=0$ line in Ser-SMM3 (\autoref{detected_SMM3}), which has a double-peaked profile centered around the source velocity. Such a line profile is typical for a rotating structure around its protostar (which could be envelope or disk in nature). Detailed modeling of line profiles is out of scope of this paper, as additional observations would be necessary in order to achieve meaningful results. For the purpose of studying the H$_{2}$S/OCS ratio, these effects are secondary and likely do not significantly affect the calculated ratio and the conclusions of this paper. For IRAS 16293 A, the column density of H$_{2}^{34}$S is not used to get the column density of H$_{2}$S, because it is computed to be partially optically thick. Meanwhile, the column density of OCS as obtained from O$^{13}$CS is within a factor of $2$ of what is obtained from OC$^{33}$S and $^{18}$OCS. For Ser-SMM3, the lack of constraints on the excitation temperature dominates the uncertainty in the H$_{2}$S/OCS ratio.

\subsection{H$_2$S/OCS ratio determination}

The column densities of H$_2$S and OCS derived on the basis of ALMA ACA observations have been used to constrain the ratio of H$_2$S to OCS (\autoref{ratio}). It was possible to compute this ratio for five out of ten sources in the considered sample. Neither H$_2$S nor OCS were detected in Emb-25 and TMC1, consequently the H$_2$S/OCS ratio could not be constrained. The non-detection of OCS in IRS7B allowed to derive only a lower limit on the H$_2$S/OCS ratio.  \autoref{ratio} also contains the best-available estimates of the H$_2$S/OCS ratio for the warm and cold components of B1-c, and for the cold component of BHR71-IRS1, although these numbers carry a higher level of uncertainty due to line opacity that could not be resolved on the basis of these observations. For the warm component of BHR71-IRS1, a lower limit on the H$_2$S/OCS ratio could be computed. For further analysis, the sample has been divided into three sub-samples: compact binary, wide binary, and single, based on the separations between components of multiple sources or closest neighbours.

\begin{table*}
    \caption{H$_2$S/OCS ratio for the studied sources, including their evolutionary class, binarity, environment, and the derived column densities of H$_2$S and OCS for the stated excitation temperatures. The H$_2$S/OCS ratios for the cold and warm components of B1-c, and the cold component of BHR71-IRS1 are the best-available estimates pending opacity issues.}
    \label{ratio}
    \centering
    \begin{adjustbox}{width=1.0\textwidth}
    \begin{tabular}{l c c c c c c r}
    \hline
    Source   & Class & Binarity & Environment & $T_{\text{ex}}$ (K) & $N$(H$_2$S) (cm$^{-2}$) & $N$(OCS) (cm$^{-2}$) & $N$(H$_2$S)/$N$(OCS)\\
\hline
    IRAS 16293-2422 A  & 0 & CB  & Clustered & 125$\pm$20  & (2.4$\pm$0.4)$\times$10$^{17}$  &  (3.6$\pm$1.4)$\times$10$^{17}$ & 0.7$\pm$0.3 \\ 
    IRAS 16293-2422 B & 0 & WB  & Clustered & 125$\pm$20  & (9.2$\pm$1.7)$\times$10$^{16}$  &  (7.0$\pm$0.8)$\times$10$^{16}$ & 1.3$\pm$0.3 \\ 
    NGC 1333-IRAS4A    & 0 & CB &  Clustered    & 150-300   &  (3.4$\pm$0.8)$\times$10$^{16}$  &  (1.8$\pm$0.2)$\times$10$^{16}$ & 1.9$\pm$0.5 \\
    RCrA IRS7B   & 0/I & WB & Clustered & 100$\pm$20 & (5.6$\pm$0.8)$\times$10$^{13}$  &   $\leq$3.6$\times$10$^{13}$  & $\geq$1.5   \\ 
    Per-B1-c    & 0 & S & Clustered    &  60  &   $>$9.7$\times$10$^{15}$  &  $>$5.0$\times$10$^{15}$ & (1.9) \\
    &  &  & & 200  &   $>$1.2$\times$10$^{16}$  &  $>$3.6$\times$10$^{15}$ & (3.3) \\
    BHR71-IRS1    & 0 & WB & Isolated & 100  &   $>$2.4$\times$10$^{16}$  &  $>$2.7$\times$10$^{15}$ & (8.9) \\ 
    &  &  &  &  250  &   $>$3.3$\times$10$^{16}$  &  (3.4$\pm$0.3)$\times$10$^{15}$ & $\geq$9.7 \\ 
    Per-emb-25    & 0/I & S & Clustered & 50-300   & $\leq$8.3$\times$10$^{13}$  &  $\leq$3.2$\times$10$^{14}$   &  --   \\
    NGC 1333-IRAS4B   & 0 & WB  & Clustered    & 100$\pm$20  &  $>$5.8$\times$10$^{15}$  &  (2.8$\pm$0.6)$\times$10$^{16}$ & $\geq$0.21 \\ 
    Ser-SMM3     & 0 & S & Clustered  &  100-250      &  (5.8$\pm$3.2)$\times$10$^{14}$  &  (8.7$\pm$4.9)$\times$10$^{14}$ & 0.7$\pm$0.5 \\
    TMC1    & I & CB   &  Clustered   & 40 & $\leq$1.5$\times$10$^{13}$  &   $\leq$2.6$\times$10$^{13}$  &  --    \\

    \hline
    Comet (67P/C-G) & & & & & & & 26.8$^{+47.5,\ }_{-21.6}$\tablefootmark{a} \\
    \hline
    &&&&& $N$(H$_2$S)/$N$(H$_2$O)& $N$(OCS)/$N$(H$_2$O) & $N$(H$_2$S)/$N$(OCS)\\
    \hline
    ISM ices (W33A) & & & & & $<$3.0$\times$10$^{-2,\ }$\tablefootmark{b} & 
    4.0$\times$10$^{-4,\ }$\tablefootmark{c} & $<$75 \\
    ISM ices (Mon R2 IRS2) & & & & & $<$4.7$\times$10$^{-3,\ }$\tablefootmark{d} & 5.5$\times$10$^{-4,\ }$\tablefootmark{c} & $<$8.5 \\
    \hline
    \end{tabular}
    \end{adjustbox}
\tablefoot{CB: close binary ($< 500$ au) that is not spatially resolved in the data used for this work; WB: wide binary ($500-5~000$ au) that is spatially resolved in the data used for this work; S: single source within $5~000$ au. ISM stands for interstellar medium. References: \tablefoottext{a}{\cite{Rubin2019}, }
\tablefoottext{b}{\cite{Vandertak2003}, \tablefoottext{c}{\cite{Palumbo1995}},
\tablefoottext{d}{\cite{Smith1991}}.}}
\end{table*}

\section{Discussion}
\label{Discussions}

\autoref{final_plot} displays the derived protostellar H$_2$S/OCS ratios, as well as the cometary (67P/Churyumov–Gerasimenko, hereafter 67P/C-G) and interstellar ice (W33A and Mon R2 IRS2) H$_2$S/OCS ratios. The derived protostellar H$_2$S/OCS ratios span a range from 0.2 to above 9.7. The ratios show a variation of approximately one order of magnitude, being the lowest in IRAS 16293 A and SMM3, and the highest in BHR71-IRS1.

In the subsections \ref{ISMices} and \ref{comet}, a comparison of the protostellar H$_2$S/OCS ratios with this ratio in interstellar and cometary ices, respectively, is made. Comets are thought to preserve the chemical composition of the Sun's birth cloud \citep{Mumma2011}. By comparing the H$_2$S/OCS ratio of comet 67P/C-G with the ratios in nascent solar-like protostellar systems, an assessment can be made whether such an inheritance is true in the case of S-bearing molecules. 

\subsection{Interstellar ices}
\label{ISMices}

Observations towards the cold, outer protostellar envelopes of high-mass protostars W33A and Mon R2 IRS2 are used to acquire the H$_2$S/OCS ratio in interstellar ices. The ratio is computed based on the ice abundances of OCS detected as an absorption feature at 4.9 $\mu$m \citep{Palumbo1995} using the Infrared Telescope Facility (IRTF) and upper limits on the H$_2$S abundance derived based on the non-detection of the 3.98 $\mu$m band. The column density of solid OCS with respect to solid H$_2$O is $N_{\text{solid}}$(OCS)/$N_{\text{solid}}$(H$_2$O) = 4$\times$10$^{-4}$ \citep{Palumbo1995} and $N_{\text{solid}}$(OCS)/$N_{\text{solid}}$(H$_2$O) = 5.5$\times$10$^{-4}$ \citep{Palumbo1997} for W33A and Mon R2 IRS2, respectively. Based on the non-detection of solid H$_2$S towards W33A in the Infrared Space Observatory (ISO) Spectra from the Short Wavelength Spectrometer (SWS), $N_{\text{solid}}$(H$_2$S)/$N_{\text{solid}}$(H$_2$O)$_{\text{solid}}$ $<$0.03 \citep{Vandertak2003}. The upper limits on the solid H$_2$S and solid H$_2$O in Mon R2 IRS2 are $<0.2\times10^{17}$ and $42.7\times10^{17}$~cm$^{-2}$ \citep{Smith1991}, respectively, yielding $N_{\text{solid}}$(H$_2$S)/$N_{\text{solid}}$(H$_2$O) $<$4.7$\times$10$^{-3}$. The H$_2$S/OCS ratio in interstellar ices is poorly constrained due to the non-detection of solid H$_2$S to date. The upper limits on the interstellar ices ratio are within the uncertainties of the cometary ices ratio. The derived protostellar ratios for all the sources are lower than the upper limit on the H$_2$S/OCS ratio determined for interstellar ices except BHR71-IRS1 with H$_2$S/OCS ratio exceeding the upper limit on the ratio for Mon R2 IRS2.

\subsection{Comet 67P/Churyumov-Gerasimenko}
\label{comet}

Comets are thought to be the most unprocessed objects in the Solar System \citep{Mumma2011}. Cometary chemical composition has been shown to be similar to a degree to that of star-forming regions \citep{Bockelee2000, Drozdovskaya2019}. Consequently, the cometary H$_2$S/OCS ratio is thought to provide an independent measurement of this ratio in interstellar ices. The H$_2$S and OCS abundances from the ESA \textit{Rosetta} mission were used to compute the H$_2$S/OCS ratio for the Jupiter-family comet 67P/C-G. The H$_2$S and OCS abundances relative to H$_2$O are 1.10$\pm$0.46\% and 0.041$^{+0.082}_{-0.020}$\%, respectively \citep{Rubin2019}. These molecules are typical constituents of comets \citep{Lis1997, Bockelee2000, Boissier2007, Mumma2011}. BHR71-IRS1 is the only protostar in the sample with a H$_2$S/OCS ratio within the uncertainties of the cometary ices ratio. The H$_2$S/OCS ratio for the other sources is at least an order of magnitude lower than for 67P/C-G, while even considering the high uncertainties on the cometary value. The availability of H$_{2}$S relative to H$_{2}$O in cometary ice ($0.0064-0.0156$) appears to be higher than in the interstellar ices towards Mon R2 IRS2 ($<0.0047$). For W33A, the currently available upper limit ($<0.03$) is less constraining and hence no conclusion can be drawn about how its ices compare to those of comet 67P/C-G. The relative ratio of H$_{2}$S to OCS is only one window onto the inventory of S-bearing molecules in gas and ice at different stages of star and planet formation, meanwhile the overall availability relative to, for example, H$_{2}$O is another window that requires dedicated exploration.

\begin{figure*}
    \centering
    \includegraphics[scale=0.3]{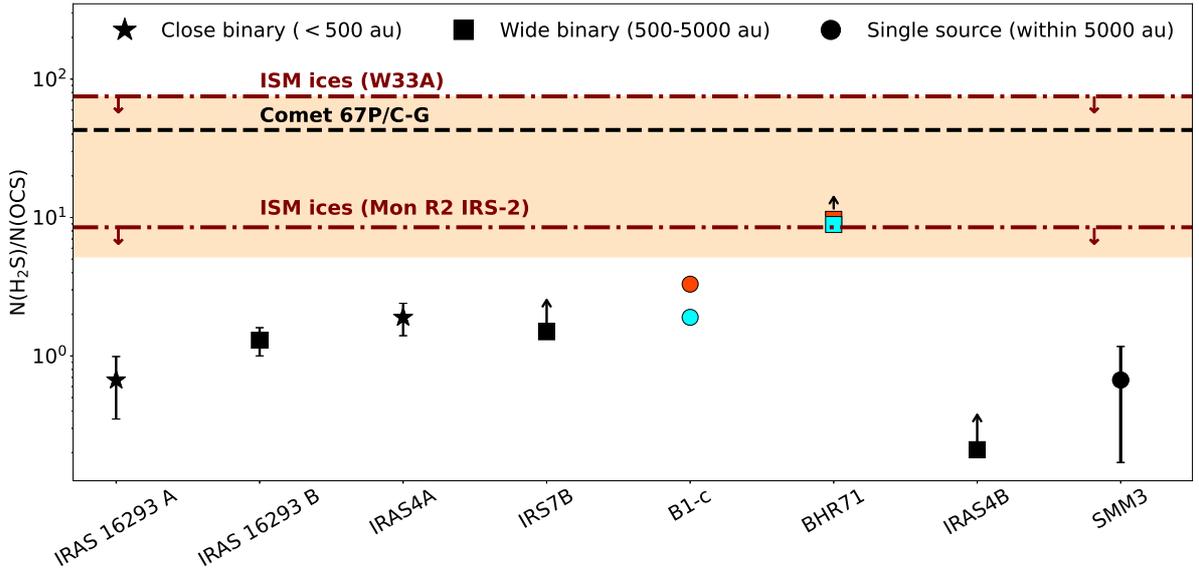}
    \caption{$N$(H$_2$S)/$N$(OCS) of the studied sources. Different symbols represent different types of sources, i.e., `star' for close binary ($<500$ au), `square' for wide binary ($500-5000$ au), and `circle' for single sources (within $5~000$ au). The upper limits on the interstellar ice (W33A and Mon R2 IRS2) ratios are shown by downward arrows. The uncertainty on the H$_2$S/OCS ratio in comet 67P/C-G is shown by the coral shaded region. The lower limit on the ratio in IRS7B is shown by an upwards arrow. The H$_2$S/OCS ratios for the cold (cyan) and warm (orange) components of B1-c, and cold (cyan) component of BHR71-IRS1 are the best-available estimates pending opacity issues. These latter three data points do not have error bars associated to them in the figure to indicate that they are merely estimates.}
    \label{final_plot}
\end{figure*}

\subsection{H$_2$S/OCS ratio as an environmental (clustered/isolated) tracer}

The measured gas-phase H$_2$S/OCS ratios in the sample of young, low-mass protostars explored in this paper are predominantly lower (by as much as an order of magnitude) than the solid-state ratio measured through direct infrared observations of interstellar ices and indirectly via comets (\autoref{final_plot}). There appears to be no correlation with binarity nor specific host cloud (\autoref{ratio}). The dependence with evolutionary stage could not be properly explored, as the sample contains only one Class I source (TMC1), which did not result in detections of neither H$_2$S nor OCS.

The highest ratio of $\geq$9.7 is found for the warm component (250 K) of BHR71-IRS1, which is a wide binary ($\sim3~200$ au; \citealt{2001ApJ...554L..91B, Parise2006, 2008ApJ...683..862C, Tobin2019}) Class 0 protostar. The ratio in BHR71-IRS1 resides within the uncertainty of the cometary ratio, but is in-between the two upper limits derived for the interstellar ices. The overall envelope mass and bolometric luminosity of BHR71-IRS1 is comparable to those of other compact binary and wide binary systems. The similarity of its gas-phase H$_2$S/OCS ratio to the ratio in ices may suggest that it is displaying the most recently thermally desorbed volatiles that have not been subjected to gas-phase processing for long. However, what makes BHR71-IRS1 stand out is that it is located in an isolated cloud, i.e., it is not associated with processes typical for clustered environments such as dynamical interactions, mechanical and chemical feedback from outflows, and enhanced irradiation. 

Possibly, isolation resulted in lower irradiation of the ice grains during the prestellar phase in BHR71-IRS1, thus converting less H$_2$S ices to OCS ices by photodissociation in the presence of CO ice. Consequently, leaving a higher H$_2$S/OCS ratio in ices, which after evaporation resulted in a higher H$_2$S/OCS ratio in the gas phase. Another reason could be more efficient hydrogenation chemistry in such a colder environment. On dust grains, hydrogenation is expected to be the most effective process leading to the formation of H$_2$S \citep{Wakelam2011, Esplugues2014}. Hence, BHR71-IRS1 may have a higher H$_2$S content and a lower OCS content, which results in a higher H$_2$S/OCS ratio. Water deuteration is also higher by a factor of $2-4$ in isolated protostars such as BHR71-IRS1 in comparison to those in clustered environments such as IRAS 16293 and IRAS4 \citep{Jensen2019}.

One alternative cause of lower H$_2$S/OCS ratios towards clustered low-mass protostars could be local temperature differences in their birth clouds, e.g., due to enhanced irradiation from the neighbouring protostars. Laboratory experiments have proven that OCS forms readily in ices when interstellar ice-analogs are irradiated by high-energy photons \citep{Ferrante2008, Garozzo2010, Jimenez2011, Chen2015}. This would lead to a lower H$_2$S/OCS ratio. 

Additionally, cosmic rays and other forms of radiation (UV and X-ray photons) are a ubiquitous source of ionization of the interstellar gas. It is a pivotal factor in the dynamical and chemical evolution of molecular clouds \citep{Padovani2018, Padovani2020}. Cosmic rays are not attenuated in the molecular clouds as strongly as UV photons \citep{Ivlev2018, Padovani2018, Silsbee2018}. Thus, dust grains in the interstellar medium can be heated by impinging cosmic rays, thereby heating up the icy grain mantles and resulting in calamitous explosions \citep{Leger1985, Ivlev2015b}, thereby activating chemistry in solids \citep{Shingledecker2017}. Magnetohydrodynamic simulations have shown a higher cosmic ray production in protostars in a clustered environment \citep{Kuffmeier2020}, which would be consistent with the lower H$_2$S/OCS ratios for such protostars found in this work. The results suggest that the H$_2$S/OCS ratio traces the environment (isolated/clustered) of the protostellar systems. However, a follow-up study is needed as the sample consisted of only one isolated source.

\section{Conclusions}
\label{Conclusions}

This work probed a sample of ten low-mass protostars for the presence of H$_2$S, OCS, and their isotopologs using ALMA ACA Band 6 observations. For 5 out of 10 protostars, the H$_2$S/OCS ratio was firmly constrained and for an additional 3, best-possible estimates were obtained. This ratio is thought to be a potential chemical and physical clock of star-forming regions, which sheds light on the sulfur depletion that transpires from the diffuse medium to the dense core stage. The main conclusions are:

\begin{itemize}
    \item Main S-bearing species, H$_2$S and OCS are detected in IRAS 16293-2422 A, IRAS 16293-2422 B, NGC 1333-IRAS4A, NGC 1333-IRAS4B, Per-B1-c, BHR71-IRS1, and Ser-SMM3. 1-$\sigma$ upper limits on the column densities of OCS are derived for RCrA IRS7B, TMC1, and Per-emb-25. 1-$\sigma$ upper limits on the column densities H$_2$S are derived for TMC1 and Per-emb-25.
    \item The gas-phase H$_2$S/OCS ratio ranges from 0.2 to above 9.7, and is typically at least one order of magnitude lower than that of ices. The lowest ratio is obtained for IRAS 16293 A and Ser-SMM3, while the highest for BHR71-IRS1. The environment of the natal cloud, prior to the onset of star formation, may have played a major role in the distribution of sulfur across various S-bearing molecules, which have resulted in an order of magnitude spread in the H$_2$S/OCS ratio.
    \item The upper limits derived for the interstellar ices (Mon R2 IRS2 and W33A) lie within the uncertainties of the cometary ices ratio, specifically that of comet 67P/Churyumov-Gerasimenko. The protostellar ratios are lower than the upper limits on the interstellar ices ratio and the cometary ices ratio by at least an order of magnitude for all sources except BHR71-IRS1. 
    \item The lower ratio in clustered protostellar regions could be due to elevated birth cloud temperatures or due to additional radiation from nearby protostars, thereby enhancing the photodissociation pathways from H$_2$S to OCS.
    \item The high H$_2$S/OCS ratio in BHR71-IRS1 could be the result of less efficient photodissociation of H$_2$S to OCS in the presence of CO ice in its isolated birth cloud or more efficient hydrogenation chemistry leading to more efficient H$_2$S formation.
\end{itemize}

Follow-up high spatial resolution observations are required towards several sources to {better constrain the spatial distribution and excitation temperatures} associated with the H$_2$S and OCS detections. Furthermore, the difference of more than tenfold in the H$_2$S/OCS ratio towards Class 0 protostars in clustered and isolated environments is a strong motivation for performing more spectroscopic observations towards such sources, thereby understanding the physical and chemical differences in the two types of environments. Observations from James Webb Space Telescope (JWST) could play an important role in constraining the H$_2$S/OCS ice ratio in low- and intermediate-mass stars. More studies of the H$_2$S/OCS ratio in a larger sample of Class 0 and Class I protostars in clustered and isolated environments should also be performed to further understand the sulfur chemistry in star-forming regions. 

\begin{acknowledgements}

The research was started as part of the Leiden/ESA Astrophysics Program for Summer Students (LEAPS) 2021. M.N.D. acknowledges the support by the Swiss National Science Foundation (SNSF) Ambizione grant no. 180079, the Center for Space and Habitability (CSH) Fellowship, and the IAU Gruber Foundation Fellowship. This paper makes use of the following ALMA data: ADS/JAO.ALMA\#2017.1.00108.S and ADS/JAO.ALMA\#2017.1.01350.S. ALMA is a partnership of ESO (representing its member states), NSF (USA) and NINS (Japan), together with NRC (Canada), MOST and ASIAA (Taiwan), and KASI (Republic of Korea), in cooperation with the Republic of Chile. The Joint ALMA Observatory is operated by ESO, AUI/NRAO and NAOJ.
This research made use of Astropy, a community-developed core Python package for Astronomy \citep{astropy:2013, astropy:2018}\footnote{\url{http://www.astropy.org}}.

The authors would like to thank Prof. Dr. Ewine van Dishoeck for useful discussions about the H$_2$S/OCS ratio and the anonymous referee for constructive feedback.
\end{acknowledgements}

\bibliographystyle{aa} 
\bibliography{mybibli.bib} 

\newpage

\onecolumn
\begin{appendix}\label{appendix} 

\section{Inventory of targeted sources and detected and undetected emission lines}
\label{A}

\begin{landscape}
\begin{table}
    \centering
\begin{adjustbox}{width=1.1\textwidth}
    \begin{tabular}{r c r c r c c c c c c c c c c c}
    \hline
    \hline
    Species & CDMS entry & Transition & Freq. & $E_{\text{up}}$ & $A_{ij}$  & IRAS 16293 A & IRAS 16293 B  & IRAS4A & IRS7B & B1-c & BHR71-IRS1 & Emb25 & IRAS4B & SMM3 & TMC1\\
    & & & (GHz) & (K) & (s$^{-1}$) & & & & & & & & & &\\
    \hline
    H$_2$S & 46519 & 2$_{2,0}$-2$_{1,1}$ & 216.710  & 84 &  4.9$\times$10$^{-5}$ & \cmark & \cmark & \cmark & \cmark & \cmark & \cmark & \xmark & \cmark & \cmark & \xmark\\
    &&&&&&&&&&&&&&&\\

    H$_2$$^{33}$S & 35503 & 2$_{2,0,1}$-2$_{1,1,2}$  & 215.494  & 84  &   2.4$\times$10$^{-5}$  & \cmark & \cmark & \cmark & \xmark & -- & -- & -- & -- & -- & -- \\

    & & 2$_{2,0,1}$-2$_{1,1,1}$  & 215.497  & 84 &   2.4$\times$10$^{-5}$  & \cmark & \cmark & \cmark & \xmark & -- & -- & -- & -- & -- & -- \\

    &  & 2$_{2,0,4}$-2$_{1,1,3}$ & 215.501 &	84 &	6.9$\times$10$^{-6}$ & \cmark & \cmark & \cmark & \xmark & -- & -- & -- & -- & -- & -- \\

    && 2$_{2,0,4}$-2$_{1,1,4}$ &  215.503 	&	84 &	4.1$\times$10$^{-5}$ & \cmark & \cmark & \cmark & \xmark & -- & -- & -- & -- & -- & -- \\

    && 2$_{2,0,2}$-2$_{1,1,3}$ & 215.504 	& 	84 &	1.7$\times$10$^{-5}$ & \cmark & \cmark & \cmark & \xmark & -- & -- & -- & -- & -- & -- \\

    && 2$_{2,0,2}$-2$_{1,1,2}$ &  215.505	    & 	84 &	1.9$\times$10$^{-5}$ & \cmark & \cmark & \cmark & \xmark & -- & -- & -- & -- & -- & -- \\

    && 2$_{2,0,2}$-2$_{1,1,1}$ &  215.508	    & 	84 &	1.2$\times$10$^{-5}$ & \cmark & \cmark & \cmark & \xmark & -- & -- & -- & -- & -- & -- \\

    & & 2$_{2,0,3}$-2$_{1,1,3}$ & 215.512	    & 	84 &	2.8$\times$10$^{-5}$ & \cmark & \cmark & \cmark & \xmark & -- & -- & -- & -- & -- & -- \\

    && 2$_{2,0,3}$-2$_{1,1,2}$ & 215.513 	& 	84	&	1.1$\times$10$^{-5}$ & \cmark & \cmark & \cmark & \xmark & -- & -- & -- & -- & -- & -- \\

    && 2$_{2,0,3}$-2$_{1,1,4}$ & 215.513 	& 	84 &	9.1$\times$10$^{-6}$ & \cmark & \cmark & \cmark & \xmark & -- & -- & -- & -- & -- & -- \\
    &&&&&&&&&&&&&&&\\

    H$_2$$^{34}$S & 36504 & 2$_{2,0}$-2$_{1,1}$ & 214.377  & 84  &   4.7$\times$10$^{-5}$  & \cmark & \cmark & \cmark & \xmark & -- & -- & --  & -- & -- & --\\
    &&&&&&&&&&&&&&&\\

     HDS & 35502 & 7$_{3,4}$-7$_{3,5}$ & 234.046 & 417 &	6.6$\times$10$^{-6}$ & -- & -- & -- & -- & \xmark & \xmark & \xmark & \xmark & \xmark & \xmark\\

     && 12$_{5,7}$-12$_{5,8}$ & 234.528 & 1148 & 6.3$\times$10$^{-6}$ & -- & -- & -- & -- & \xmark & \xmark & \xmark& \xmark & \xmark & \xmark\\
    && 14$_{2,12}$-13$_{4,9}$ & 214.325 & 1277 &	5.0$\times$10$^{-8}$ & \xmark & \xmark & \xmark & \xmark  & -- & --& --& --& --& --\\

    &&&&&&&&&&&&&&\\

     HD$^{34}$S & 37503 & 7$_{3,4}$-7$_{3,5}$ & 232.964 & 416 &	6.5$\times$10$^{-6}$ & \xmark & \xmark & \xmark & \xmark & --& --& --& --& --& --\\
     & & 12$_{5,7}$-12$_{5,8}$ & 233.088 & 1145&	6.2$\times$10$^{-6}$ & \xmark & \xmark & \xmark & \xmark & --& \xmark & --& --& \xmark & \xmark\\
    &&&&&&&&&&&&&&\\

     OC$^{33}$S & 61503 & 18-17 & 216.147 & 99  &   2.9$\times$10$^{-5}$  & \cmark & \cmark & \cmark & \xmark &  \xmark & \xmark & \xmark & \cmark & \xmark & \xmark\\
    &&&&&&&&&&&&&&\\

     O$^{13}$CS & 61502 & 19-18 & 230.318  & 110 &   3.5$\times$10$^{-5}$  & \cmark & \cmark & \cmark & \xmark & -- & -- & -- & -- & -- & --\\
    &&&&&&&&&&&&&&\\

     $^{18}$OCS & 62506 & 19-18 & 216.753  & 104  &   2.9$\times$10$^{-5}$ & \cmark & \cmark & \cmark  & \xmark & \xmark & \xmark & \xmark & \xmark & \xmark & \xmark\\
    &&&&&&&&&&&&&&\\

     OCS, $v$=0 & 60503 & 19-18 & 231.061  & 111  &   3.6$\times$10$^{-5}$  & \cmark & \cmark & \cmark & \xmark &  \cmark & \cmark & \xmark & \cmark & \cmark & \xmark \\
    &&&&&&&&&&&&&&\\

    OCS, $v_2$=1$^{-}$ & 60504 & 19-18 & 231.342  & 860  &   3.5$\times$10$^{-5}$  & \cmark & \cmark & \cmark & \xmark &  \xmark & \xmark & \xmark & \xmark & \xmark & \xmark\\

    OCS, $v_2$=1$^{+}$ &  & 19-18 & 231.584  & 860  &   3.5$\times$10$^{-5}$  & \cmark & \cmark & \cmark & \xmark &  -- & -- & -- & -- & -- & --\\

    &&&&&&&&&&&&&&\\

    $^{18}$OC$^{34}$S & 64511 & 21-20 & 233.479  & 123  &   3.7$\times$10$^{-5}$  & \xmark & \xmark & \xmark & \xmark & \xmark & \xmark & \xmark & \xmark & \xmark & \xmark\\

    $^{18}$O$^{13}$CS & 63503 & 19-18 & 216.229  & 104  &   2.9$\times$10$^{-5}$  & -- & -- & -- & -- & \xmark & \xmark & \xmark & \xmark & \xmark & \xmark\\

    \hline
    \hline
    \end{tabular}
    \end{adjustbox}
    \caption{Lines of H$_2$S, OCS, and their isotopologs towards the targeted sources. The columns show 1) the CDMS entry, 2) the quantum numbers associated to the transition, 3) the rest frequency of the line, 4) the upper energy level of the transition, 5) the Einstein A coefficient of the transition, and the line evaluation done for different protostars given in columns 6-14.\\ Note: \cmark represents \textbf{detected lines}, \xmark  represents \textbf{undetected lines} and -- represents \textbf{unobserved lines} (i.e., not covered in the spectral windows observed for that source).}
    \label{line_emissions}
\end{table}
\end{landscape}

\section{IRAS 16293-2422 B}\label{a_IRAS16293B}

The synthetic spectral fitting of the detected species towards IRAS 16293 B is presented thoroughly in \Cref{IRAS16293B}. An extensive study of the effect of source size on column density and H$_2$S/OCS is also presented in \Cref{IRAS16293B}. The excitation temperature and FHWM assumed for fitting synthetic spectra to the detected lines towards IRAS 16293 B are 125 K and 1 km s$^{-1}$, respectively, based on the ALMA-PILS observations of IRAS 16293 \citep{Drozdovskaya2018}. 

\subsection{Detected lines in IRAS 16293-2422 B}

\begin{figure}[H]
\centering
\begin{minipage}[b]{6.2in}%

    \subfigure{\includegraphics[width=2.0in]{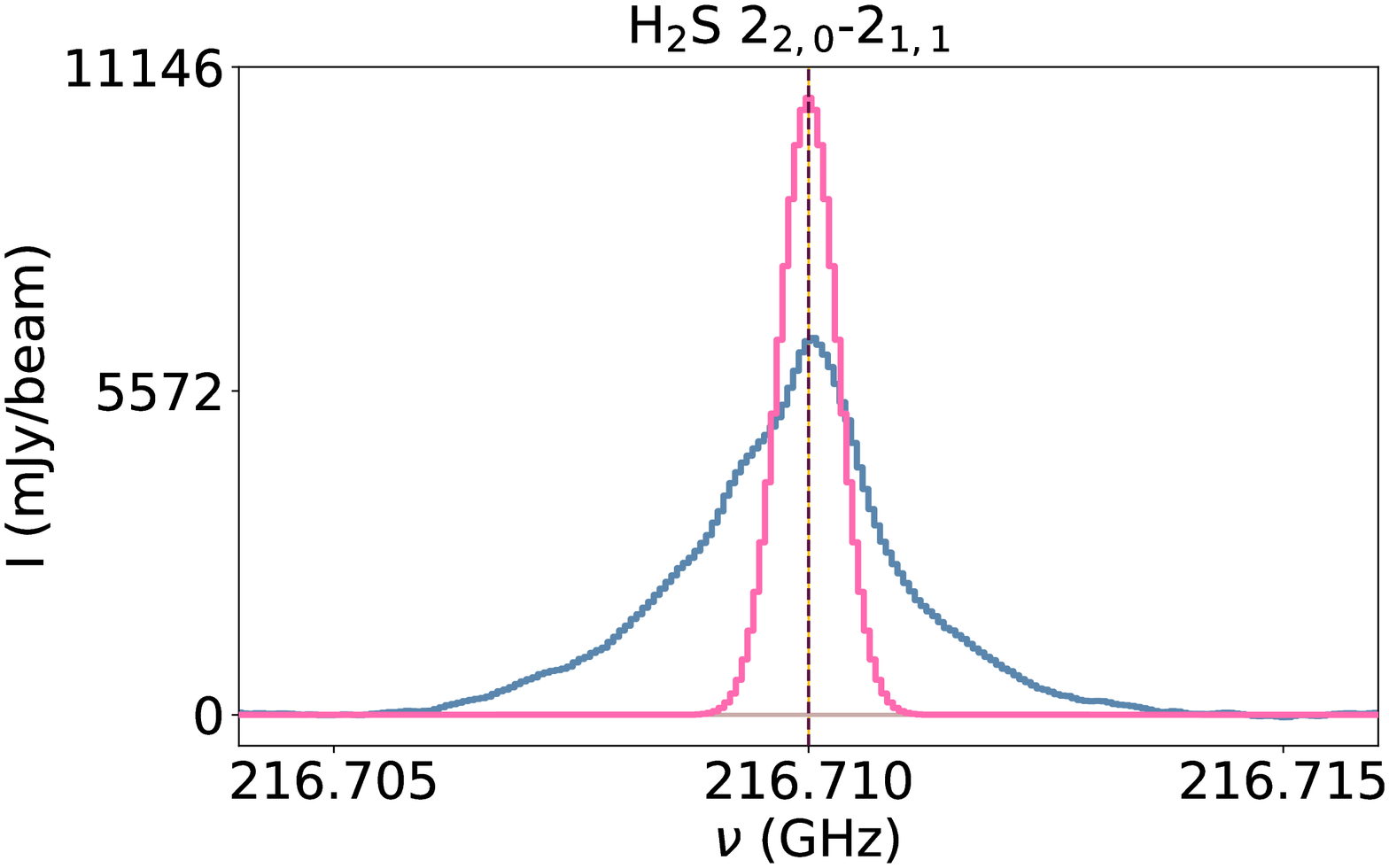}}
    \subfigure{\includegraphics[width=2.0in]{images/detected/IRAS16293B/H2S-34_214376_9236.eps}}\\
    \subfigure{\includegraphics[width=6.0in]{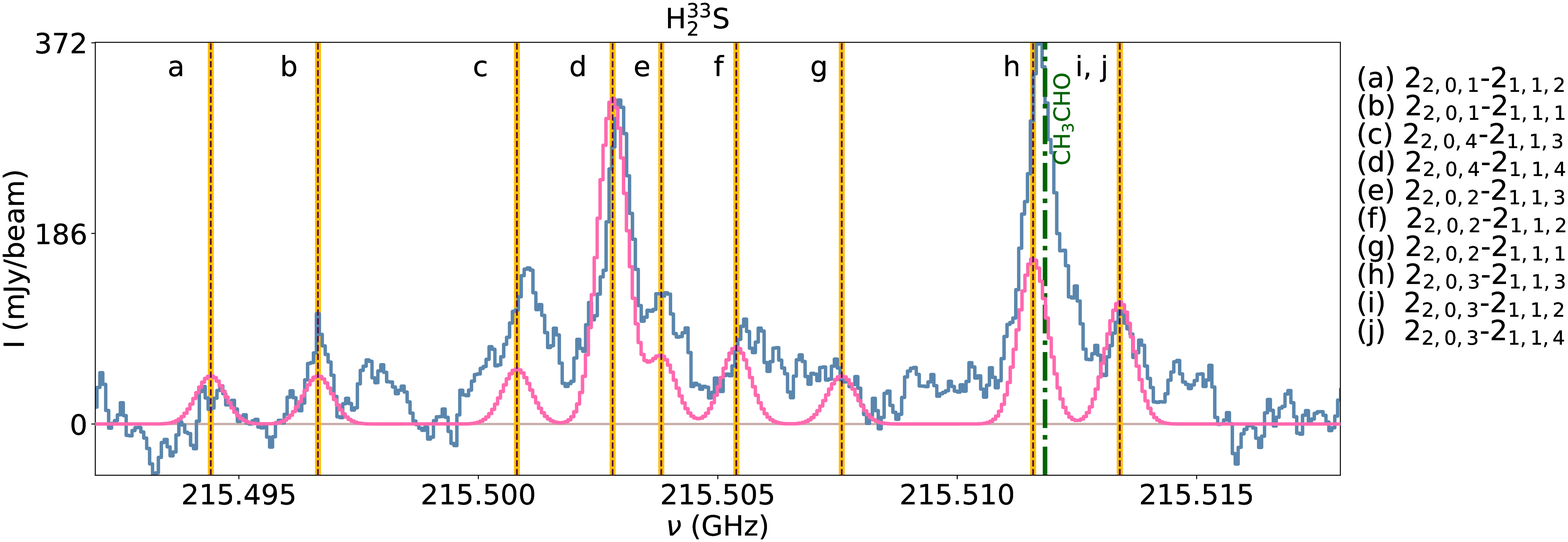}}\\
    \subfigure{\includegraphics[width=2.0in]{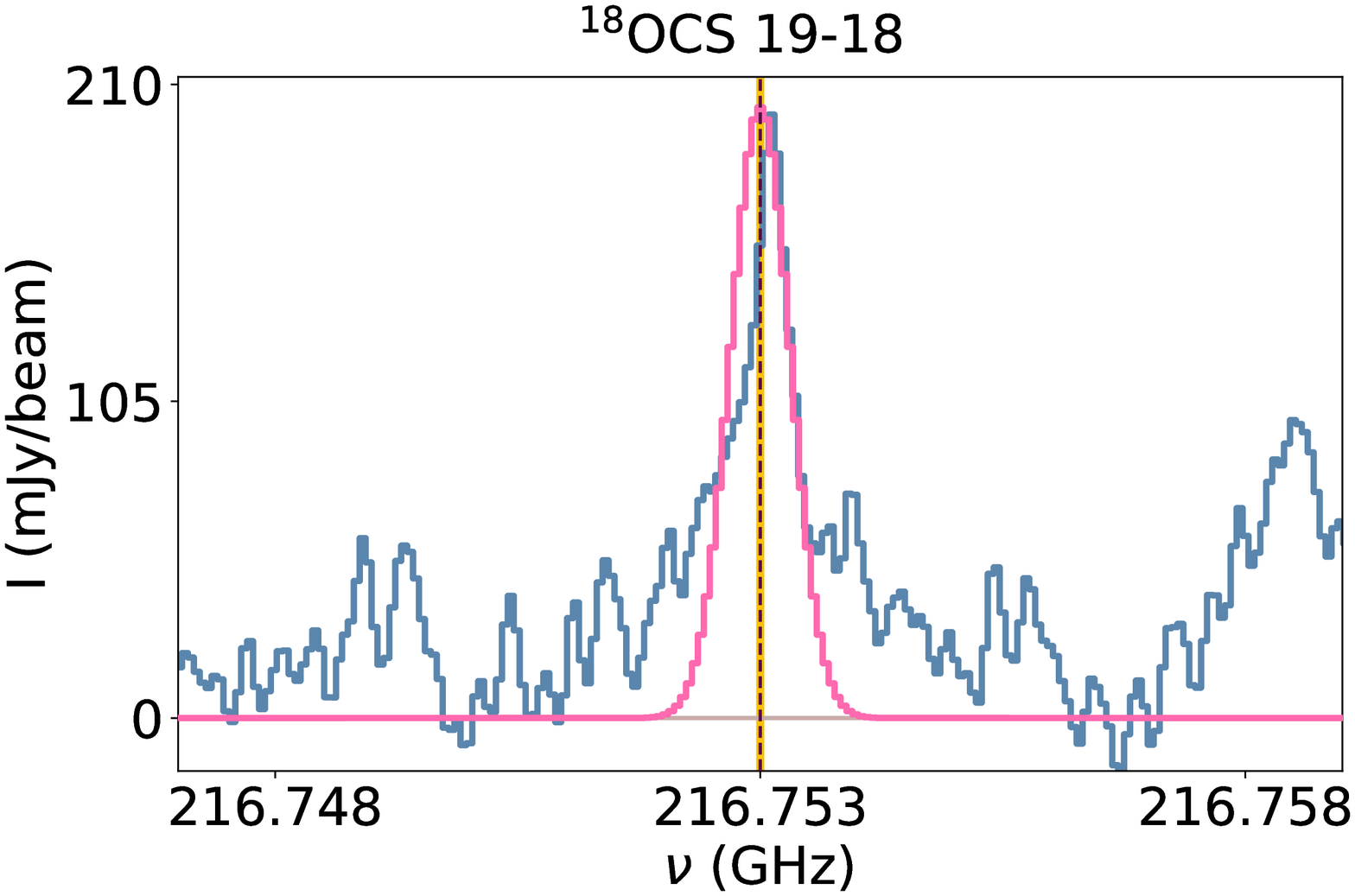}}
    \subfigure{\includegraphics[width=2.0in]{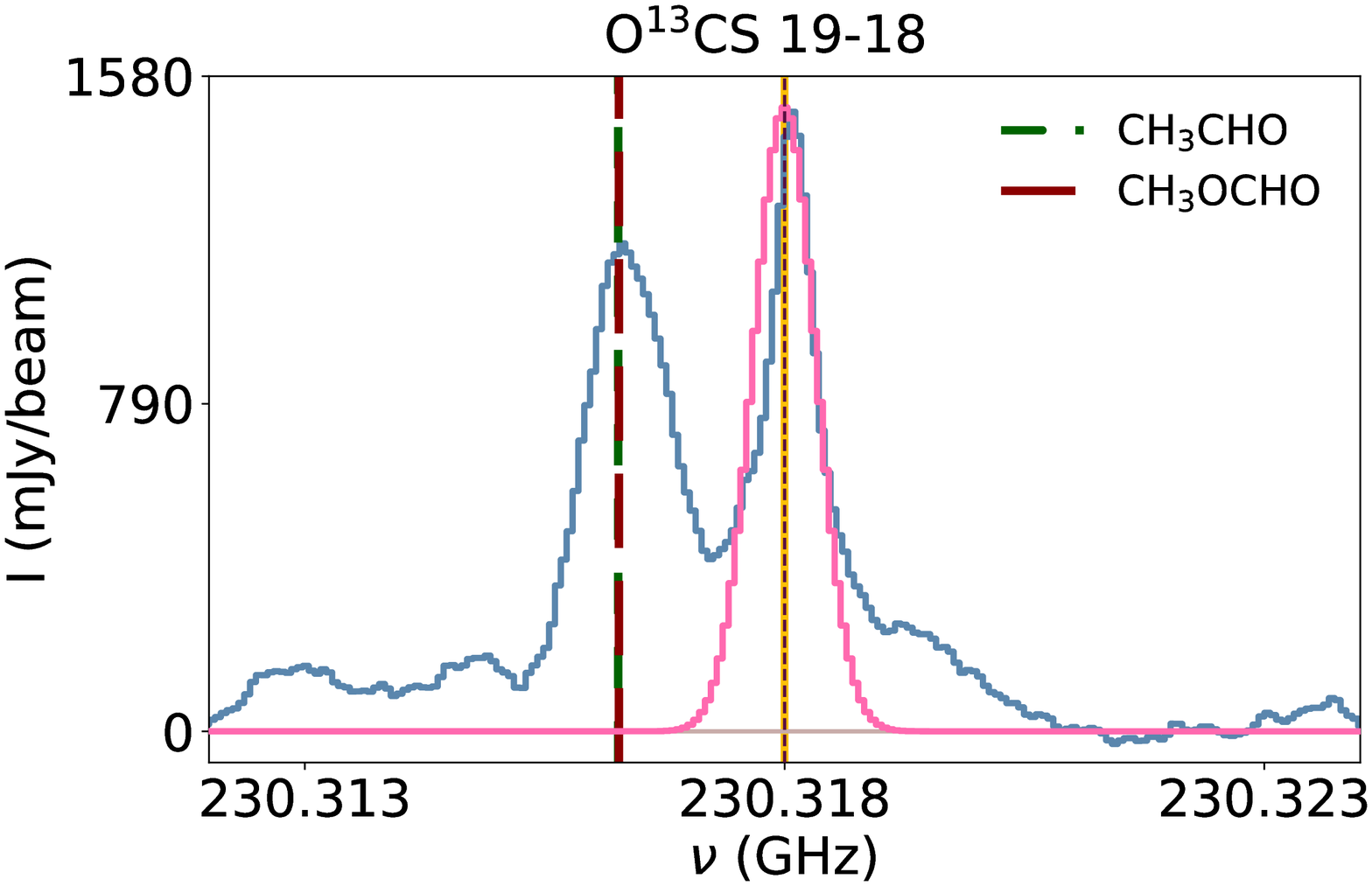}}
    \subfigure{\includegraphics[width=2.0in]{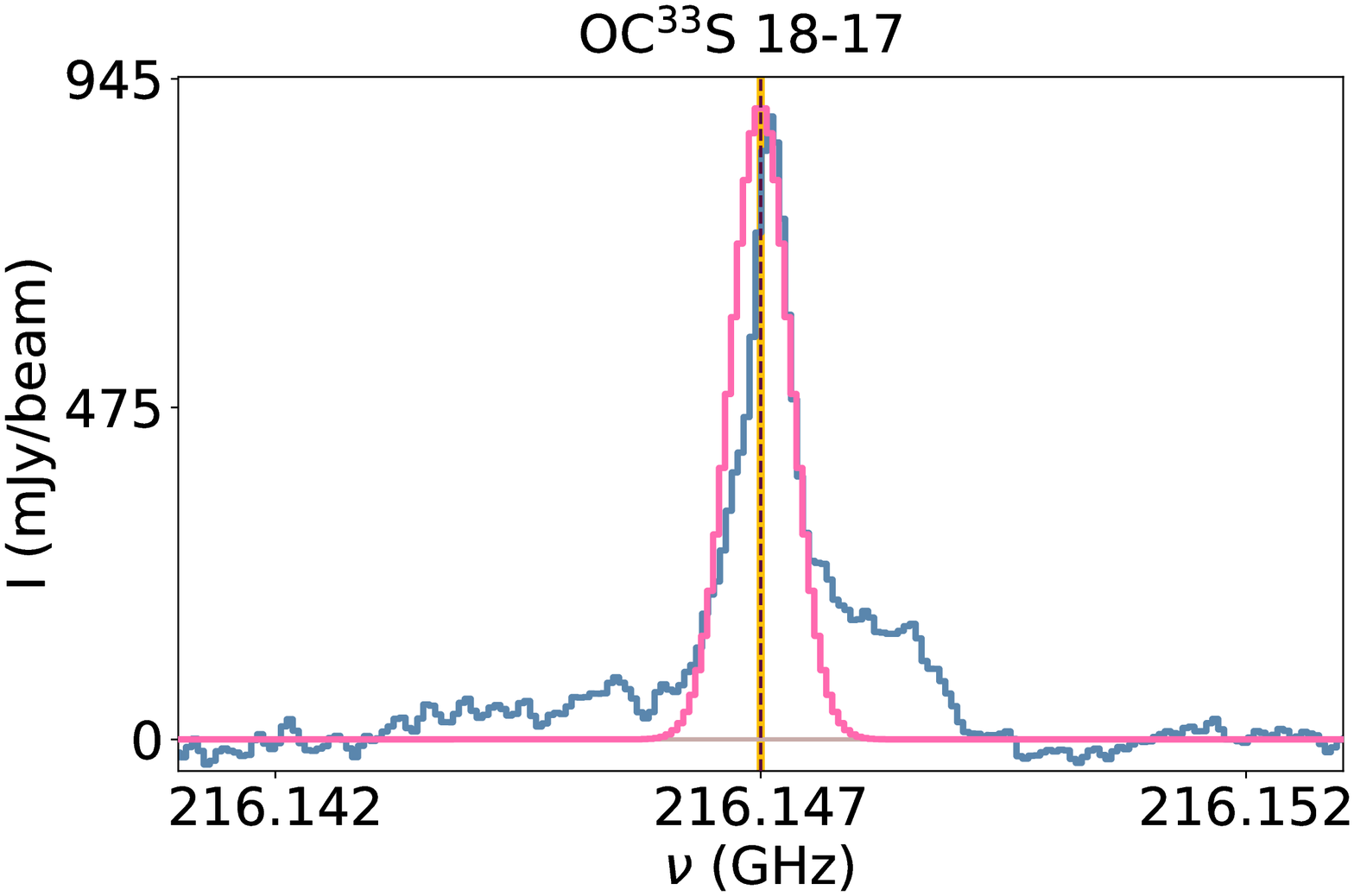}}\\ 
    \subfigure{\includegraphics[width=2.0in]{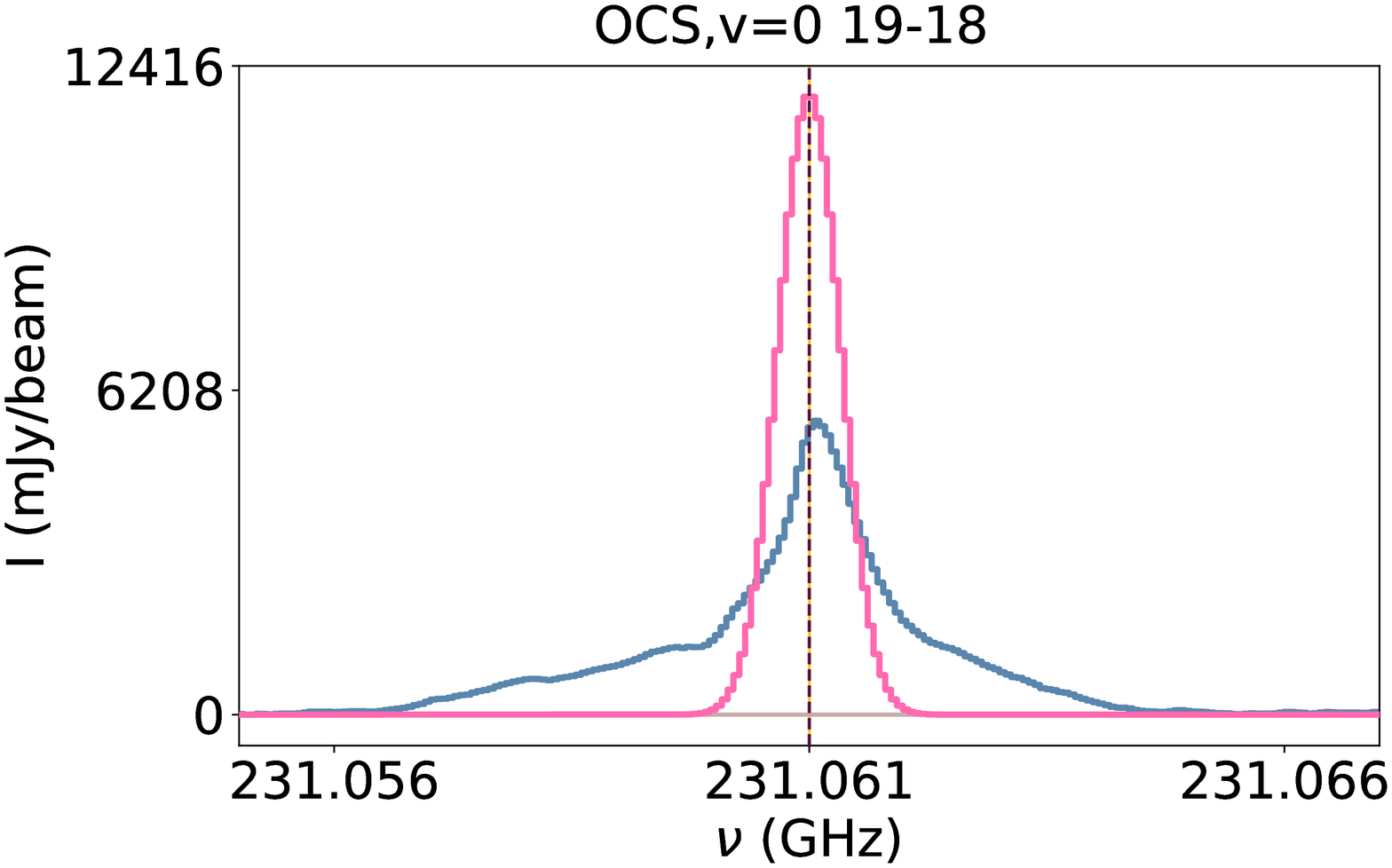}}
    \subfigure{\includegraphics[width=2.0in]{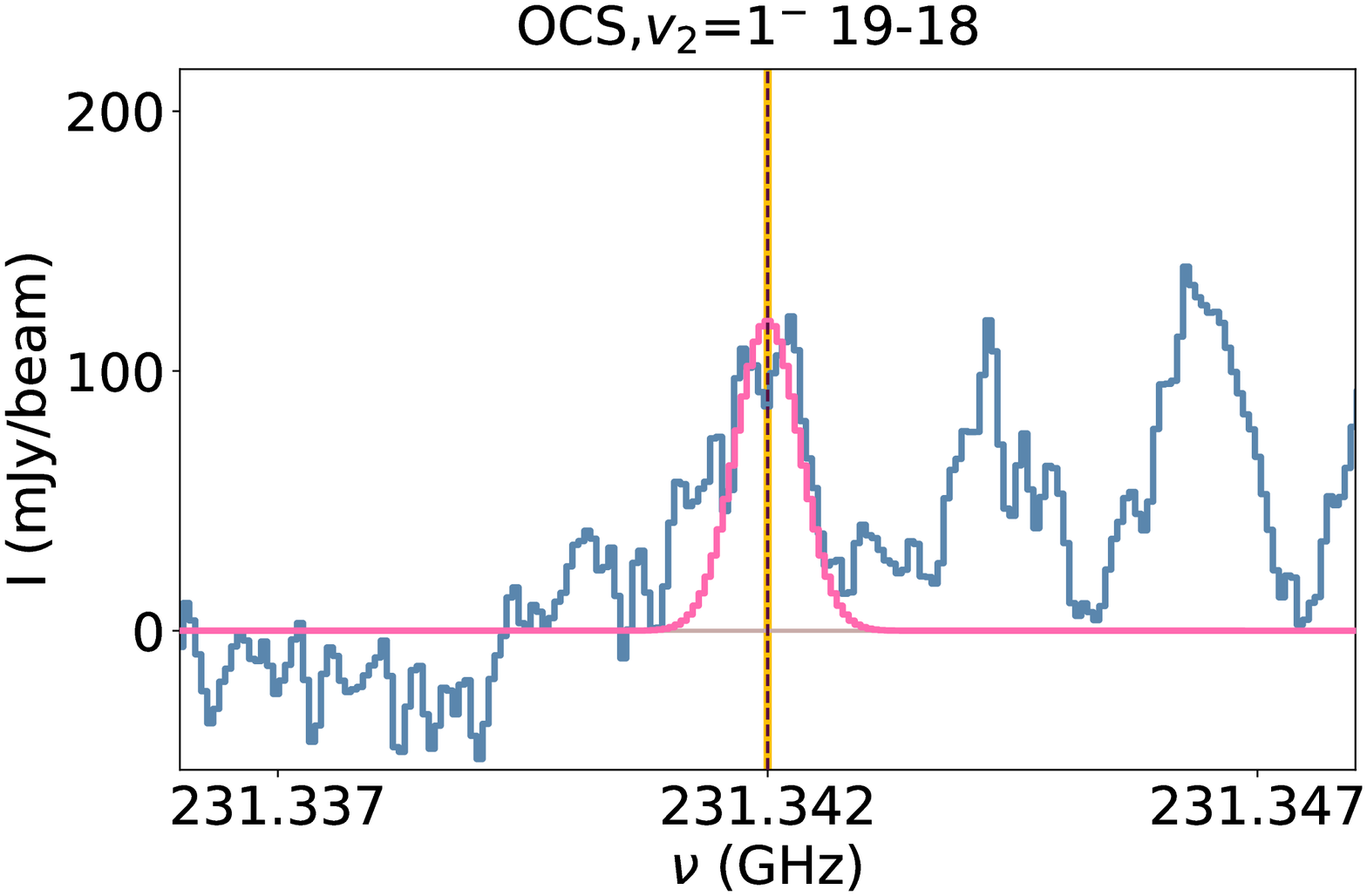}}
    \subfigure{\includegraphics[width=2.0in]{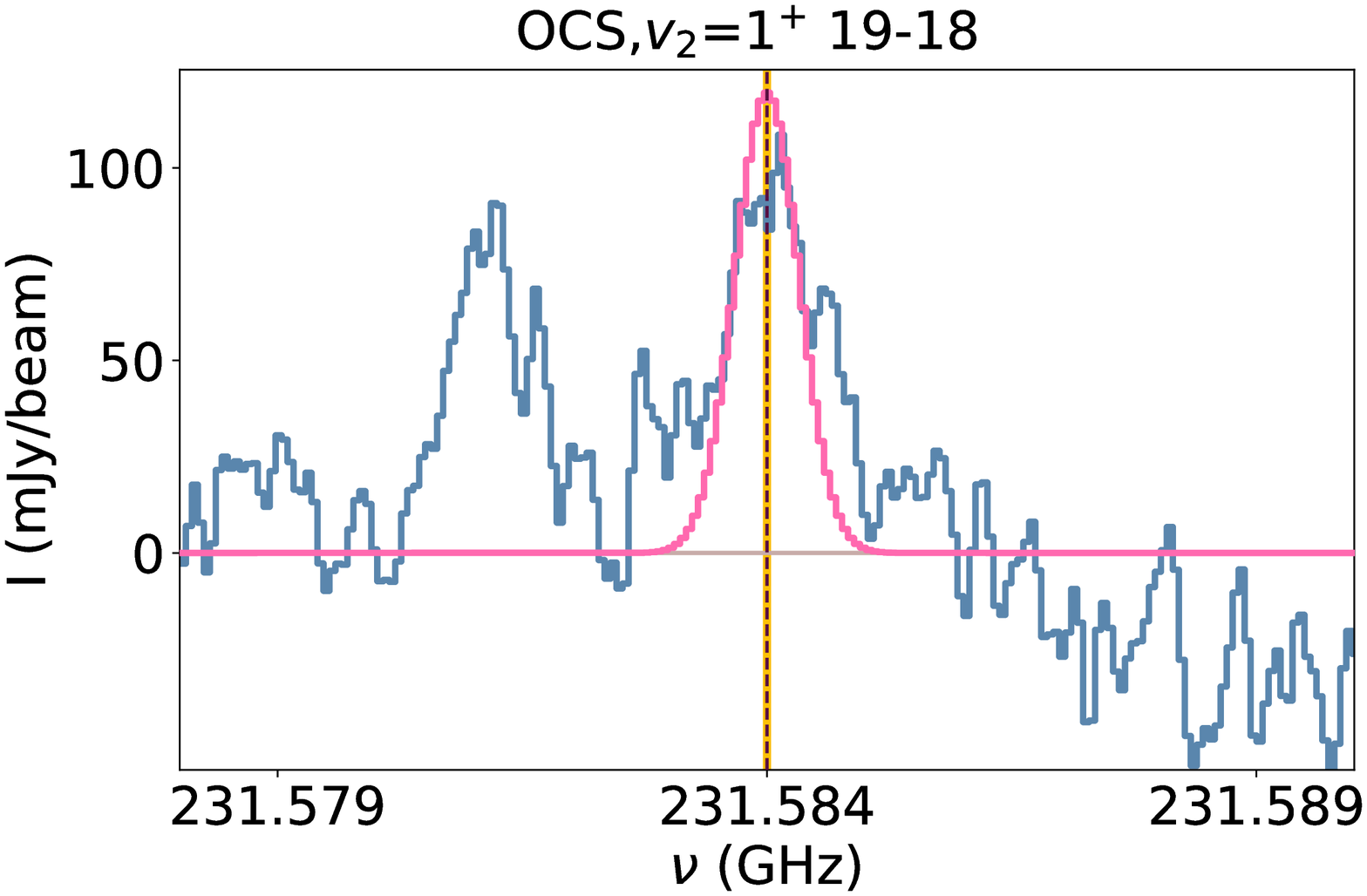}}\\ 
\end{minipage}
\caption{Observed spectra (in blue), rest frequency of the detected line (brown dashed line), spectroscopic uncertainty on the rest frequency of the detected line (yellow shaded region), blending species (green dash-dotted line and red dashed line), and fitted synthetic spectra (in pink) plotted for the sulfur-bearing species detected towards IRAS 16293-2422 B. $\overline{N\text{(H}_2\text{S)}}$ and $\overline{N\text{(OCS)}}$ are used for the synthetic spectra of the optically thick lines. For the displayed fits, a source size of 2$\arcsec$ is assumed.}
\label{detected_IRAS16293B}
\end{figure}

\subsection{Undetected lines in IRAS 16293-2422 B}

\begin{figure}[H]
\centering
\begin{minipage}[b]{6.2in}%

    \subfigure{\includegraphics[width=2.0in]{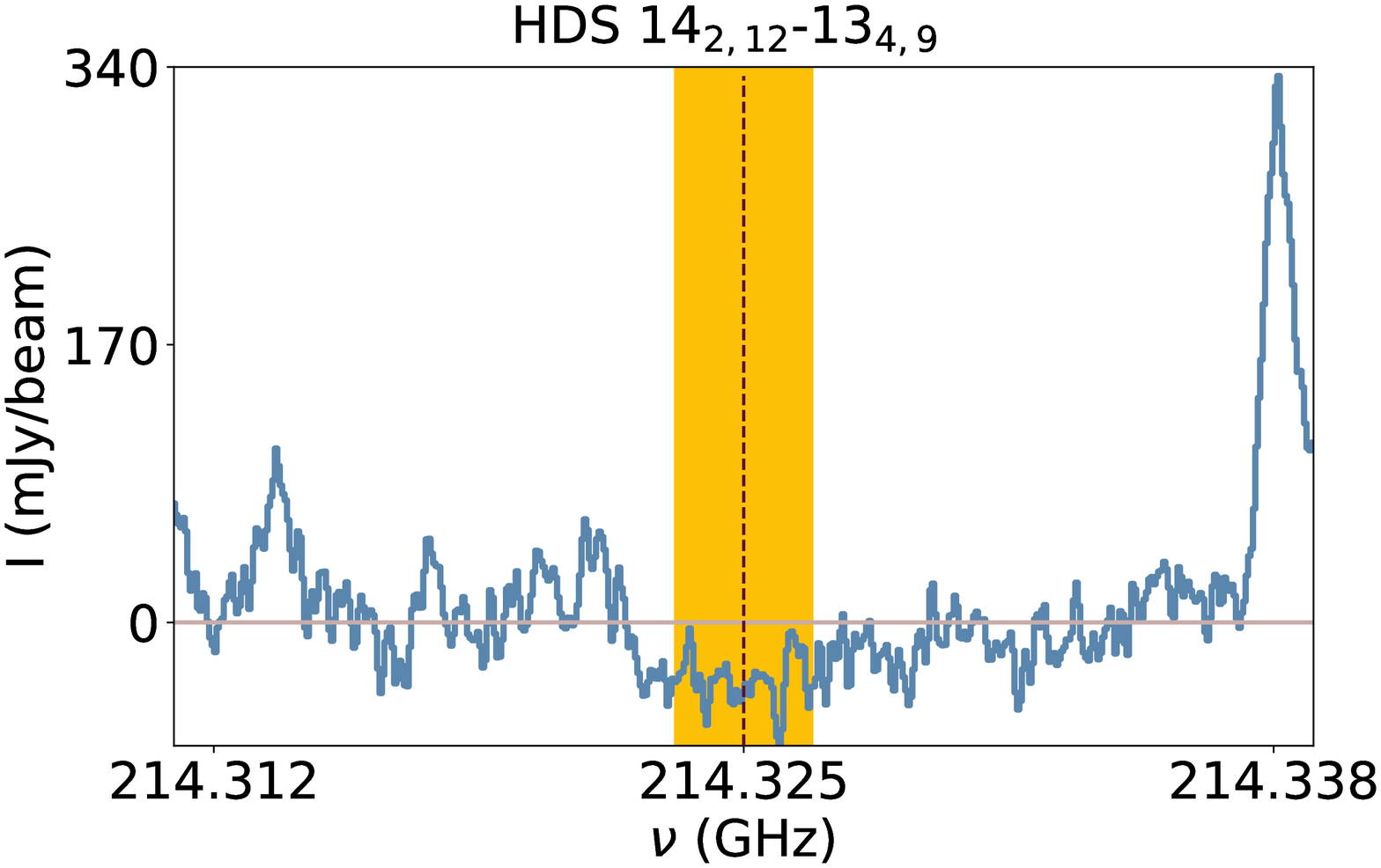}}
    \subfigure{\includegraphics[width=2.0in]{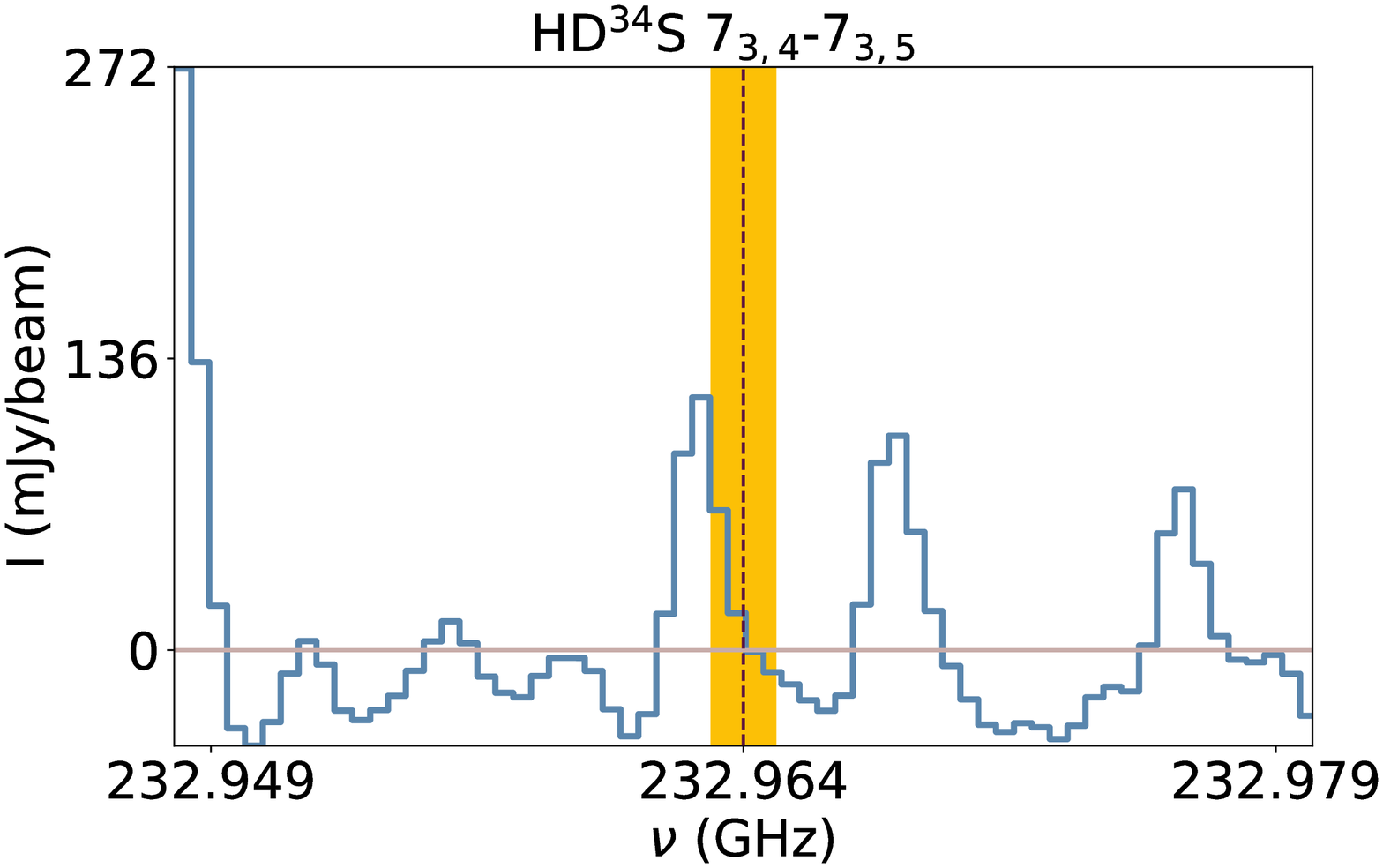}}
    \subfigure{\includegraphics[width=2.0in]{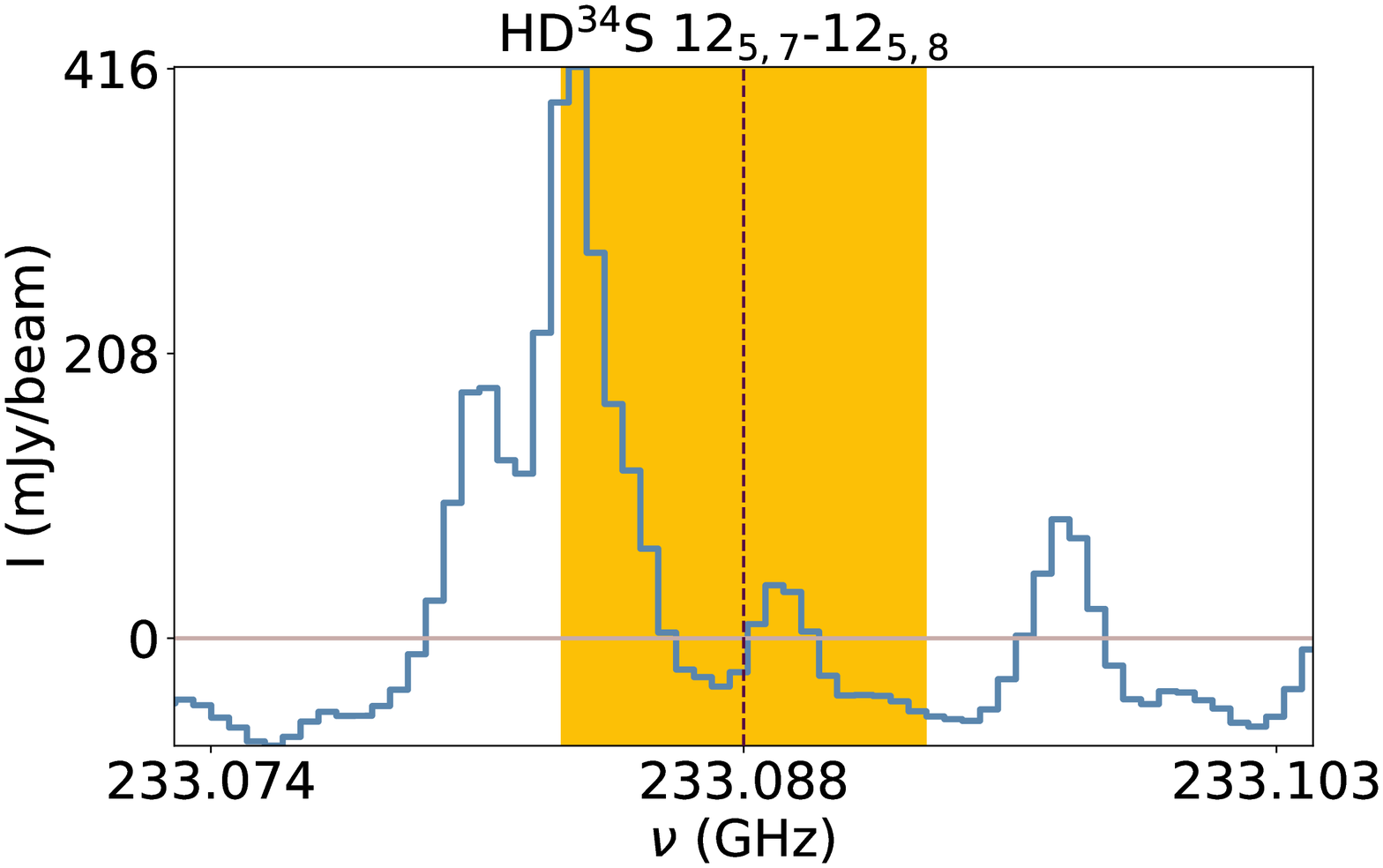}}\\
    \subfigure{\includegraphics[width=2.0in]{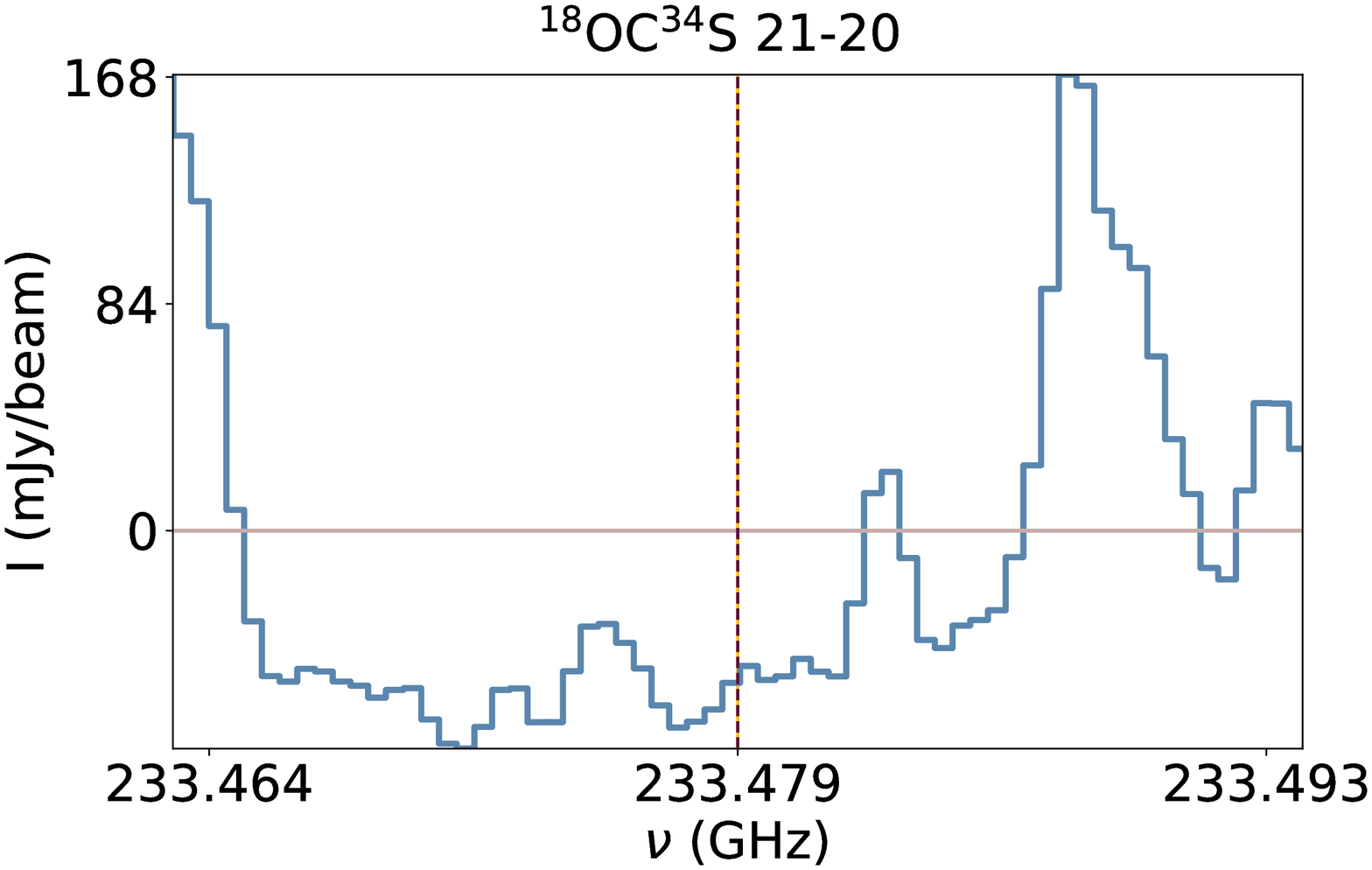}}

\end{minipage}
\caption{Observed spectra (in blue), rest frequency of the undetected line (brown dashed line), and spectroscopic uncertainty on the rest frequency of the undetected line (yellow shaded region) plotted for the sulfur-bearing species undetected towards IRAS 16293-2422 B.}
\label{undetected_IRAS16293B}
\end{figure}

\section{IRAS 16293-2422 A}
\label{a_IRAS16293A}

\begin{figure}[H]
\centering
\begin{minipage}{6.2in}%
    \subfigure{\includegraphics[width=2.0in]{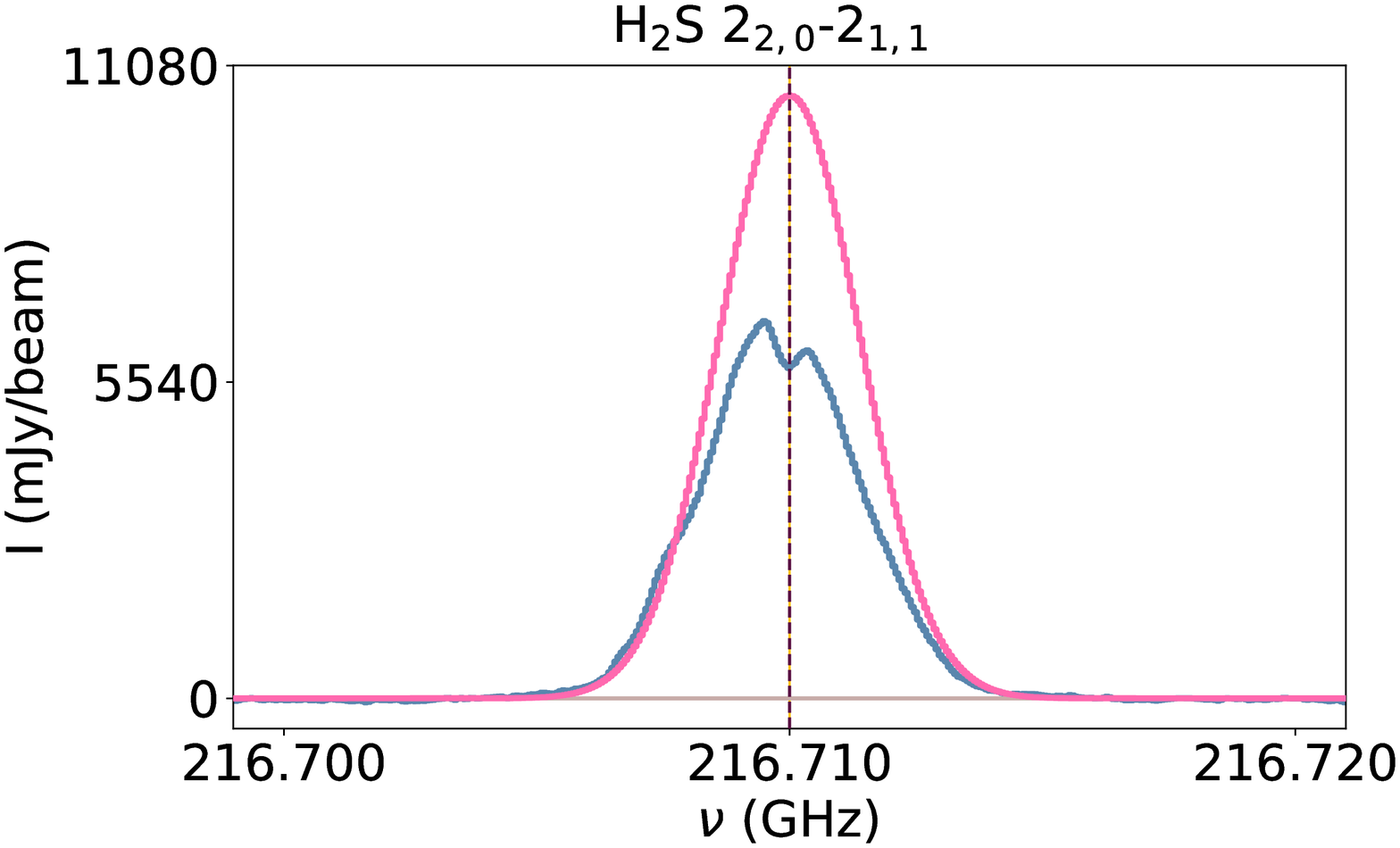}}
    \subfigure{\includegraphics[width=2.0in]{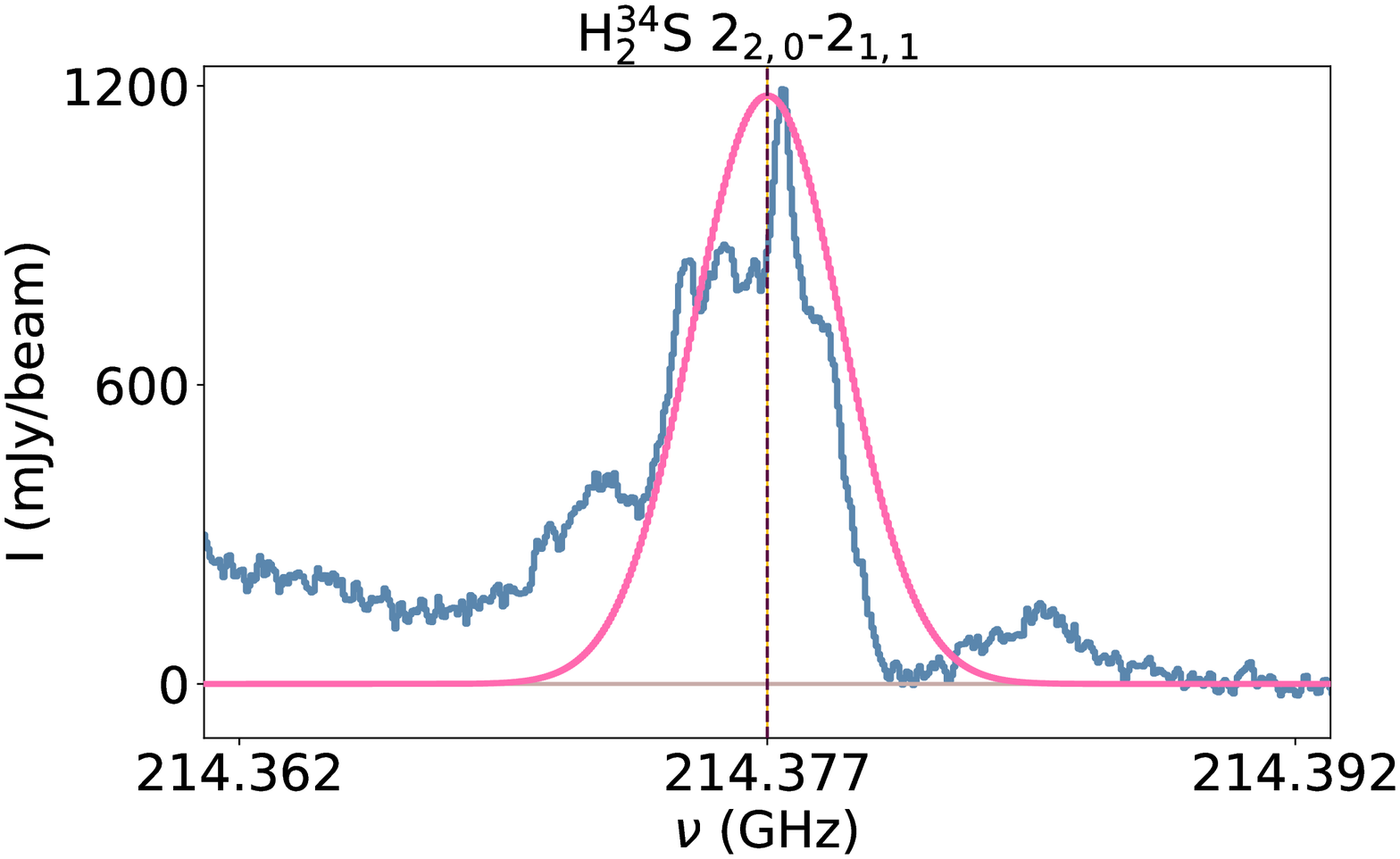}}\\
    \subfigure{\includegraphics[width=6.0in]{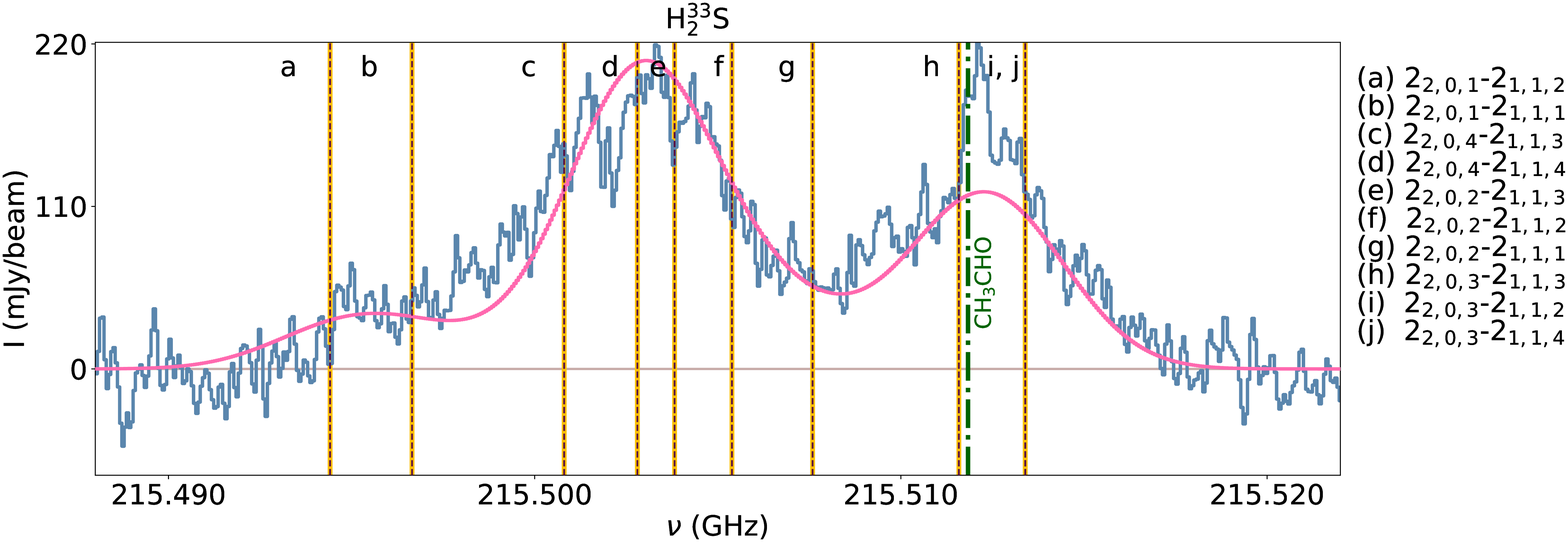}}\\
    \subfigure{\includegraphics[width=2.0in]{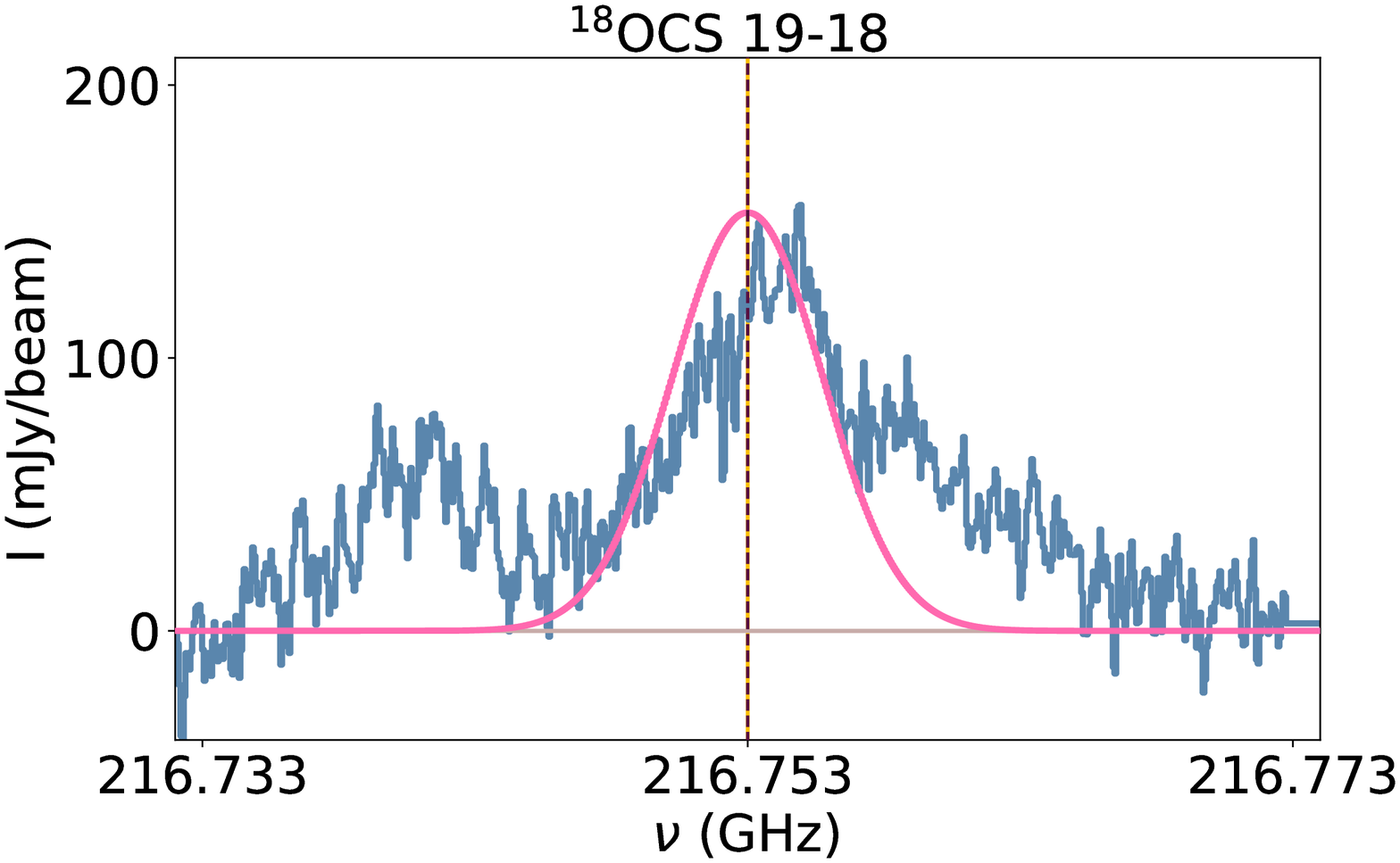}}
    \subfigure{\includegraphics[width=2.0in]{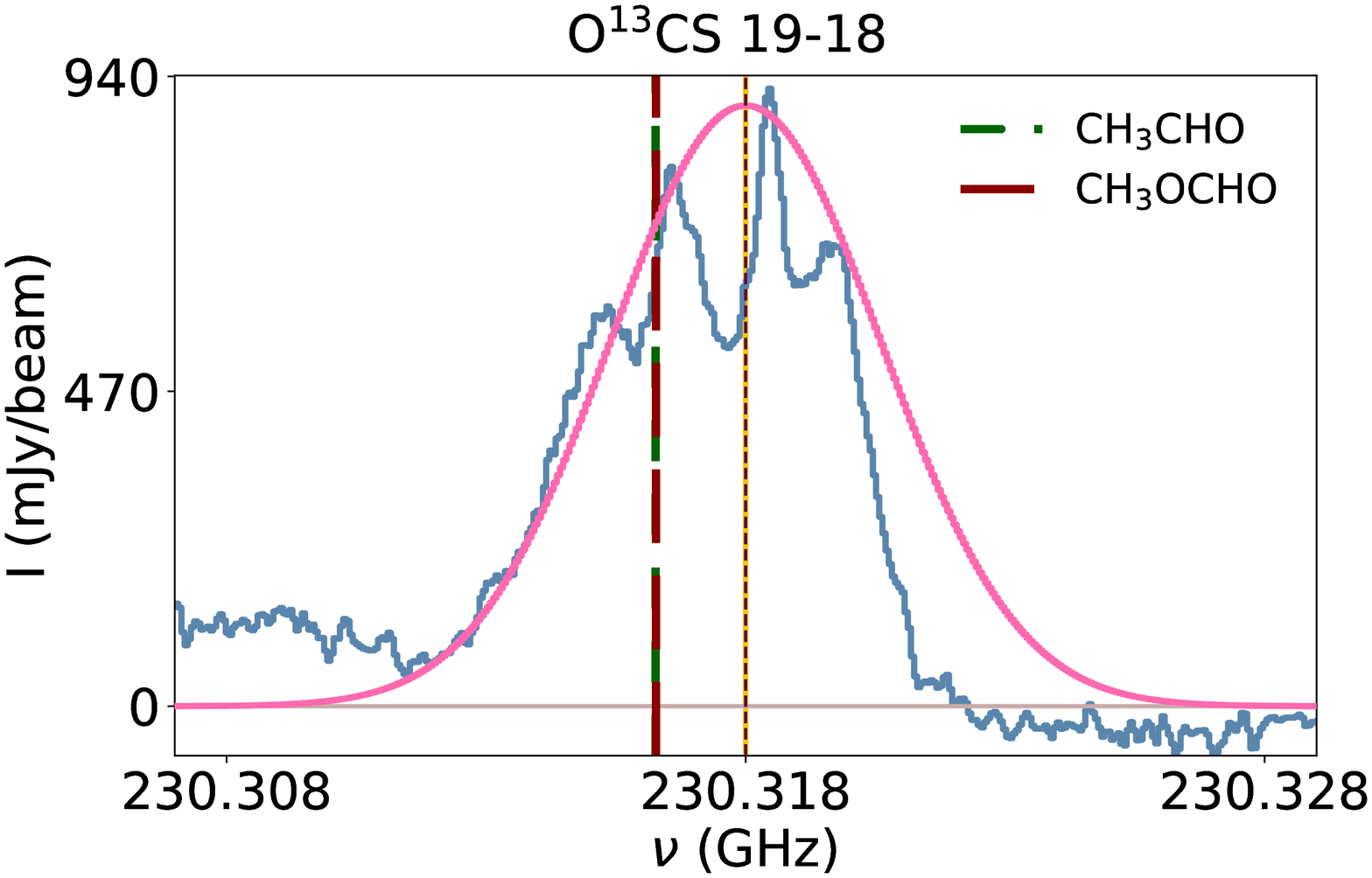}}
    \subfigure{\includegraphics[width=2.0in]{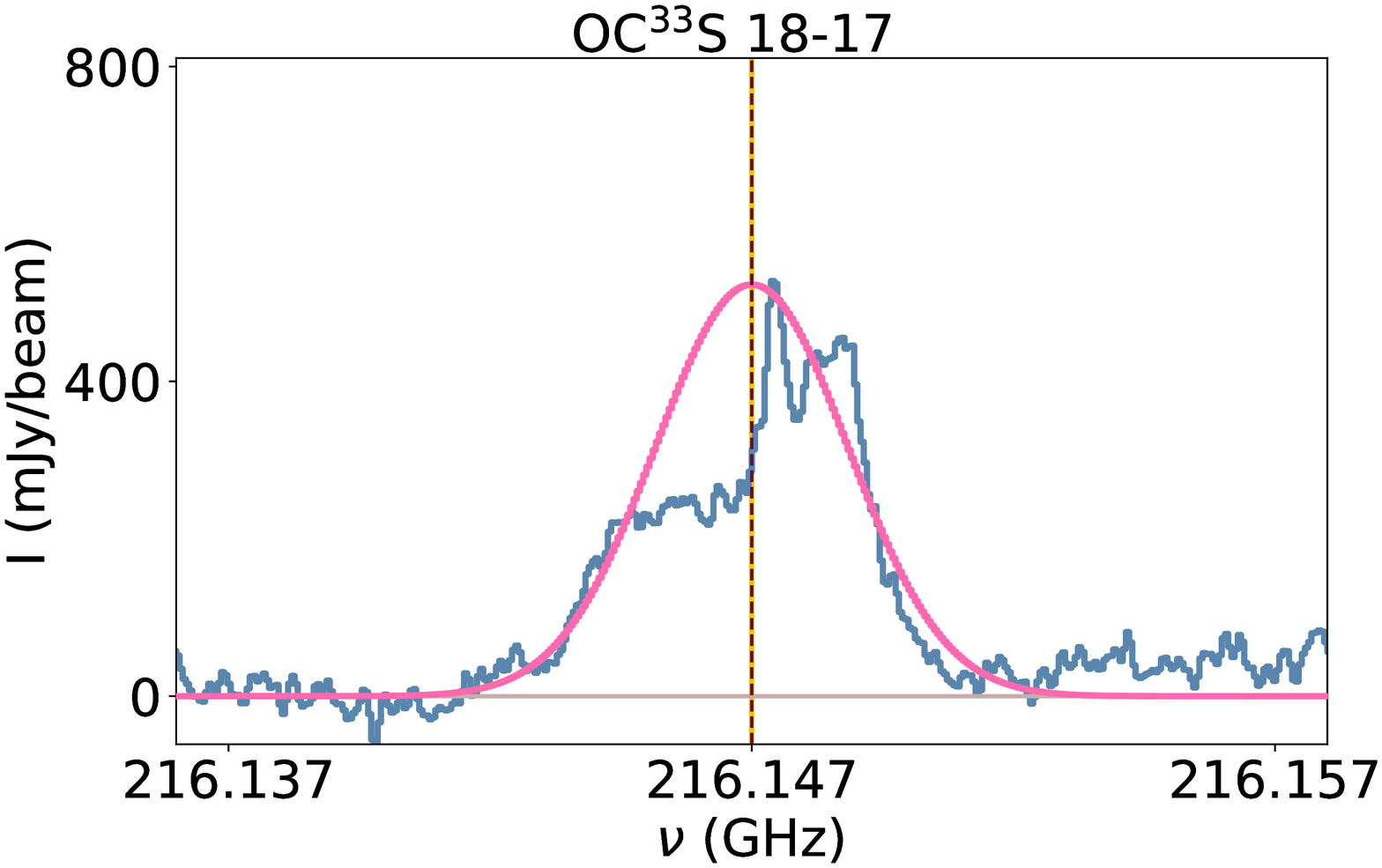}}\\ 
    \subfigure{\includegraphics[width=2.0in]{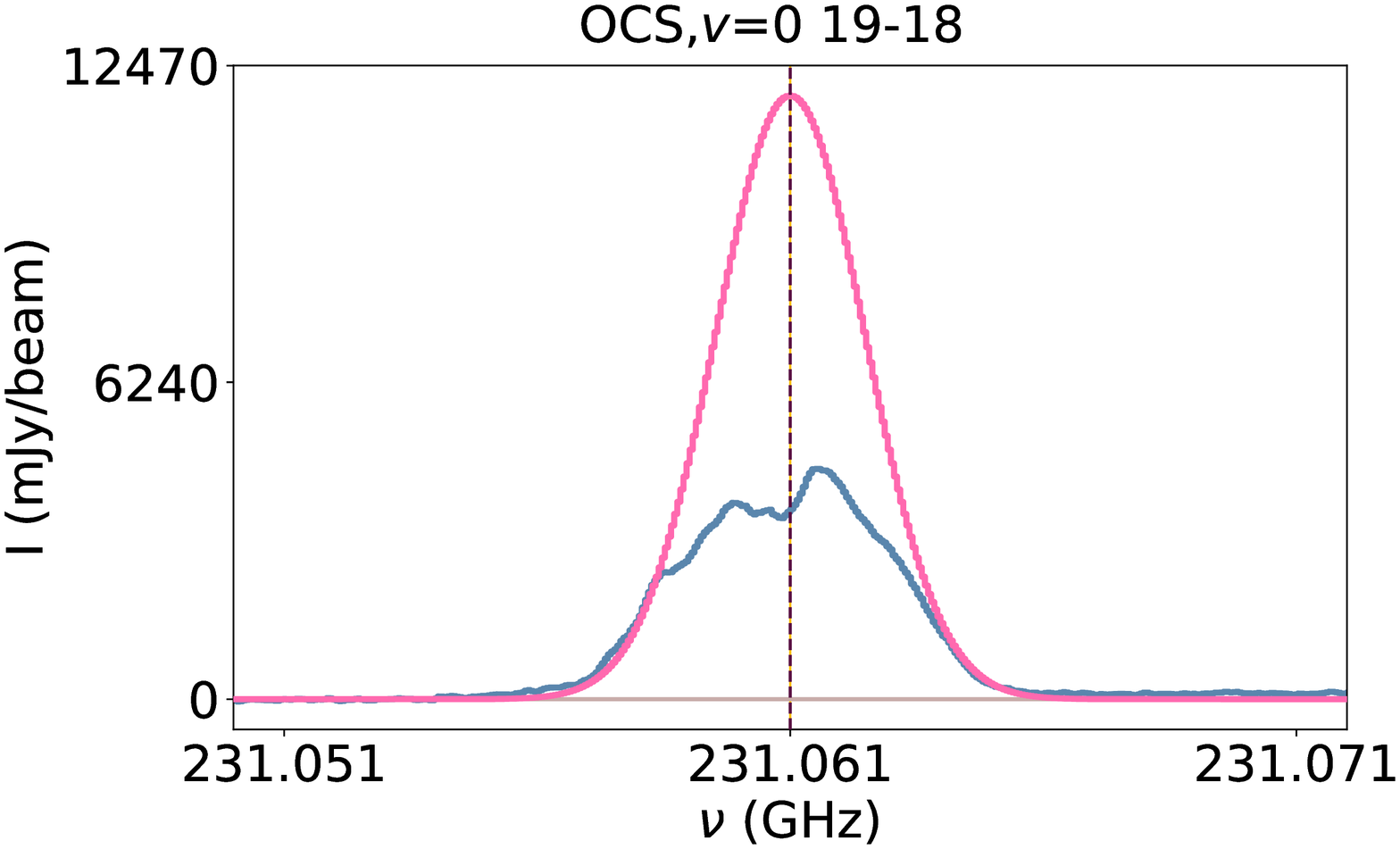}}
    \subfigure{\includegraphics[width=2.0in]{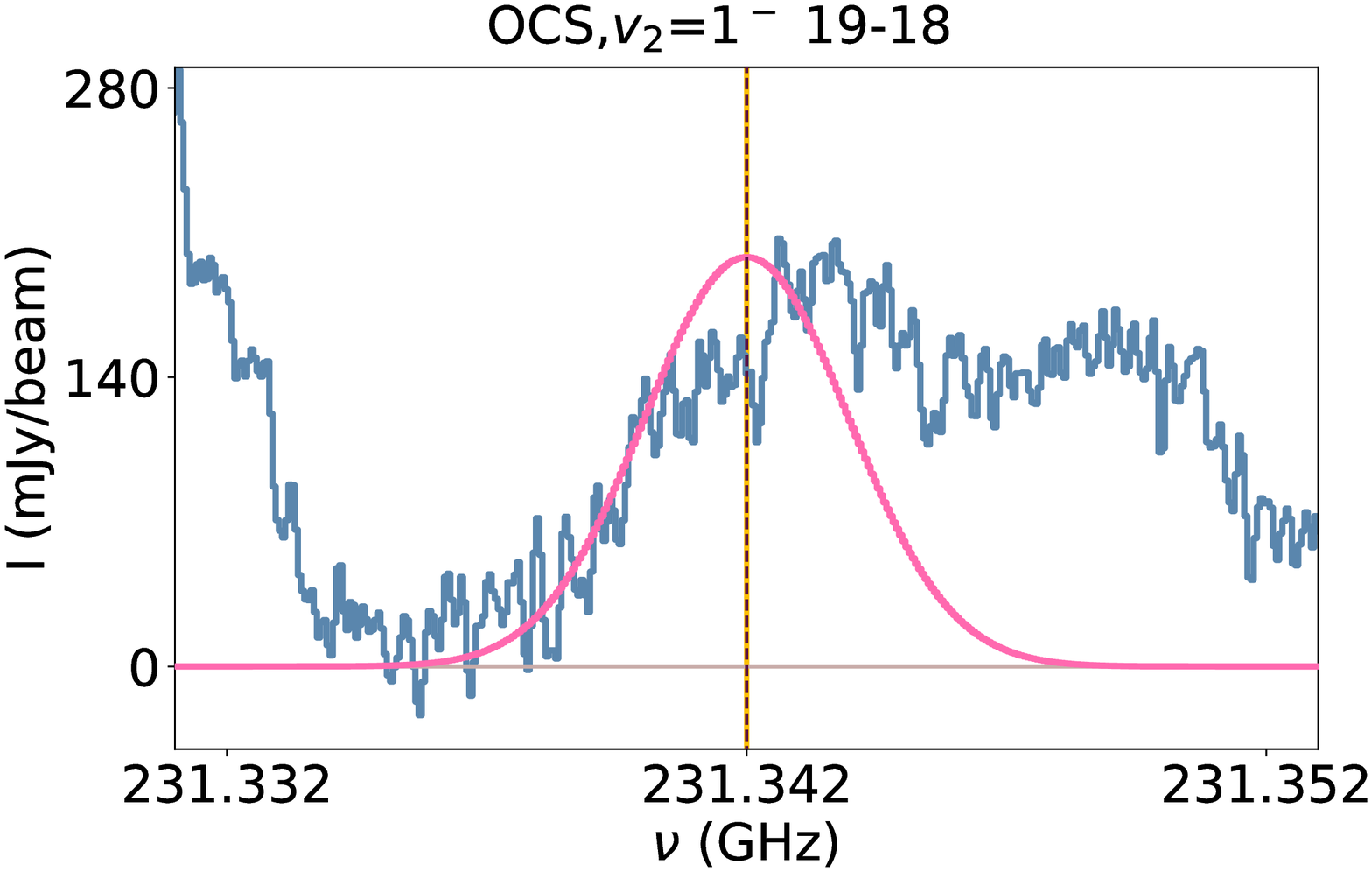}}
    \subfigure{\includegraphics[width=2.0in]{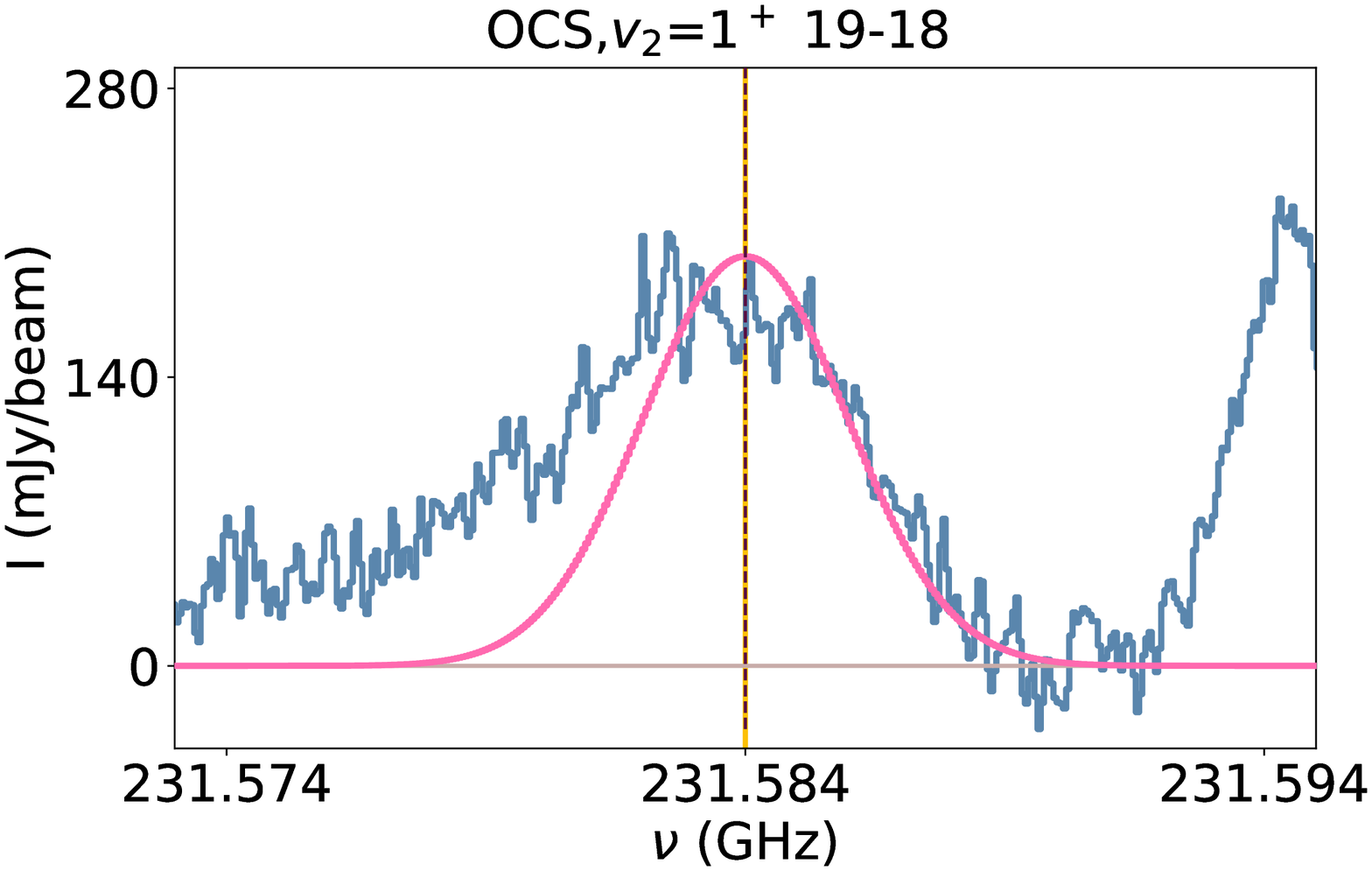}}\\ 
\end{minipage}
\caption{Observed spectra (in blue), rest frequency of the detected line (brown dashed line), spectroscopic uncertainty on the rest frequency of the detected line (yellow shaded region), blending species (green dash-dotted line and red dashed line), and fitted synthetic spectra (in pink) plotted for the sulfur-bearing species detected towards  IRAS 16293-2422 A. $\overline{N\text{(H}_2\text{S)}}$ and $\overline{N\text{(OCS)}}$ are for the synthetic spectra of the optically thick lines.}
\label{detected_IRAS16293A}
\end{figure}

\begin{figure}[H]
\centering
\begin{minipage}{6.2in}%
    \subfigure{\includegraphics[width=2.0in]{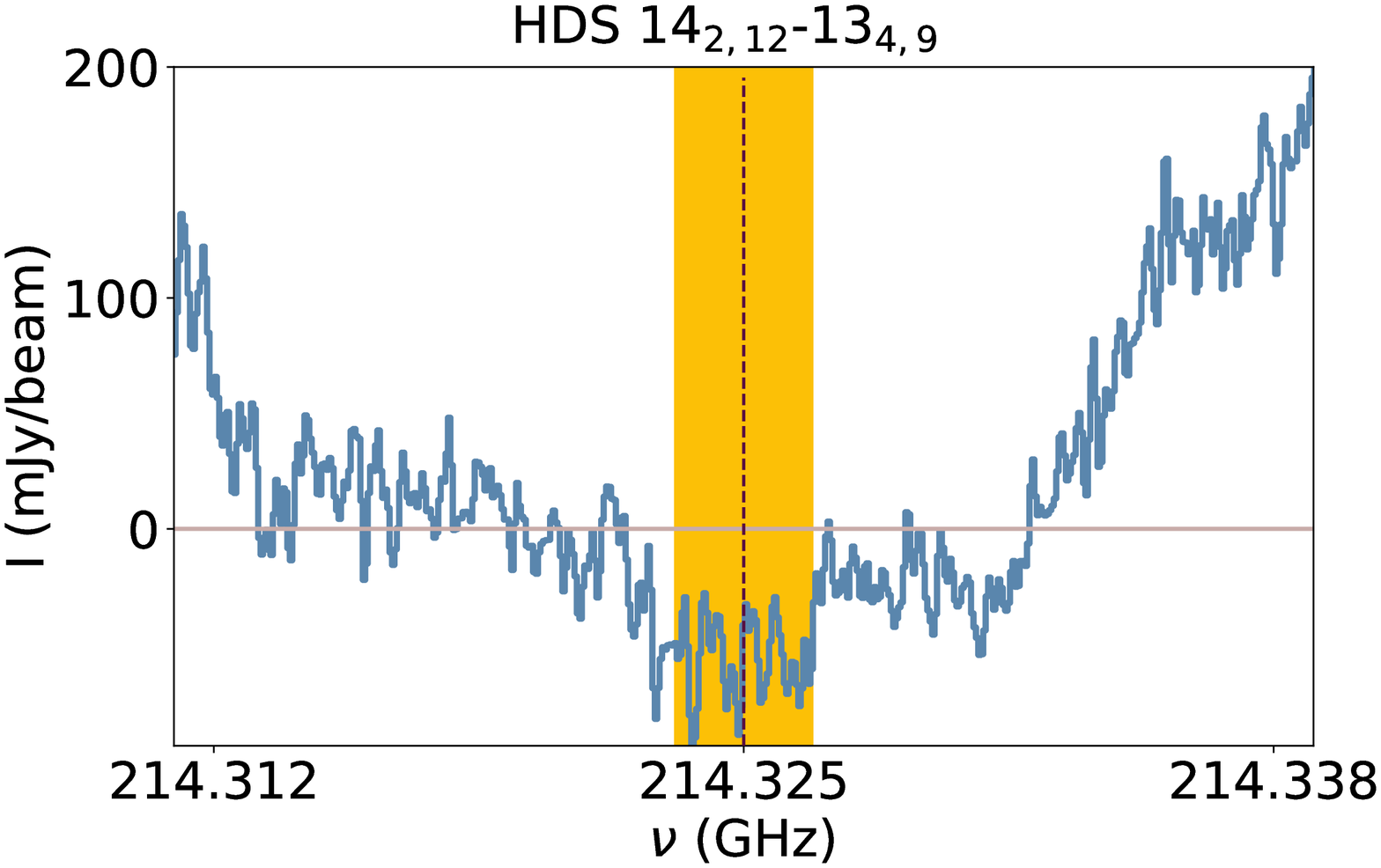}}
    \subfigure{\includegraphics[width=2.0in]{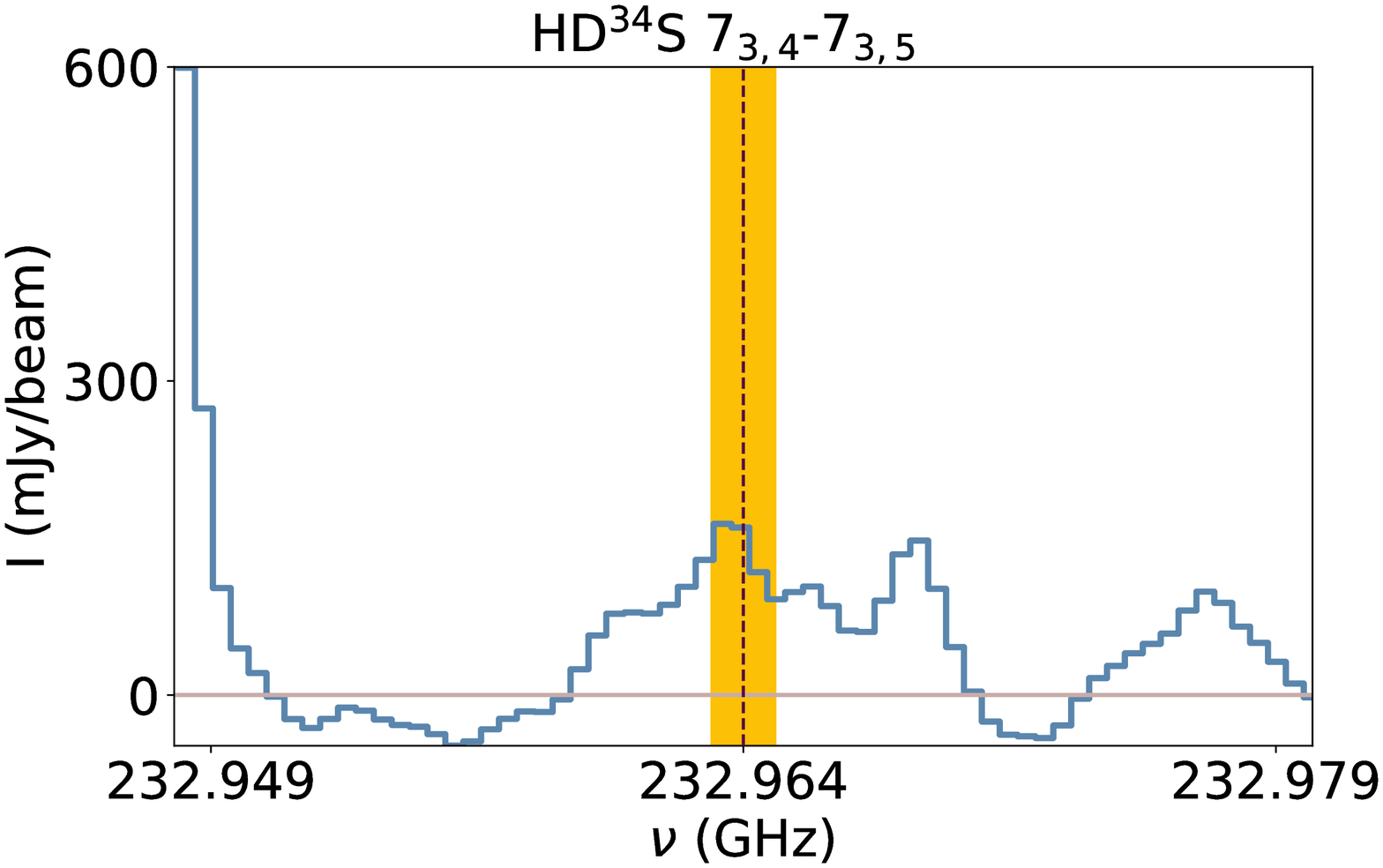}}
    \subfigure{\includegraphics[width=2.0in]{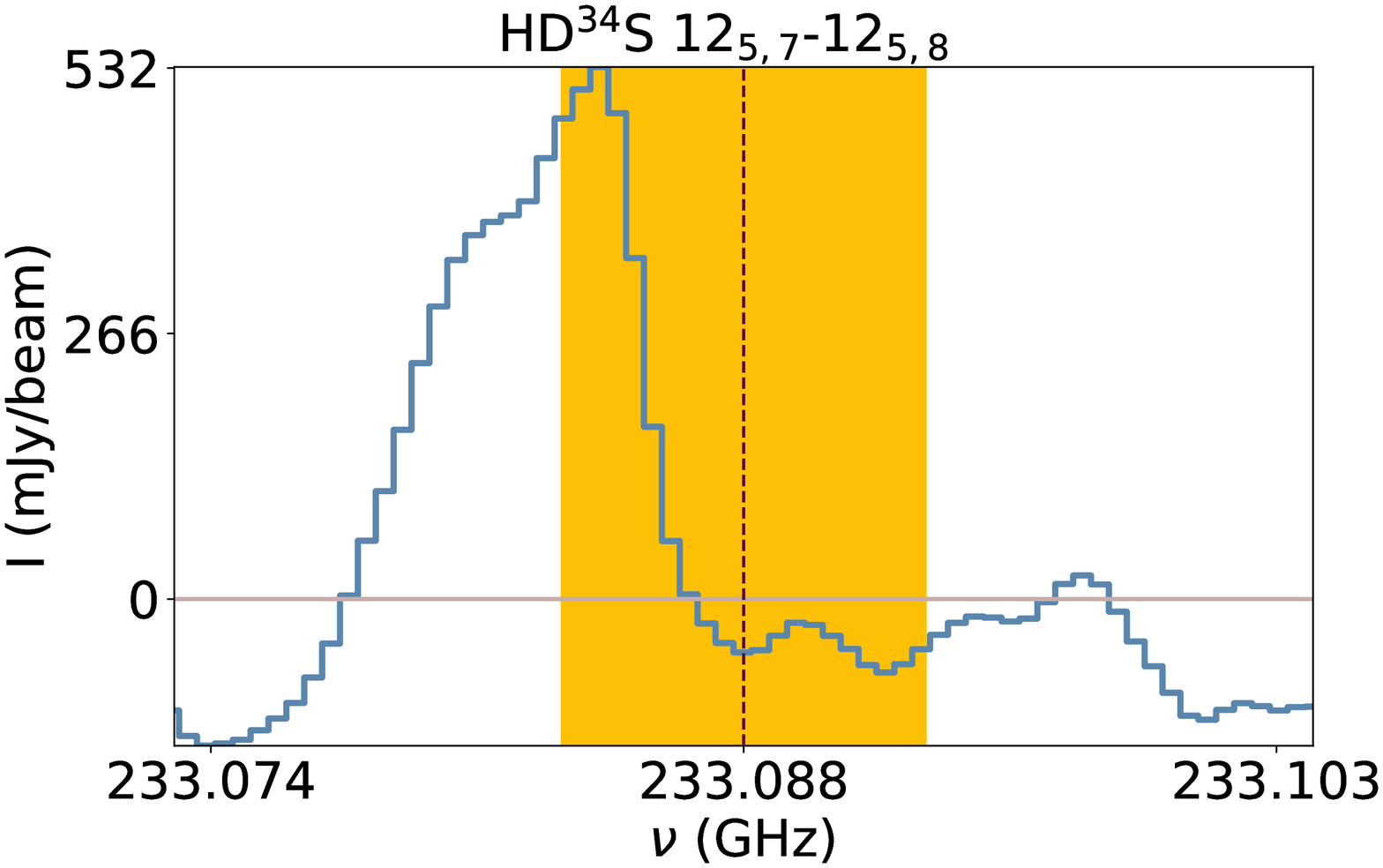}}\\
    \subfigure{\includegraphics[width=2.0in]{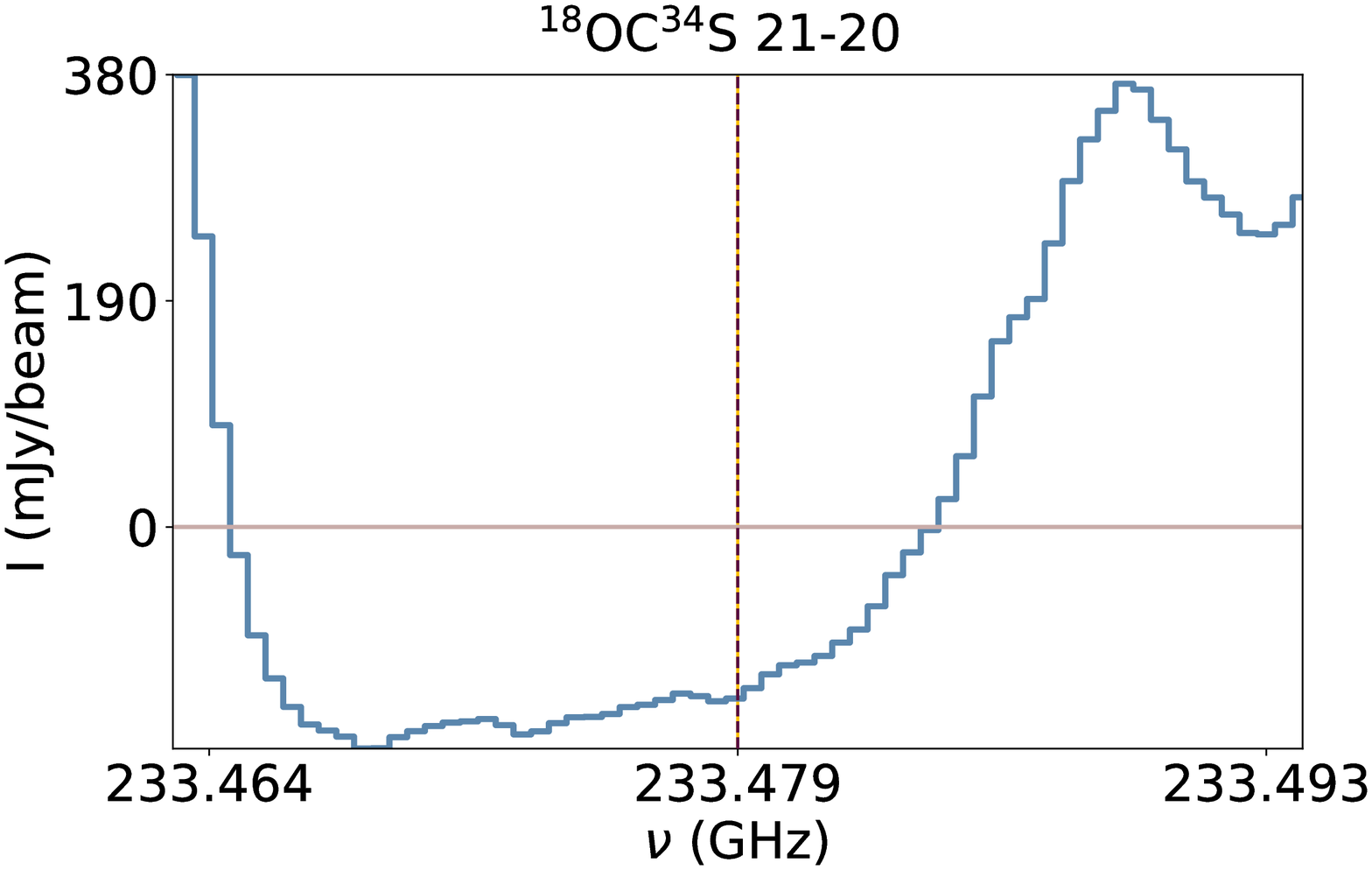}}
\end{minipage}
\caption{Observed spectra (in blue), rest frequency of the undetected line (brown dashed line), and spectroscopic uncertainty on the rest frequency of the undetected line (yellow shaded region) plotted for the sulfur-bearing species undetected towards IRAS 16293-2422 A.}
\label{undetected_IRAS16293A}
\end{figure}

The strengths of the lines detected towards IRAS 16293 A are comparable to those towards IRAS 16293 B. The lines of H$_{2}$S, 2$_{2,0}$-2$_{1,1}$ and OCS, $v$=0, J = 19-18 are bright and optically thick ($\tau\gg$1). The line of H$_2^{34}$ S is marginally optically thick ($\tau$ = 0.1). For the analysis of IRAS 16293 A, we have assumed $T_{\text{ex}}$=125 K, which is typical for the hot inner regions of this source (\citealt{Calcutt2018, Manigand2020}), and a source size of 2$''$ in diameter (based on the investigation of IRAS 16293 B; \Cref{IRAS16293B}). The fitted FWHM and $N$, and the computed $\tau$ are given in \autoref{results_IRAS16293A} for the detected S-bearing molecules. Due to the high spectral resolution of the data, all detected lines are spectrally resolved to be double- or multi- peaked. This is possibly caused by the multiple components of the compact binary IRAS 16293 A \citep{Maureira2020}. One of the detected H$_2^{33}$S lines (at 215.512 GHz) is contaminated by the CH$_3$CHO molecule. The line of HD$^{34}$S is heavily blended with that of CH$_3$CN, $v_8$=1 at 232.965 GHz. Having $^{32}$S/$^{34}$S=22 (\citealt{Wilson1999}) and D/H $\sim$0.04 (incl. statistical correction; \citealt{Drozdovskaya2018}), it is most likely that all the emission seen around the rest frequency of HD$^{34}$S stems from CH$_3$CN. The spectra of detected and undetected molecules can be found in \autoref{detected_IRAS16293A} and \autoref{undetected_IRAS16293A}, respectively.

\begin{table}[H]
    \centering
    \caption{Synthetic fitting of the detected S-bearing species towards IRAS 16293-2422 A for an excitation temperature ($T_{\text{ex}}$) of 125 K and a source size of 2$''$. Directly across from a specific minor isotopolog under ``Derived $N$ of isotopologs’ follows the column density of the main isotopolog upon the assumption of the standard isotopic ratio. In bold in the same column is the average column density of the main isotopolog based on all the available minor isotopologs (only if the minor isotopolog is optically thin and including the uncertainties).}
    \label{results_IRAS16293A}
    \begin{adjustbox}{width=1\textwidth}
    \begin{tabular}{r r c r  c c  c c c c  r}
    \hline
    \hline
    Species & Transition & Frequency & $E_{\text{up}}$ & $A_{ij}$ &
    Beam size & FWHM &
    $N$ & Derived $N$ & $\tau$\\

    &  &  &  & & & & & {of isotopologs} &  \\
    & & (GHz) & (K) & (s$^{-1}$) & ($''$) & (km s$^{-1}$) & (cm$^{-2}$) & (cm$^{-2}$) &  \\
\hline
& & & &  & & & & &\\

H$_2$S & 2$_{2,0}$-2$_{1,1}$ & 216.710 & 84 & 4.9$\times$10$^{-5}$  & 6.0 & 4.5  & op. thick  & \textbf{$\overline{N(\text{H}_2\text{S})}$=(2.4$\pm$0.4)$\times$10$^{17}$} & 6.0 \\

&&&&&&&\\

H$_2$ $^{33}$S & 2$_{2,0,1}$-2$_{1,1,2}$
& 215.494 & 84  & 2.4$\times$10$^{-5}$ &  6.0 & 6.0  & 2.0$_{-0.5}^{+0.2}\times$10$^{15}$ & $N$(H$_2$S)=2.5$_{-0.6}^{+0.3}\times$10$^{17}$ &  0.002\\
& 2$_{2,0,1}$-2$_{1,1,1}$ & 215.497 & 84 & 2.4$\times$10$^{-5}$  &  & & &  & 0.002  \\
& 2$_{2,0,4}$-2$_{1,1,3}$ & 215.501 & 84 & 6.9$\times$10$^{-6}$ &  &  &  &  & 0.002   \\
& 2$_{2,0,4}$-2$_{1,1,4}$ & 215.503 & 84 & 4.1$\times$10$^{-5}$ &  &   &  & & 0.010   \\
& 2$_{2,0,2}$-2$_{1,1,3}$ & 215.504 & 84 & 1.7$\times$10$^{-5}$ &  &  & &  & 0.003   \\
& 2$_{2,0,2}$-2$_{1,1,2}$ & 215.505 & 84 & 1.9$\times$10$^{-5}$ &  &  & &  & 0.003   \\
& 2$_{2,0,2}$-2$_{1,1,1}$ & 215.508 & 84 & 1.2$\times$10$^{-5}$   &  &  & &  & 0.002    \\
& 2$_{2,0,3}$-2$_{1,1,3}$ & 215.512 & 84 & 2.8$\times$10$^{-5}$ &  &  & &  & 0.007   \\
& 2$_{2,0,3}$-2$_{1,1,2}$ & 215.513 & 84 & 1.1$\times$10$^{-5}$ &  &  & &  & 0.003  \\
& 2$_{2,0,3}$-2$_{1,1,4}$ & 215.513 & 84 & 9.1$\times$10$^{-6}$ &  &  & &  & 0.002   \\

&&&&&&&&&\\

H$_2$ $^{34}$S & 2$_{2,0}$-2$_{1,1}$ & 214.377 & 84  & 4.7$\times$10$^{-5}$ &  6.0 & 7.0 & $>$6.6$\times$10$^{15}$ & & 0.1\\
&&&&&&&&&\\

&&&&&&&&&\\

OCS, $v$=0& 19-18 & 231.061 & 111  & 3.6$\times$10$^{-5}$  & 5.6  & 4.5  & op. thick & \textbf{$\overline{N\text{(OCS)}}$= (3.6$\pm$1.4)$\times$10$^{17}$} & 9.00 \\
&&&&&&&&&\\

OC$^{33}$S & 18-17 & 216.147 & 99 & 2.9$\times$10$^{-5}$ & 6.0 & 6.0  & 1.9$_{-0.3}^{+0.1}\times$10$^{15}$ &  $N$(OCS)=2.4$_{-0.2}^{+0.1}\times$10$^{17}$ & 0.05\\

&&&&&&&&&\\

O$^{13}$CS  & 19-18 & 230.318 & 110  & 3.5$\times$10$^{-5}$  & 5.7 & 8.0 & 3.8$_{-0.2}^{+0.3}\times$10$^{15}$ & $N$(OCS)=2.6$_{-0.1}^{+0.2}\times$10$^{17}$ & 0.08 \\

&&&&&&&&&\\

$^{18}$OCS  & 19-18 & 216.753 & 104  & 3.0$\times$10$^{-5}$ & 6.0 &  9.0 & 8.5$_{-2.0}^{+0.5}\times$10$^{14}$ &  $N$(OCS)=4.7$_{-1.1}^{+0.3}\times$10$^{17}$ & 0.01\\

&&&&&&&&&\\
OCS, $v_2$=1$^-$  & 19-18 & 231.342 & 860 & 3.5$\times$10$^{-5}$ & 5.6  & 6.0  & 2.5$_{-1.8}^{+1.5}\times$10$^{17}$ & & 0.02\\
OCS, $v_2$=1$^+$  & 19-18 & 231.584 & 860  &  3.5$\times$10$^{-5}$ &  5.6  & & & &  0.02\\
\hline
\hline
    \end{tabular}
    \end{adjustbox}

\end{table}

\subsection{Detected lines in IRAS 16293-2422 A}

\subsection{Undetected lines in IRAS 16293-2422 A}

\newpage

\section{NGC 1333-IRAS4A}
\label{IRAS4A}

Line detections in IRAS4A are nearly the same as in IRAS 16293 A and IRAS 16293 B. The main sulfur species, H$_{2}$S, 2$_{2,0}$-2$_{1,1}$ and OCS, $v$=0, $J = 19-18$, are bright and optically thick, $\tau$>1 for $50\leq T_{\text{ex}}<200$ K or partially optically thick, $0.1\leq\tau<1$ for $T_{\text{ex}}$ $\geq$200 K. The two protostars in IRAS4A, 4A1 and 4A2, could not be disentangled due to insufficient angular resolution of the ALMA ACA observations towards this source in comparison to the separation between the two components (1.8$''$, i.e., 540 au; \citealt{Sahu2019}). Due to a lack of small-scale constraints on the distribution of S-bearing molecules in IRAS4A, a range of excitation temperature between 150 and 300 K is considered (\citealt{Lopez2017}). Temperatures of 50 and 100 K have been ruled out on the grounds that the observed signal at the frequency corresponding to the OCS, $v_2$=1$^{\pm}$ is incompatible with these lowest temperatures for column densities in the physical range of values. Looking at high spatial resolution maps from \citet{Sahu2019}, a 2$''$ source size is assumed to cover the circumbinary envelope around 4A1 and 4A2. The spectra fitted to the line profiles have a FWHM of 1.8 km s$^{-1}$. The computed column densities of the detected species are reported in Table \ref{results3_IRAS4A} for excitation temperatures in the $150-300$~K range. The lines of H$_{2}^{33}$S and HD$^{34}$S face the same blending effects as in IRAS 16293 A and IRAS 16293 B.

\subsection{Results of the synthetic fitting of the line profiles towards NGC 1333-IRAS4A for $T_{\text{ex}}=150-300$ K}

\begin{table}[H]
    \centering
    \caption{Synthetic fitting of the detected S-bearing species towards NGC 1333-IRAS4A for a range of excitation temperatures between 150 and 300 K, a FWHM of 1.8$''$, and a source size of 2$''$. Directly across from a specific minor isotopolog under ``Derived $N$ of isotopologs’ follows the column density of the main isotopolog upon the assumption of the standard isotopic ratio. In bold in the same column is the average column density of the main isotopolog based on all the available minor isotopologs (only if the minor isotopolog is optically thin and including the uncertainties).}
    \label{results3_IRAS4A}
    \begin{adjustbox}{width=1\textwidth}
    \begin{tabular}{c c c c c c c  c c  r}
    \hline
    \hline
    Species & Transition & Frequency & $E_{\text{up}}$ & $A_{ij}$ & Beam size &
    $N$ & Derived $N$ & $\tau$ \\

    &  &  &  & & & & {of isotopologs} &  \\
    & & (GHz) & (K) & (s$^{-1}$) & ($''$) & (cm$^{-2}$) & (cm$^{-2}$) &  \\
\hline
& & & & &  &  &  &  \\

H$_2$S & 2$_{2,0}$-2$_{1,1}$ & 216.710 & 84 & 4.9$\times$10$^{-5}$  & 6.1 &  op. thick  & \textbf{$\overline{N(\text{H}_2\text{S})}$=(2.6-4.2)$\times$10$^{16}$}  & 0.4-1.2 \\
&&&&&&&\\

H$_2$ $^{33}$S 
& 2$_{2,0,1}$-2$_{1,1,2}$  & 215.494 & 84  & 2.4$\times$10$^{-5}$ &  6.8  &  3$\times$10$^{14}$ & $N$(H$_2$S)=(3.8-8.0)$\times$10$^{16}$ &  0.0003-0.0007\\
&2$_{2,0,1}$-2$_{1,1,1}$  & 215.497 & 84 & 2.4$\times$10$^{-5}$  &  &  &  & 0.0003-0.0007  \\
& 2$_{2,0,4}$-2$_{1,1,3}$ & 215.501 & 84 & 6.9$\times$10$^{-6}$ &  &  &  & 0.0004-0.0008   \\
& 2$_{2,0,4}$-2$_{1,1,4}$ & 215.503 & 84 & 4.1$\times$10$^{-5}$ &  &  &  & 0.0020-0.0080  \\
& 2$_{2,0,2}$-2$_{1,1,3}$ & 215.504 & 84 & 1.7$\times$10$^{-5}$ &  &  &  & 0.0005-0.0010   \\
& 2$_{2,0,2}$-2$_{1,1,2}$ & 215.505 & 84 & 1.9$\times$10$^{-5}$ &  &  &  & 0.0005-0.0010  \\
& 2$_{2,0,2}$-2$_{1,1,1}$ & 215.508 & 84 & 1.2$\times$10$^{-5}$  &  &  &  & 0.0003-0.0007    \\
& 2$_{2,0,3}$-2$_{1,1,3}$ & 215.512 & 84 & 2.8$\times$10$^{-5}$ &  &  &  & 0.0010-0.0020  \\
& 2$_{2,0,3}$-2$_{1,1,2}$ & 215.513 & 84 & 1.1$\times$10$^{-5}$ &  &  &  & 0.0005-0.0010   \\
& 2$_{2,0,3}$-2$_{1,1,4}$ & 215.513 & 84 & 9.1$\times$10$^{-6}$ &  &  &  & 0.0004-0.0008   \\

&&&&&&&\\

H$_2$ $^{34}$S & 2$_{2,0}$-2$_{1,1}$ & 214.377 & 84  & 4.7$\times$10$^{-5}$ &  6.8 &  6.3$\times$10$^{14}$ & $N$(H$_2$S)=(1.4-2.9)$\times$10$^{16}$ & 0.01-0.03\\

&&&&&&&\\

OCS, $v$=0 & 19-18 & 231.061 & 111  & 3.6$\times$10$^{-5}$  & 6.2 & op. thick  & \textbf{$\overline{N\text{(OCS)}}$= (1.6-2.0)$\times$10$^{16}$} & 0.5-1.1\\

&&&&&&&\\

OC$^{33}$S & 18-17 & 216.147 & 99 & 2.9$\times$10$^{-5}$ & 6.8 &  1.0$\times$10$^{14}$ &  $N$(OCS)=(1.3-1.9)$\times$10$^{16}$ & 0.003-0.007 \\

&&&&&&&\\

O$^{13}$CS & 19-18 & 230.318 & 110  & 3.5$\times$10$^{-5}$  & 6.3 & 1.4$\times$10$^{14}$ & $N$(OCS)=(0.1-1.3)$\times$10$^{16}$ & 0.005-0.010 \\

&&&&&&&\\

$^{18}$OCS & 19-18 & 216.753 & 104  & 3.0$\times$10$^{-5}$ & 6.1 &  3.8$\times$10$^{13}$ &  $N$(OCS)=(2.3-3.0)$\times$10$^{16}$ & 0.001-0.002\\

&&&&&&&\\

OCS, $v_2$=1$^-$  & 19-18 & 231.342 & 860 & 3.5$\times$10$^{-5}$ & 5.6  & (0.4-1.2)$\times$10$^{15}$ & & 0.001-0.002\\
OCS, $v_2$=1$^+$  & 19-18 & 231.584 & 860  &  3.5$\times$10$^{-5}$ &  5.6  & & & 0.001-0.004\\
\hline
\hline
    \end{tabular}
    \end{adjustbox}

\end{table}

\subsection{Detected lines in NGC 1333-IRAS4A}

\begin{figure}[H]
\centering
\begin{minipage}{6.2in}%
    \subfigure{\includegraphics[width=2.0in]{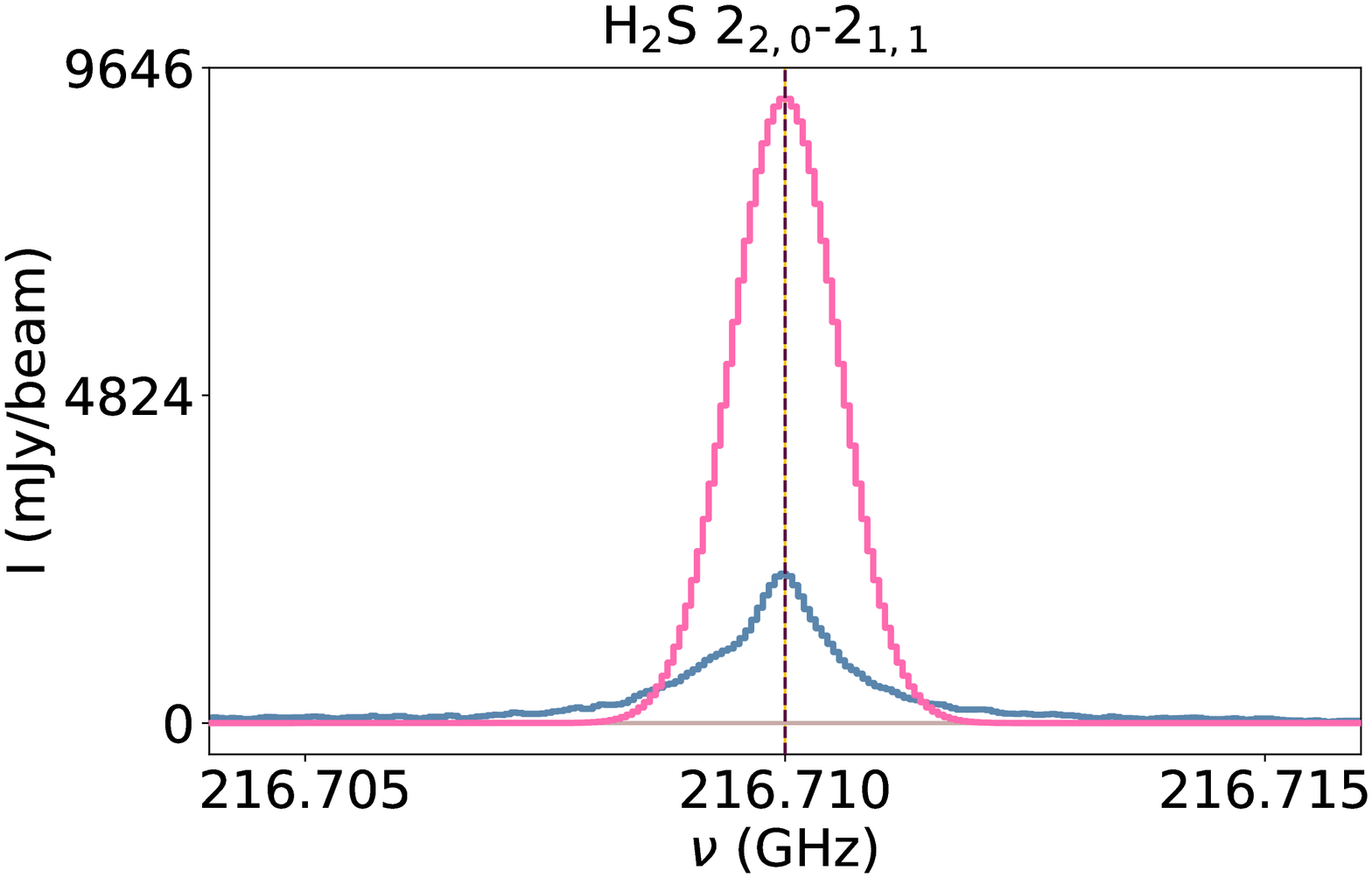}}
    \subfigure{\includegraphics[width=2.0in]{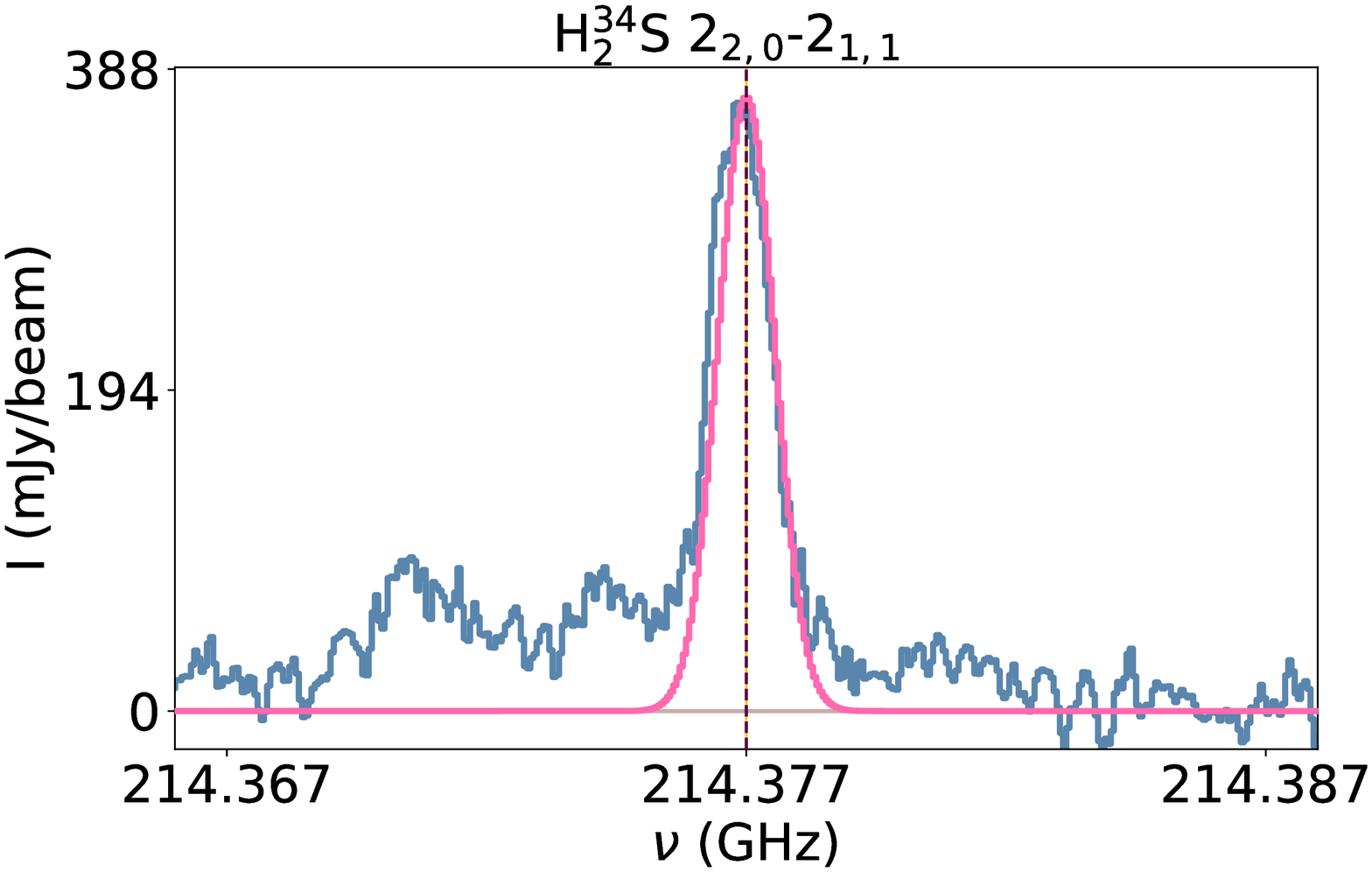}}\\
    \subfigure{\includegraphics[width=6.0in]{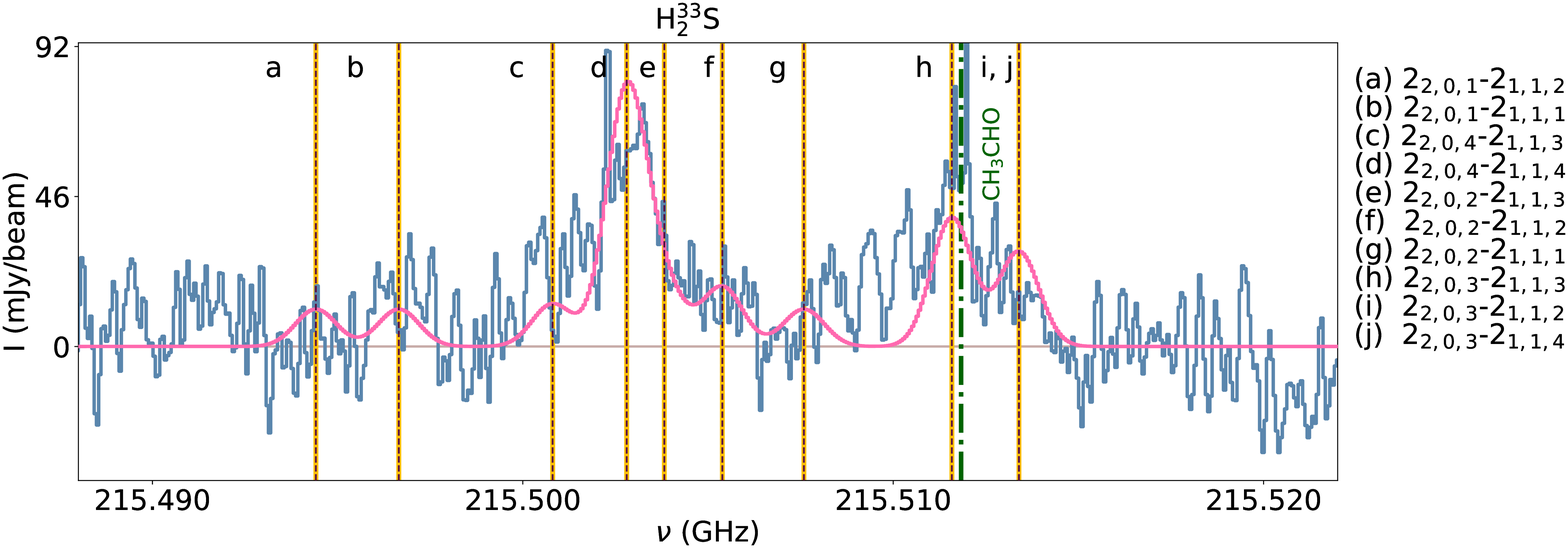}}\\
    \subfigure{\includegraphics[width=2.0in]{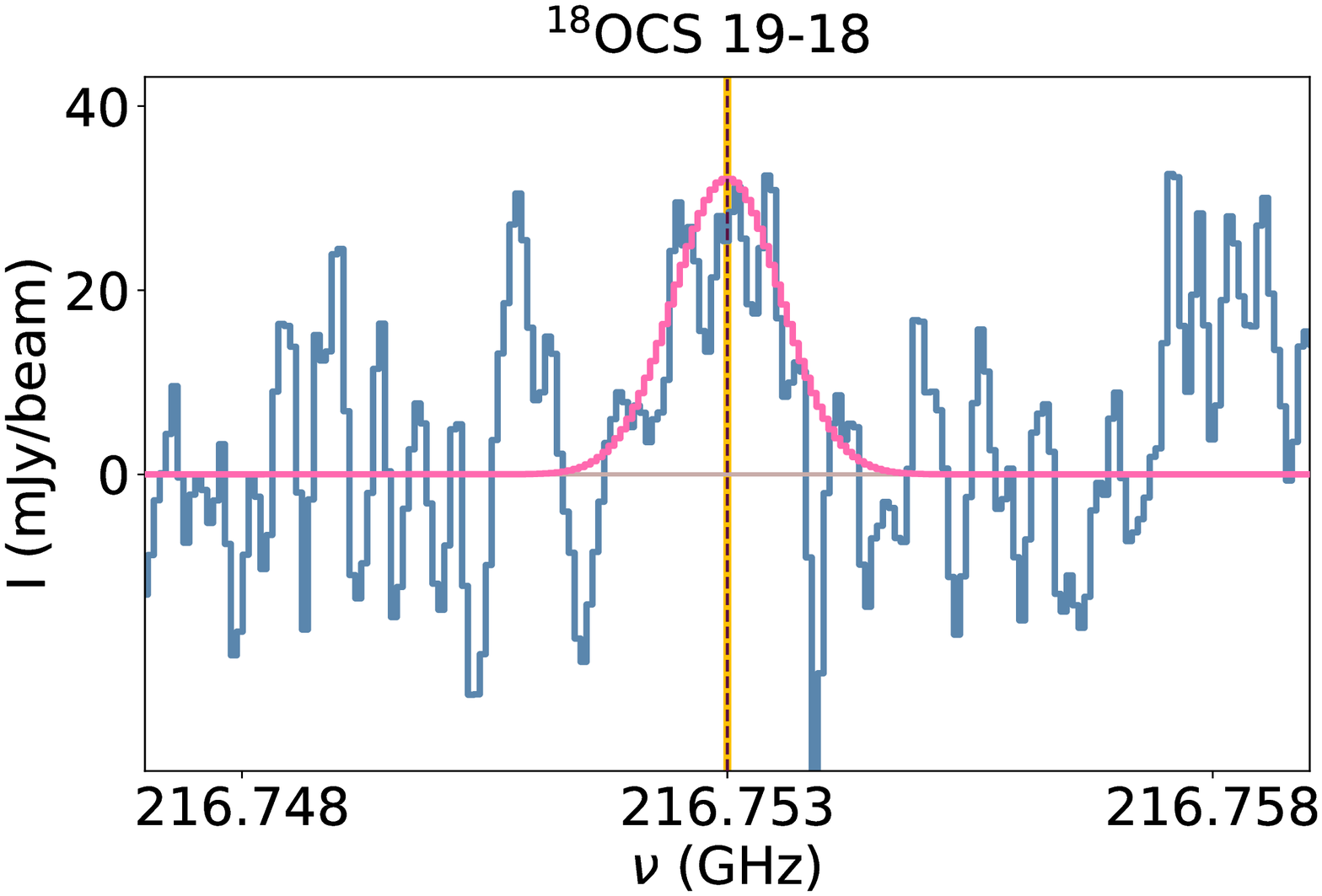}}
    \subfigure{\includegraphics[width=2.0in]{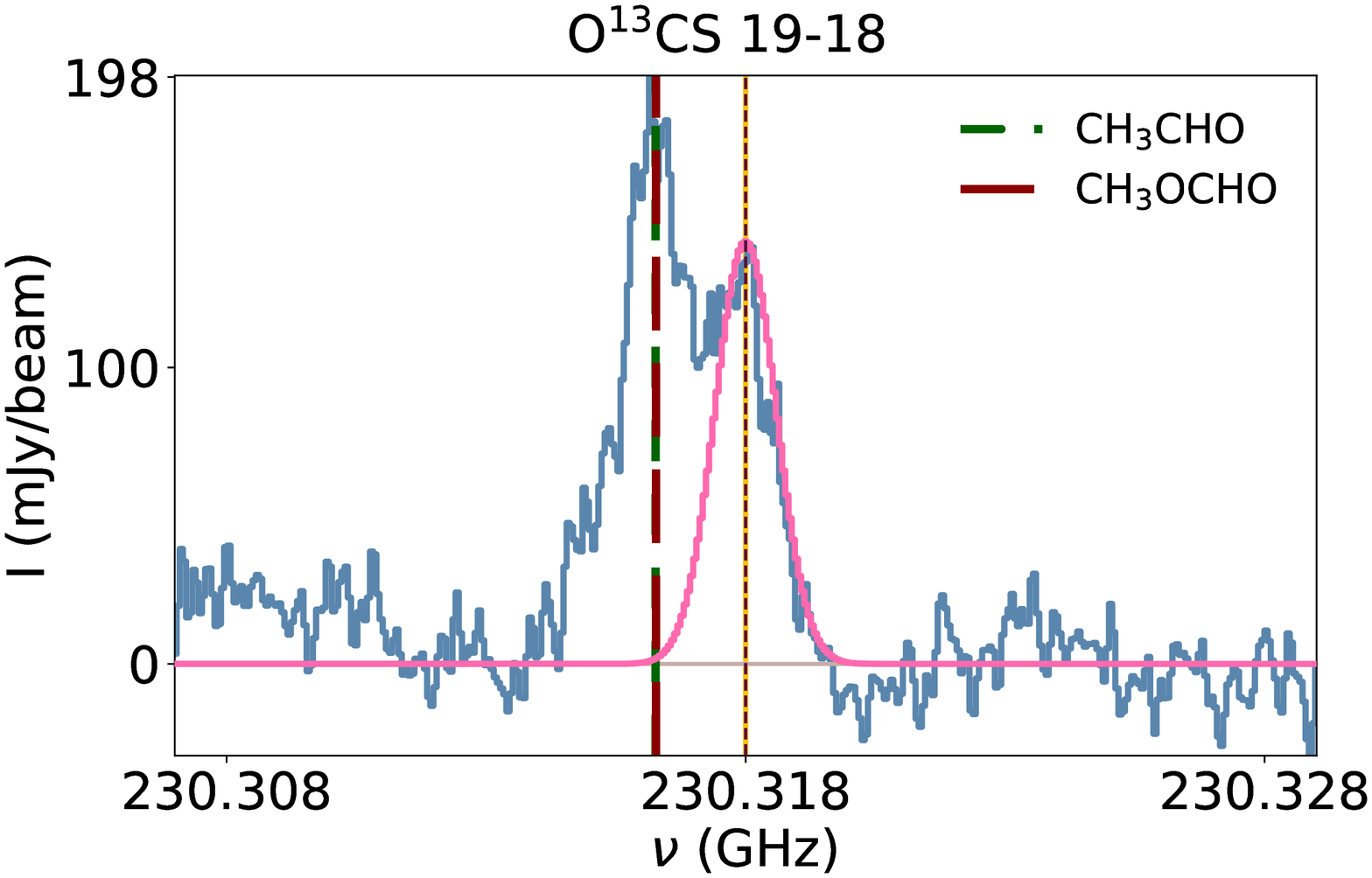}}
    \subfigure{\includegraphics[width=2.0in]{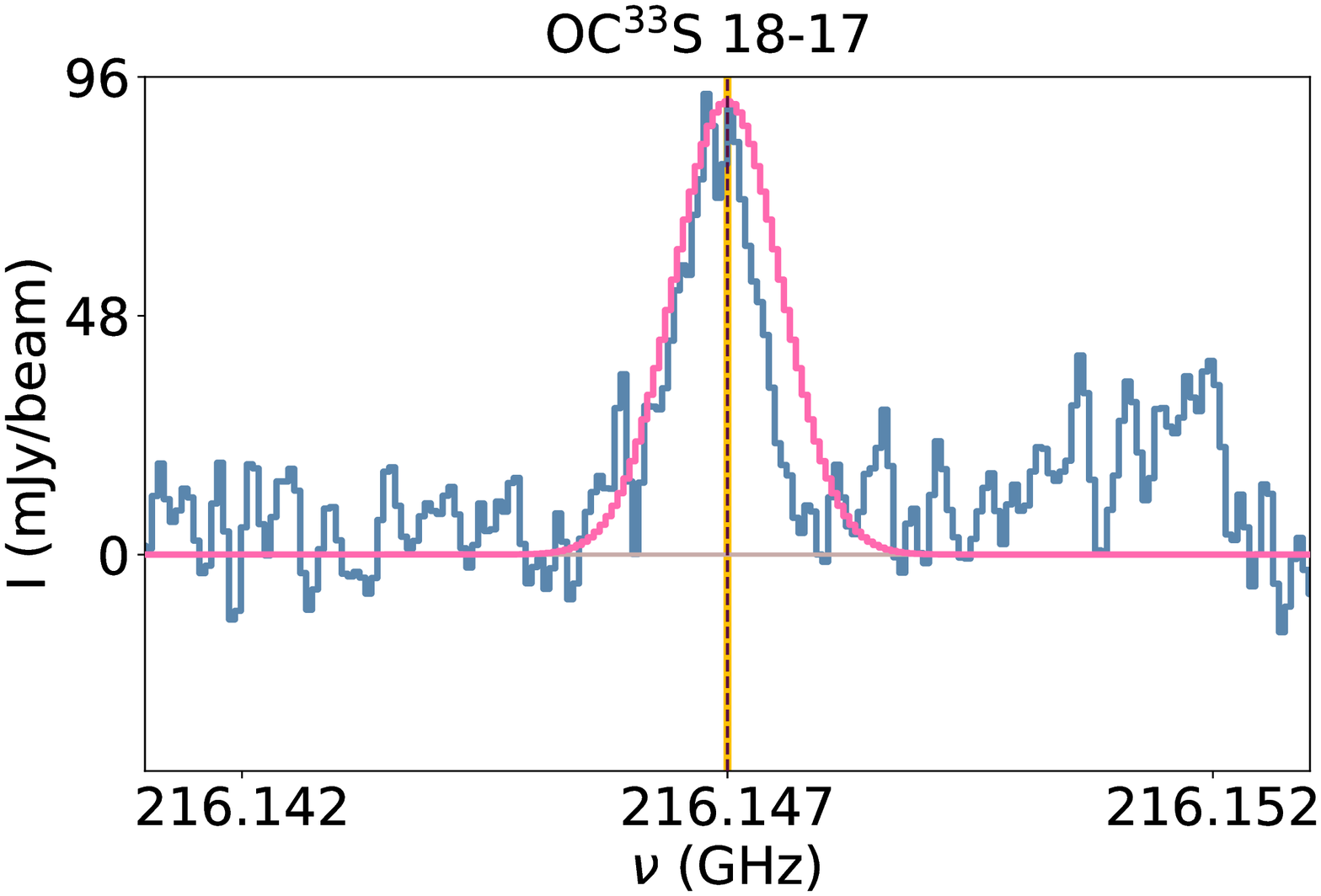}}\\ 
    \subfigure{\includegraphics[width=2.0in]{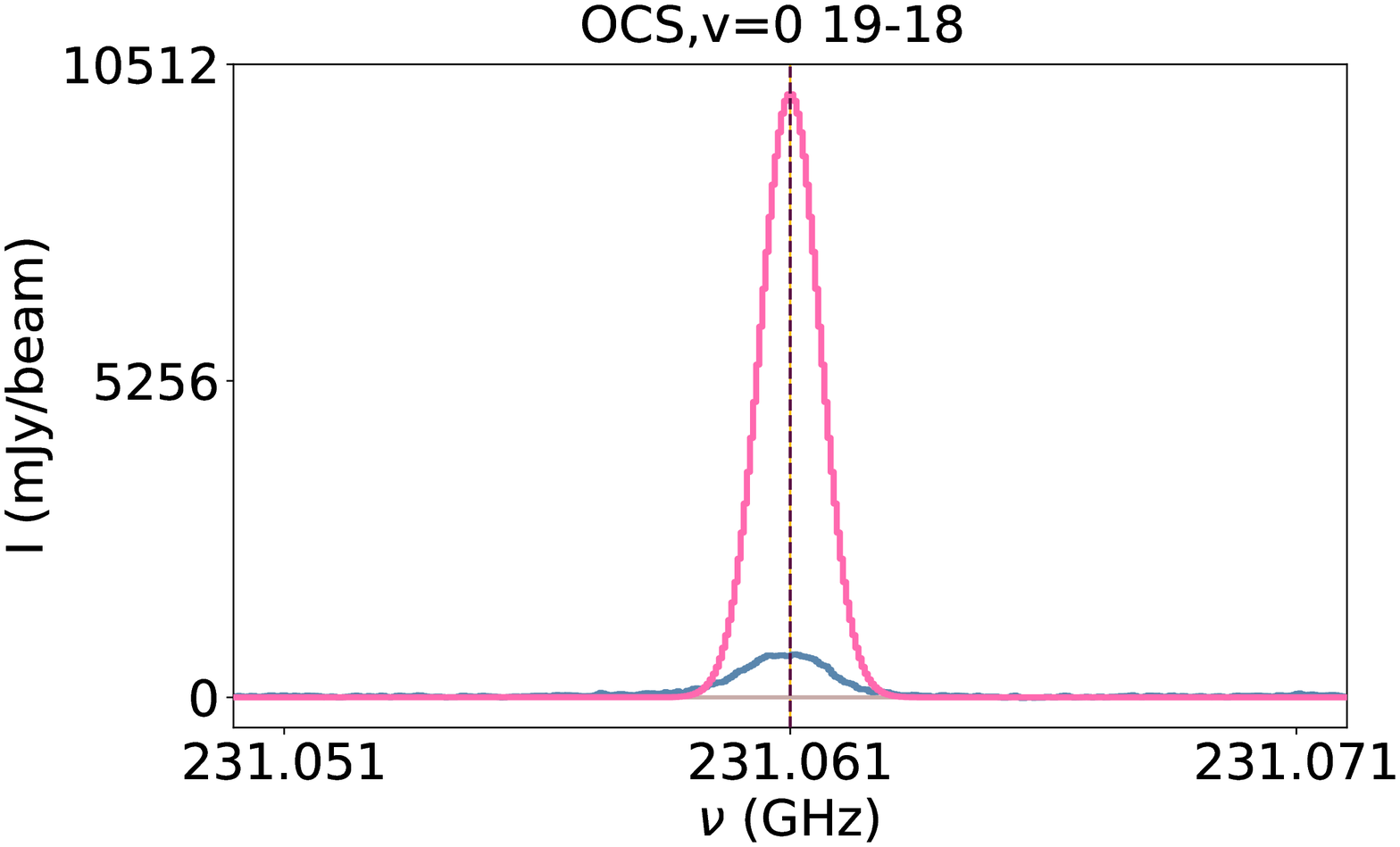}}
    \subfigure{\includegraphics[width=2.0in]{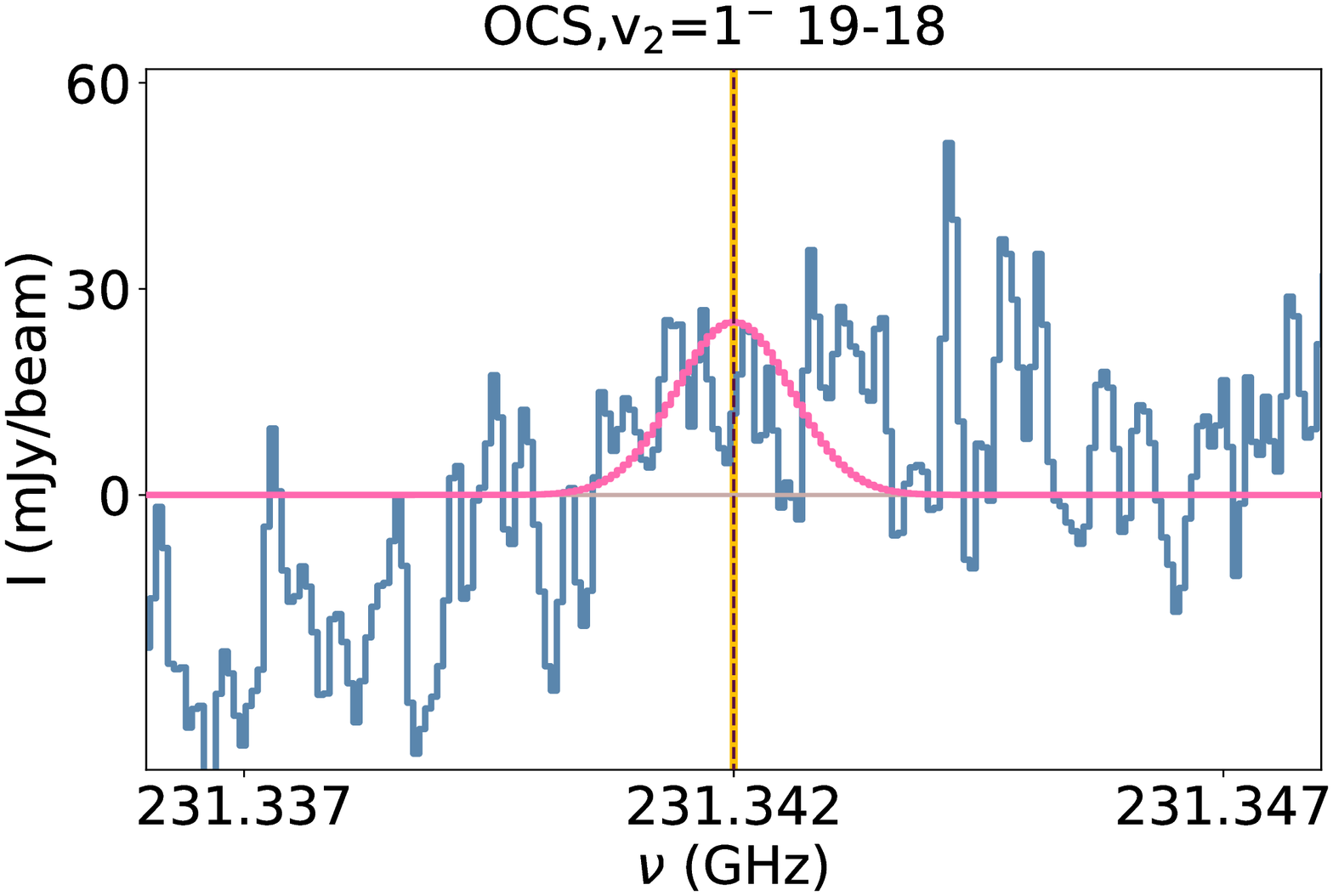}}
    \subfigure{\includegraphics[width=2.0in]{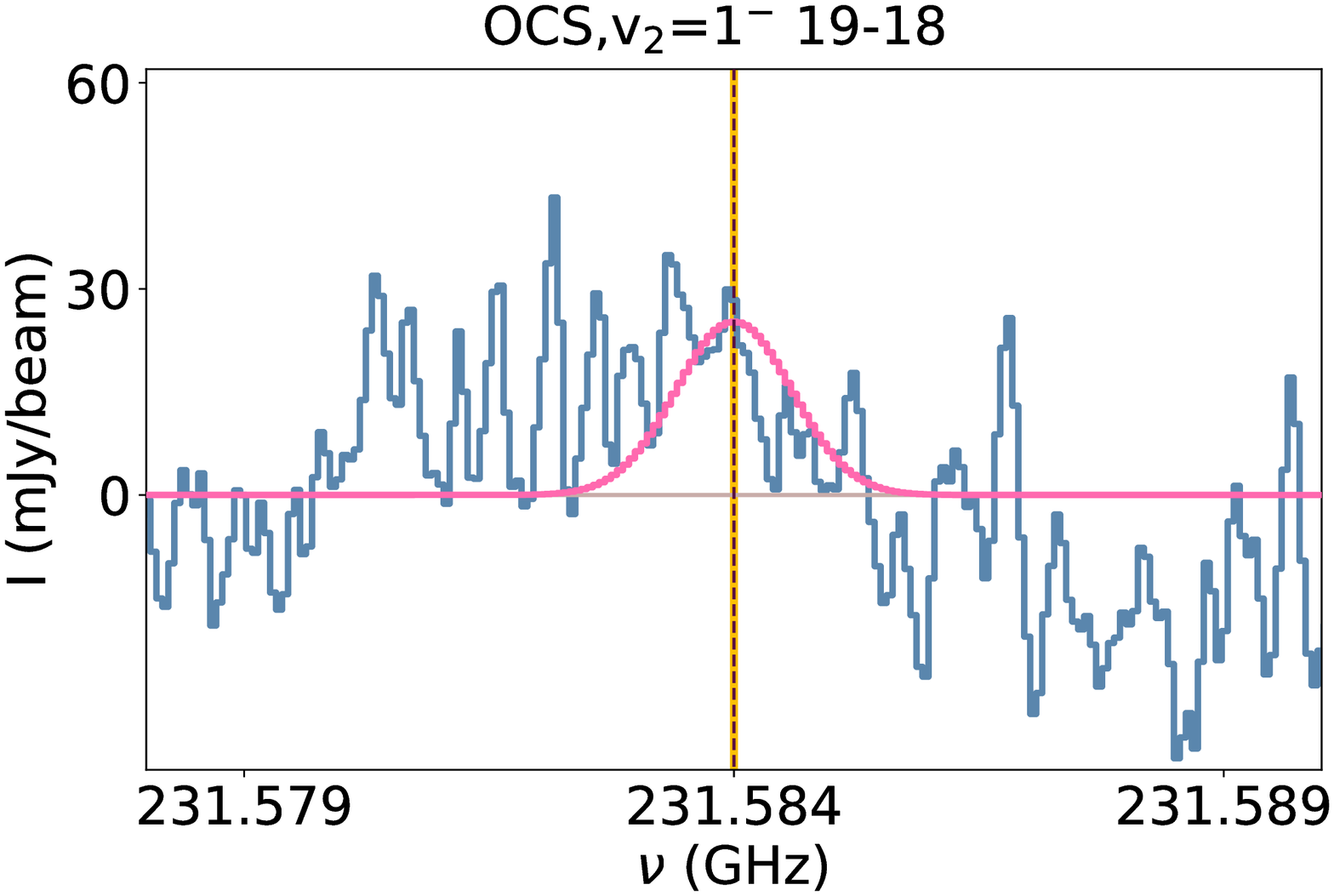}}\\ 
\end{minipage}
\caption{Observed spectra (in blue), rest frequency of the detected line (brown dashed line), spectroscopic uncertainty on the rest frequency of the detected line (yellow shaded region), blending species (green dash-dotted line and red dashed line), and fitted synthetic spectra (in pink) plotted for the sulfur-bearing species detected towards NGC 1333-IRAS4A for $T_{\text{ex}}=150$ K. $\overline{N\text{(H}_2\text{S)}}$ and $\overline{N\text{(OCS)}}$ are used to fit the synthetic spectra of the optically thick lines.}
\label{detected_IRAS4A}
\end{figure}

\subsection{Undetected lines in NGC 1333-IRAS4A}

\begin{figure}[H]
\centering
\begin{minipage}{6.2in}%
    \subfigure{\includegraphics[width=2.0in]{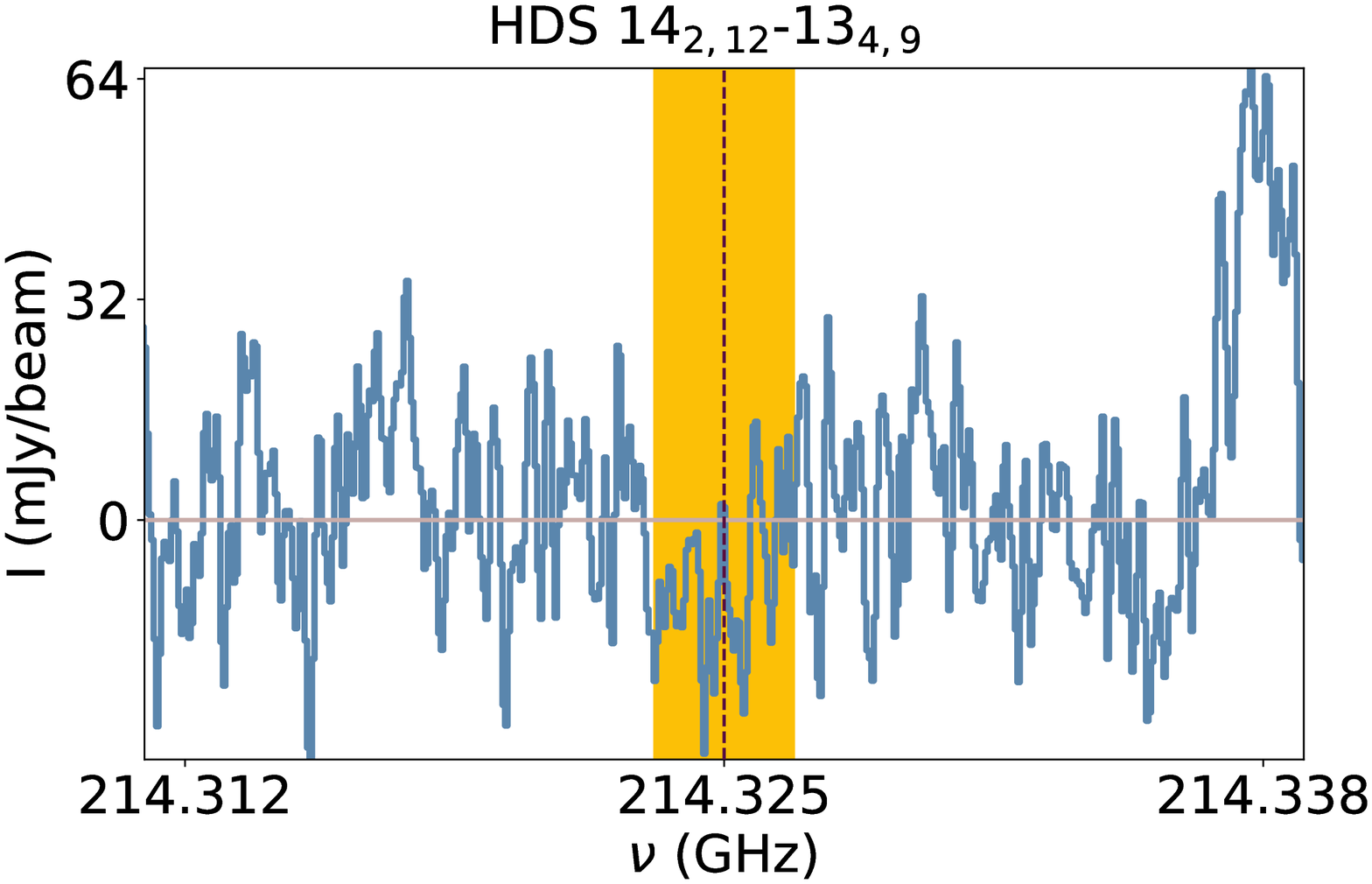}}
    \subfigure{\includegraphics[width=2.0in]{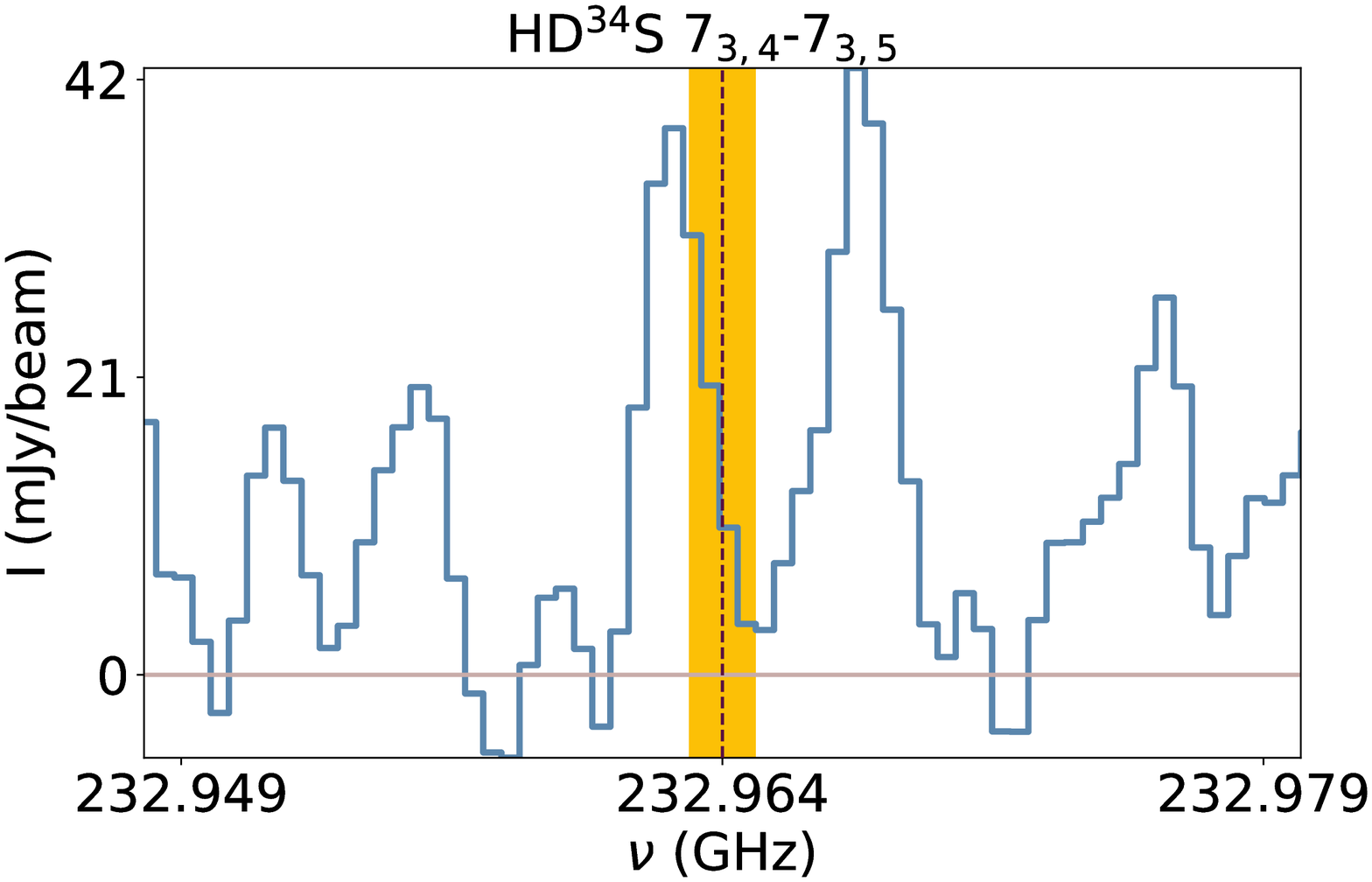}}
    \subfigure{\includegraphics[width=2.0in]{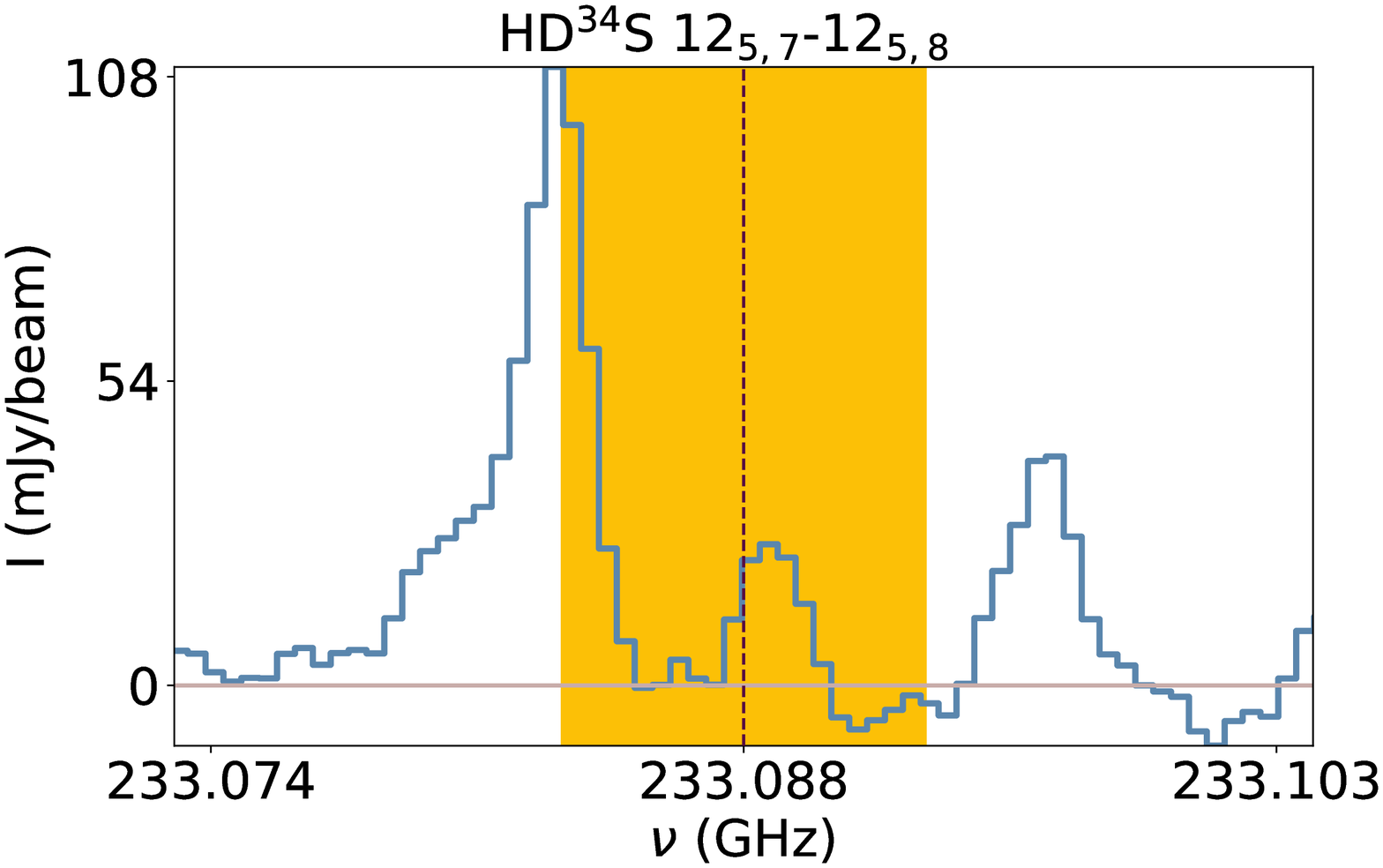}}\\
    \subfigure{\includegraphics[width=2.0in]{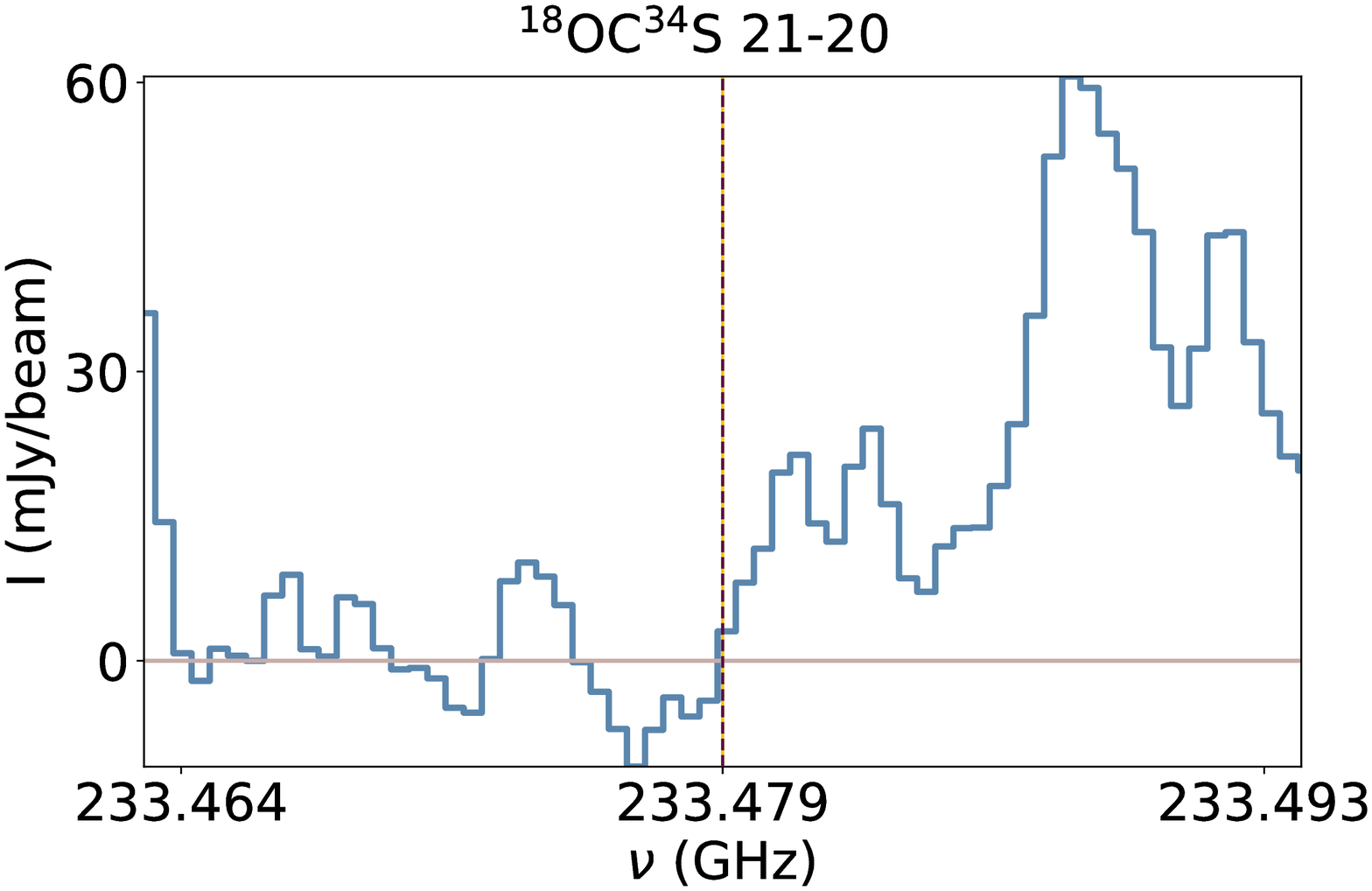}}
\end{minipage}
\caption{Observed spectra (in blue), rest frequency of the undetected line (brown dashed line), and spectroscopic uncertainty on the rest frequency of the undetected line (yellow shaded region) plotted for the sulfur-bearing species undetected towards NGC 1333-IRAS4A.}
\label{undetected_IRAS4A}
\end{figure}

\section{RCrA IRS7B}

\begin{figure}[H]
\centering
\begin{minipage}[b]{4.0in}%
    \subfigure{\includegraphics[width=3.0in]{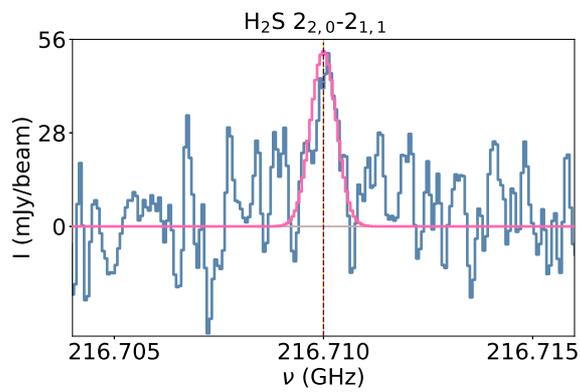}}
\end{minipage}
\caption{Observed spectrum (in blue), rest frequency of the detected line (brown dashed line), spectroscopic uncertainty on the rest frequency of the detected line (yellow shaded region), and fitted synthetic spectrum (in pink) plotted for the sulfur-bearing species detected towards RCrA IRS7B.}
\label{detected_IRS7B}
\end{figure}

H$_2$S, 2$_{2,0}$-2$_{1,1}$ is the only emission line detected towards IRS7B (\autoref{detected_IRS7B}). The synthetic spectrum for this line is best-fitted for a FWHM of 1 km s$^{-1}$, an excitation temperature of 100 K, and a source size of 1.5$''$, as determined from ALMA 12m observations (project id: 2017.1.00108.S; PI: M. N. Drozdovskaya) towards IRS7B. The upper limit on the column density of OCS, $v$=0 is stated in \autoref{ratio} and \ref{results_IRS7B}. IRS7B shows a weak hot corino activity resulting in few detections of complex organic molecules and their low abundances \citep{Lindberg2015}. The chemical composition in the envelope of IRS7B could be affected by the enhanced UV radiation (photodissociation) from the nearby Herbig Ae star RCrA located at a distance of 39$''$ ($6~630$ au) NW of IRS7B \citep{Watanabe2012}.

\begin{table}[H]
    \centering
    \caption{Synthetic fitting of the detected S-bearing species towards RCrA IRS7B for an excitation temperature of 100 K, a FWHM of 1 km s$^{-1}$, and a source size of 1.5$''$.}
    \label{results_IRS7B}
    \begin{tabular}{c c c c r r c r c r}
    \hline
    \hline
    Species & Transition & Frequency & $E_{\text{up}}$ & $A_{ij}$ &
    Beam size & $N$ & $\tau$ \\
    &&&&&&& \\
    & & (GHz) & (K) & (s$^{-1}$) & ($''$) & (cm$^{-2}$) &   \\
\hline
& & & & &  & & &  \\
H$_2$S & 2$_{2,0}$-2$_{1,1}$ & 216.710 & 84 & 4.9$\times$10$^{-5}$  & 6.5 &  5.5$_{-0.6}^{+0.9}\times$10$^{13}$ & 0.01\\
OCS, $v$=0 & 19-18 & 231.061 & 111  & 3.6$\times$10$^{-5}$  & 6.2 & $\leq$3.6$\times$10$^{13}$  & --\\
\hline
\hline
    \end{tabular}
\end{table}

\subsection{Detected lines in RCrA IRS7B}

\subsection{Undetected lines in RCrA IRS7B}

\begin{figure}[H]
\centering
\begin{minipage}[b]{6.2in}%
    \subfigure{\includegraphics[width=4.0in]{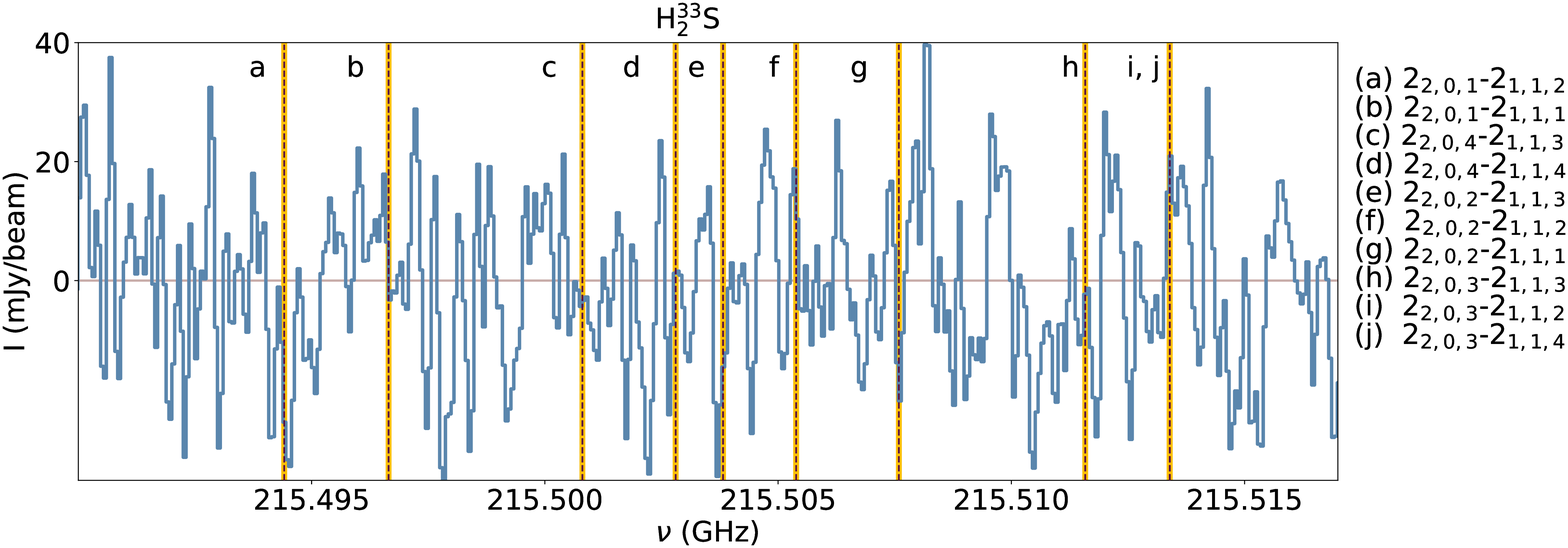}}
    \subfigure{\includegraphics[width=2.0in]{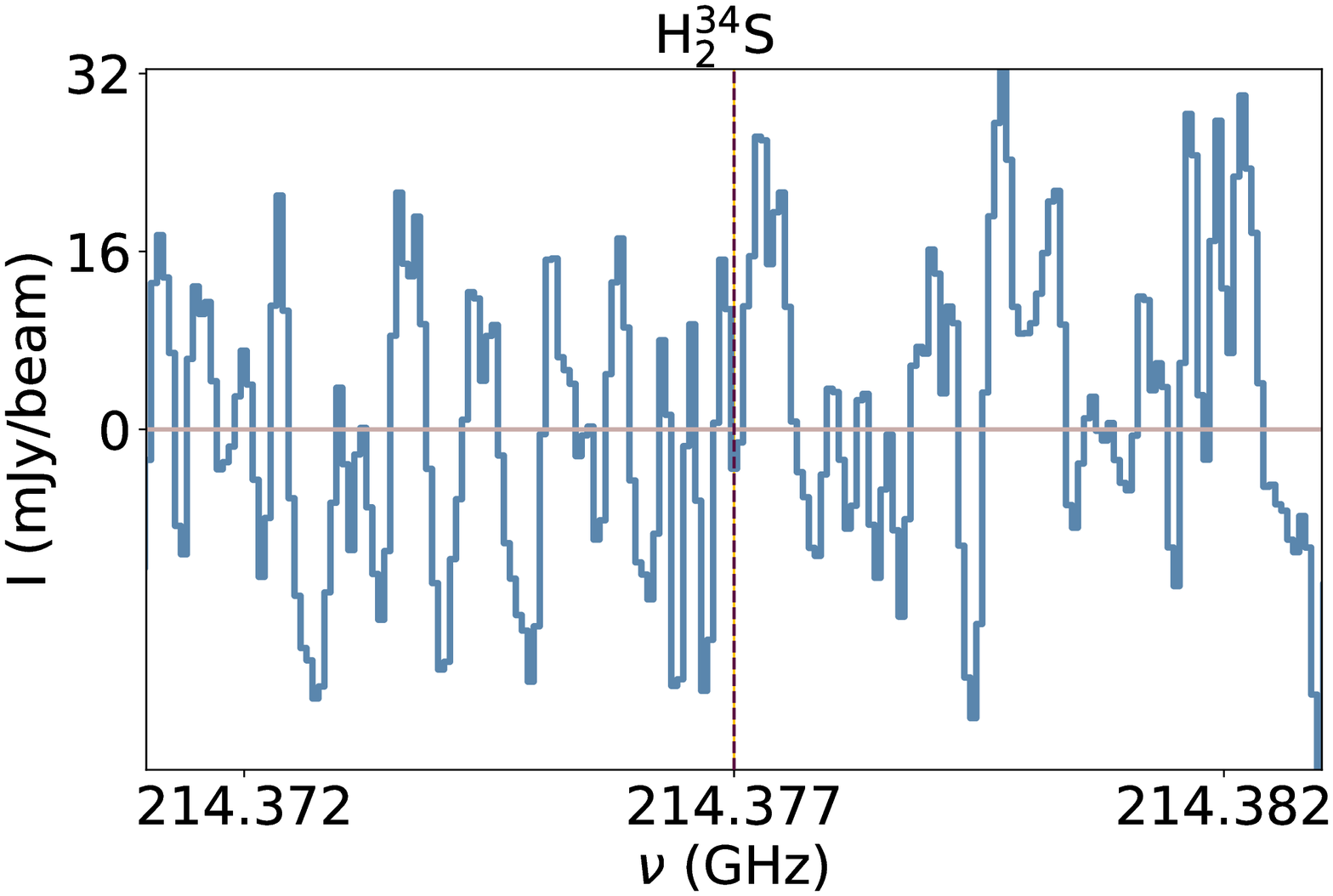}}\\
    \subfigure{\includegraphics[width=2.0in]{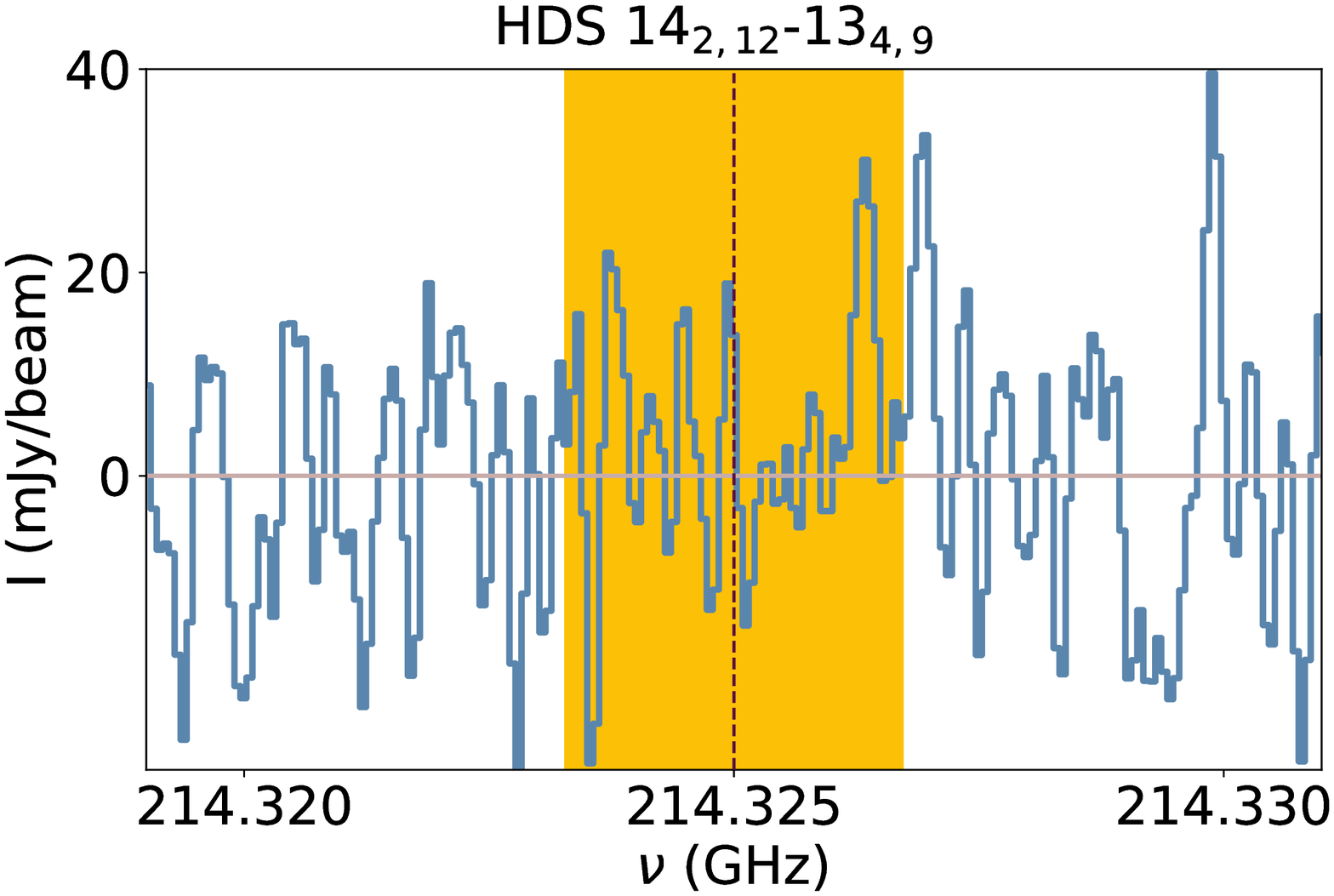}}
    \subfigure{\includegraphics[width=2.0in]{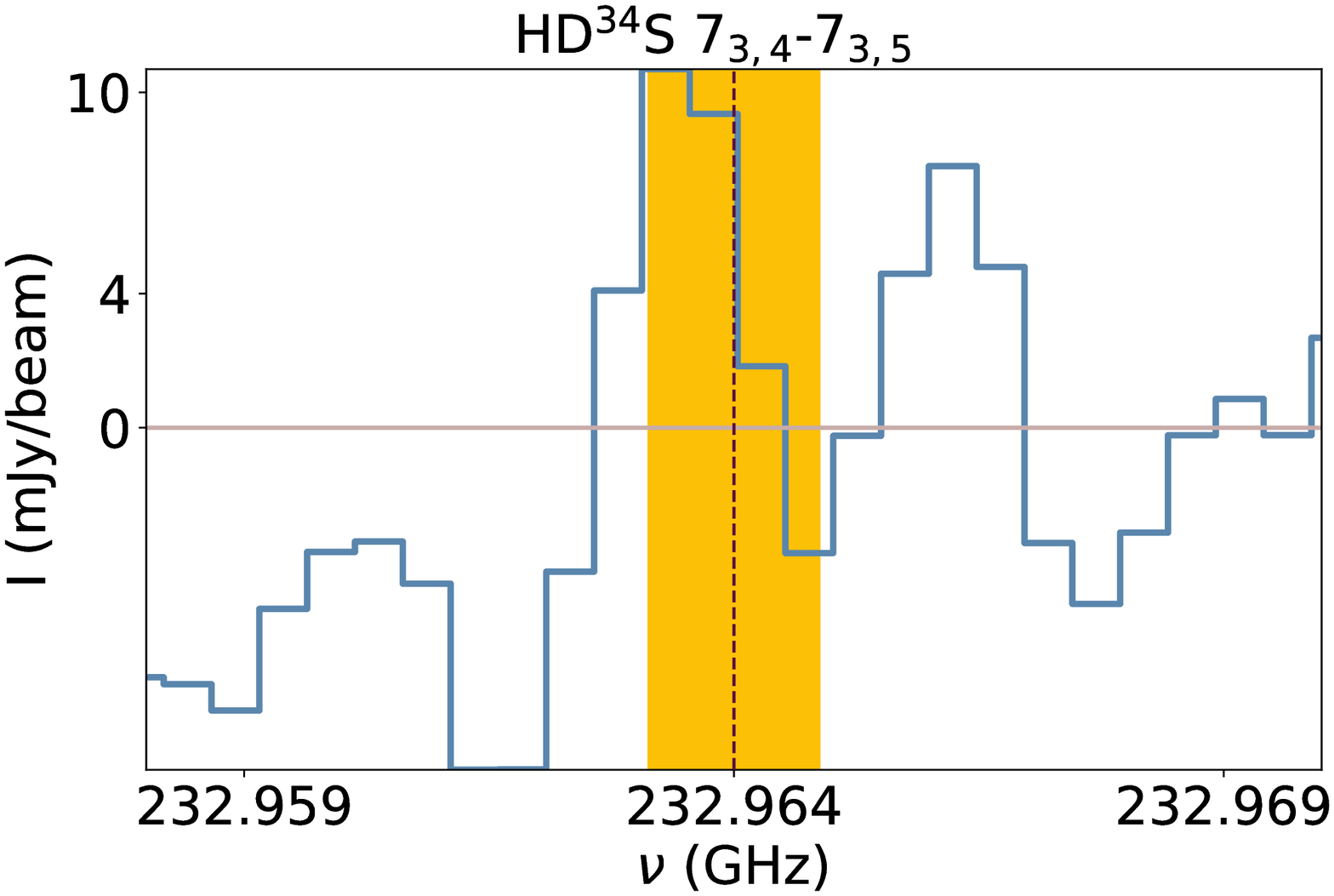}}
    \subfigure{\includegraphics[width=2.0in]{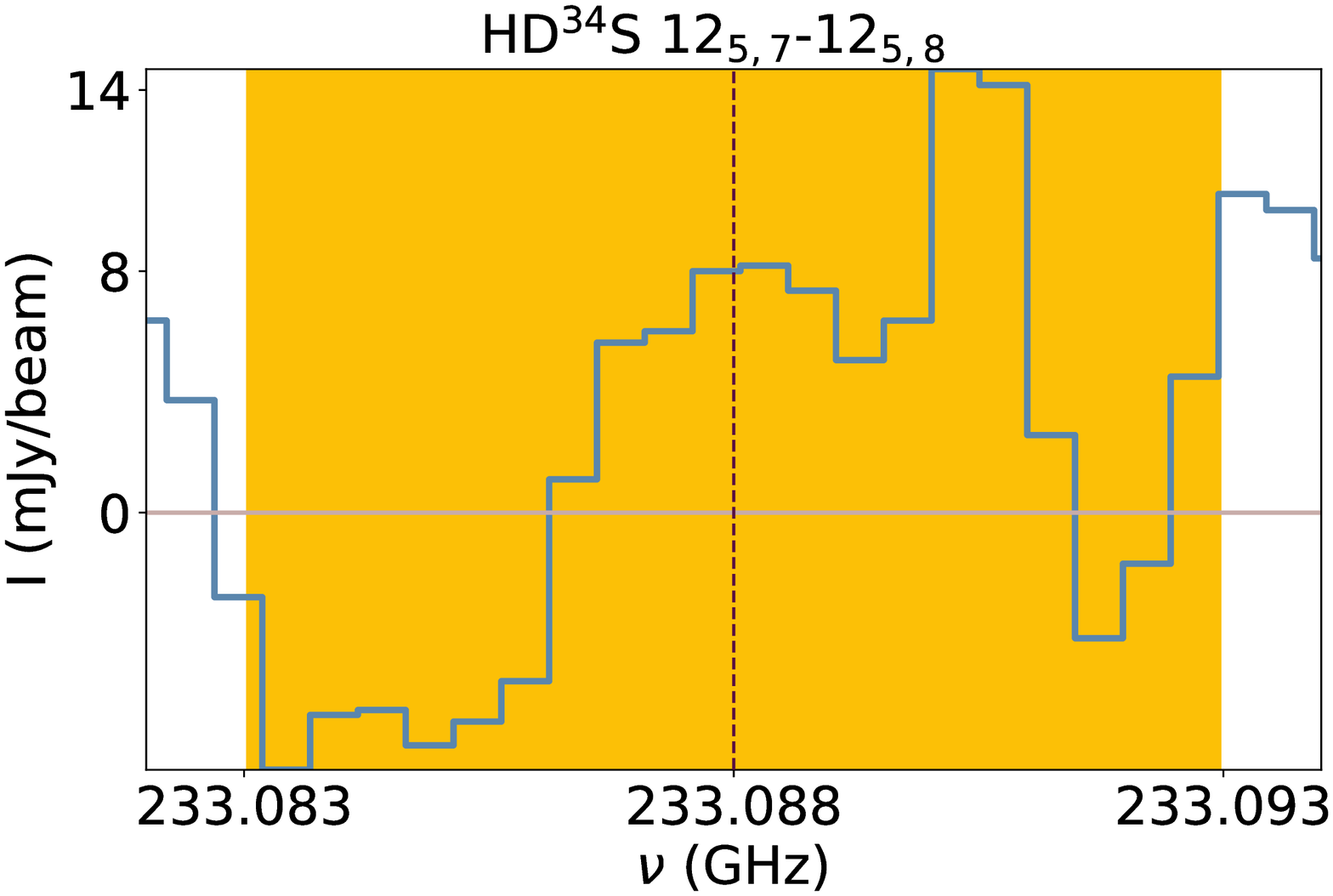}}\\
    \subfigure{\includegraphics[width=2.0in]{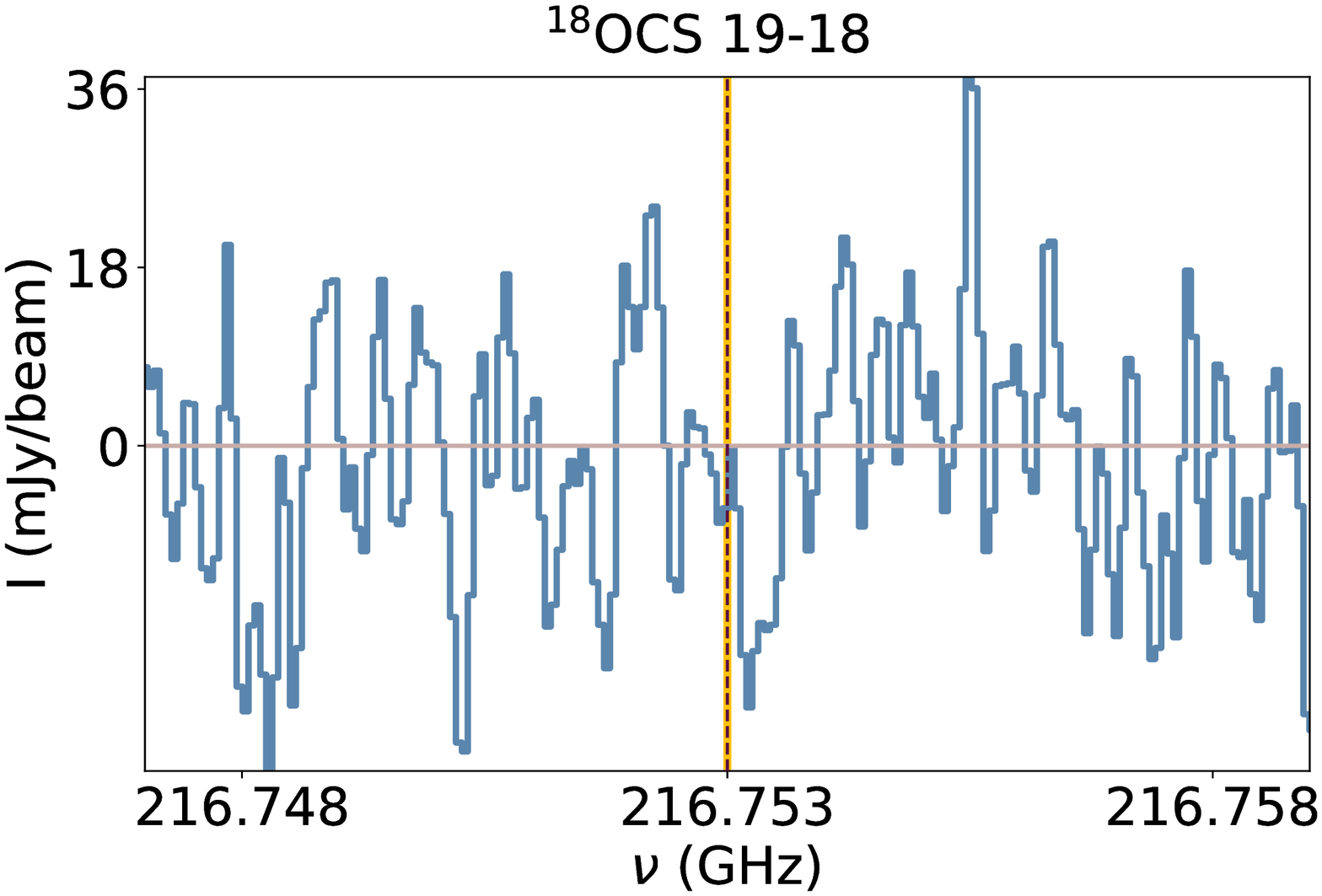}}
    \subfigure{\includegraphics[width=2.0in]{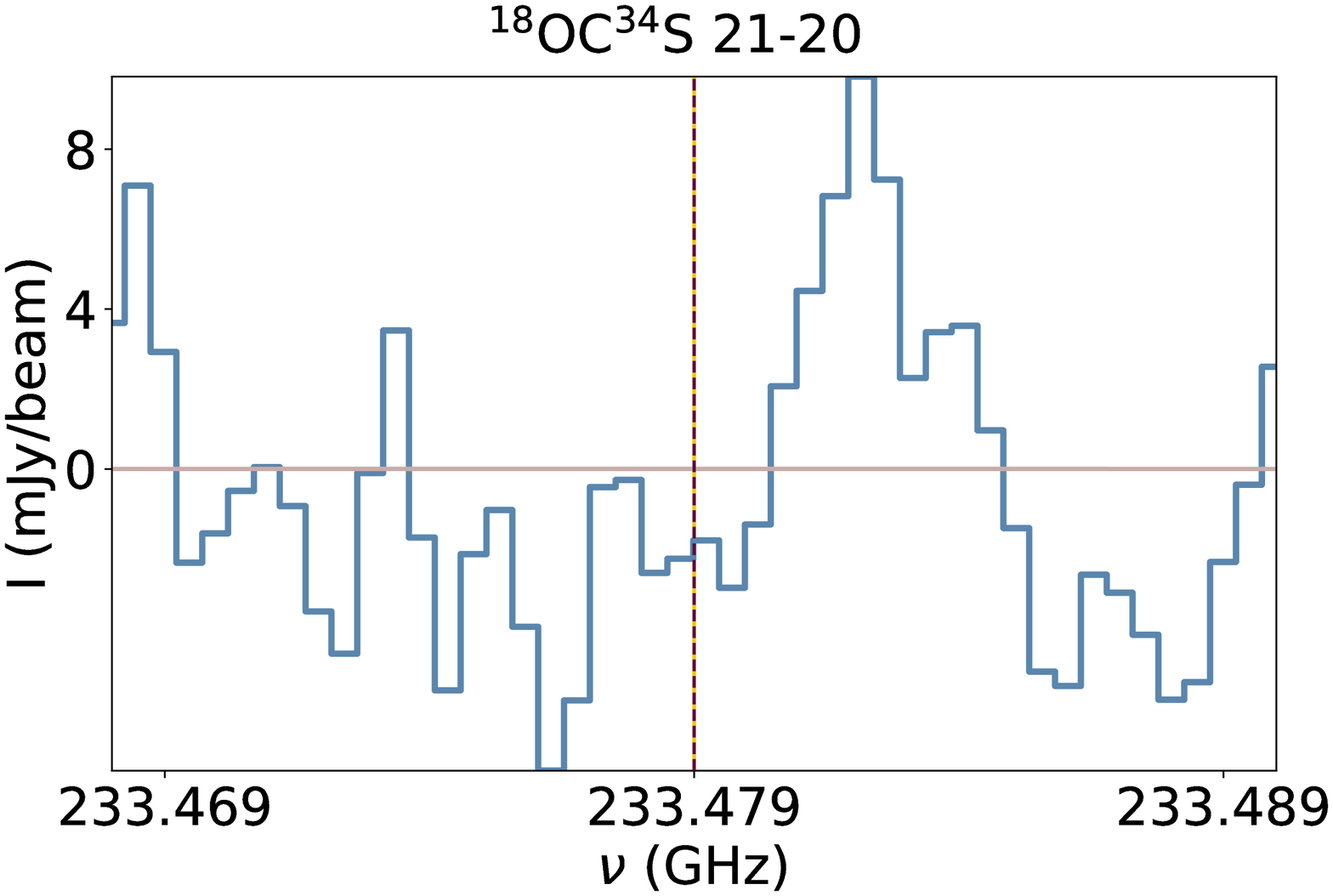}}
    \subfigure{\includegraphics[width=2.0in]{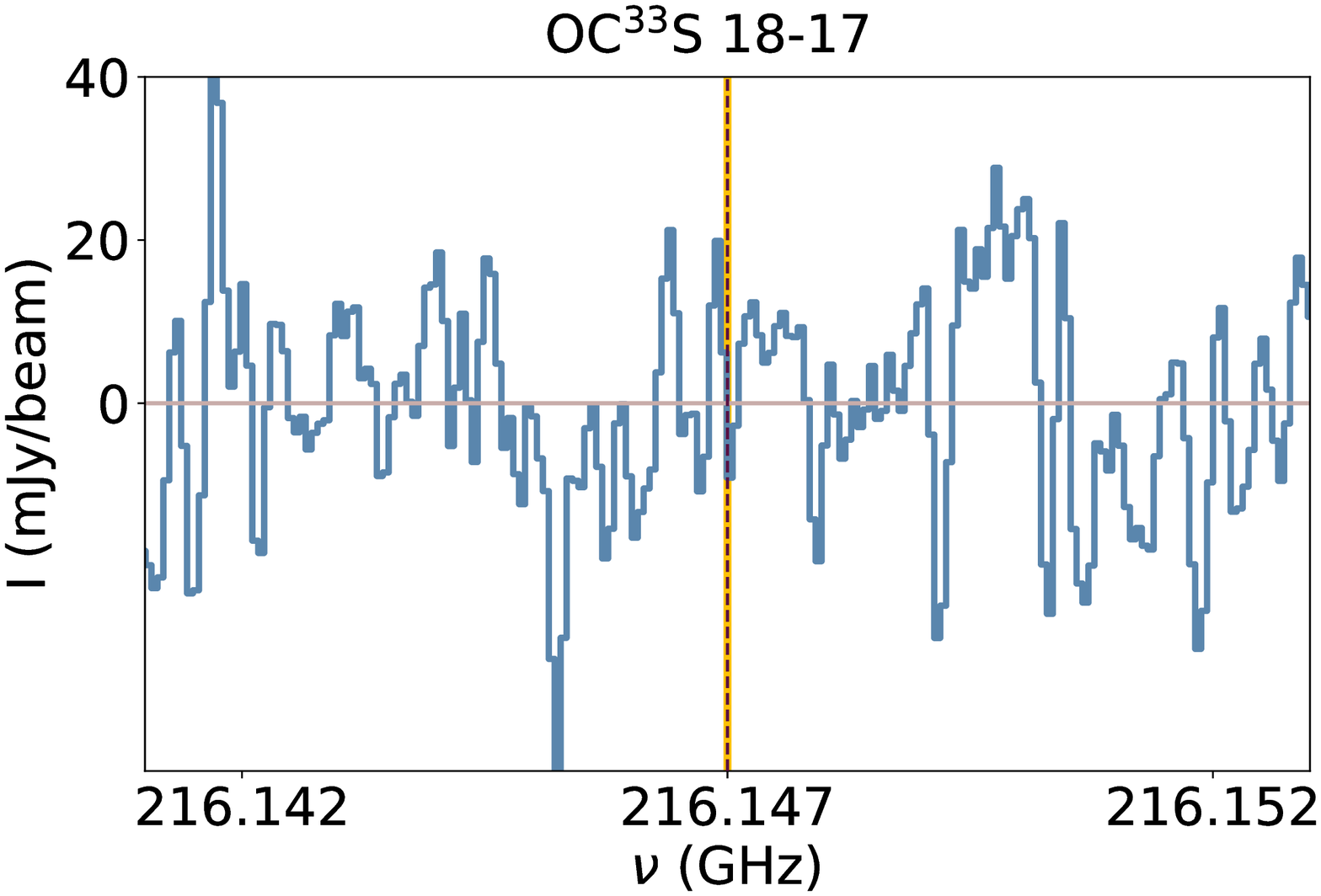}}\\
    \subfigure{\includegraphics[width=2.0in]{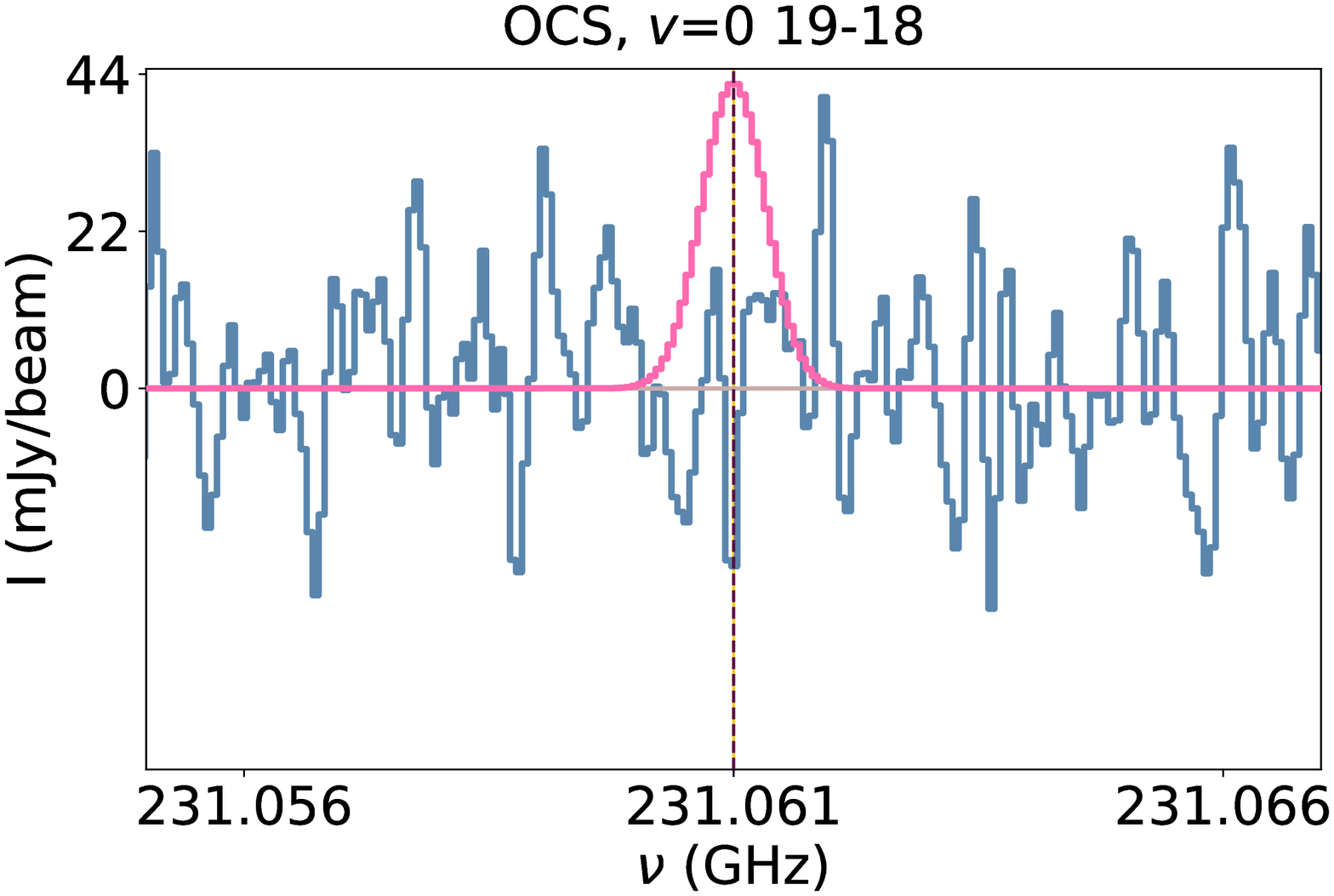}}
    \subfigure{\includegraphics[width=2.0in]{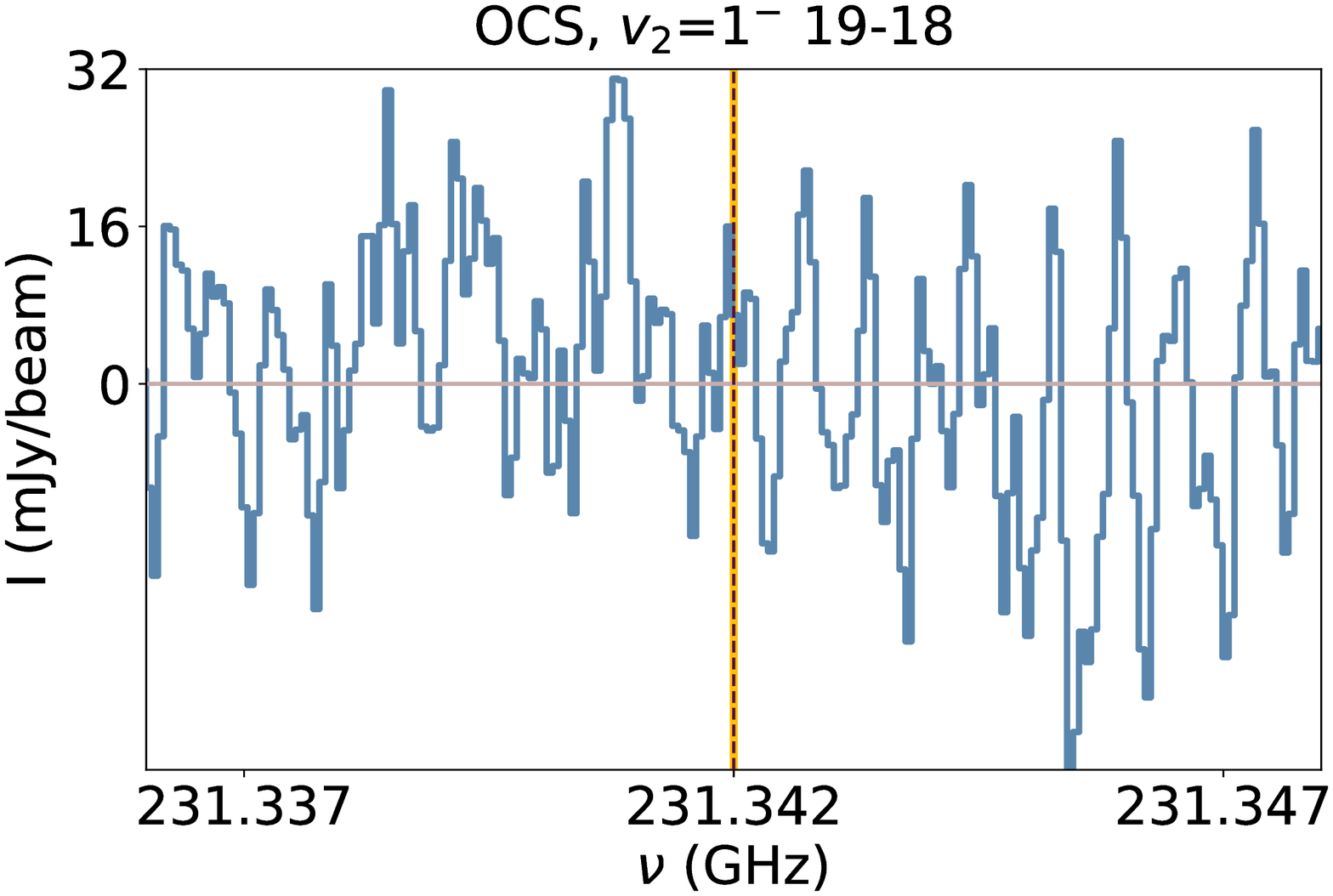}}
    \subfigure{\includegraphics[width=2.0in]{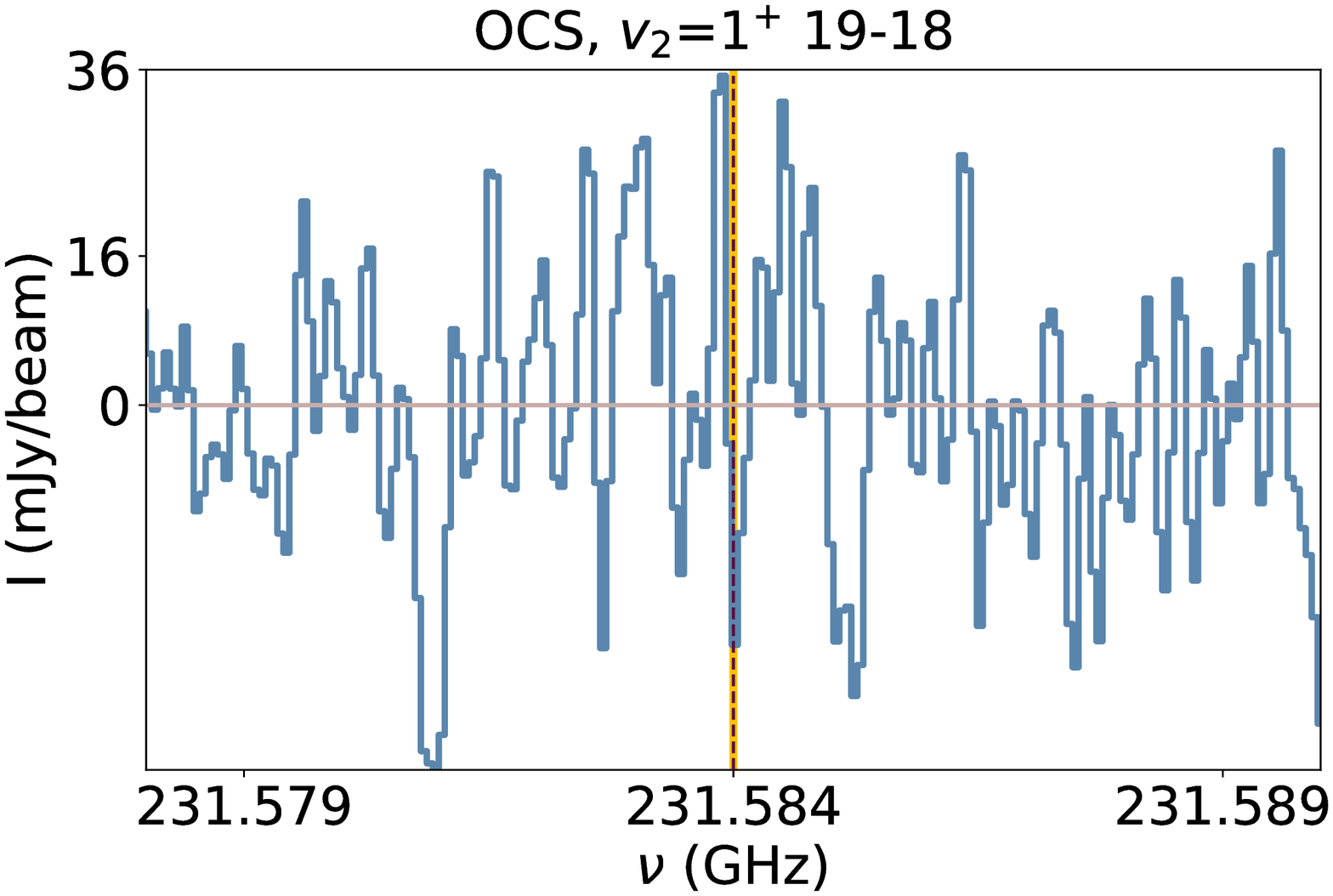}}\\
    \subfigure{\includegraphics[width=2.0in]{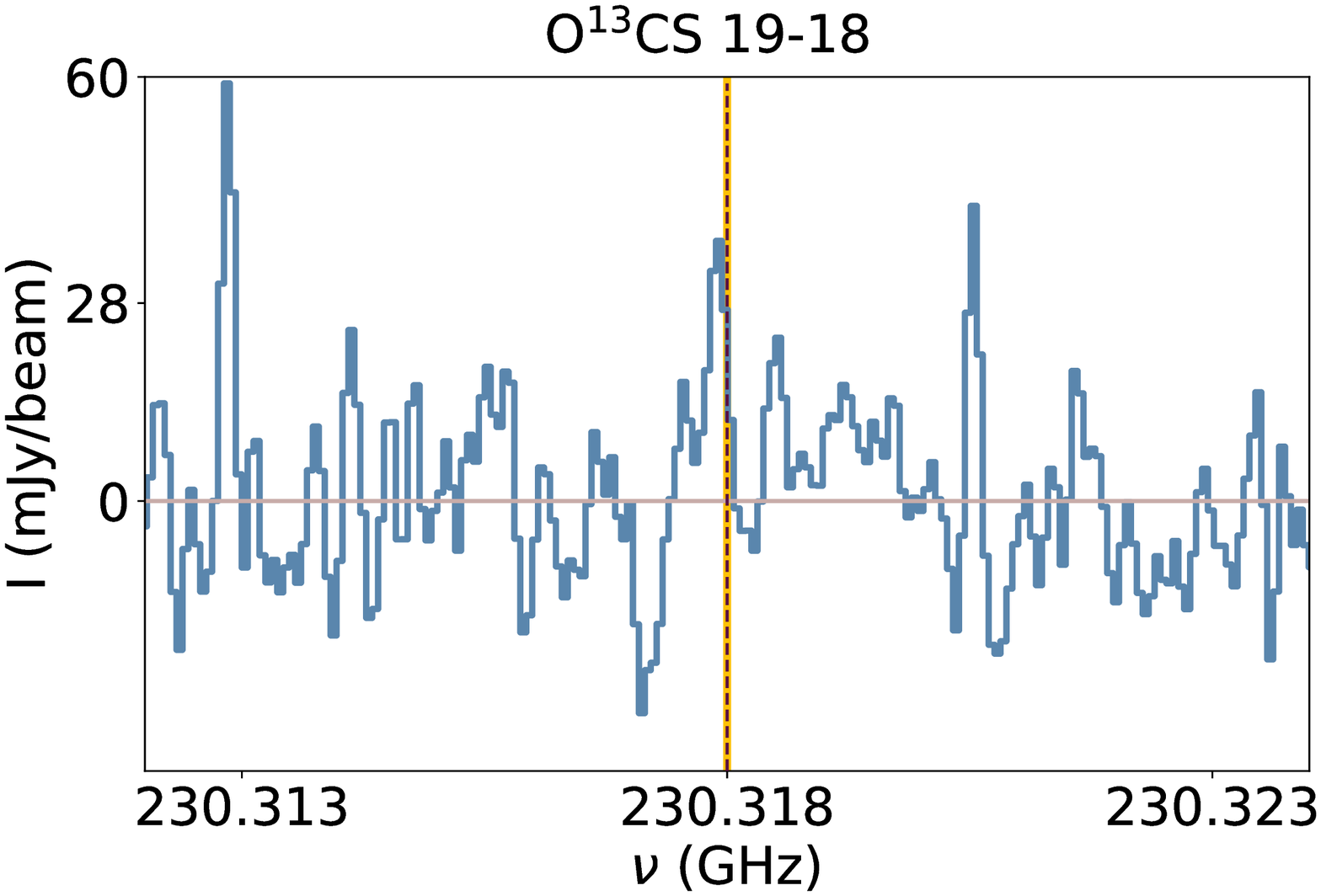}}
\end{minipage}
\caption{Observed spectra (in blue), rest frequency of the undetected line (brown dashed line), and spectroscopic uncertainty on the rest frequency of the undetected line (yellow shaded region) plotted for the sulfur-bearing species undetected towards RCrA IRS7B.  Synthetic spectrum (in pink) fitted to the OCS, $v$=0 line with the 1-$\sigma$ upper limit on its column density.}
\label{undetected_IRS7B}
\end{figure}

\section{B1-c}

H$_2$S, 2$_{2,0}$-2$_{1,1}$ and OCS, $v$=0, $J = 19-18$ lines are detected towards B1-c. Unfortunately, no optically thin isotopologs of H$_2$S and OCS were observed towards B1-c in this observational data set. Synthetic spectra for the detected H$_2$S and OCS lines are fitted for two components of B1-c, a cold component ($T_{\text{ex}}=60$ K) and a warm central component ($T_{\text{ex}}=200$ K) for a source size of 0.45$''$ (as assumed in \citealt{vanGelder2020} for the same observations). The best-fitting FWHM is 2.2 km s$^{-1}$. The fitting parameters, computed N and $\tau$ values are tabulated in \autoref{results_cold_B1c} and \ref{results_hot_B1c}. For the cold component, the H$_2$S line is optically thick ($\tau$=2.0); whereas, the OCS, $v$=0, $J=19-18$ line is partially optically thick ($\tau=0.7$). For the warm component, both H$_2$S and OCS, $v$=0 lines are partially optically thick ($\tau=0.26$ and $\tau=0.10$, respectively). The undetected lines are shown in \autoref{undetected_B1c}. 

\begin{table}[H]
    \centering
    \caption{Synthetic fitting of the detected S-bearing species towards B1-c for the cold component ($T_{\text{ex}}=60$ K), a FWHM of 2.2 km s$^{-1}$, and a source size of 0.45$''$.}
    \label{results_cold_B1c}
    \begin{tabular}{r r c r c c  c c c r}
    \hline
    \hline
    Species & Transition & Frequency & $E_{\text{up}}$ & $A_{ij}$ &
    Beam size & $N$  & $\tau$ \\
    &&&&&& & \\
    & & (GHz) & (K) & (s$^{-1}$) & ($''$) & (cm$^{-2}$)  &  \\
\hline
& & & & &  & & &  \\
H$_2$S & 2$_{2,0}$-2$_{1,1}$ & 216.710 & 84 & 4.9$\times$10$^{-5}$  & 6.5 &   $>$9.7$\times$10$^{15,}$ \tablefootmark{*} & 2.0\\
OCS, $v$=0 & 19-18 & 231.061 & 111  & 3.6$\times$10$^{-5}$  & 6.1  &  $>$5.0$\times$10$^{15,}$ \tablefootmark{**} &    0.7  \\
\hline
\hline
    \end{tabular}
    \tablefoot{\tablefoottext{*}{optically thick}, \tablefoottext{**}{partially optically thick}}
\end{table}

\begin{table}[H]
    \centering
    \caption{Synthetic fitting of the detected S-bearing species towards B1-c for the warm component ($T_{\text{ex}}=200$ K), a FWHM of 2.2 km s$^{-1}$, and a source size of 0.45$''$.}
    \label{results_hot_B1c}
    \begin{tabular}{r r c r c c c c c r}
    \hline
    \hline
    Species & Transition & Frequency & $E_{\text{up}}$ & $A_{ij}$ &
    Beam size & $N$ & $\tau$ \\
    &&&&&& & \\
    & & (GHz) & (K) & (s$^{-1}$) & ($''$) & (cm$^{-2}$) &  \\
\hline
& & & & &  & & &  \\
H$_2$S & 2$_{2,0}$-2$_{1,1}$ & 216.710 & 84 & 4.9$\times$10$^{-5}$  & 6.5 &  $>$1.2$\times$10$^{16, }$ \tablefootmark{*}  & 0.26\\
OCS, $v$=0 & 19-18  & 231.061 & 111  & 3.6$\times$10$^{-5}$  & 6.1 & $>$3.6$\times$10$^{15,}$ \tablefootmark{*}  & 0.10 \\
\hline
\hline
    \end{tabular}
    \tablefoot{\tablefoottext{*}{partially optically thick}}
\end{table}

\subsection{Detected lines in B1-c}

\begin{figure}[H]
\centering
\begin{minipage}[b]{6.2in}%
    \subfigure{\includegraphics[width=3.0in]{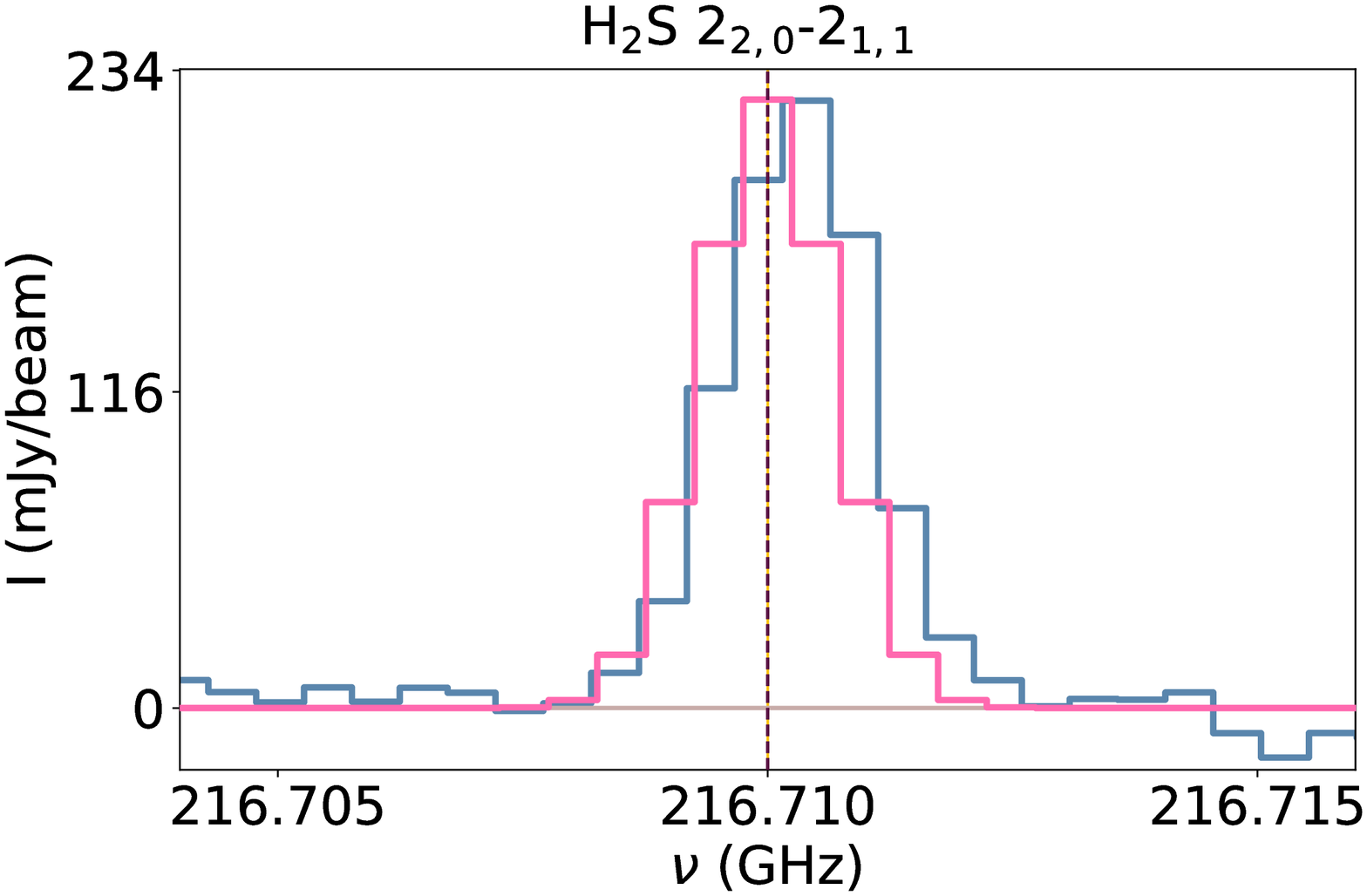}}
    \subfigure{\includegraphics[width=3.0in]{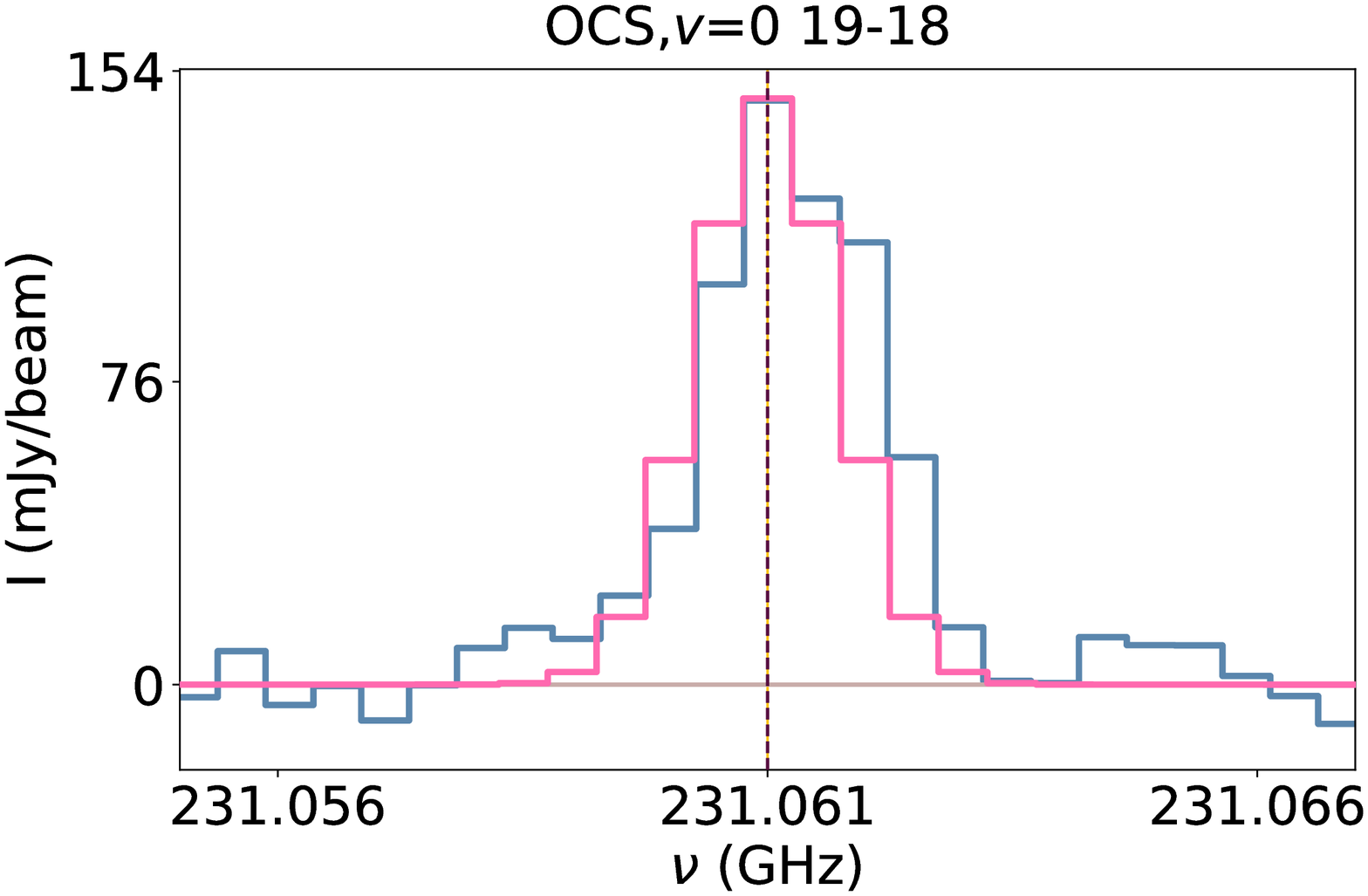}}
\end{minipage}
\caption{Observed spectra (in blue), rest frequency of the detected line (brown dashed line), spectroscopic uncertainty on the rest frequency of the detected line (yellow shaded region), and fitted synthetic spectra (in pink) plotted for the sulfur-bearing species detected towards the cold component ($T_{\text{ex}}=60$ K) of B1-c.}
\label{detected_cold_B1c}
\end{figure}

\begin{figure}[H]
\centering
\begin{minipage}[b]{6.2in}%

    \subfigure{\includegraphics[width=3.0in]{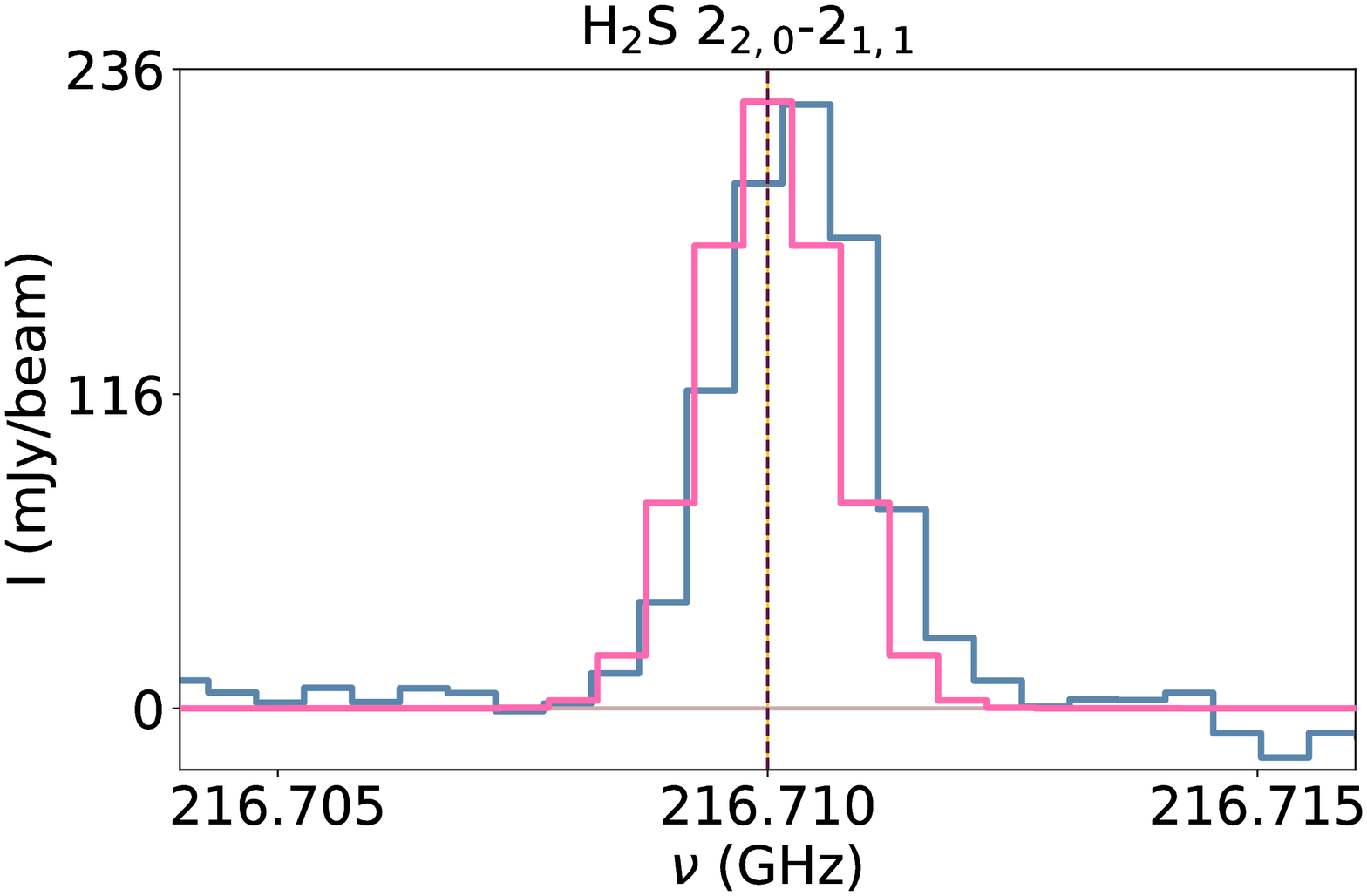}}
    \subfigure{\includegraphics[width=3.0in]{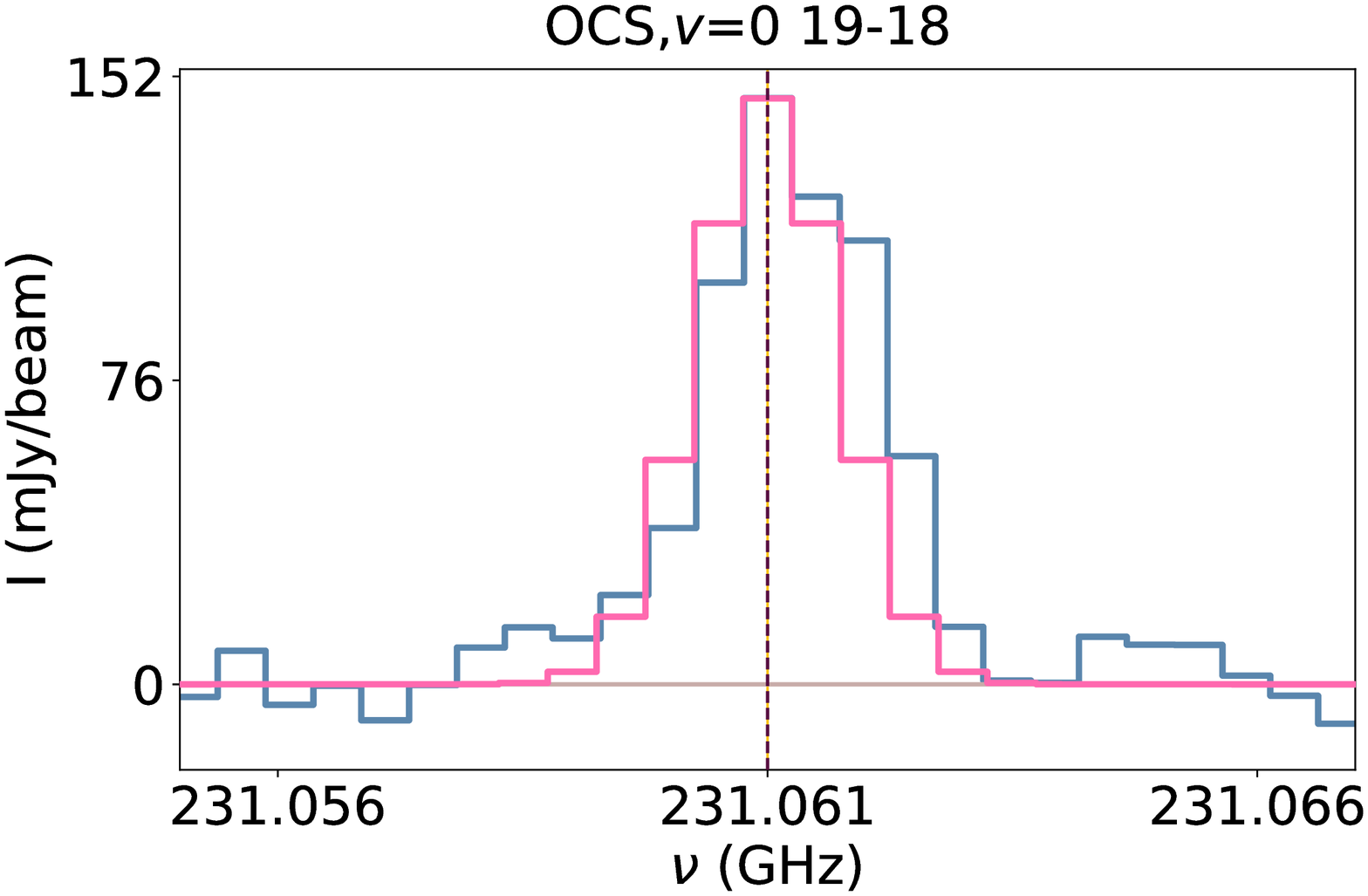}}
\end{minipage}
\caption{Observed spectra (in blue), rest frequency of the detected line (brown dashed line), spectroscopic uncertainty on the rest frequency of the detected line (yellow shaded region), and fitted synthetic spectra (in pink) plotted for the sulfur-bearing species detected towards the warm component ($T_{\text{ex}}=200$ K) of B1-c.}
\label{detected_warm_B1c}
\end{figure}

\subsection{Undetected lines in B1-c}

\begin{figure}[H]
\centering
\begin{minipage}[b]{6.2in}%
    \subfigure{\includegraphics[width=2.0in]{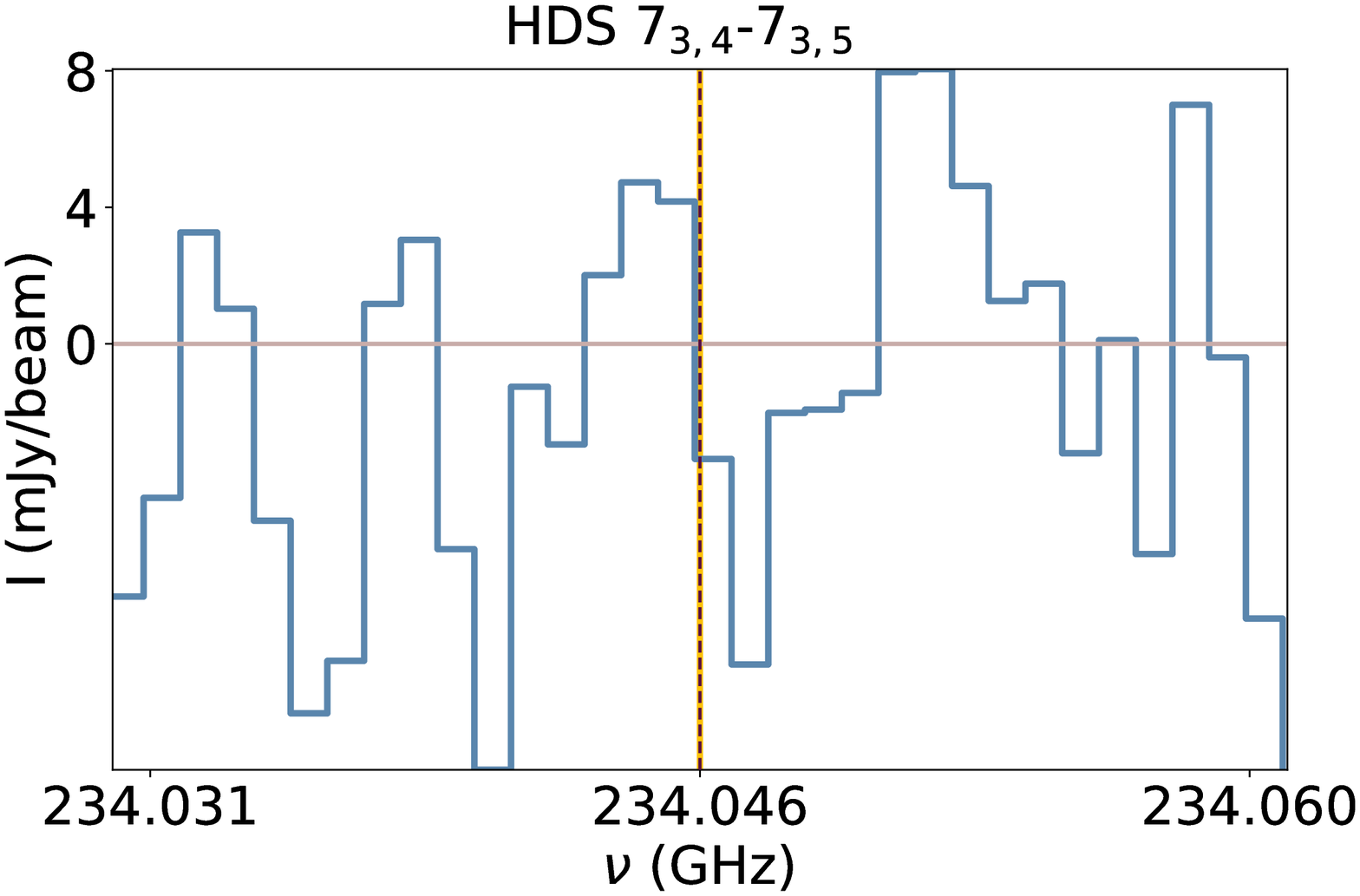}}
    \subfigure{\includegraphics[width=2.0in]{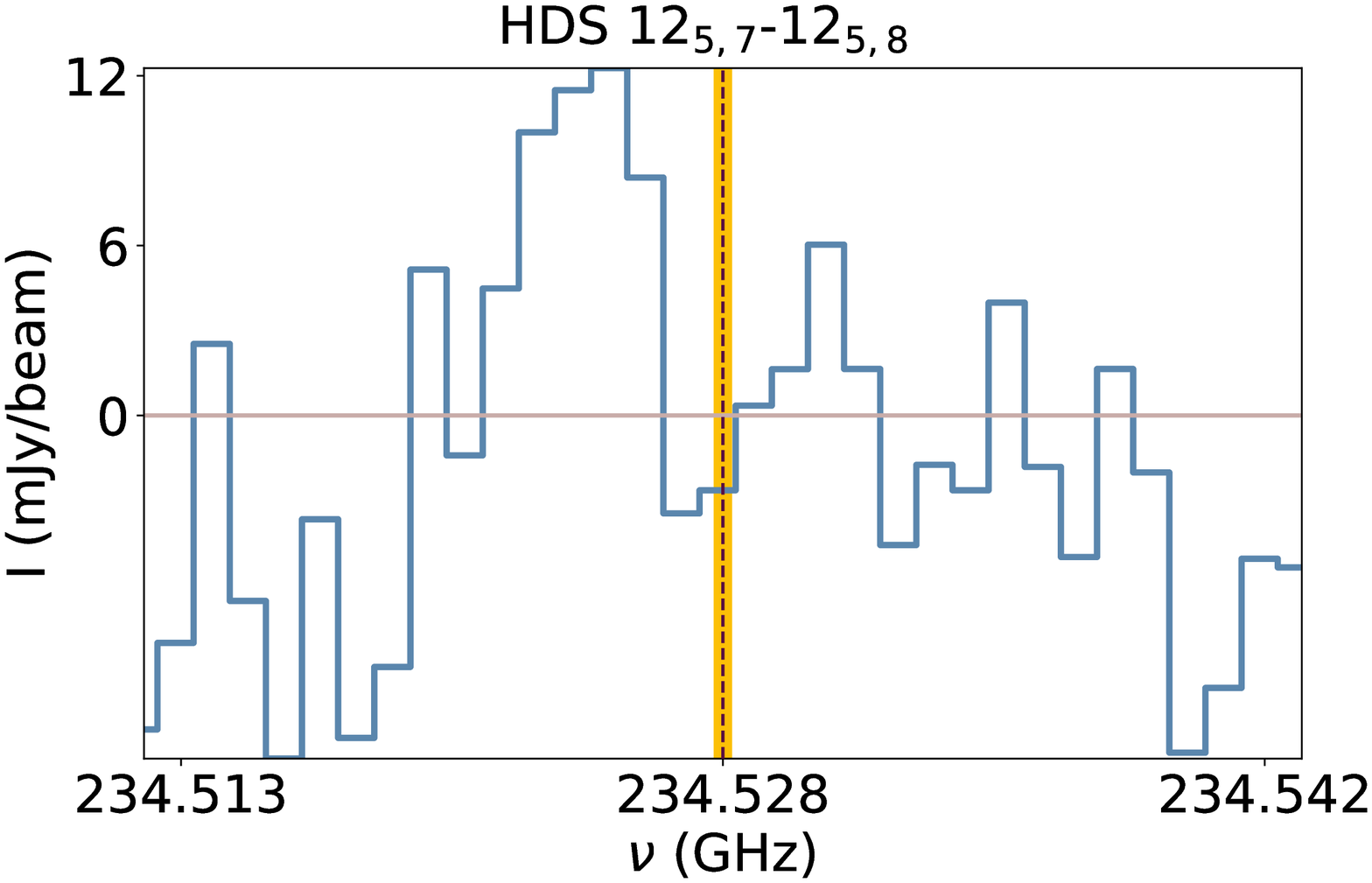}}
    \subfigure{\includegraphics[width=2.0in]{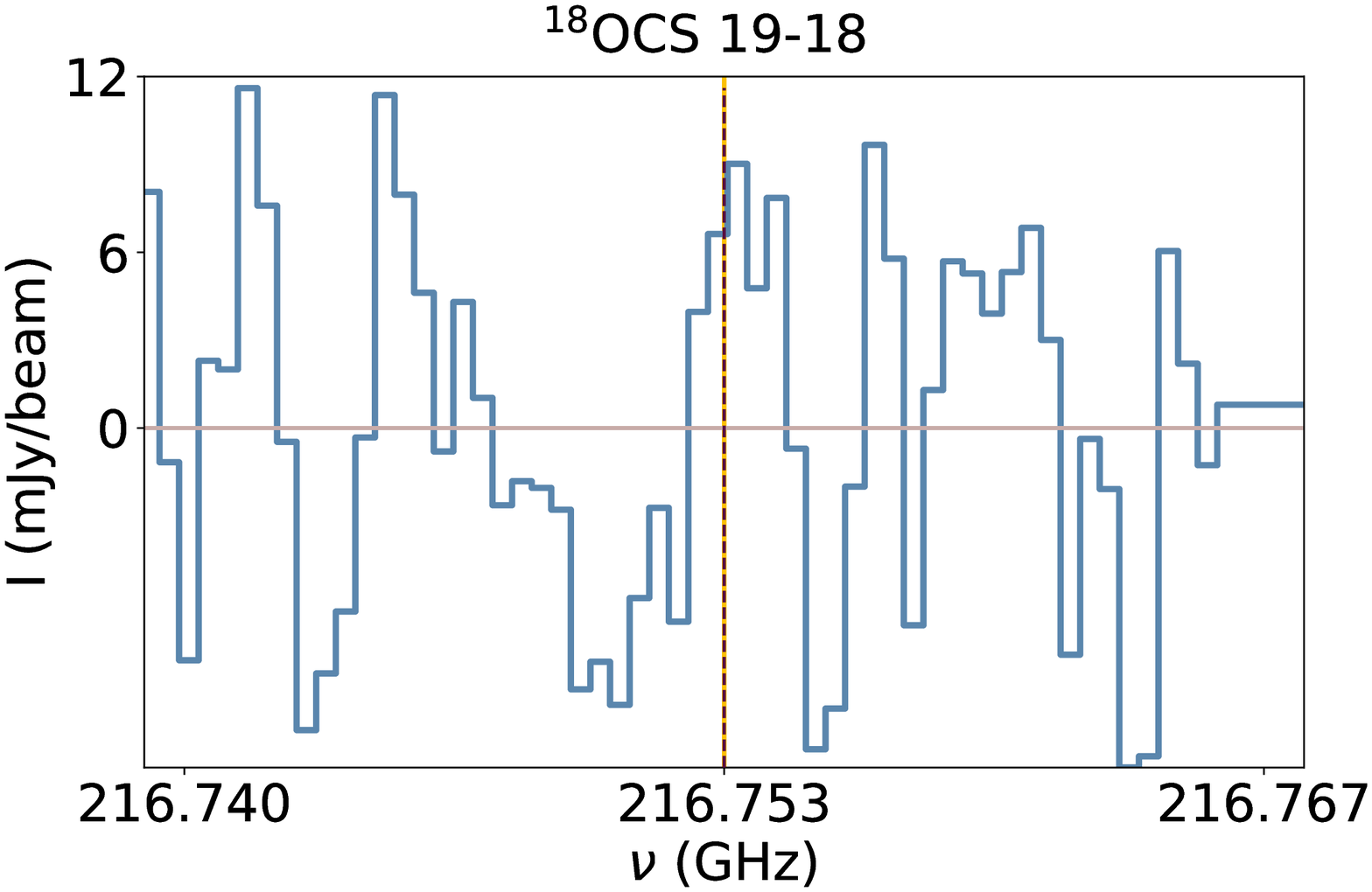}}\\
    \subfigure{\includegraphics[width=2.0in]{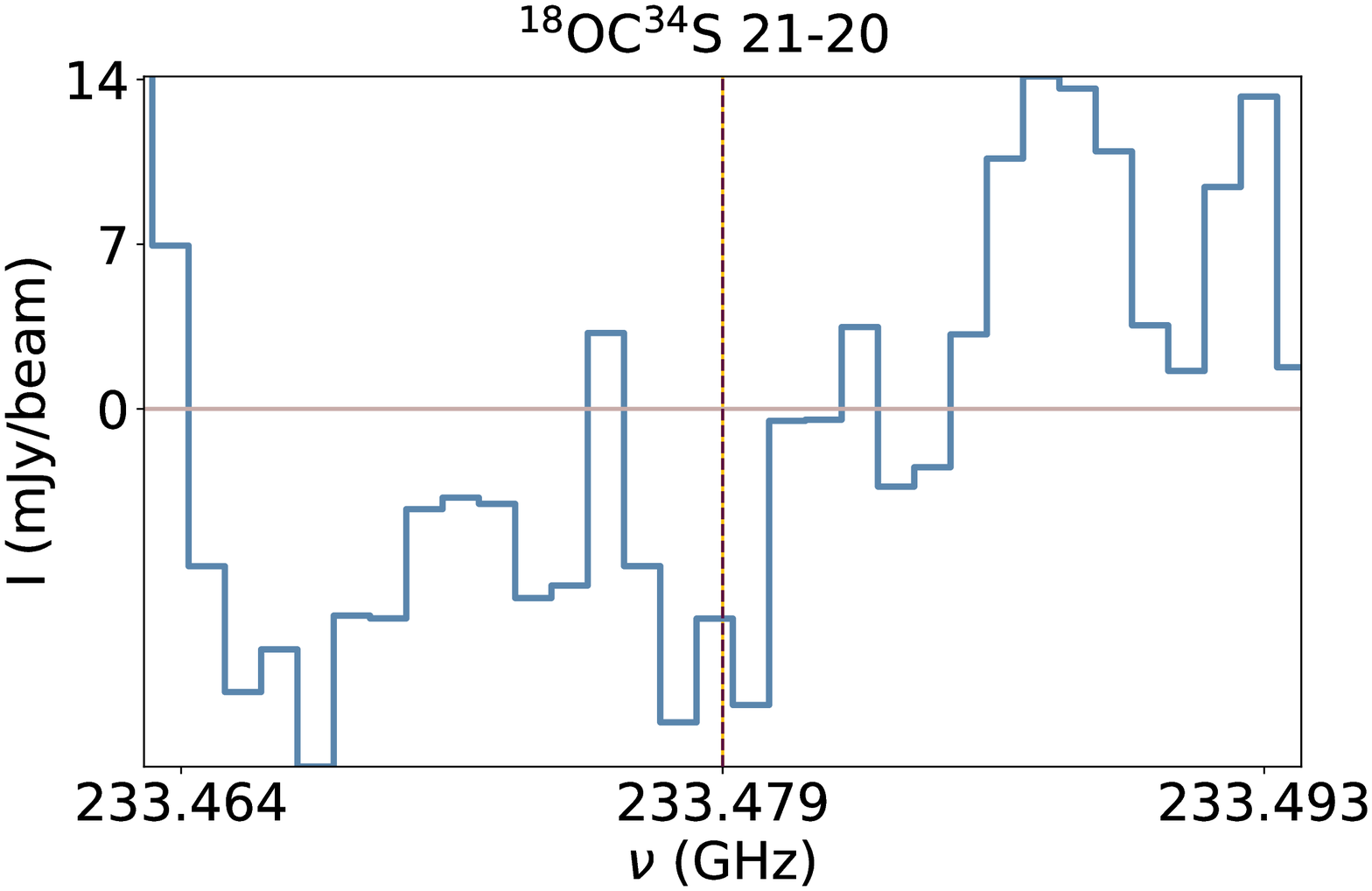}}
    \subfigure{\includegraphics[width=2.0in]{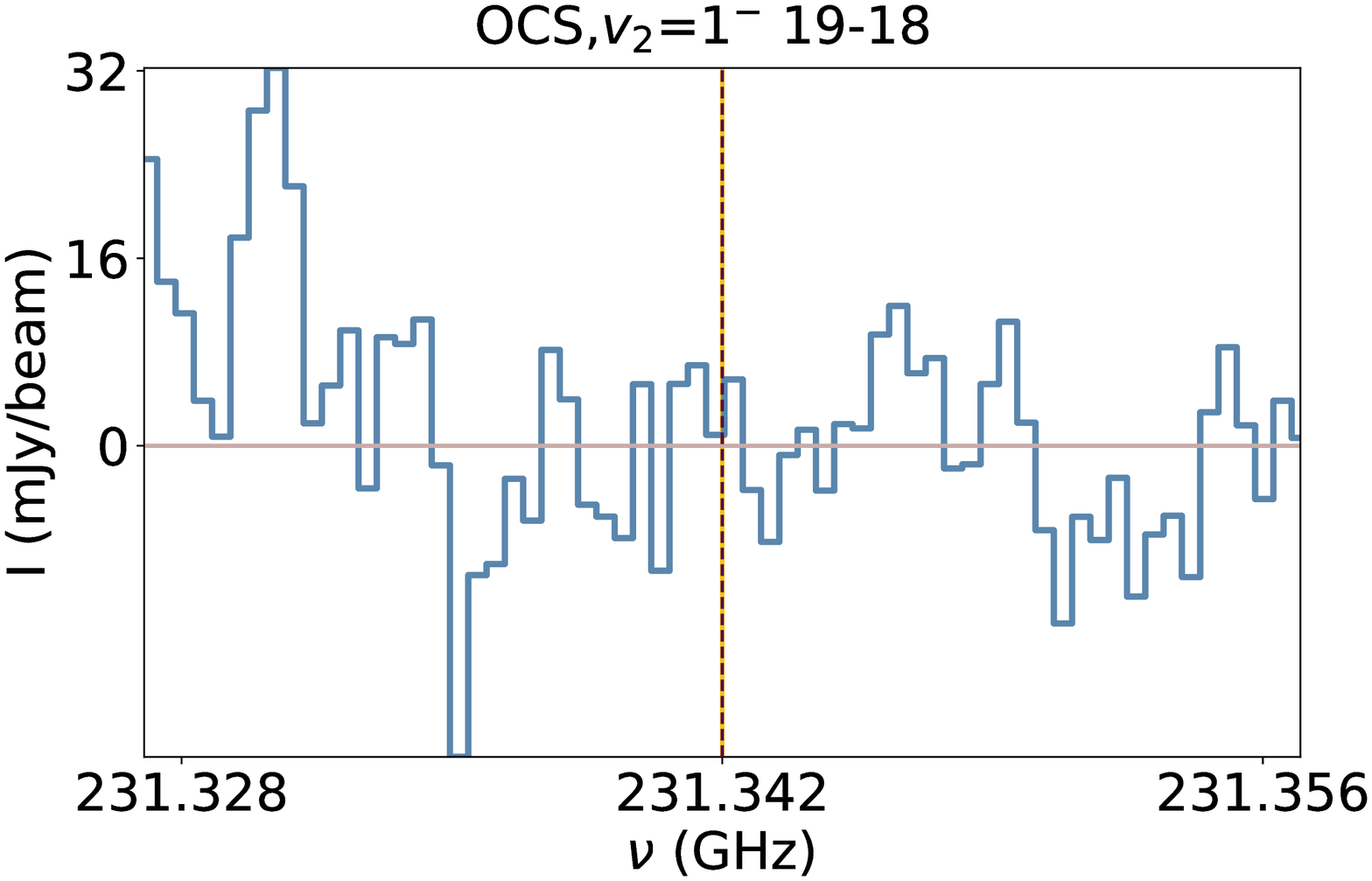}}
    \subfigure{\includegraphics[width=2.0in]{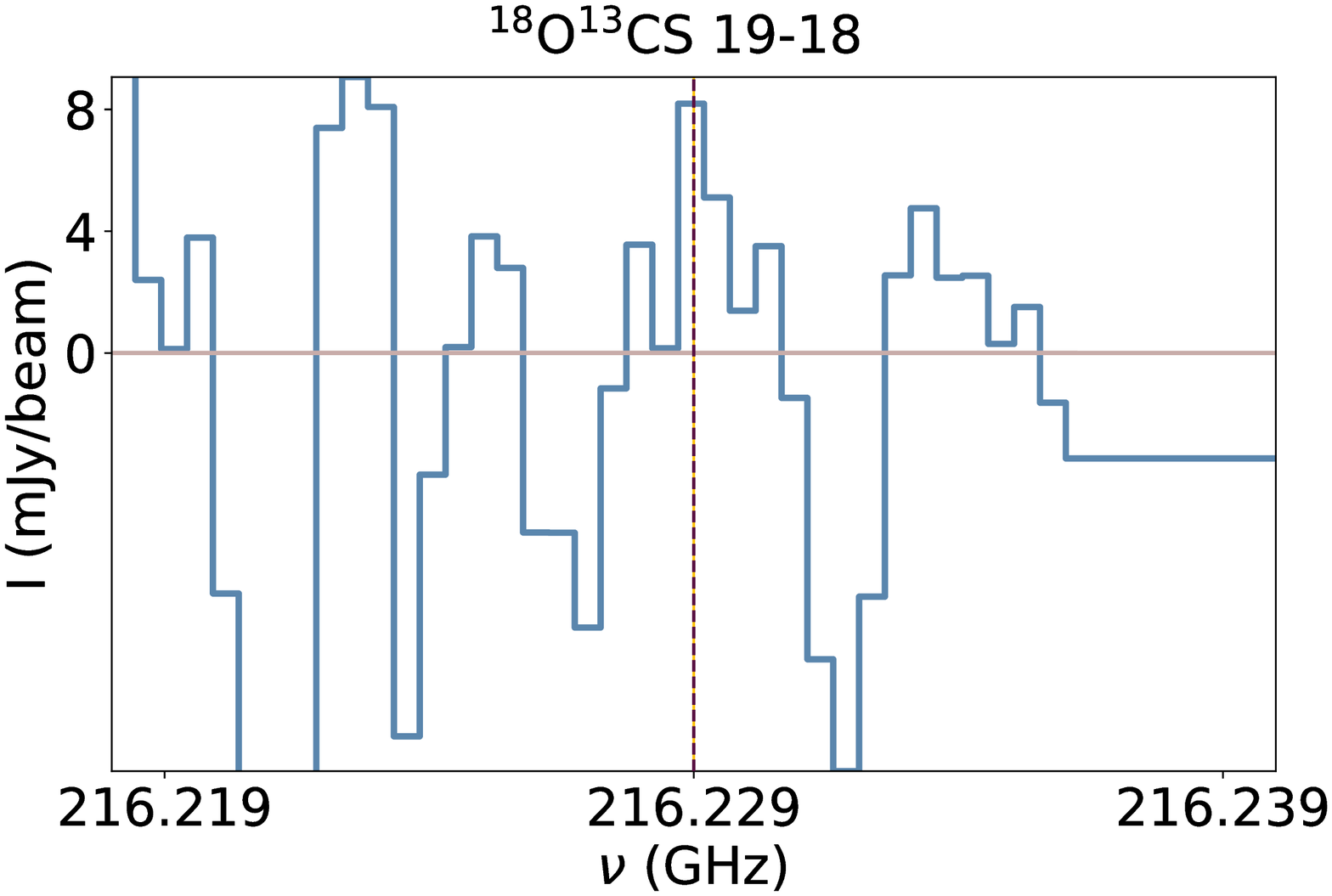}}
    \subfigure{\includegraphics[width=2.0in]{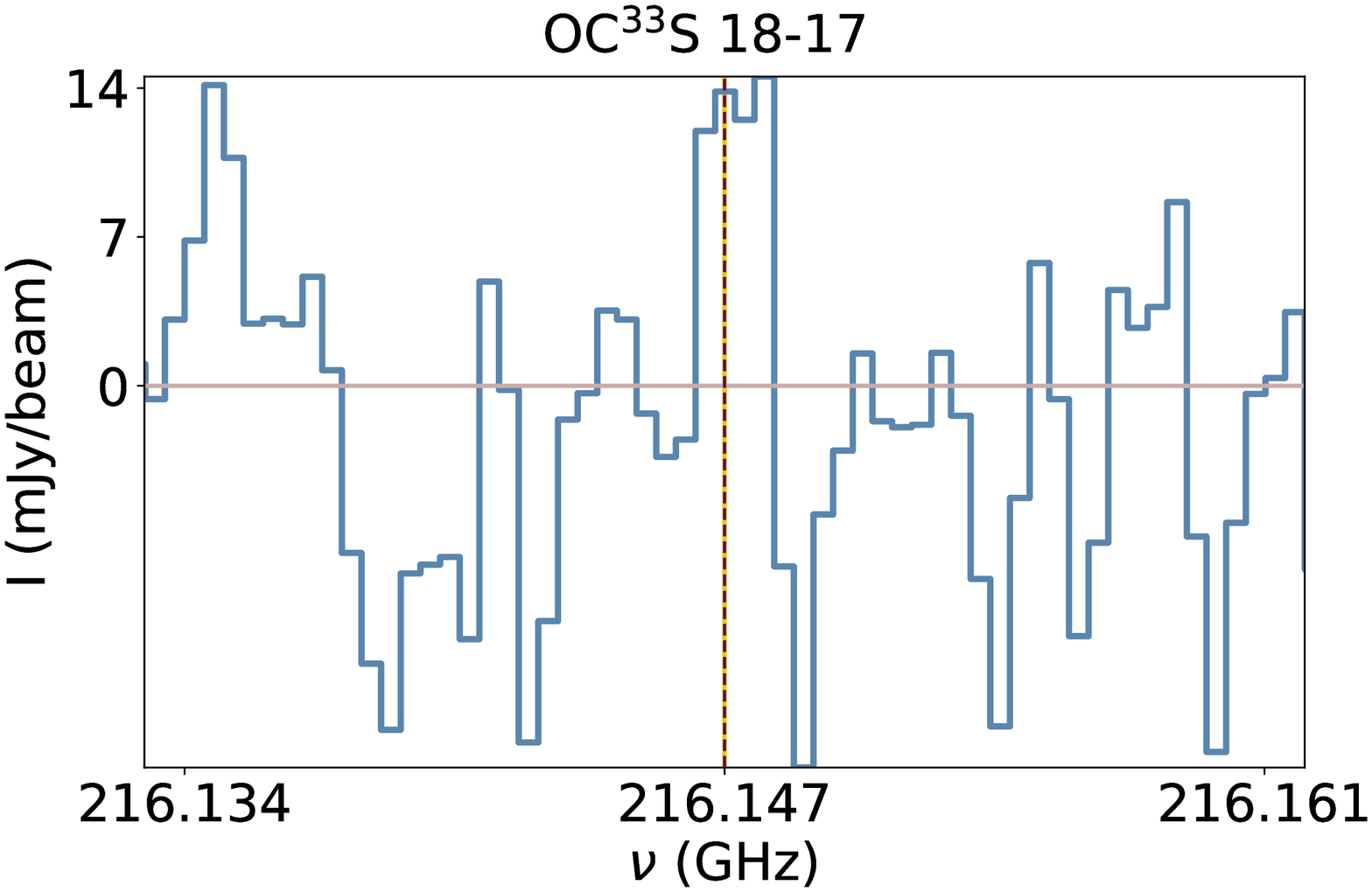}}    
\end{minipage}
\caption{Observed spectra (in blue), rest frequency of the undetected line (brown dashed line), and spectroscopic uncertainty on the rest frequency of the undetected line (yellow shaded region) plotted for the sulfur-bearing species undetected towards B1-c.}
\label{undetected_B1c}
\end{figure}

\section{BHR71-IRS1}

\begin{figure}[H]
\centering
\begin{minipage}[b]{6.2in}%
    \subfigure{\includegraphics[width=3.0in]{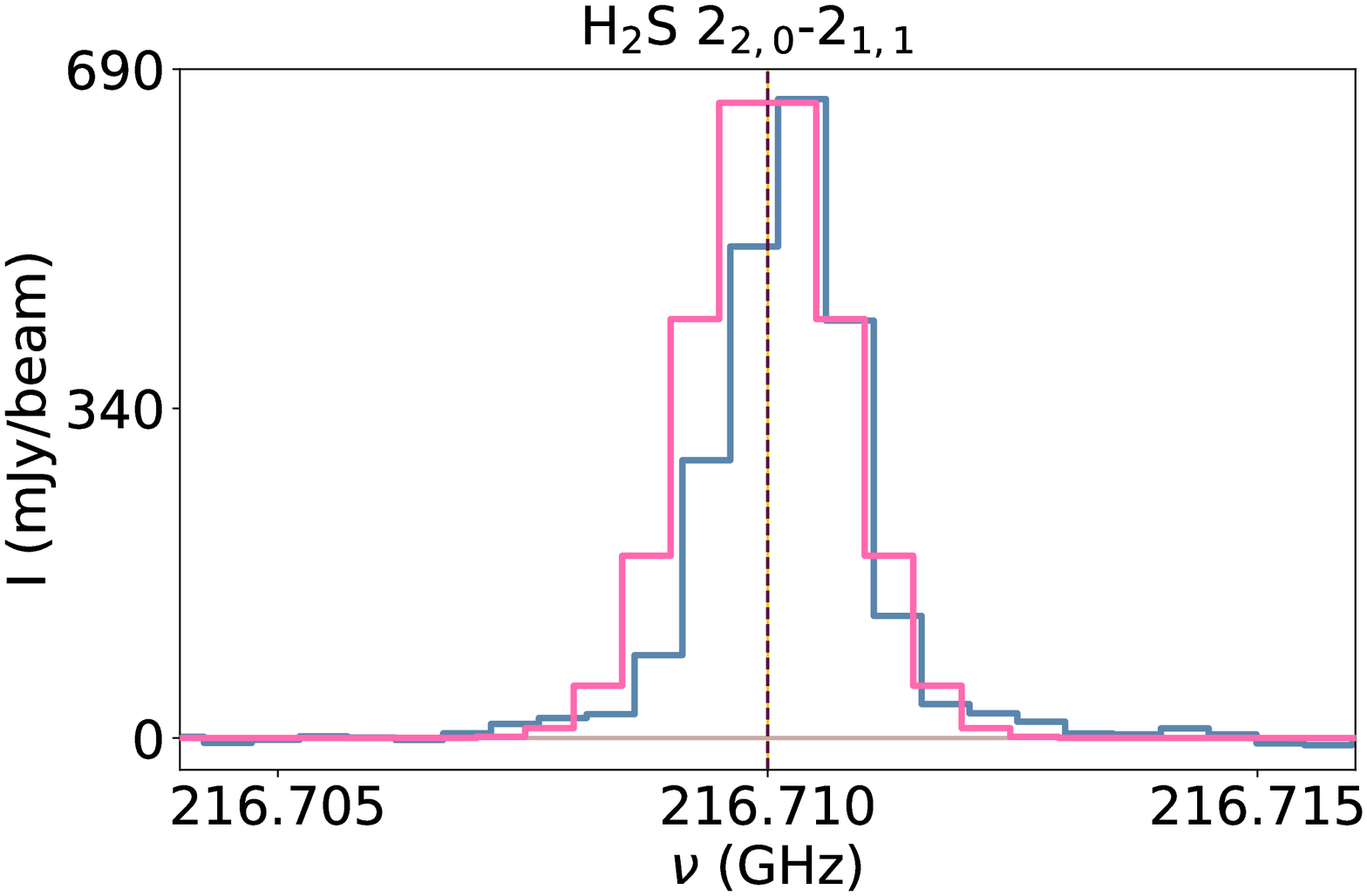}}    \subfigure{\includegraphics[width=3.0in]{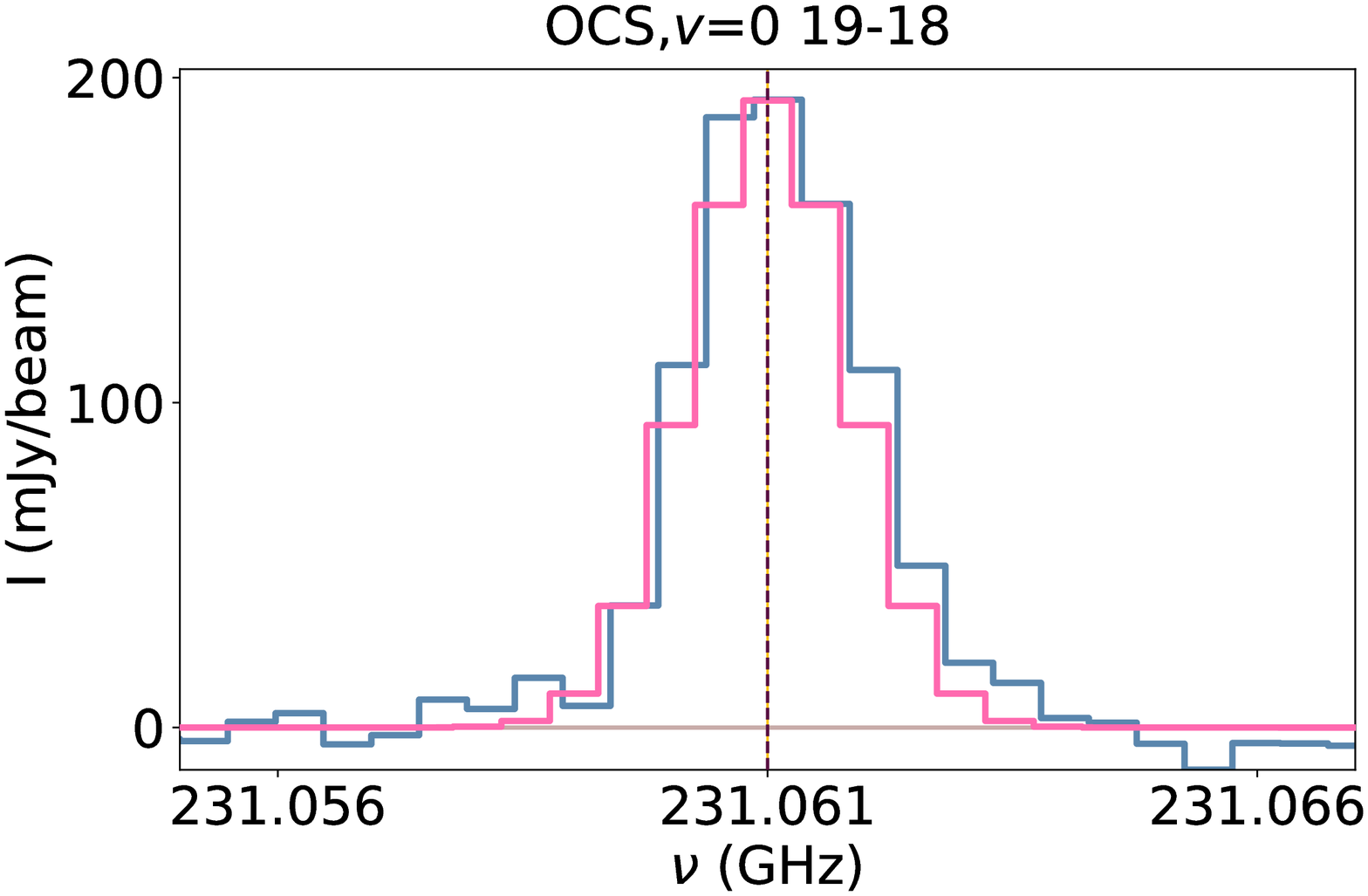}}
\end{minipage}
\caption{Observed spectra (in blue), rest frequency of the detected line (brown dashed line), spectroscopic uncertainty on the rest frequency of the detected line (yellow shaded region), and fitted synthetic spectra (in pink) plotted for the sulfur-bearing species detected towards the cold component ($T_{\text{ex}}=100$ K) of BHR71-IRS1.}
\label{detected_cold_BHR71}
\end{figure}

\begin{figure}[H]
\centering
\begin{minipage}[b]{6.2in}%

    \subfigure{\includegraphics[width=3.0in]{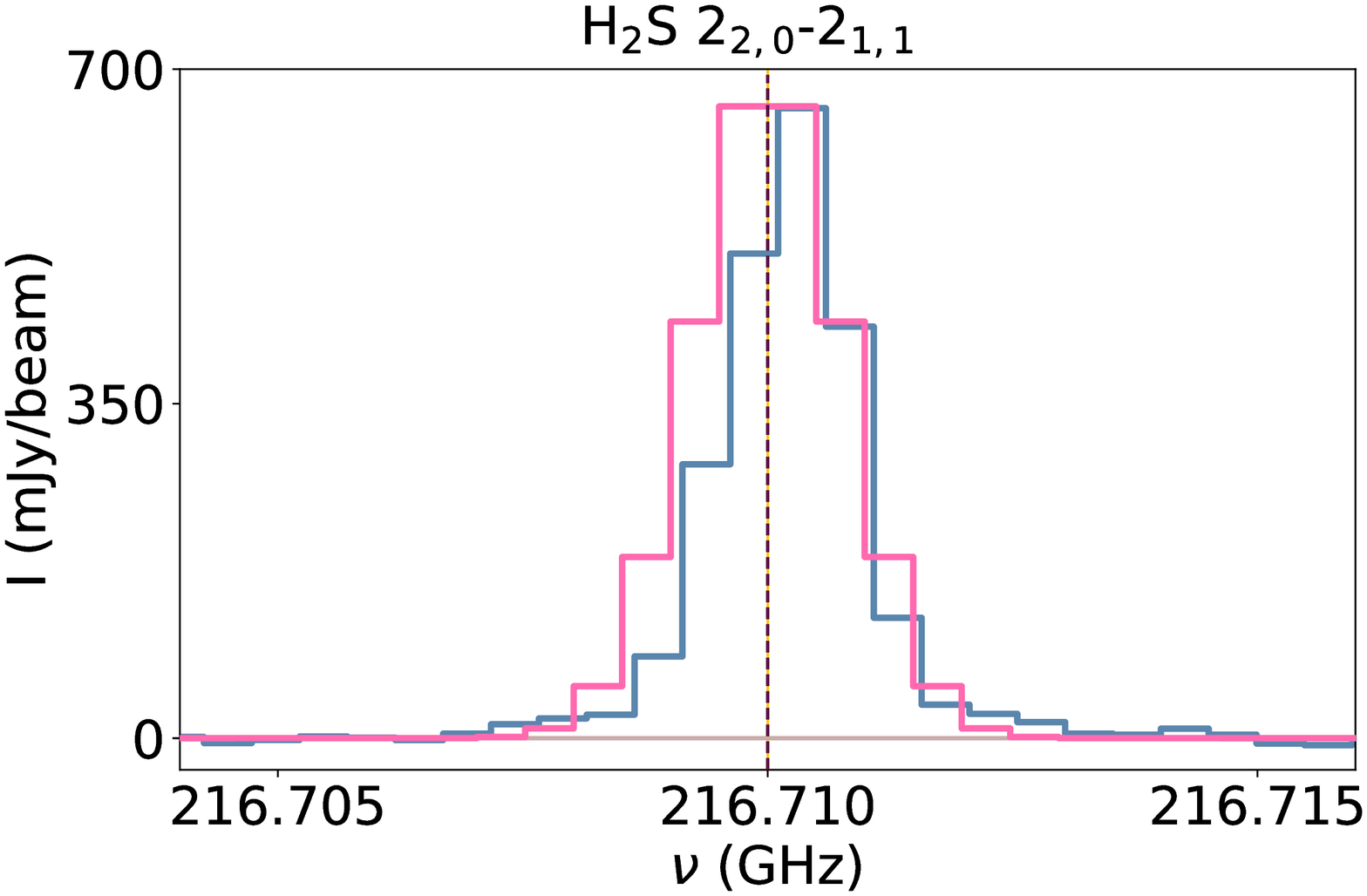}}
    \subfigure{\includegraphics[width=3.0in]{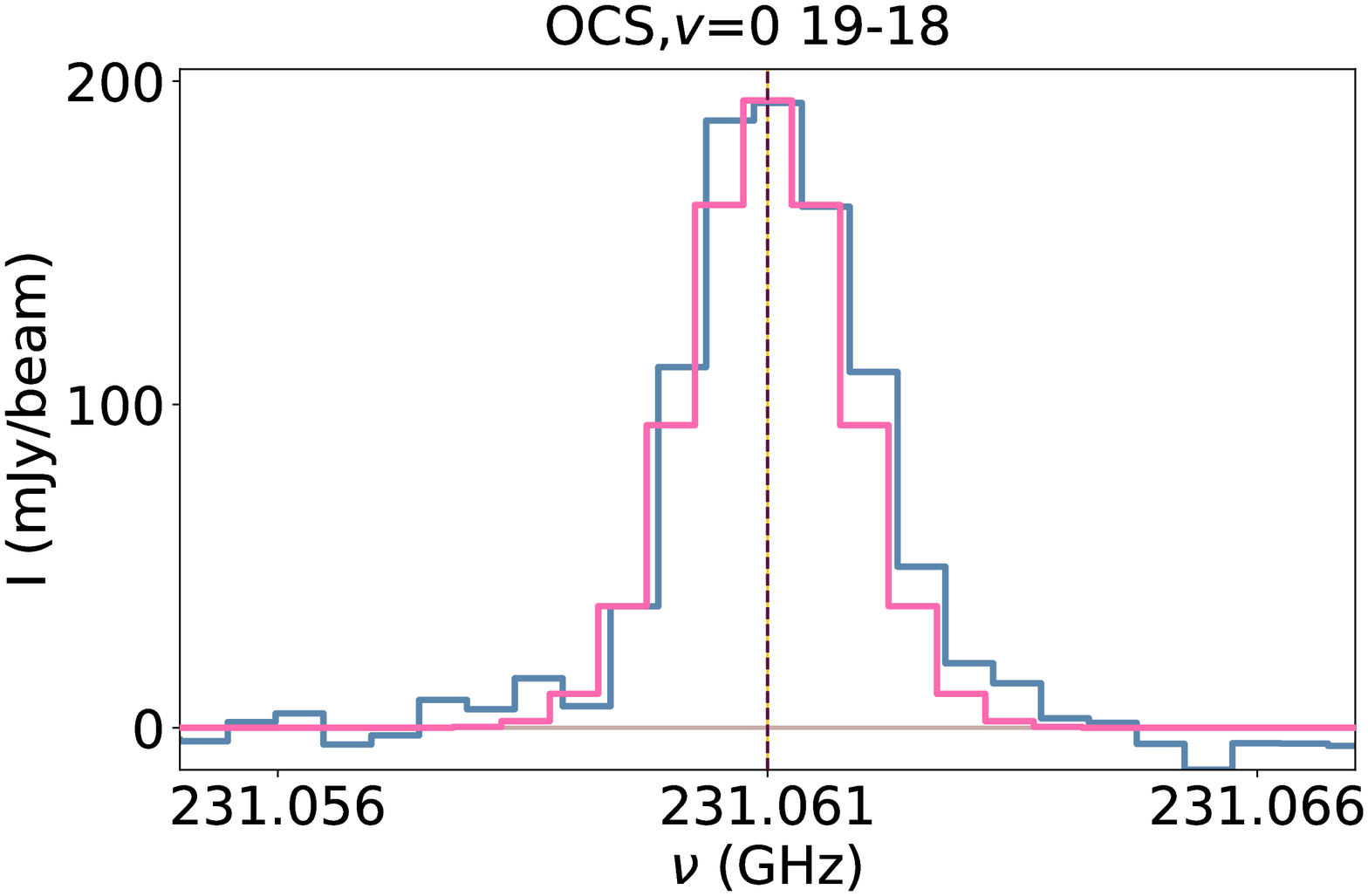}}
\end{minipage}
\caption{Observed spectra (in blue), rest frequency of the detected line (brown dashed line), spectroscopic uncertainty on the rest frequency of the detected line (yellow shaded region), and fitted synthetic spectra (in pink) plotted for the sulfur-bearing species detected towards the warm component ($T_{\text{ex}}=250$ K) of BHR71-IRS1.}
\label{detected_warm_BHR71}
\end{figure}

\begin{figure}[H]
\centering
\begin{minipage}[b]{6.2in}%
    \subfigure{\includegraphics[width=2.0in]{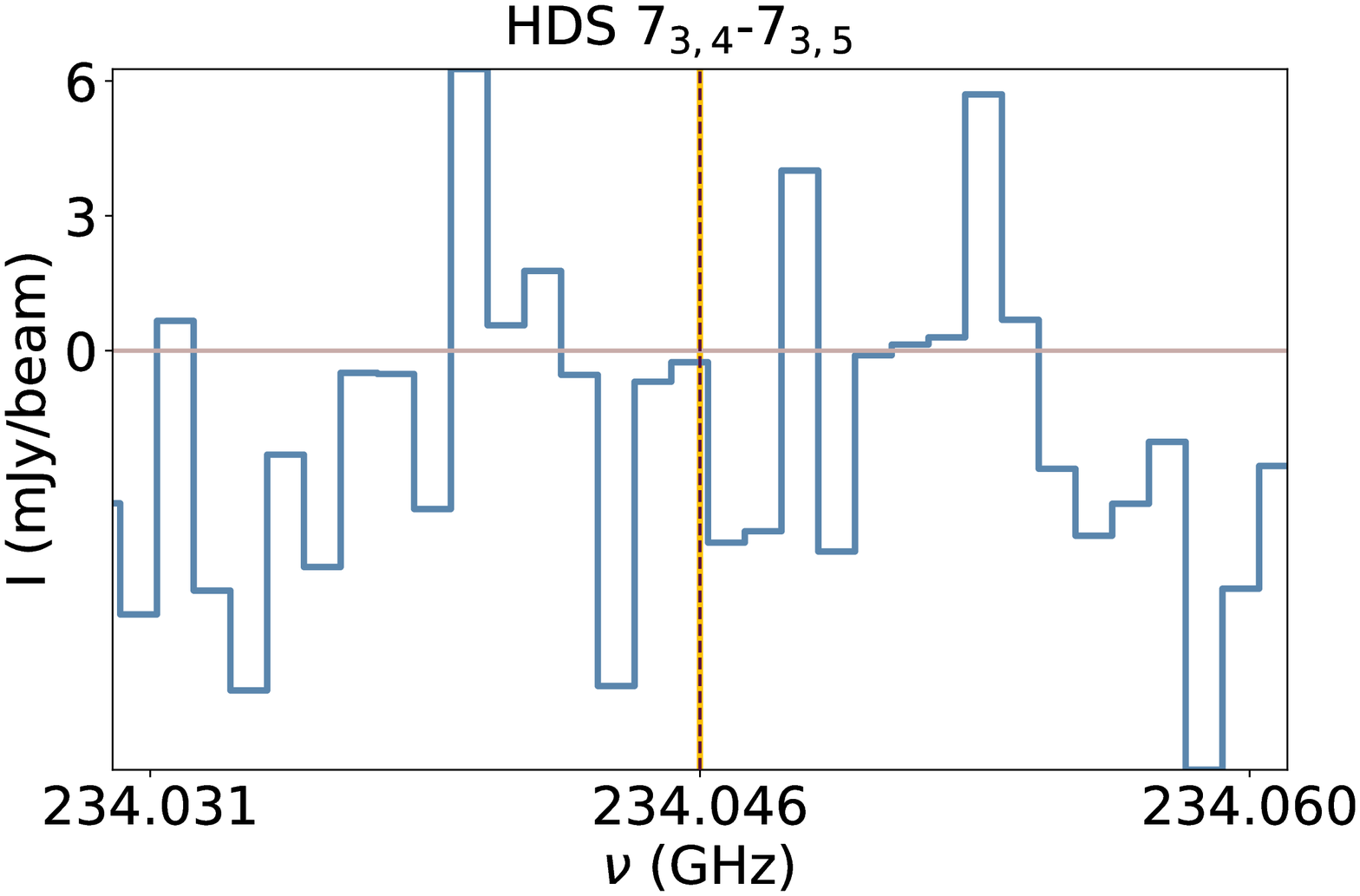}}
    \subfigure{\includegraphics[width=2.0in]{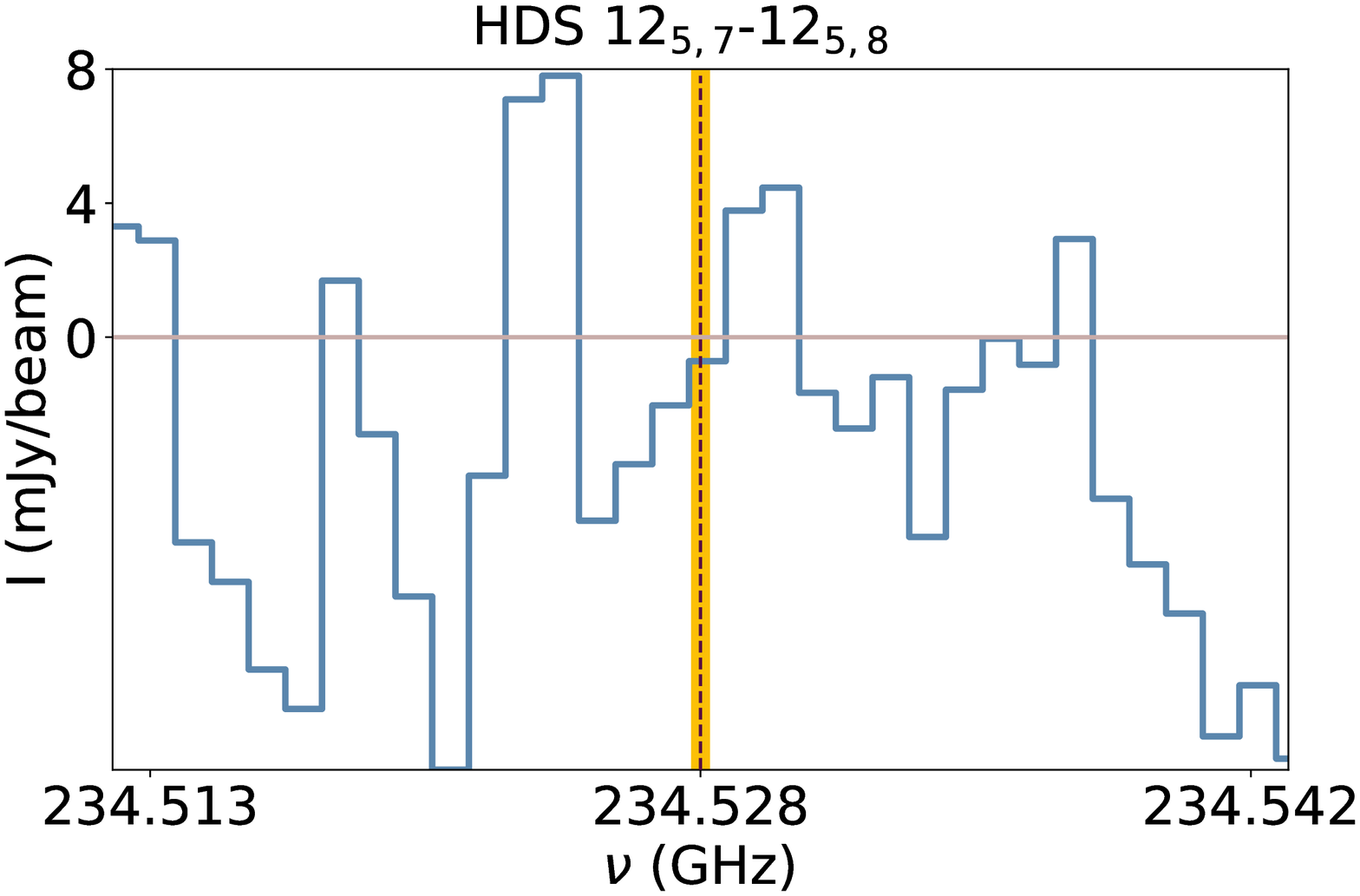}}
    \subfigure{\includegraphics[width=2.0in]{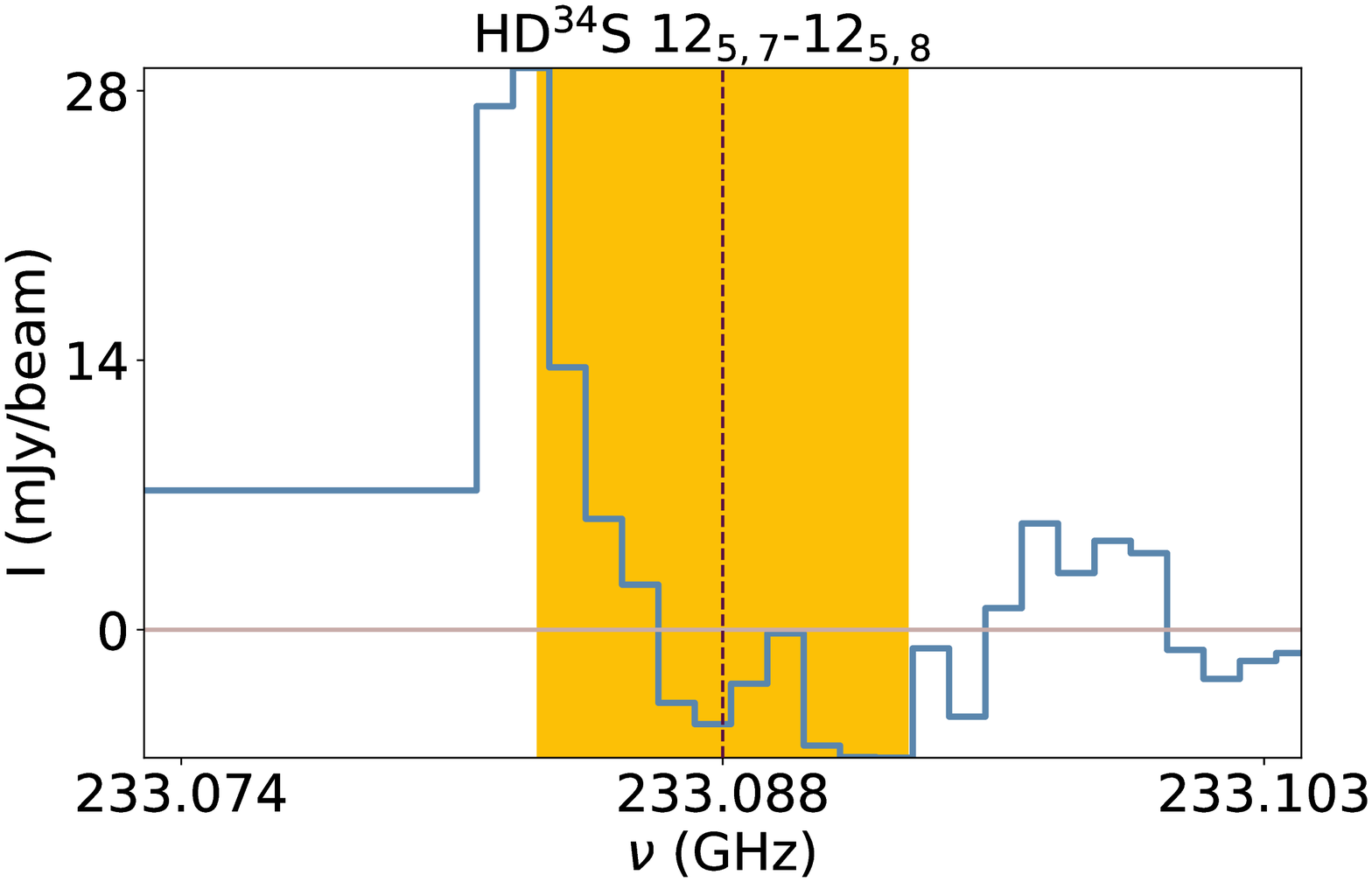}}\\
    \subfigure{\includegraphics[width=2.0in]{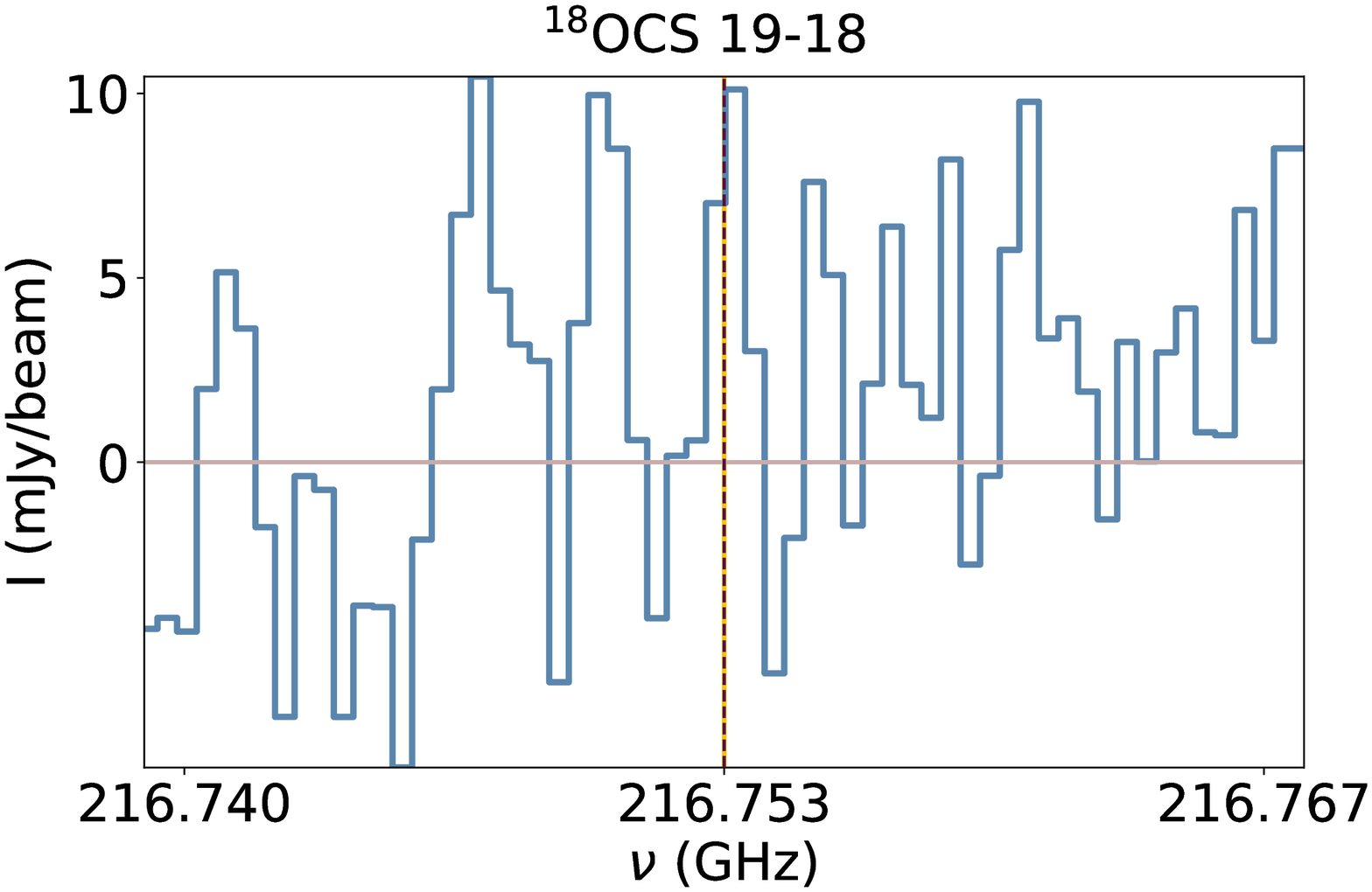}}
    \subfigure{\includegraphics[width=2.0in]{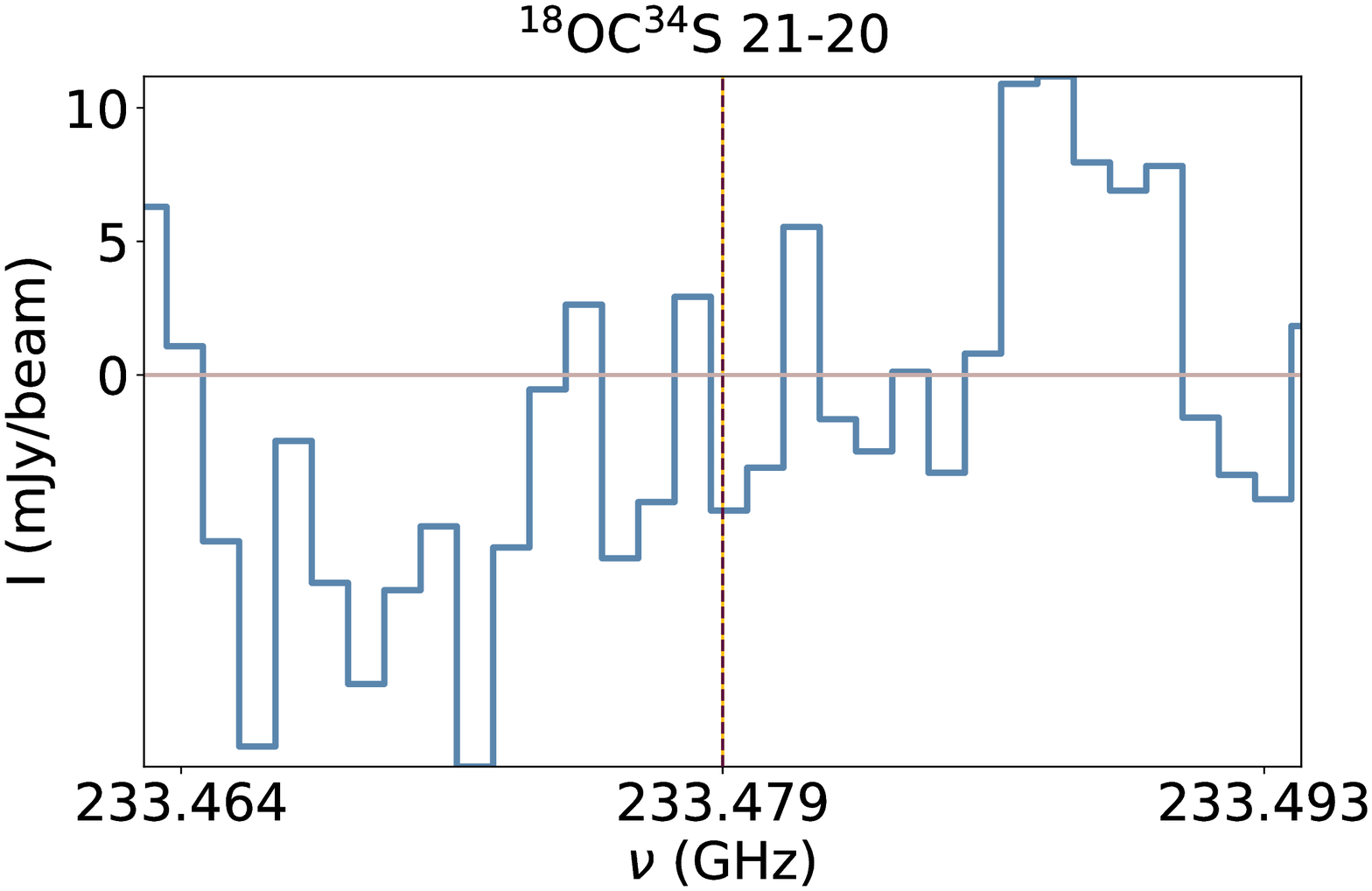}}
    \subfigure{\includegraphics[width=2.0in]{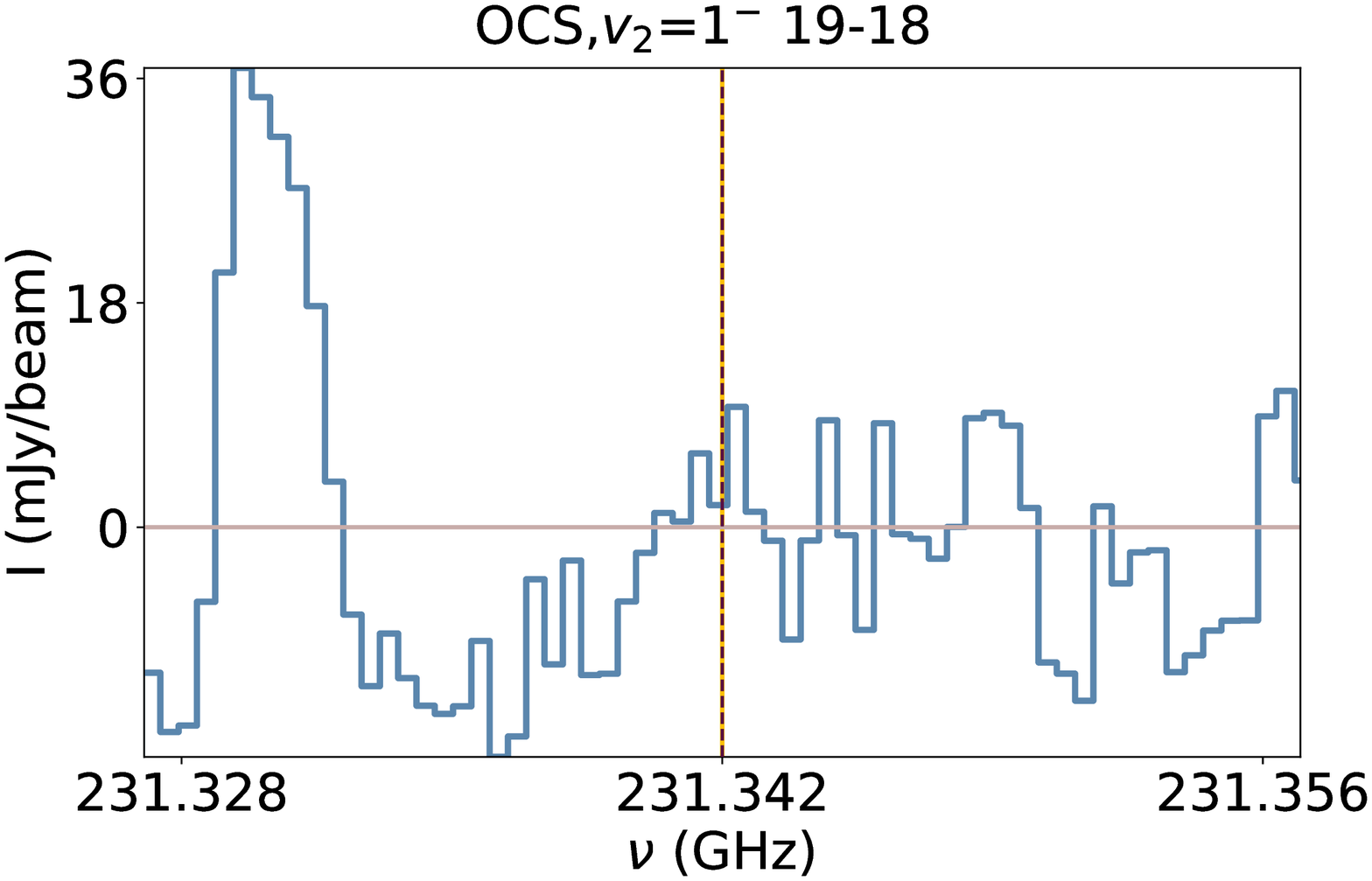}}
    \subfigure{\includegraphics[width=2.0in]{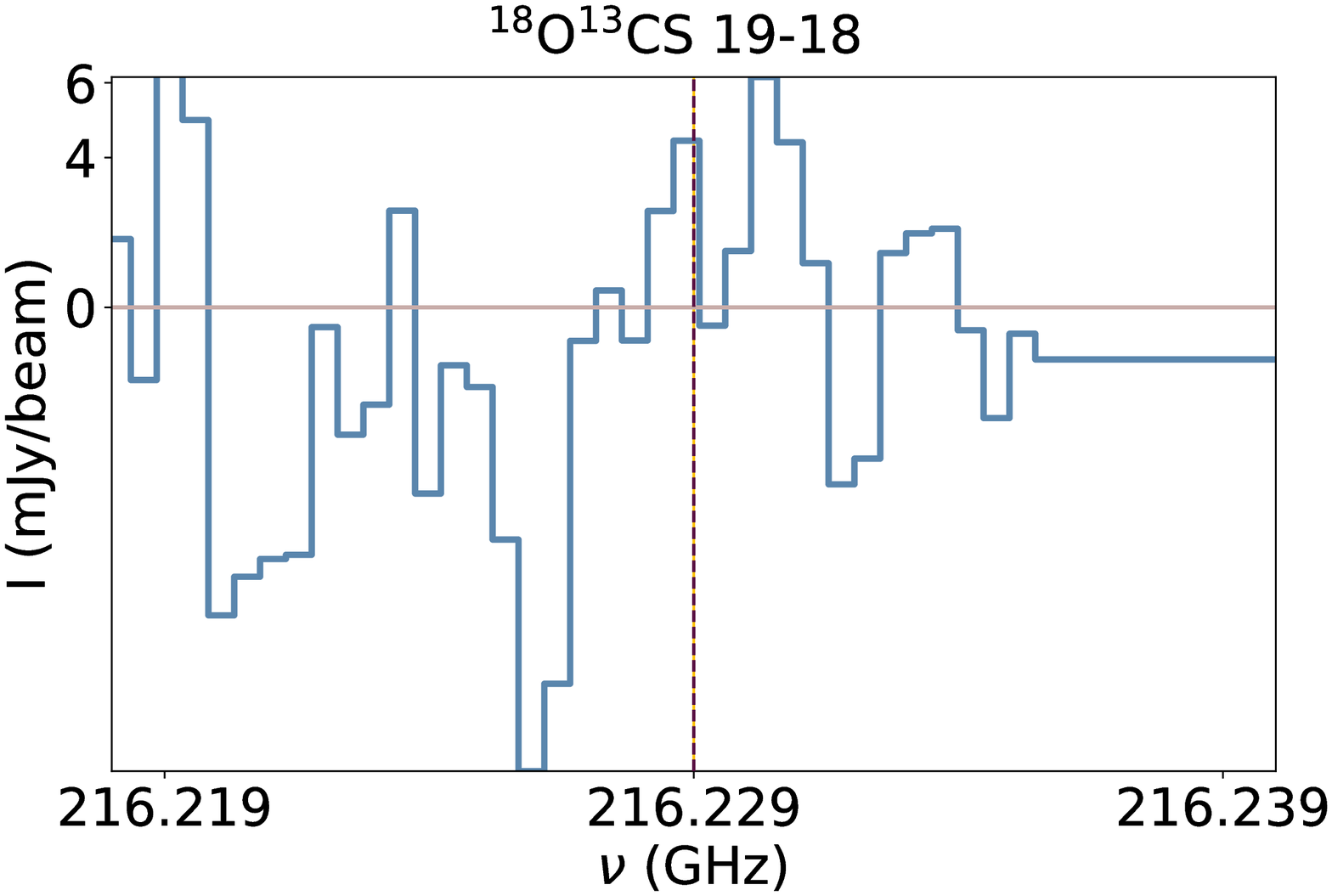}}
    \subfigure{\includegraphics[width=2.0in]{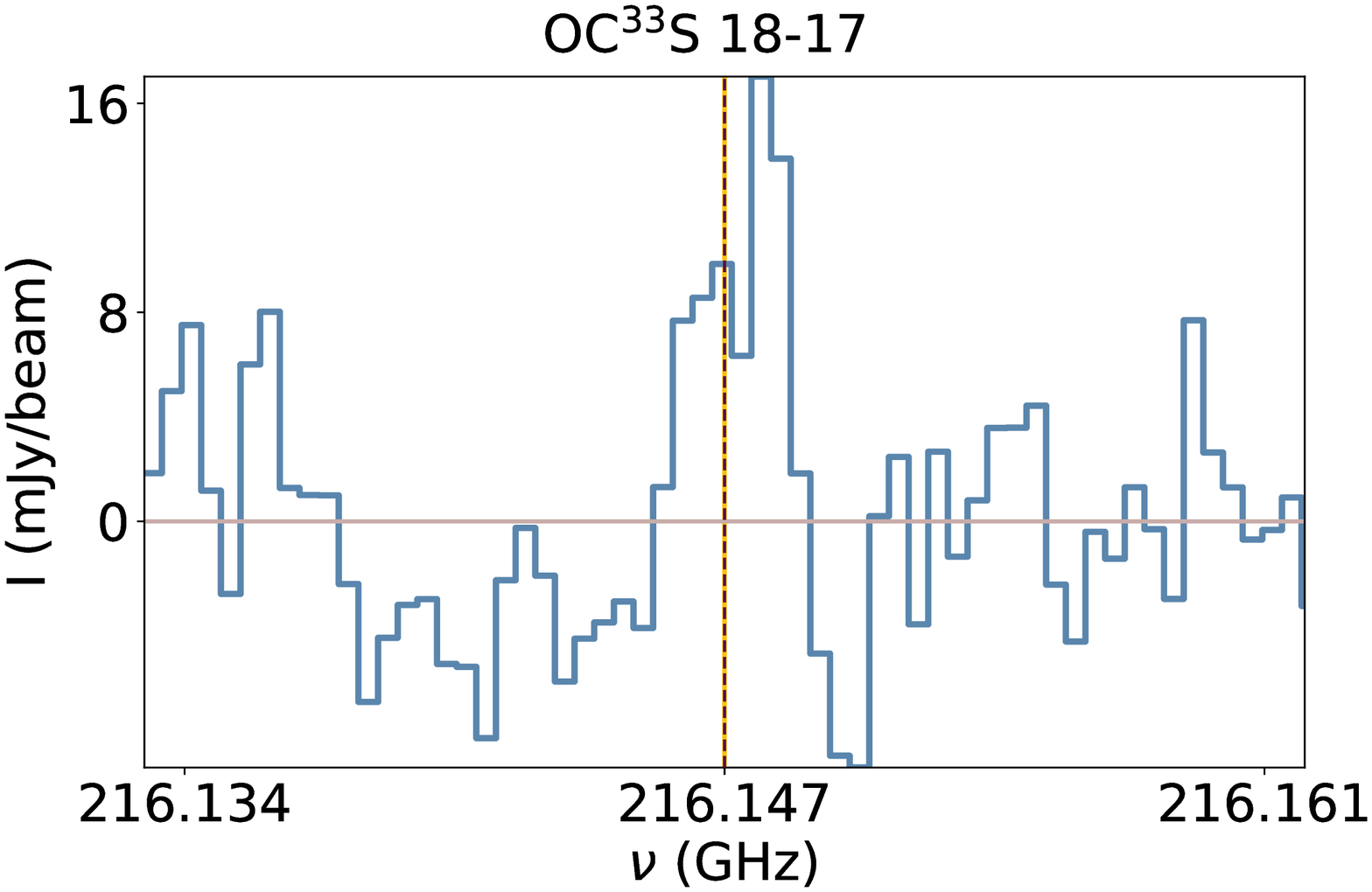}}
\end{minipage}
\caption{Observed spectra (in blue), rest frequency of the undetected line (brown dashed line), and spectroscopic uncertainty on the rest frequency of the undetected line (yellow shaded region) plotted for the sulfur-bearing species undetected towards BHR71-IRS1.}
\label{undetected_BHR71}
\end{figure}

H$_2$S, 2$_{2,0}$-2$_{1,1}$ and OCS, $v=0$, $J = 19-18$ emission lines are detected towards BHR71-IRS1 (\autoref{detected_cold_BHR71}, \ref{detected_warm_BHR71}). As for B1-c, two excitation temperatures (100 and 250 K) probing two different regions are inputted into the modeling of synthetic spectra for BHR71-IRS1. The excitation temperatures are taken from the detection of methanol at $T_{\text{ex}}=100$ K and gauche-ethanol at $T_{\text{ex}}=250$ K \citep{Yang2020}. \cite{Yang2020} fitted CS for a source size of 0.32$''$; however, in order to cover a larger area of the envelope a source size of 0.6$''$, and a FWHM of 2.5 km s$^{-1}$ is assumed here. The H$_2$S line of the colder component seems to be optically thick ($\tau=1.8$), while the OCS, $v$=0 line is partially optically thick for this component ($\tau=0.23$). H$_2$S is partially optically thin for the warmer component ($\tau = 0.39$). The spectra of the undetected lines are shown in \autoref{undetected_BHR71}. 

\begin{table}[H]
    \centering
    \caption{Synthetic fitting of the detected S-bearing species towards BHR71-IRS1 for an excitation temperature of 100 K, a FWHM of 2.5 km s$^{-1}$, and a source size of 0.6$''$.}
    \label{results_cold_BHR71}
    \begin{tabular}{r r c r c c  c c c r}
    \hline
    \hline
    Species & Transition & Frequency & $E_{\text{up}}$ & $A_{ij}$ &
    Beam size & $N$  & $\tau$ \\
    &&&&&& & \\
    & & (GHz) & (K) & (s$^{-1}$) & ($''$) & (cm$^{-2}$) &  \\
\hline
& & & & &  & & &  \\
H$_2$S & 2$_{2,0}$-2$_{1,1}$ & 216.710 & 84 & 4.9$\times$10$^{-5}$  & 6.2 &   $>$2.4$\times$10$^{16,}$ \tablefootmark{*} & 1.80\\

OCS, $v$=0 & 19-18 & 231.061 & 111  & 3.6$\times$10$^{-5}$  & 5.8  &  $>$2.7$\times$10$^{15,}$ \tablefootmark{**} &  0.23 \\
\hline
\hline
    \end{tabular}
    \tablefoot{\tablefoottext{*}{optically thick}, \tablefoottext{**}{partially optically thick}}
\end{table}

\begin{table}[H]
    \centering
    \caption{Synthetic fitting of the detected S-bearing species towards BHR71-IRS1 for an excitation temperature of 250 K, a FWHM of 2.5 km s$^{-1}$, and a source size of 0.6$''$.}
    \label{results_hot_BHR71}
    \begin{tabular}{r r c r c c  c c c r}
    \hline
    \hline
    Species & Transition & Frequency & $E_{\text{up}}$ & $A_{ij}$ &
    Beam size & $N$ &  $\tau$ \\
    &&&&&&& \\
    & & (GHz) & (K) & (s$^{-1}$) & ($''$) & (cm$^{-2}$) &  \\
\hline
& & & & &  &  &  \\
H$_2$S & 2$_{2,0}$-2$_{1,1}$ & 216.710 & 84 & 4.9$\times$10$^{-5}$  & 6.2 &  $>$3.3$\times$10$^{16,}$ \tablefootmark{*} & 0.39\\

OCS, $v$=0 & 19-18  & 231.061 & 111  & 3.6$\times$10$^{-5}$  & 5.8 & 3.4$^{+0.3}_{-0.3}\times$10$^{15}$ & 0.08 \\
\hline
\hline
    \end{tabular}
    \tablefoot{\tablefoottext{*}{partially optically thick}}
\end{table}

\subsection{Detected lines in BHR71-IRS1}
\label{shiftBHR71}

\subsection{Undetected lines in BHR71-IRS1}

\section{Per-emb-25}

\begin{figure}[H]
\centering
\begin{minipage}[b]{6.2in}%
    \subfigure{\includegraphics[width=2.0in]{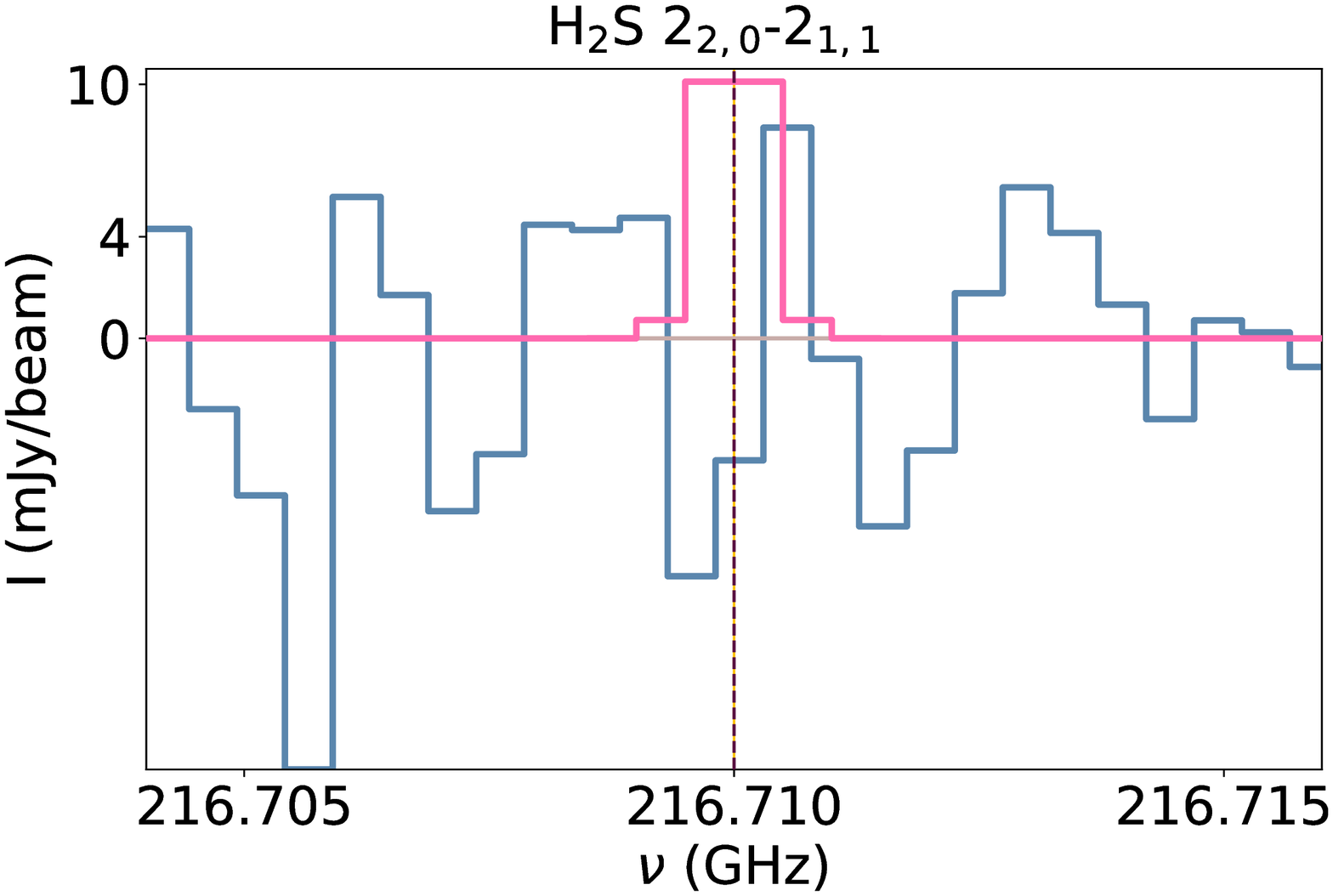}}
    \subfigure{\includegraphics[width=2.0in]{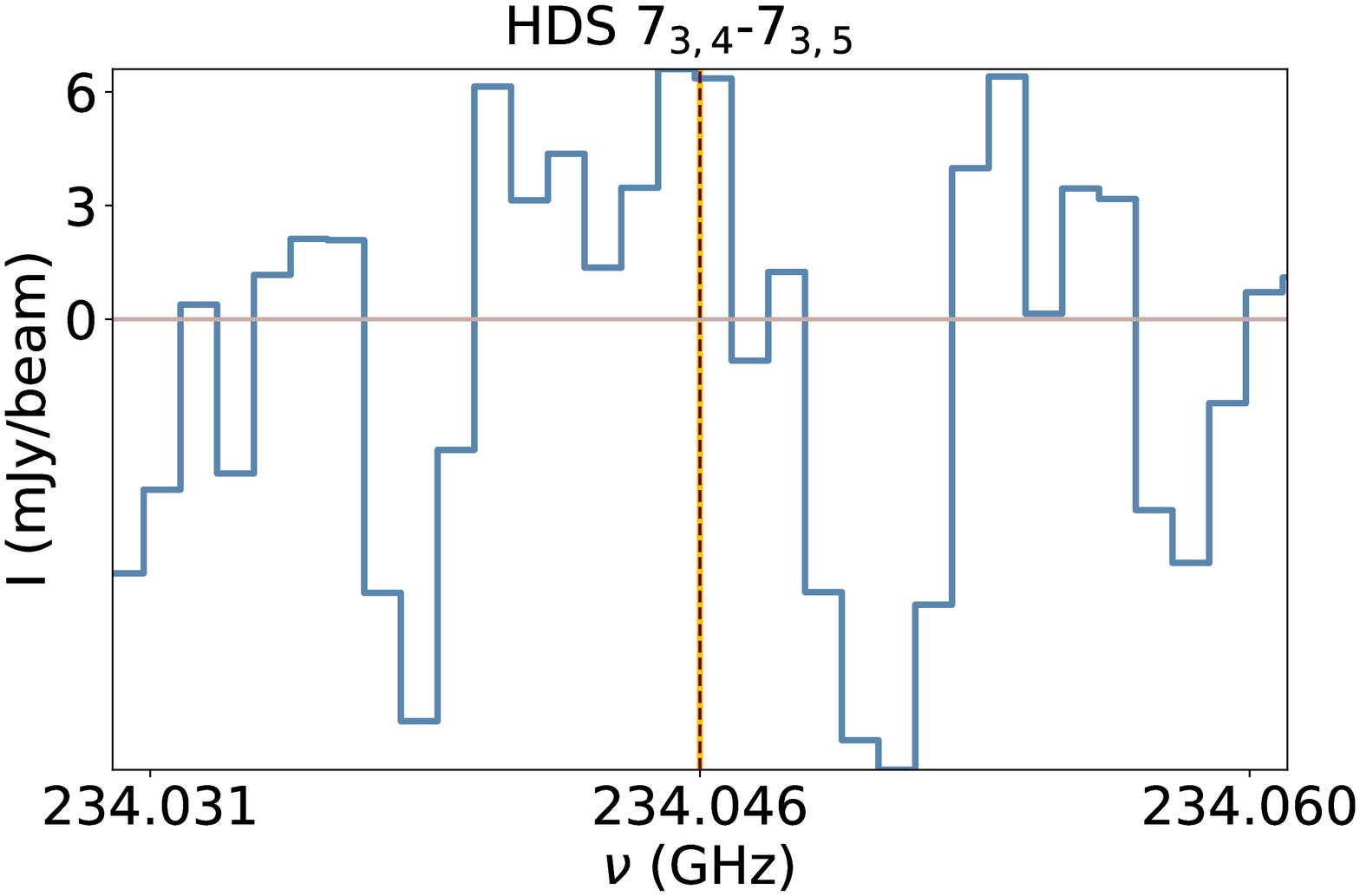}}
    \subfigure{\includegraphics[width=2.0in]{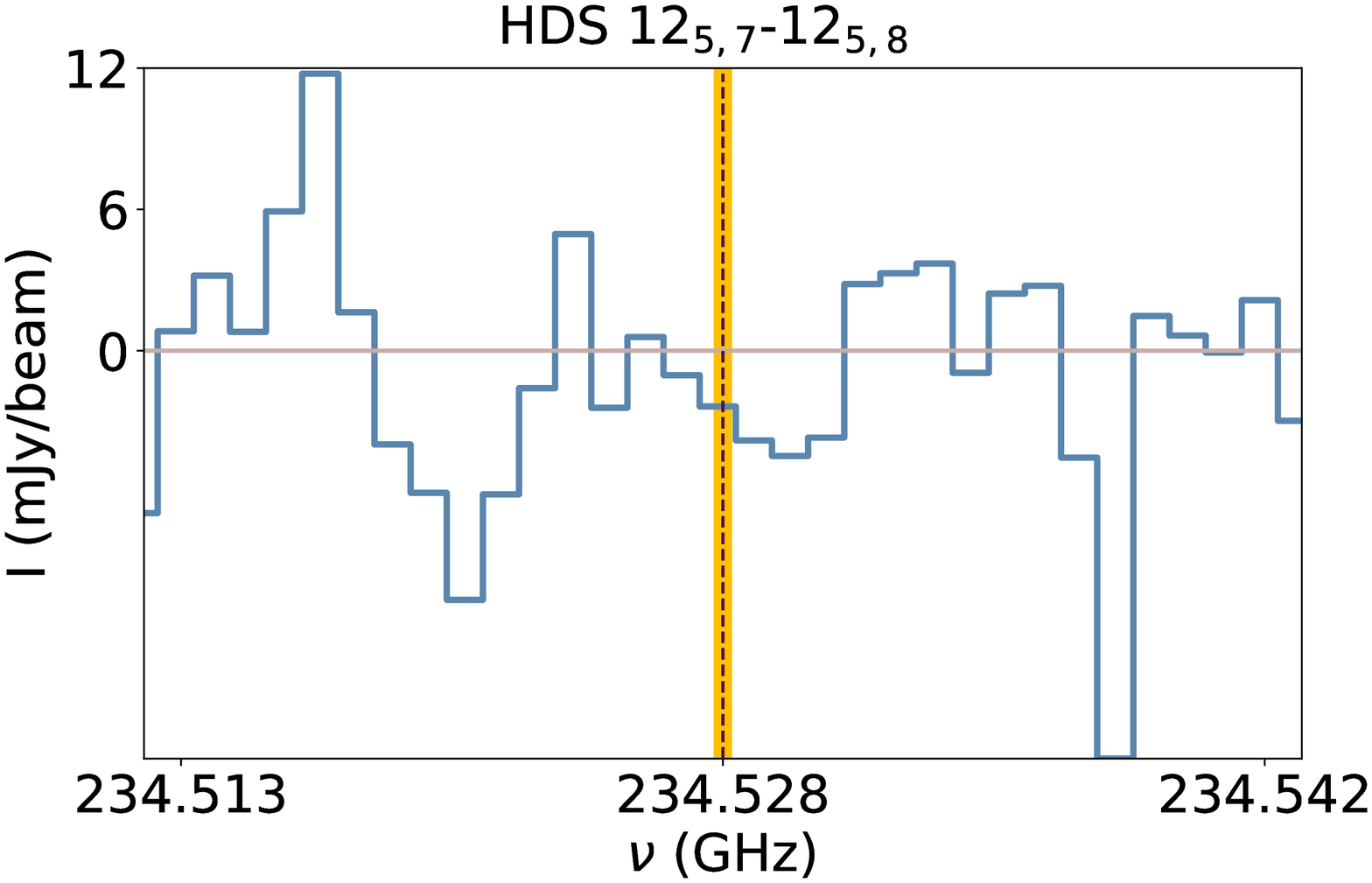}}\\
    \subfigure{\includegraphics[width=2.0in]{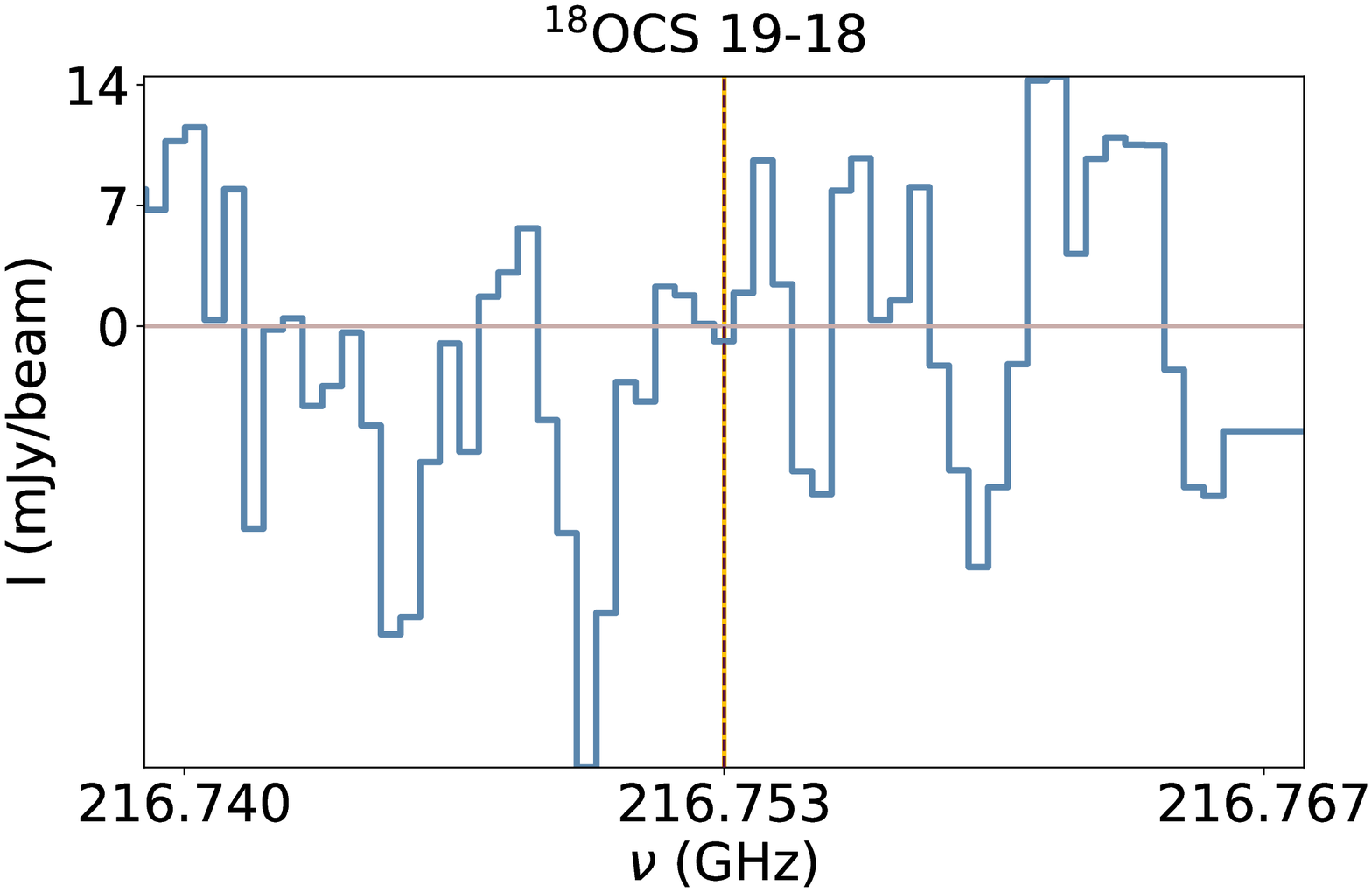}}
    \subfigure{\includegraphics[width=2.0in]{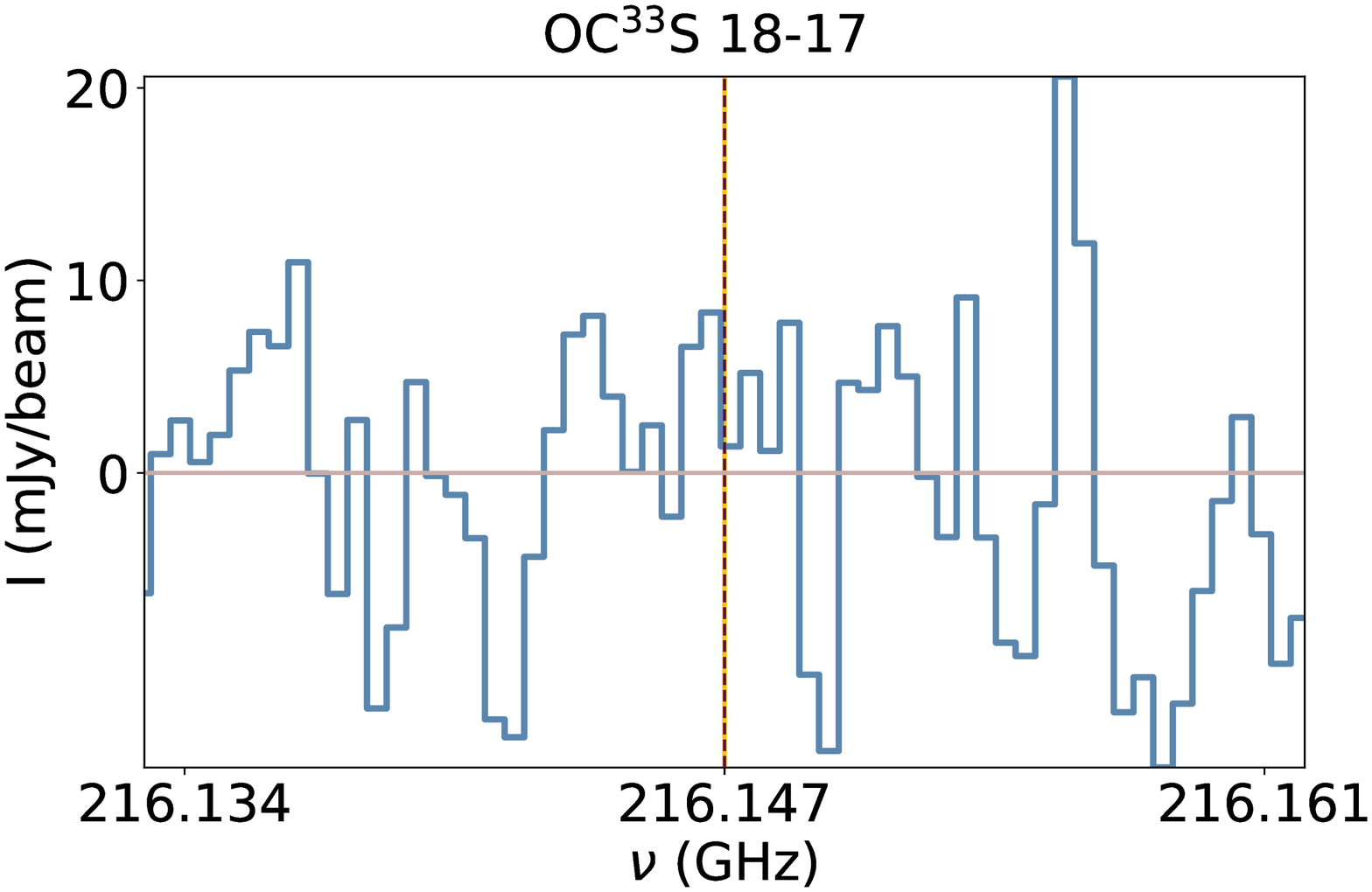}}
    \subfigure{\includegraphics[width=2.0in]{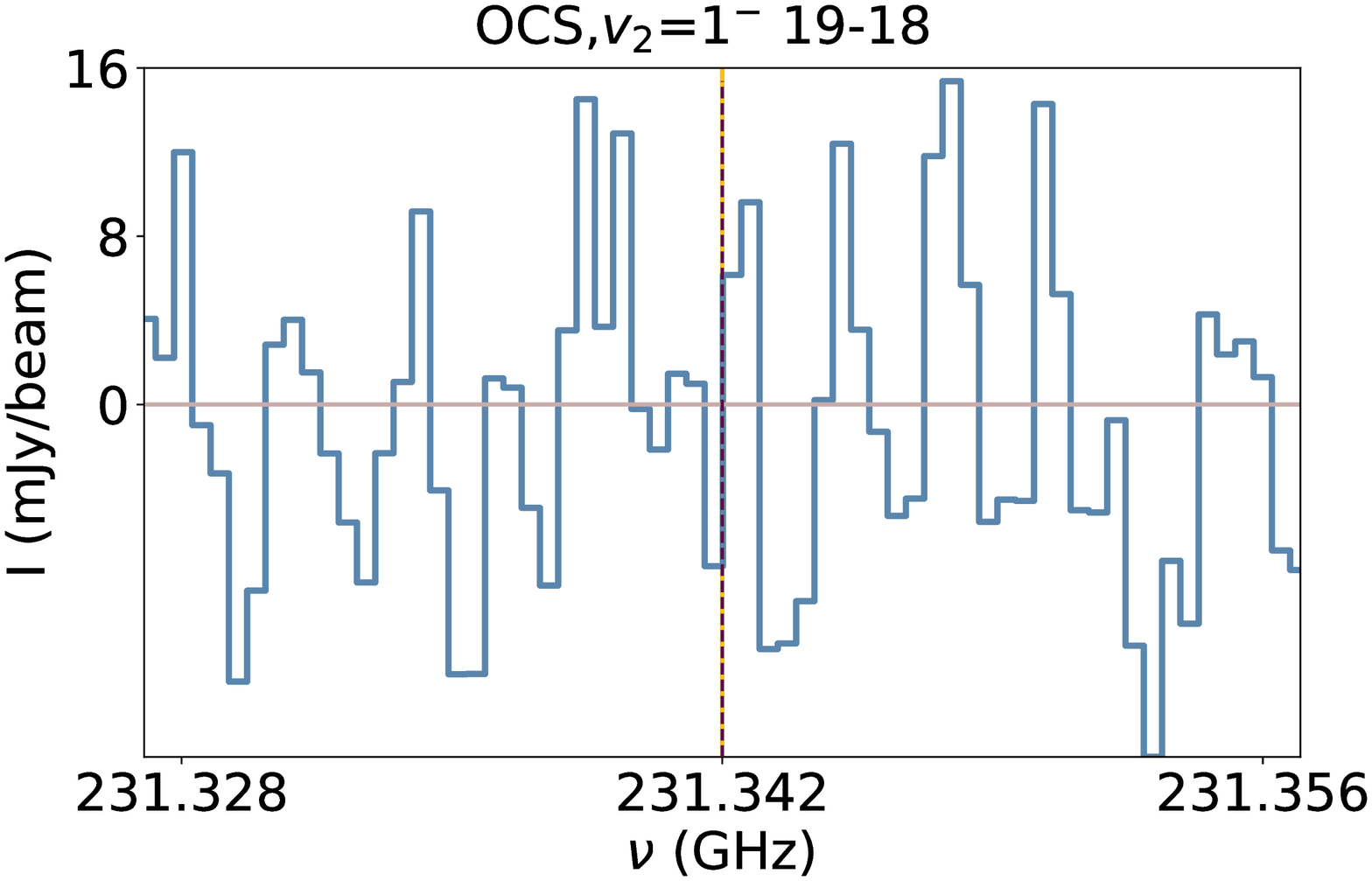}}\\    
    \subfigure{\includegraphics[width=2.0in]{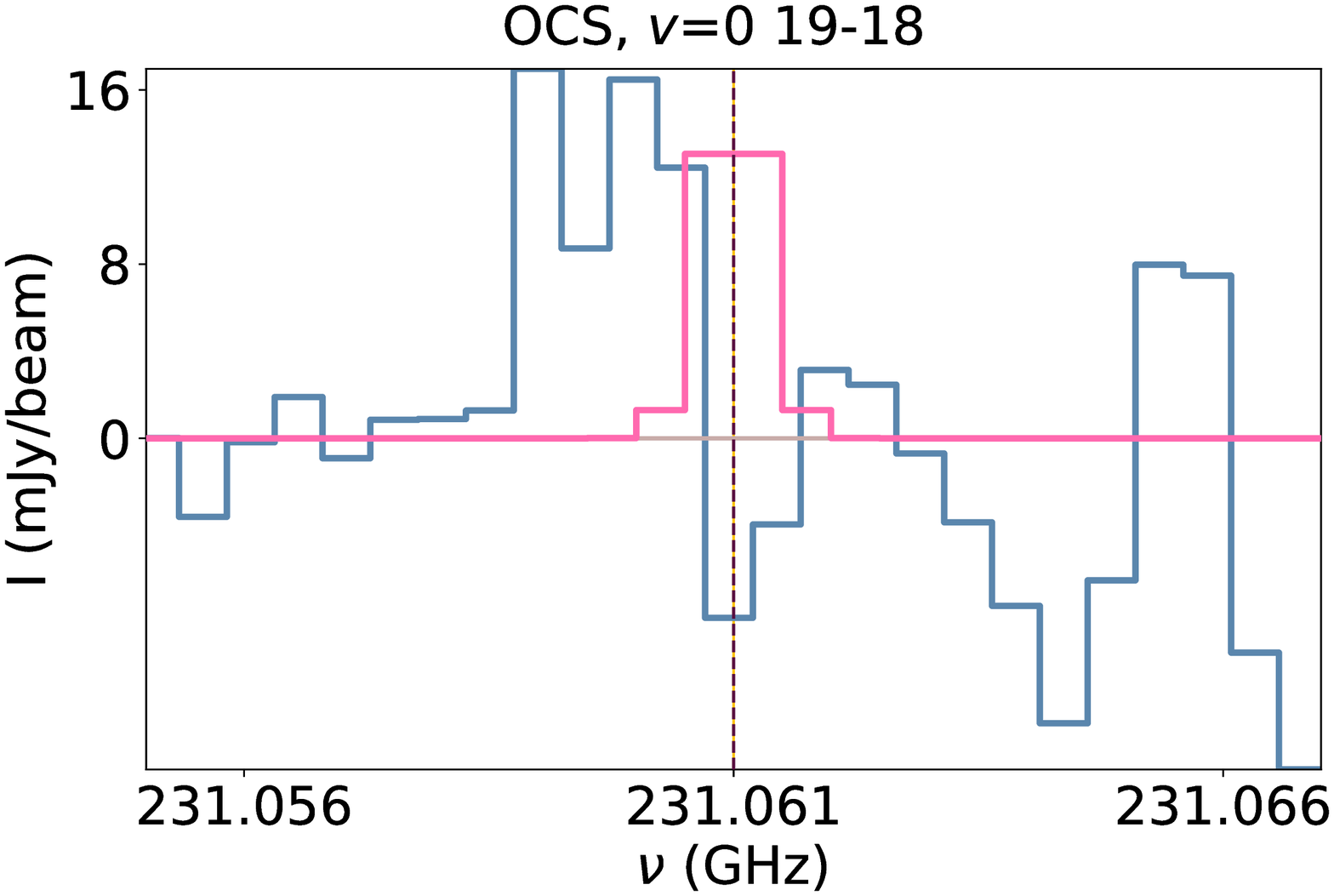}}
    \subfigure{\includegraphics[width=2.0in]{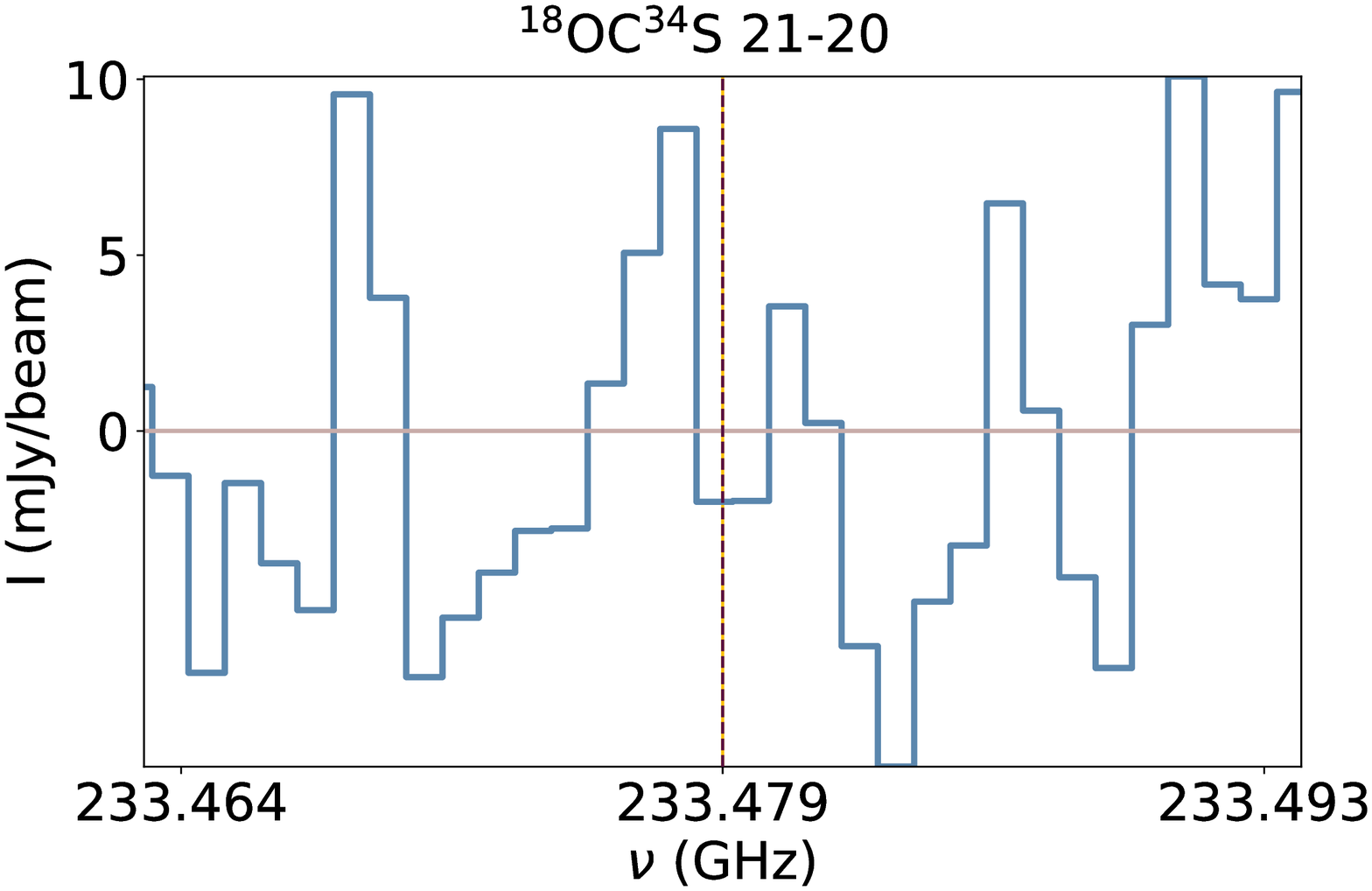}}
    \subfigure{\includegraphics[width=2.0in]{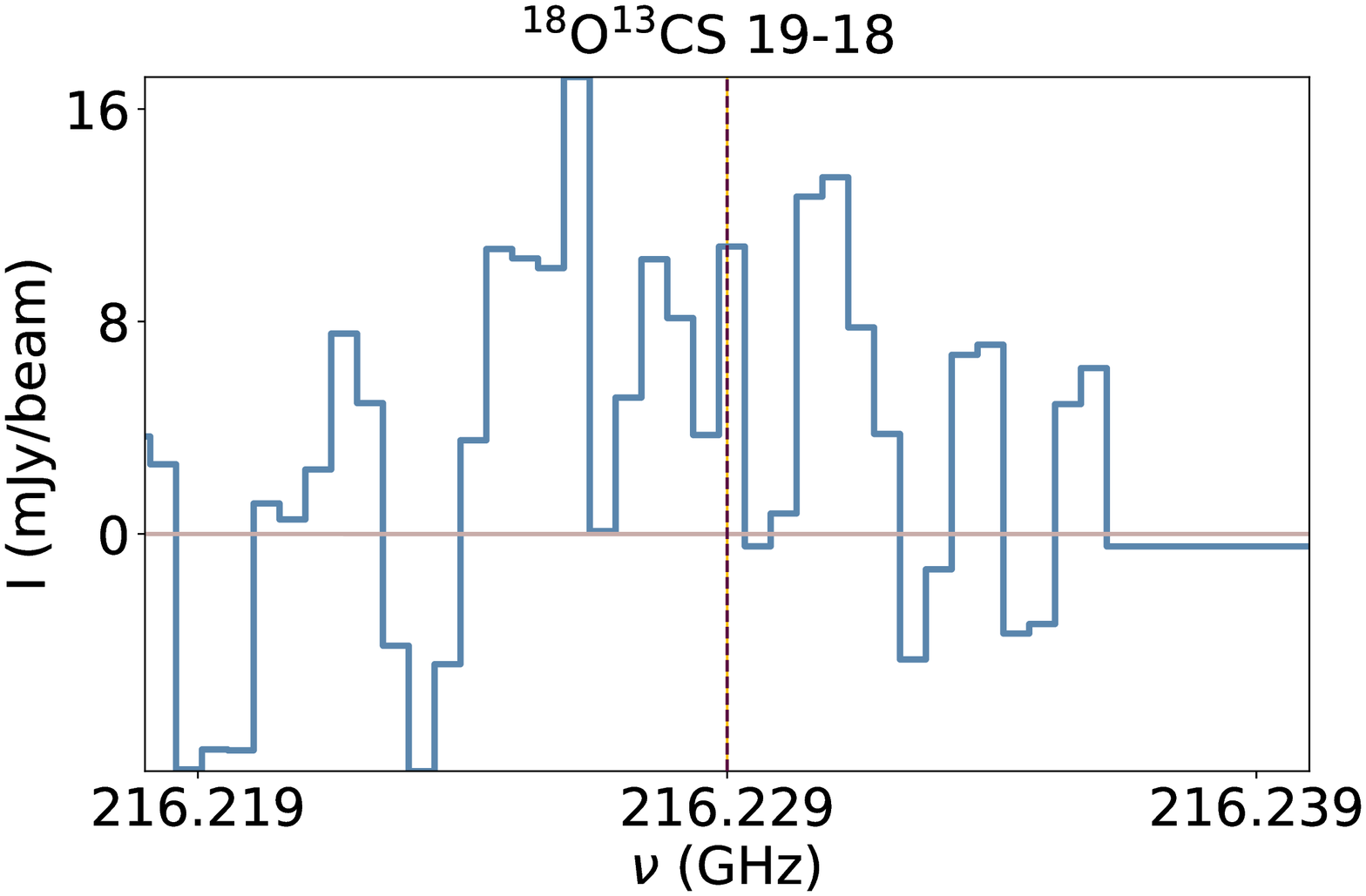}}
\end{minipage}
\caption{Observed spectra (in blue), rest frequency of the undetected line (brown dashed line), and spectroscopic uncertainty on the rest frequency of the undetected line (yellow shaded region) plotted for the sulfur-bearing species undetected towards Per-emb-25. Synthetic spectra (in pink) fitted to H$_2$S and OCS, $v$=0 lines with the 1-$\sigma$ upper limit on their column densities.}
\label{undetected_Emb25}
\end{figure}

No S-bearing species are detected towards Emb-25. Consequently, 1-$\sigma$ upper limits on the column densities of H$_2$S and OCS, $v$=0 are computed assuming a source size of 1$''$ \citep{Yang2021}, a FWHM of 1 km s$^{-1}$, and excitation temperatures in the range of $50-300$ K. The upper limits on H$_2$S and OCS, $v$=0 column densities are $\leq8.3\times10^{13}$ and $\leq3.2\times10^{14}$ cm$^{-2}$, respectively (\autoref{ratio} and \ref{results_Emb25}). The spectra of all the undetected species are presented in \autoref{undetected_Emb25}.

\subsection{1-$\sigma$ upper limits on the column densities of H$_2$S and OCS, $v$=0 of Per-emb-25}

\begin{table}[H]
    \centering
    \caption{Synthetic fitting of the non-detected main S-bearing species towards Per-emb-25 for a range of excitation temperatures between 50 and 300 K, a FWHM of 1 km s$^{-1}$, and a source size of 1$''$.}
    \label{results_Emb25}
    \begin{tabular}{r r c r  c c  c c  r}
    \hline
    \hline
    Species & Transition & Frequency & $E_{\text{up}}$ & $A_{ij}$ &
    Beam size & source size & $N$\\

& & (GHz) & (K) & (s$^{-1}$) & ($''$)& ($''$) & (cm$^{-2}$) \\
\hline
H$_2$S & 2$_{2,0}$-2$_{1,1}$ & 216.710 & 84 & 4.9$\times$10$^{-5}$  & 6.5 & 2.0 & $\leq$8.3$\times$10$^{13}$ \\

OCS, $v$=0 & 19-18 & 231.061 & 111  & 3.6$\times$10$^{-5}$  & 6.0 & 0.5 & $\leq$3.2$\times$10$^{13}$ \\
\hline
\hline
\end{tabular}
\end{table}

\subsection{Undetected lines in Per-emb-25}

\section{NGC 1333-IRAS4B}

\begin{figure}[H]
\centering
\begin{minipage}[b]{6.2in}%

    \subfigure{\includegraphics[width=2.0in]{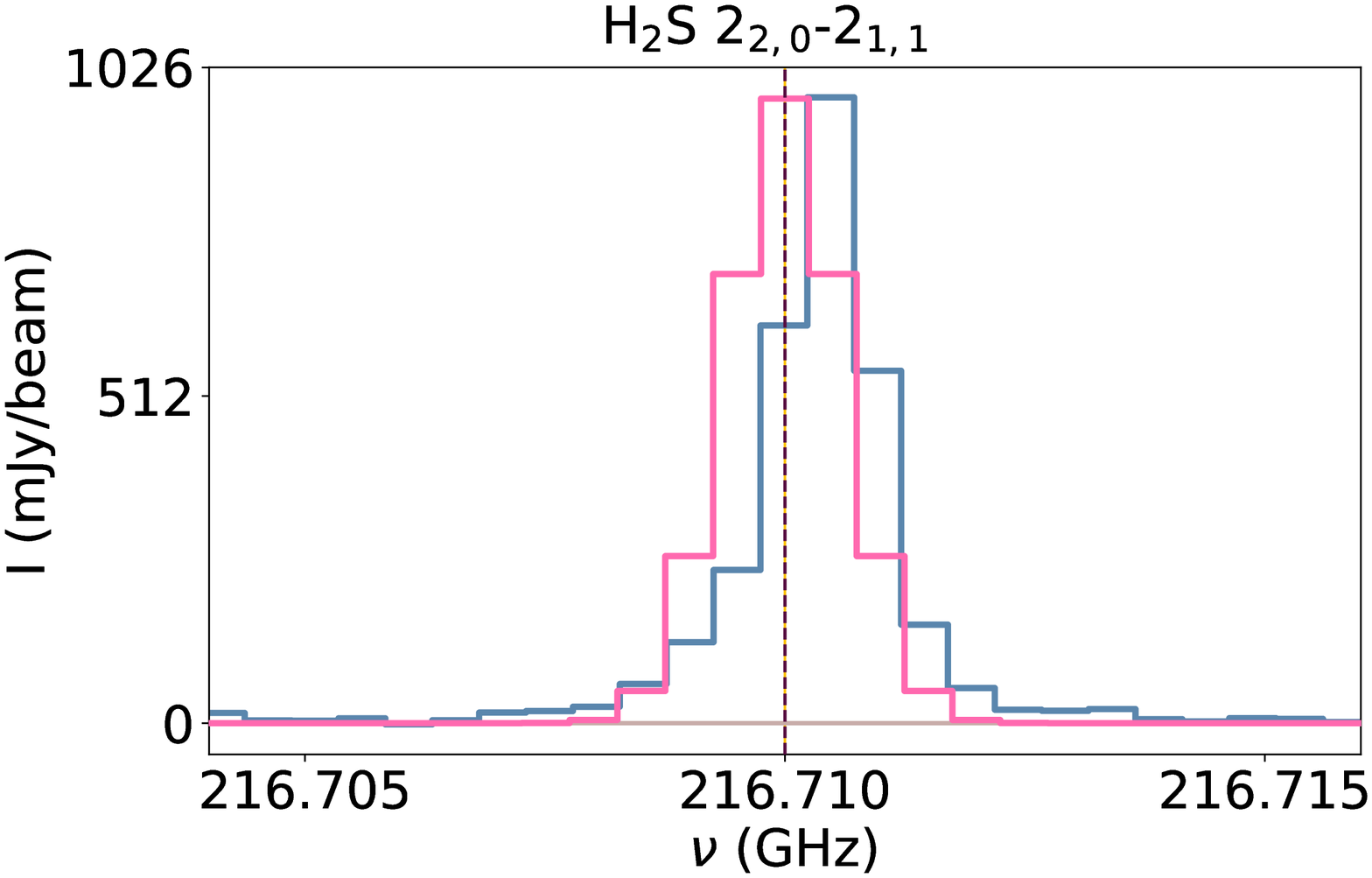}}
    \subfigure{\includegraphics[width=2.0in]{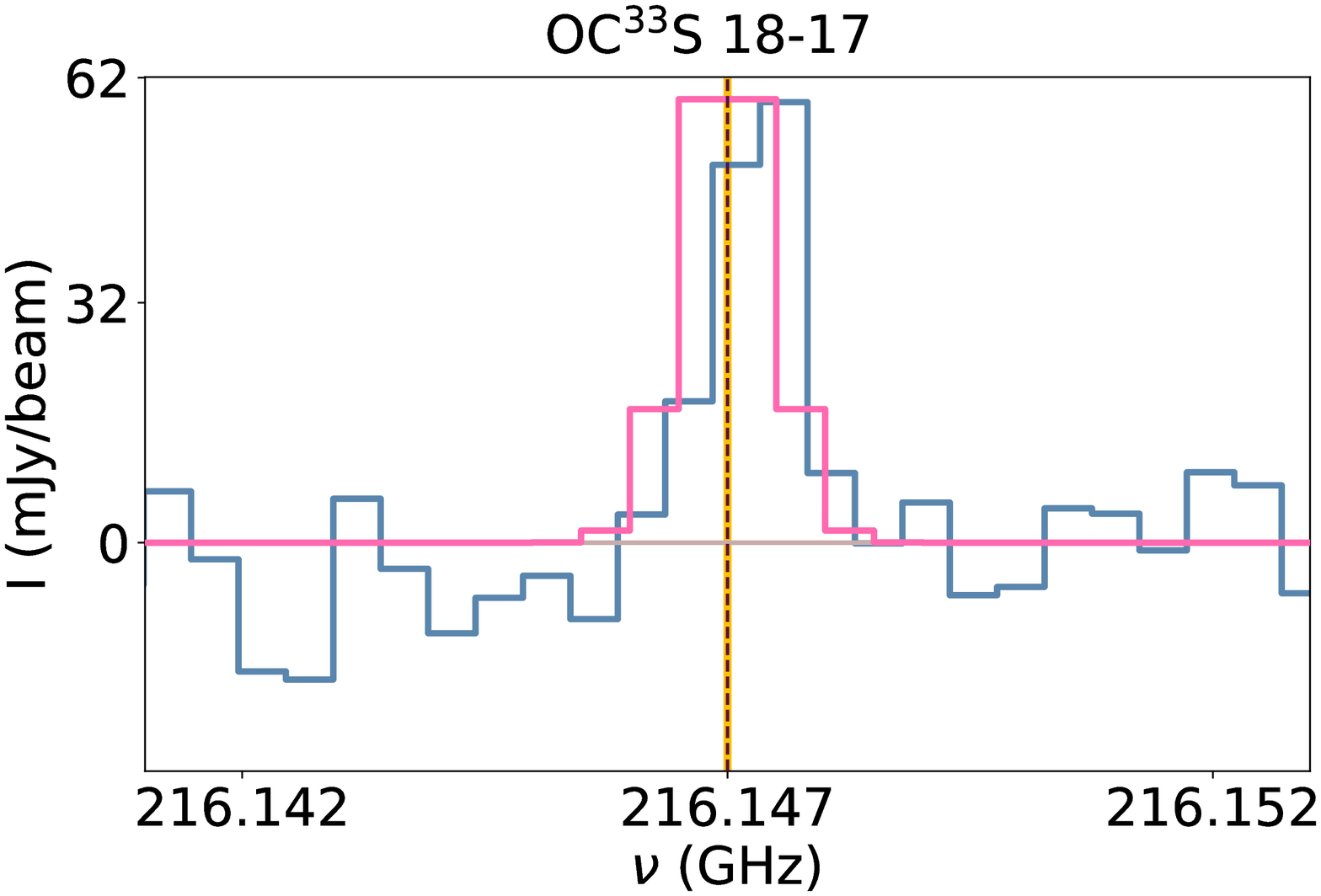}}
    \subfigure{\includegraphics[width=2.0in]{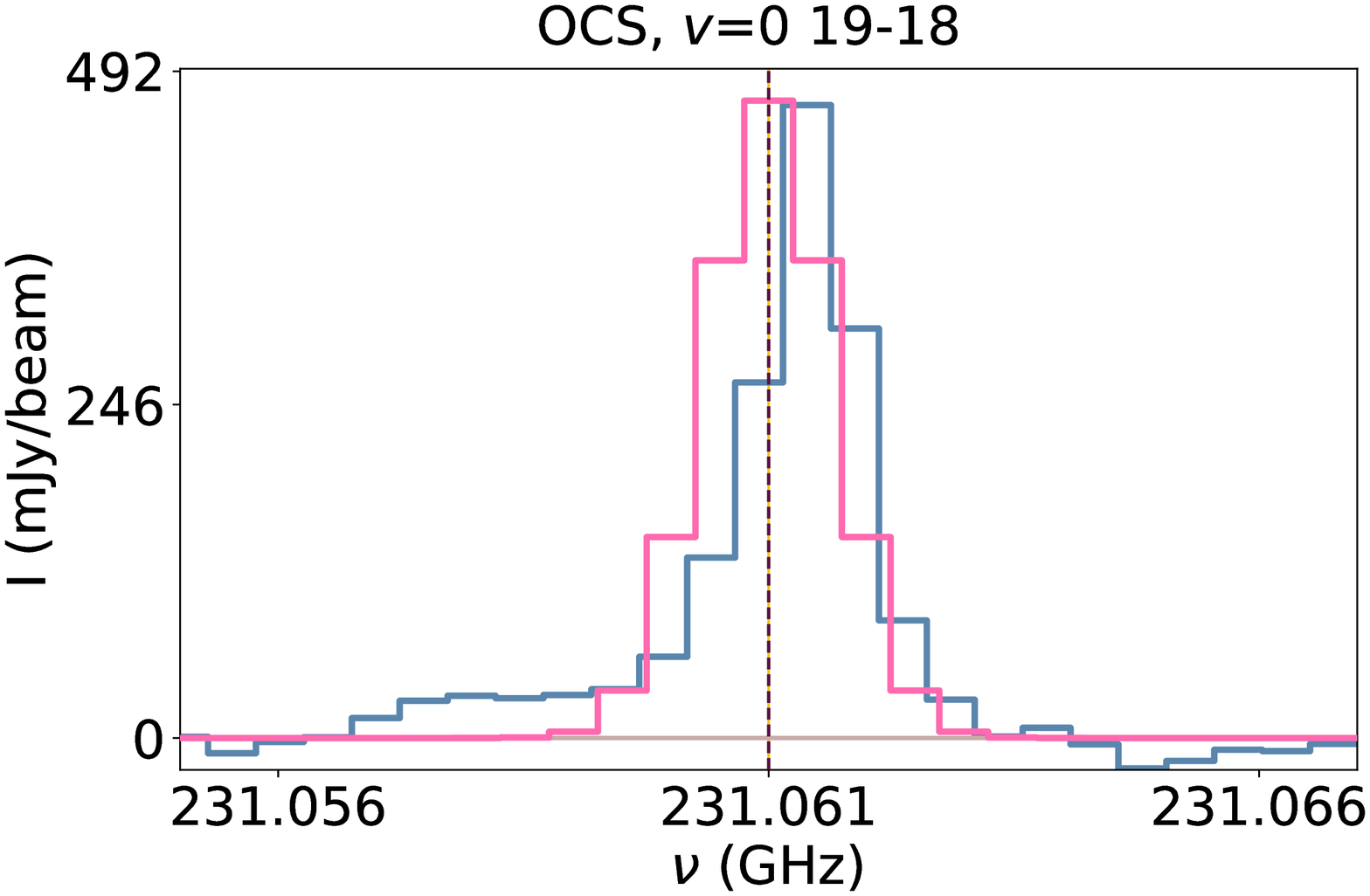}}
\end{minipage}
\caption{Observed spectra (in blue), rest frequency of the detected line (brown dashed line), spectroscopic uncertainty on the rest frequency of the detected line (yellow shaded region), and fitted synthetic spectra (in pink) plotted for the sulfur-bearing species detected towards NGC 1333-IRAS4B for $T_{\text{ex}}=100$~K and a source size of $1\arcsec$.}
\label{detected_IRAS4B}
\end{figure}

\begin{figure}[H]
\centering
\begin{minipage}[b]{6.2in}%

    \subfigure{\includegraphics[width=2.0in]{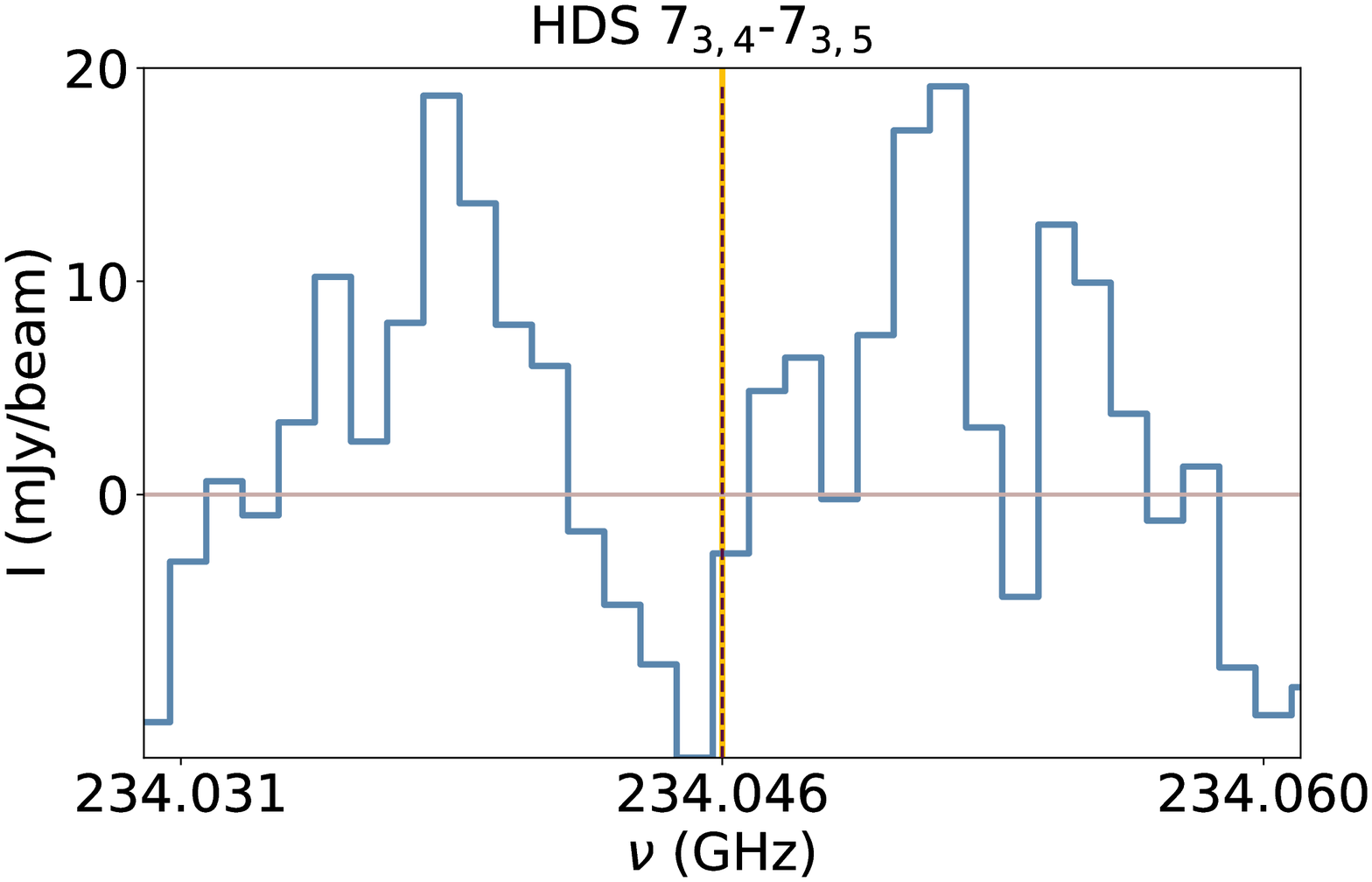}}
    \subfigure{\includegraphics[width=2.0in]{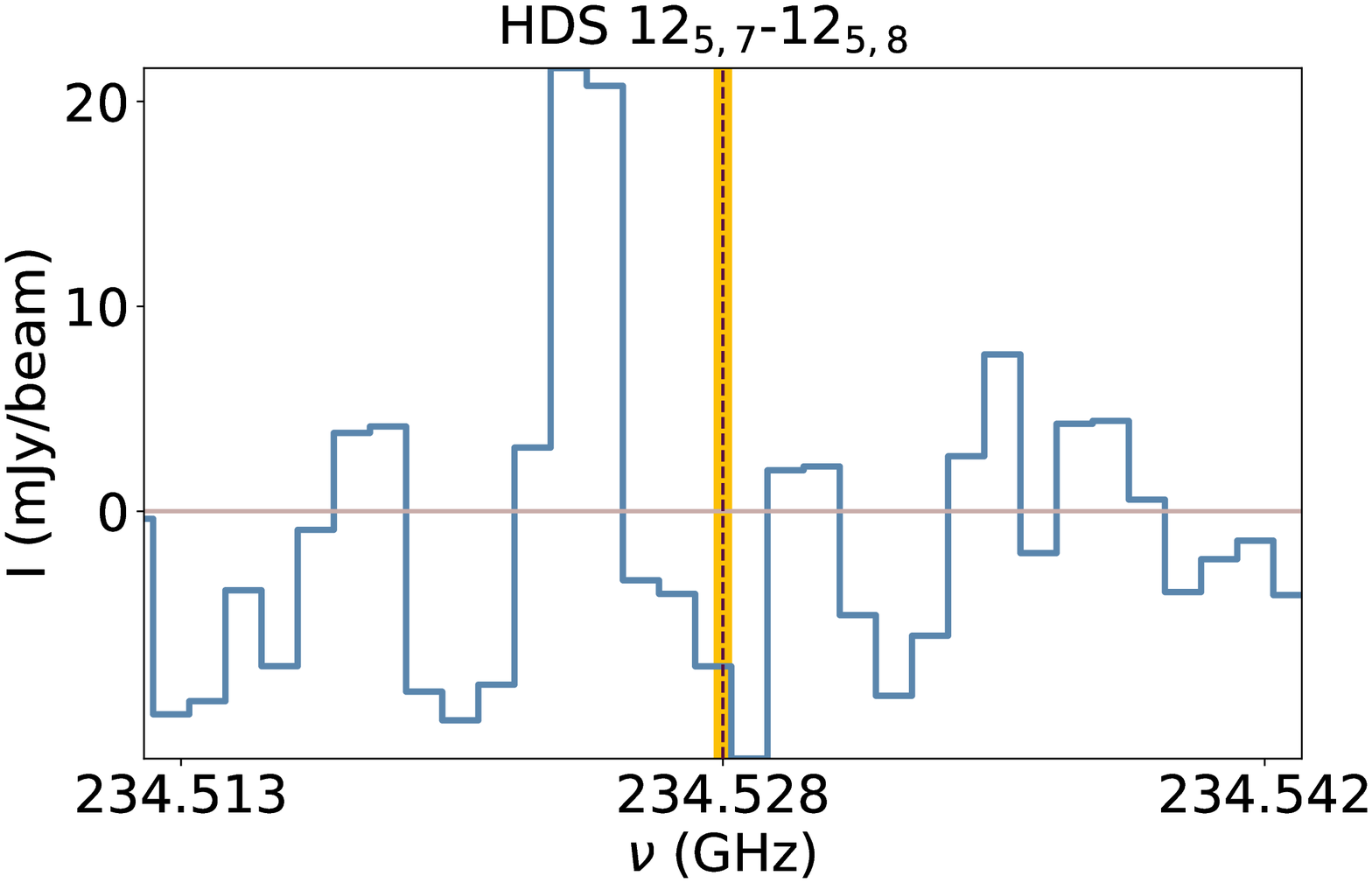}}
    \subfigure{\includegraphics[width=2.0in]{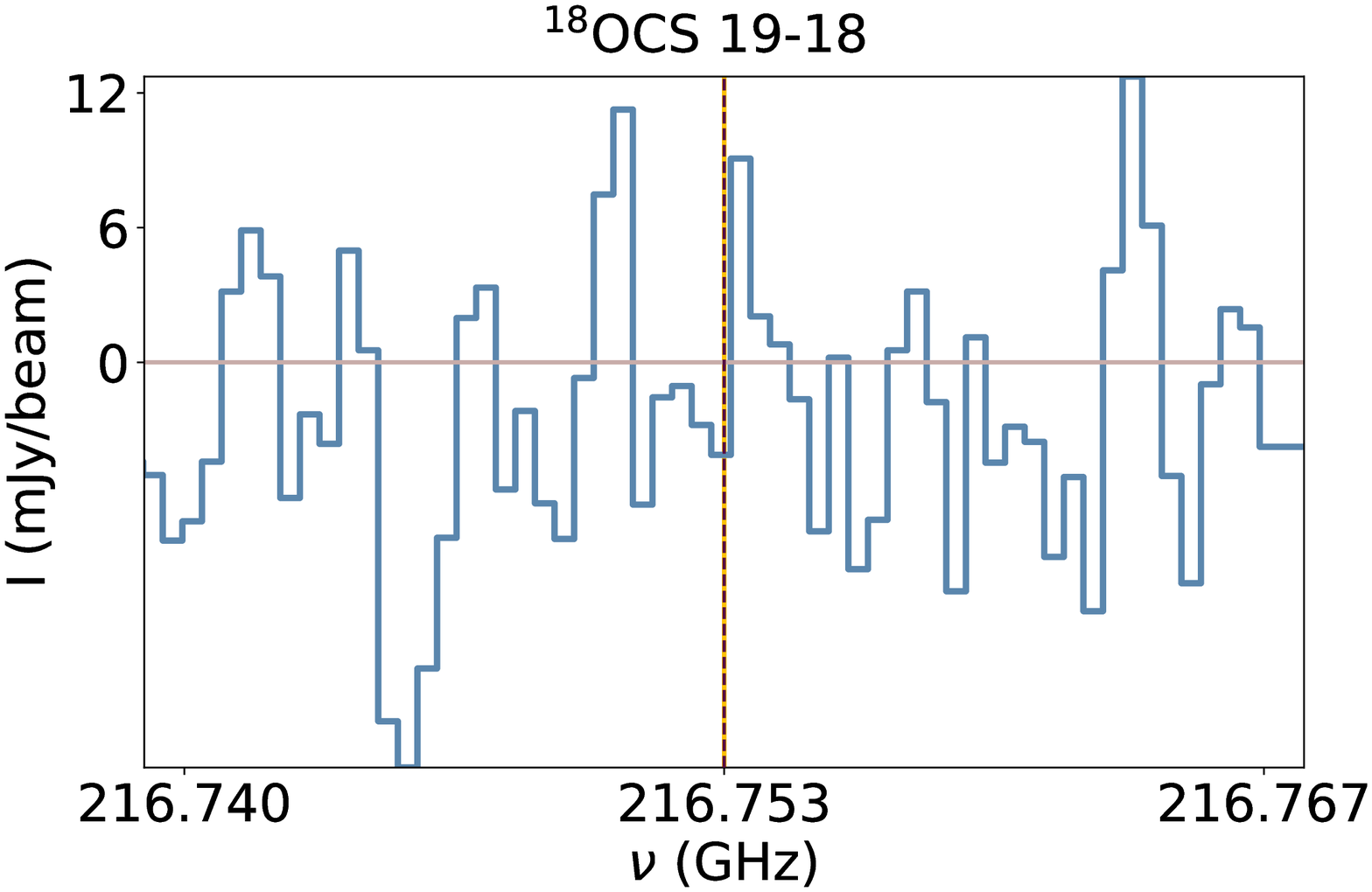}}\\
    \subfigure{\includegraphics[width=2.0in]{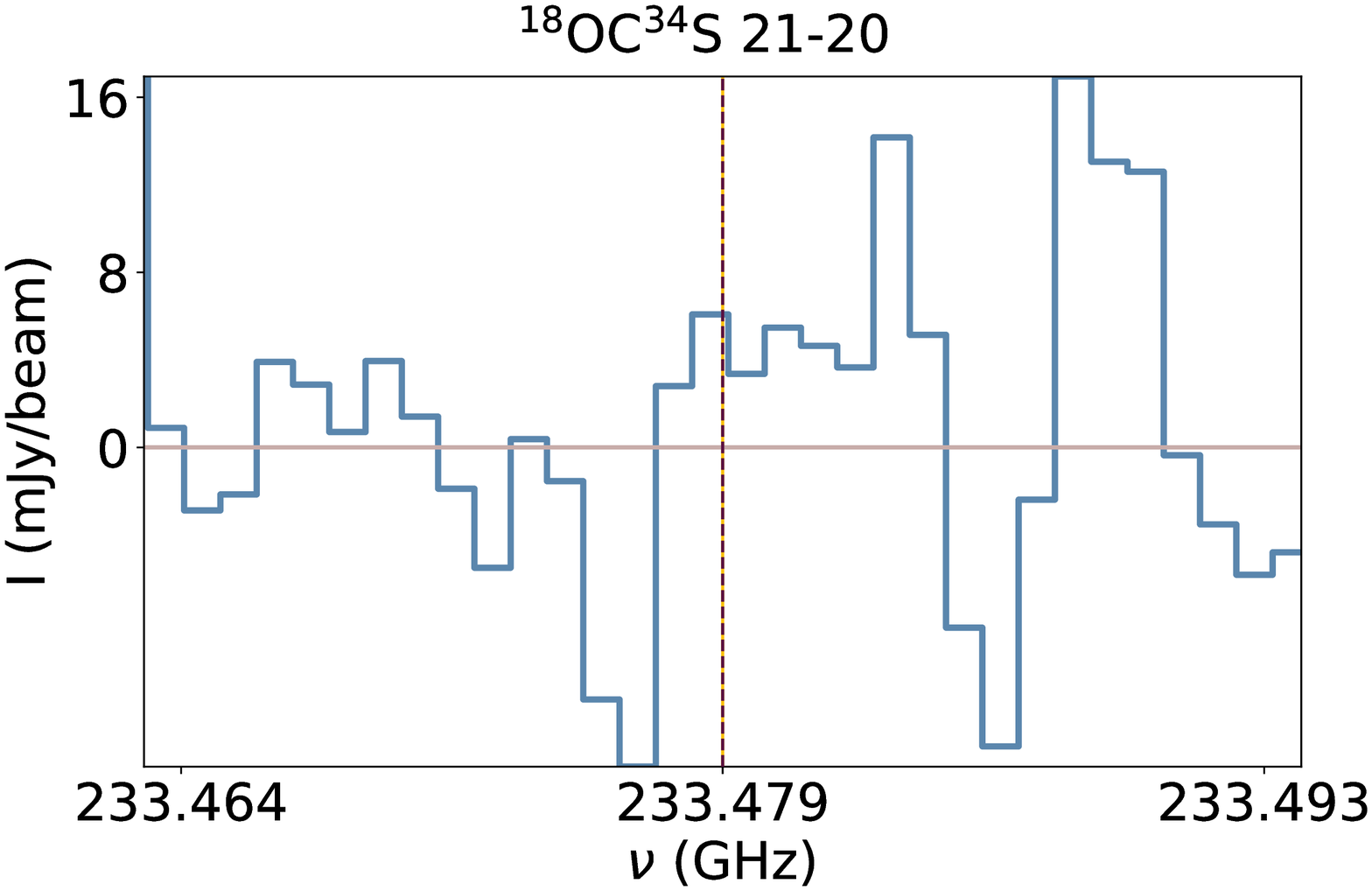}}
    \subfigure{\includegraphics[width=2.0in]{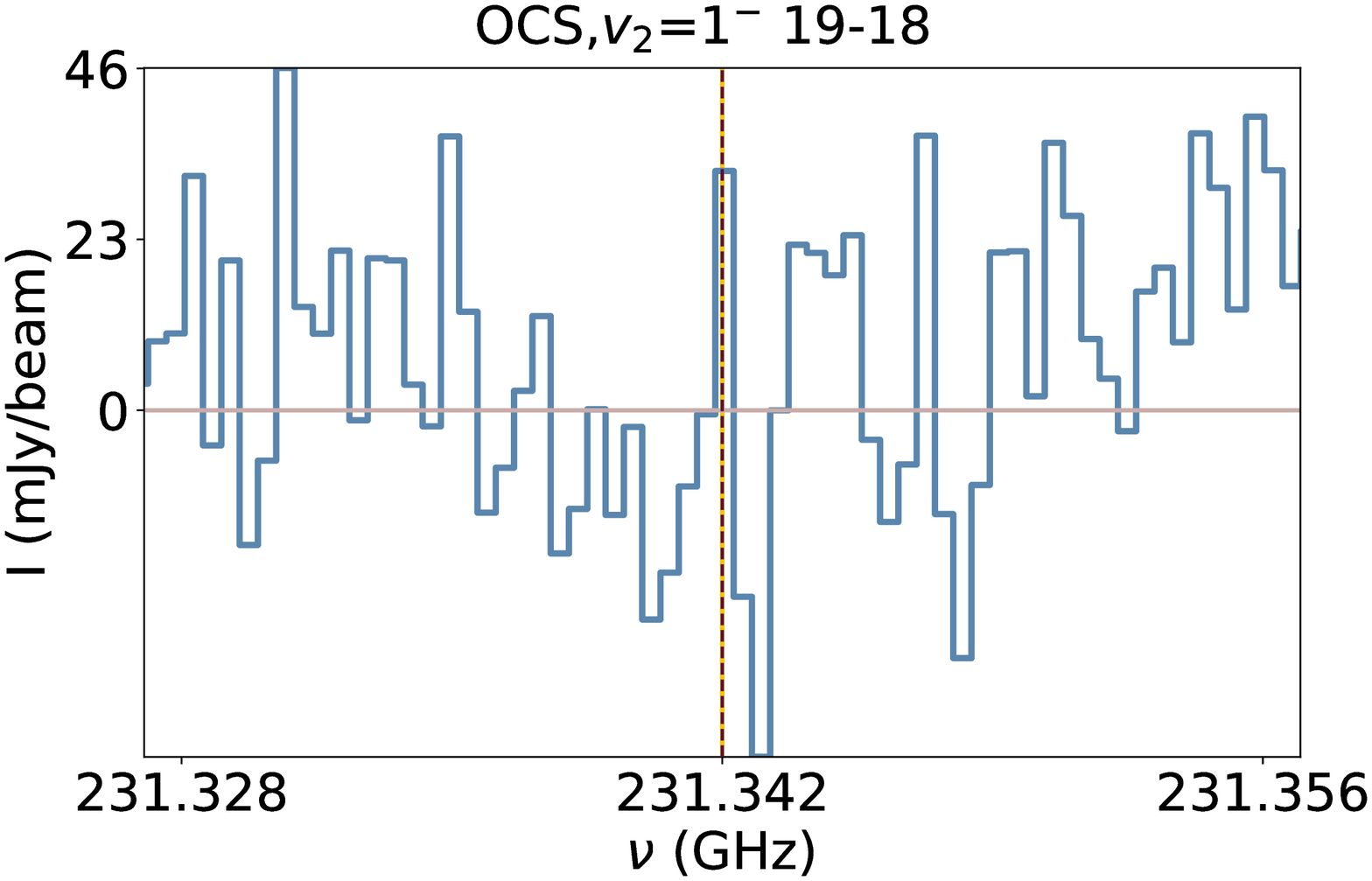}}
    \subfigure{\includegraphics[width=2.0in]{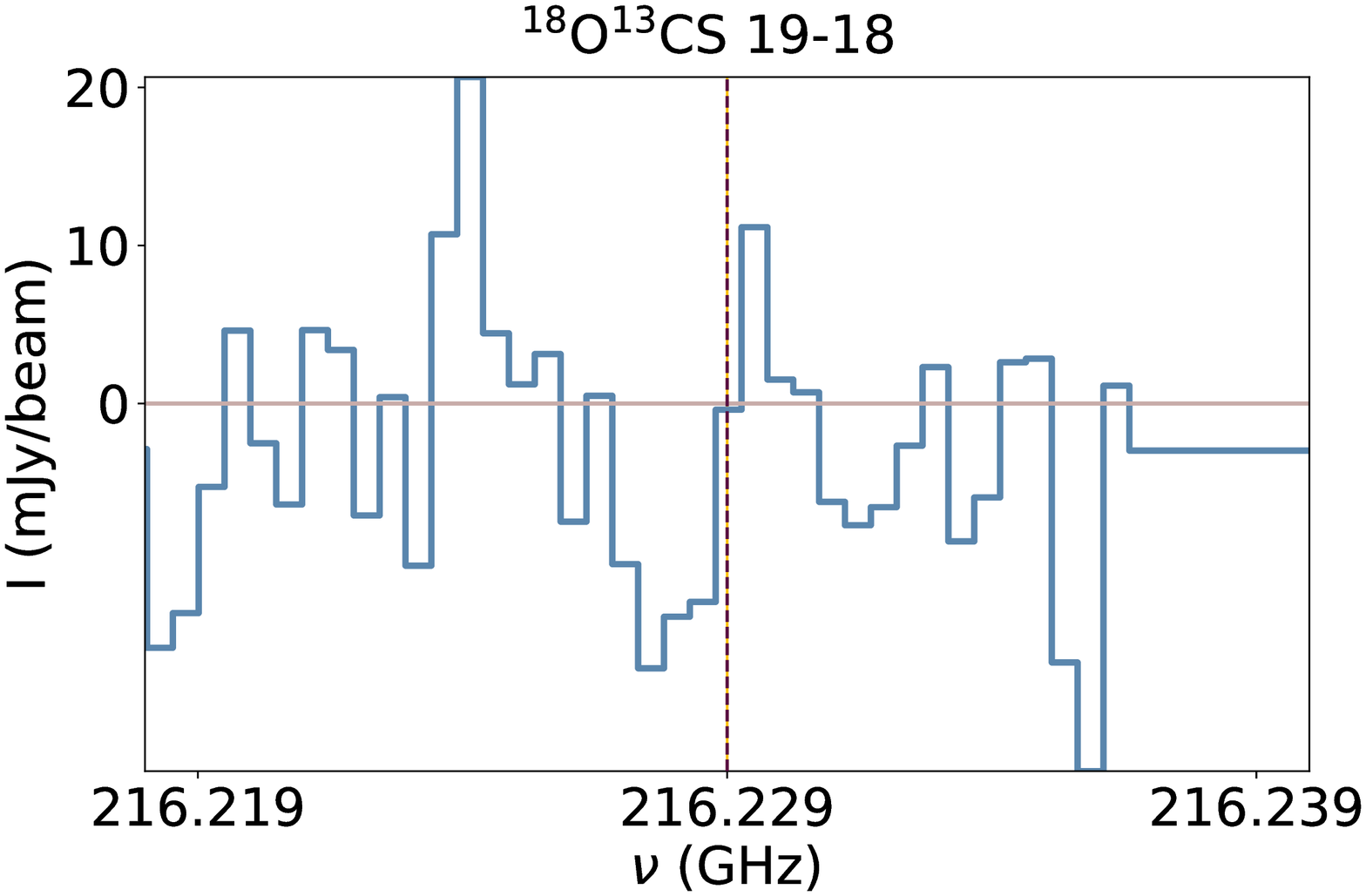}}
\end{minipage}
\caption{Observed spectra (in blue), rest frequency of the undetected line (brown dashed line), and spectroscopic uncertainty on the rest frequency of the undetected line (yellow shaded region) plotted for the sulfur-bearing species undetected towards NGC 1333-IRAS4B.}
\label{undetected_IRAS4B}
\end{figure}

H$_2$S, OCS, $v$=0, $J = 19-18$, and OC$^{33}$S, $J = 18-17$ emission lines are detected in IRAS4B. Initially, an excitation temperature and a source size of 210 K and 2$''$, respectively, were assumed for the modeling based on the rotation temperature of the detected CH$_{3}$OH lines towards this source \citep{Yang2021}. In this case, the computed $\tau$ values indicate that all the lines are optically thin (\autoref{results_IRAS4B_tex210ss2}). However, there is a mismatch by an order of magnitude in the column density of OCS as determined from its $v$=0 line and the OC$^{33}$S line. It is unlikely that the isotopic ratio of $^{32}$S/$^{33}$S deviates from the canonical value for this one source in the well-studied NGC~1333. It is more likely that the assumed source size and excitation temperature are not appropriate for OCS (and H$_{2}$S by extension). An exploration of the parameter space demonstrated that an excitation temperature of $100$~K and a source size of $1\arcsec$ are more appropriate for these S-bearing molecules (\autoref{results_IRAS4B}). In this case, the OCS, $v$=0 line is partially optically thick and thus the column density determined from its optically thin minor OC$^{33}$S isotopolog is more reliable. The fitted synthetic spectra and spectra of undetected species are shown in \autoref{detected_IRAS4B} and \autoref{undetected_IRAS4B}, respectively. \autoref{detected_IRAS4B} shows that the lines of S-bearing molecules deviate from the literature $v_{\text{LSR}}=7.4$~km/s value that was determined based on CO observations \cite{Kristensen2012}. A single $v_{\text{LSR}}$ value does not fit all the emission lines detected in this work. This discrepancy can only be understood with a more thorough characterisation of the structure of this source on large and small spatial scales.

\begin{table}[H]
    \centering
    \caption{Synthetic fitting of the detected S-bearing species towards NGC 1333-IRAS4B for an excitation temperature of 210 K and a source size of 2$''$. Directly across from a specific minor isotopolog under ``Derived $N$ of isotopologs’ follows the column density of the main isotopolog upon the assumption of the standard isotopic ratio.}
    \label{results_IRAS4B_tex210ss2}
    \begin{tabular}{r r c r c c c c c c r}
    \hline
    \hline
    Species & Transition & Frequency & $E_{\text{up}}$ & $A_{ij}$ &
    Beam size & FWHM & $N$ & Derived $N$ & $\tau$ \\
    &&&&&&&& of isotopologs & \\
    & & (GHz) & (K) & (s$^{-1}$) & ($''$) & km s$^{-1}$ & (cm$^{-2}$) & (cm$^{-2}$) &  \\
\hline
& & & & &  &  &  &\\
H$_2$S & 2$_{2,0}$-2$_{1,1}$ & 216.710 & 84 & 4.9$\times$10$^{-5}$  & 6.5 & 2.0 & 2.5$^{+0.3}_{-0.3}\times$10$^{15}$ & -- & 0.050\\
OC$^{33}$S & 18-17 & 216.147 & 99 & 2.9$\times$10$^{-5}$ & 6.5 & 1.5 &  7.6$^{+3.4}_{-1.3}\times$10$^{13}$ &  $N$(OCS)=9.5$^{+4.3}_{-1.6}\times$10$^{15}$ & 0.004 \\
OCS, $v$=0 & 19-18 & 231.061 & 111  & 3.6$\times$10$^{-5}$  & 6.1 & 2.0 & 5.7$^{+0.5}_{-0.3}\times$10$^{14}$ & -- &  0.020 \\

\hline
\hline
    \end{tabular}
\end{table}

\begin{table}[H]
    \centering
    \caption{Synthetic fitting of the detected S-bearing species towards NGC 1333-IRAS4B for an excitation temperature of 100 K and a source size of 1$''$. Directly across from a specific minor isotopolog under ``Derived N of isotopologs’ follows the column density of the main isotopolog upon the assumption of the standard isotopic ratio.}
    \label{results_IRAS4B}
    \begin{tabular}{r r c r c c c c c c r}
    \hline
    \hline
    Species & Transition & Frequency & $E_{\text{up}}$ & $A_{ij}$ &
    Beam size & FWHM & $N$ & Derived $N$ & $\tau$ \\
    &&&&&&&& of isotopologs & \\
    & & (GHz) & (K) & (s$^{-1}$) & ($''$) & km s$^{-1}$ & (cm$^{-2}$) & (cm$^{-2}$) &  \\
\hline
& & & & &  &  &  &\\
H$_2$S & 2$_{2,0}$-2$_{1,1}$ & 216.710 & 84 & 4.9$\times$10$^{-5}$  & 6.5 & 2.0 & $>$5.8$\times$10$^{15}$ & -- & 0.56\\
OC$^{33}$S & 18-17 & 216.147 & 99 & 2.9$\times$10$^{-5}$ & 6.5 & 1.5 &  2.2$^{+0.5}_{-0.4}\times$10$^{14}$ &  $N$(OCS)=2.75$^{+0.65}_{-0.55}\times$10$^{16}$ & 0.03 \\
OCS, $v$=0 & 19-18 & 231.061 & 111  & 3.6$\times$10$^{-5}$  & 6.1 & 2.0 & $>$1.9$\times$10$^{15}$ & -- &  0.20 \\
\hline
\hline
    \end{tabular}
\end{table}

\subsection{Detected lines in NGC 1333-IRAS4B}
\label{shiftIRAS4B}

\subsection{Undetected lines in NGC 1333-IRAS4B}

\section{Ser-SMM3}

The excitation temperature was difficult to estimate for SMM3, as no molecular tracer of the inner envelope has been detected. As a result, the column density of the detected H$_2$S, 2$_{2,0}$-2$_{1,1}$ and OCS, $v$=0, $J = 19-18$ lines is evaluated for a range of excitation temperatures between 100 and 250 K. The OCS, $v$=0 line shows a double peaked feature and the synthetic spectrum is fitted covering both peaks. The best-fitting models to H$_2$S and OCS, $v$=0 have line widths of 2.5 and 3.0 km s$^{-1}$, and source sizes of 2.0$''$ and 0.5$''$, respectively. Different source sizes for H$_{2}$S and OCS were assumed in this source, because 12m ALMA data at a spatial resolution of $0.5\arcsec$ show very different spatial distributions for these molecules. The H$_{2}$S and OCS integrated intensity maps in fig.~9 of \cite{Tychoniec2021} show extended emission for H$_{2}$S covering $\sim2\arcsec$ and much more compact emission for OCS $\sim0.5\arcsec$. The lines are optically thin ($\tau<0.1$) with line opacities tabulated in \autoref{results_SMM3}.

\begin{table}[H]
    \centering
    \caption{Synthetic fitting of the detected S-bearing species towards Ser-SMM3 for a range of excitation temperatures between 100 and 250 K.}
    \label{results_SMM3}
    \begin{adjustbox}{width=1\textwidth}
        \begin{tabular}{r r c r  c c c c c  r}
    \hline
    \hline
    Species & Transition & Frequency & $E_{\text{up}}$ & $A_{ij}$ & Beam size & source size & FWHM &  $N$ & $\tau$ \\

& & (GHz) & (K) & (s$^{-1}$) & ($''$)& ($''$) & (km s$^{-1}$) & (cm$^{-2}$) &  \\
\hline

&&&&&&&&\\

H$_2$S & 2$_{2,0}$-2$_{1,1}$ & 216.710 & 84 & 4.9$\times$10$^{-5}$  & 6.0 & 2.0 & 2.5 &  5.8$^{+3.1}_{-3.2}\times$10$^{14}$ & 0.01-0.02\\
OCS, $v$=0 & 19-18 & 231.060 & 111  & 3.6$\times$10$^{-5}$  & 5.6 & 0.5 & 3.0 & 8.7$^{+4.9}_{-4.9}\times$10$^{14}$ &  0.02-0.05 \\

\hline
\hline
\end{tabular}
\end{adjustbox}
\end{table}

\subsection{Detected lines in Ser-SMM3}

\begin{figure}[H]
\begin{minipage}{3.1in}
    \subfigure{\label{100K_H2S}\includegraphics[width=1.5in]{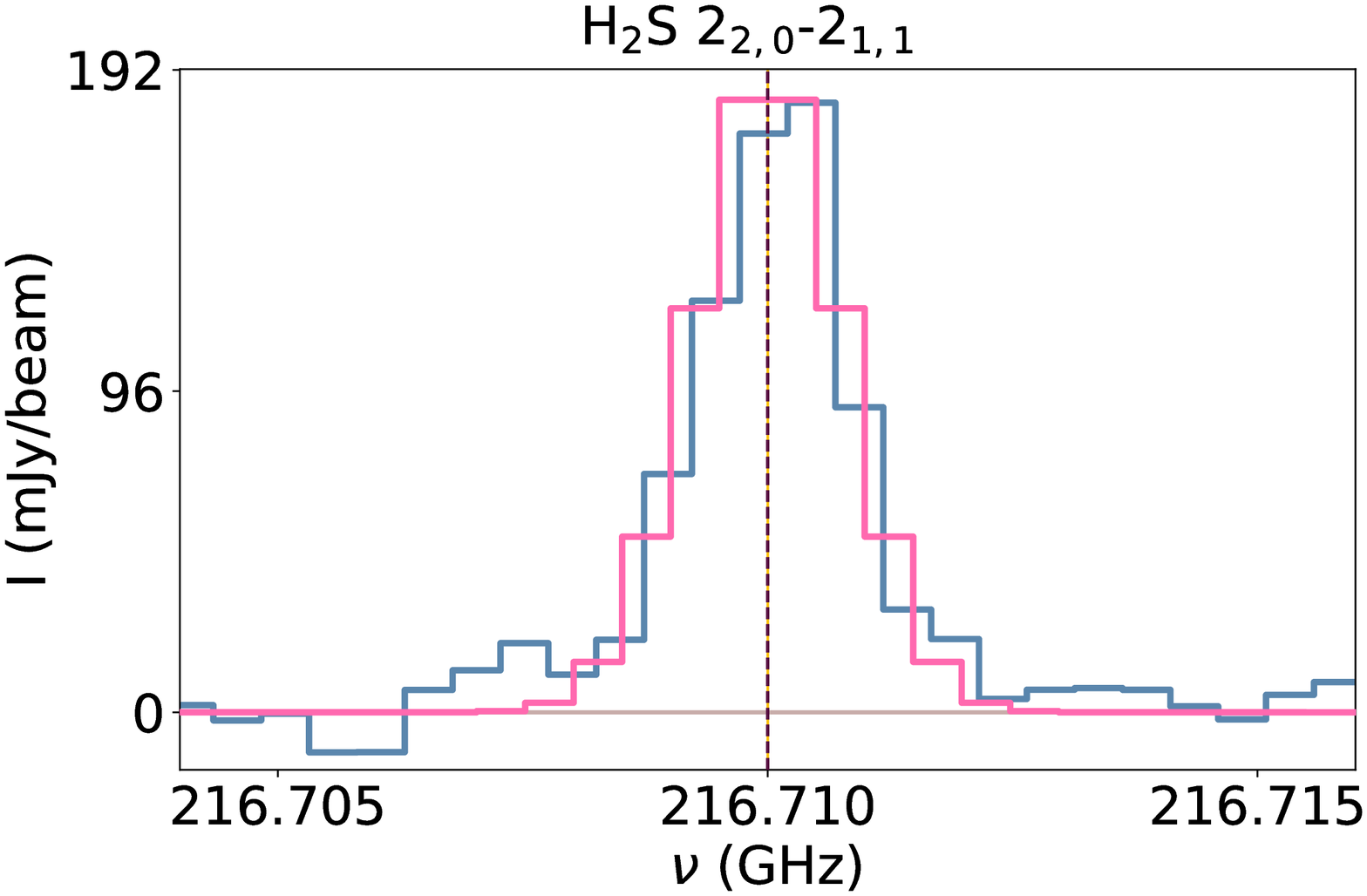}}
  \subfigure{\label{100K_OCS}\includegraphics[width=1.5in]{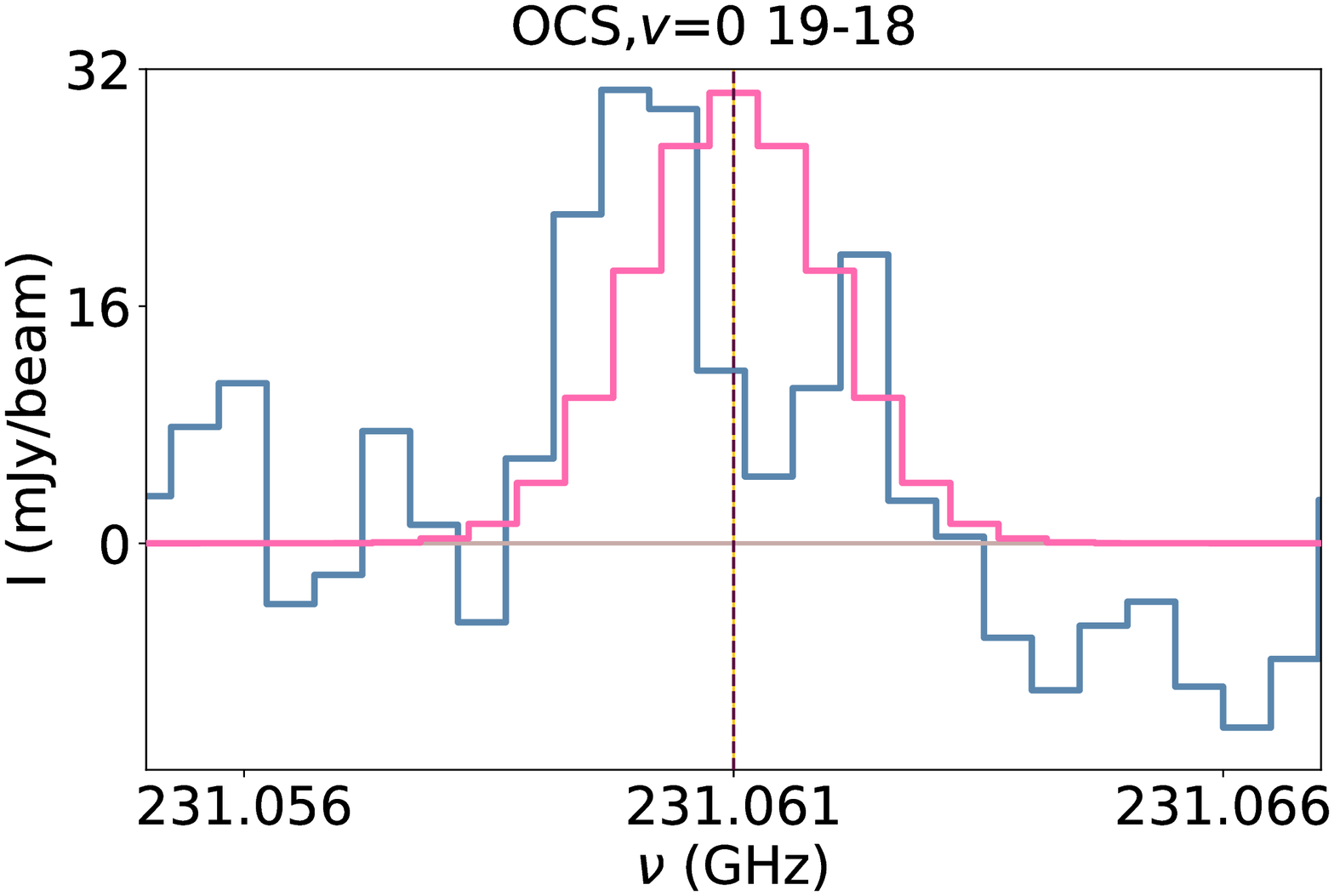}}
    \caption*{$T_{\text{ex}} = 100$ K}
\end{minipage}
\begin{minipage}{3.1in}
    \subfigure{\label{150K_H2S}\includegraphics[width=1.5in]{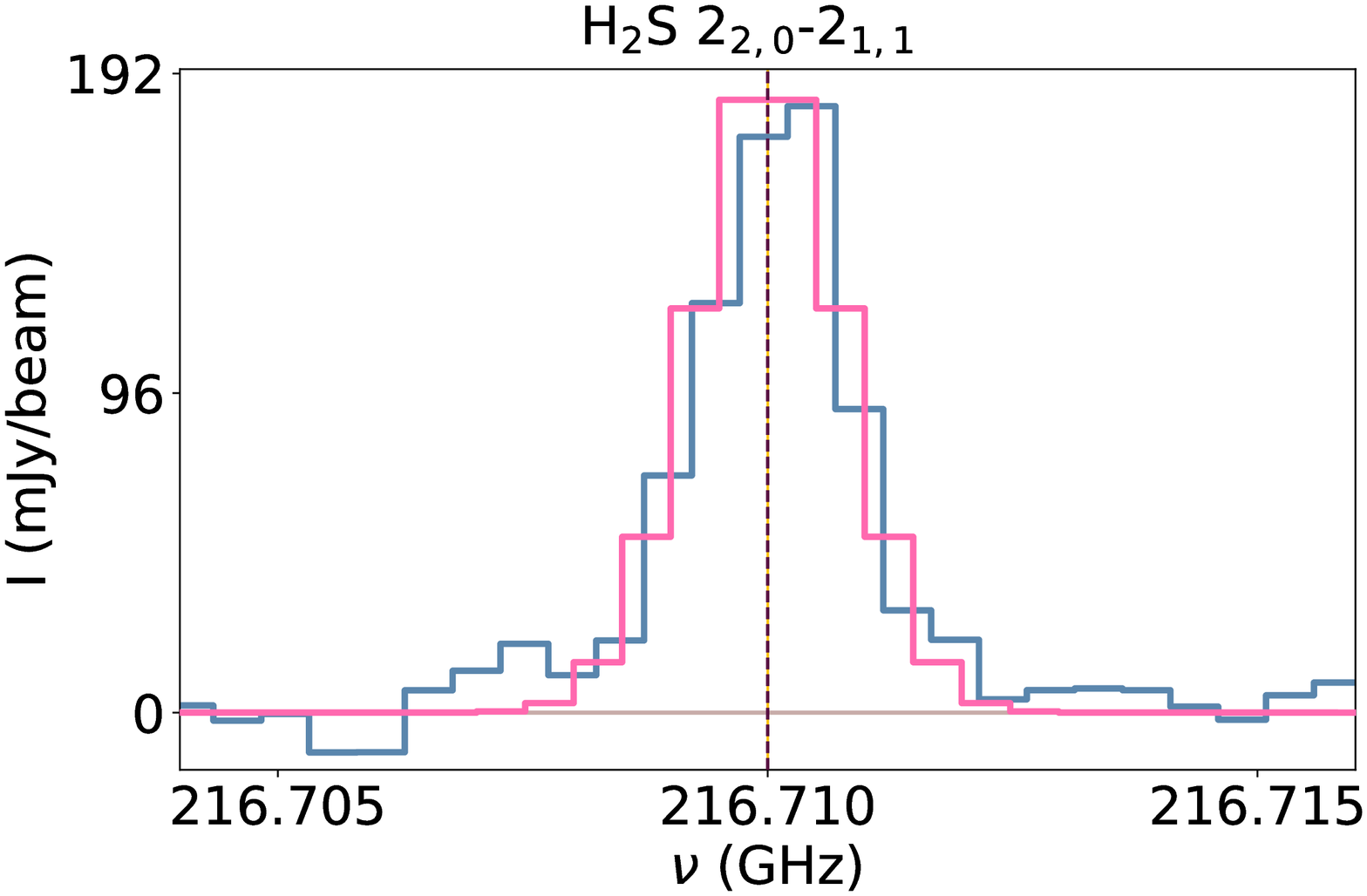}}    \subfigure{\label{150K_OCS}\includegraphics[width=1.5in]{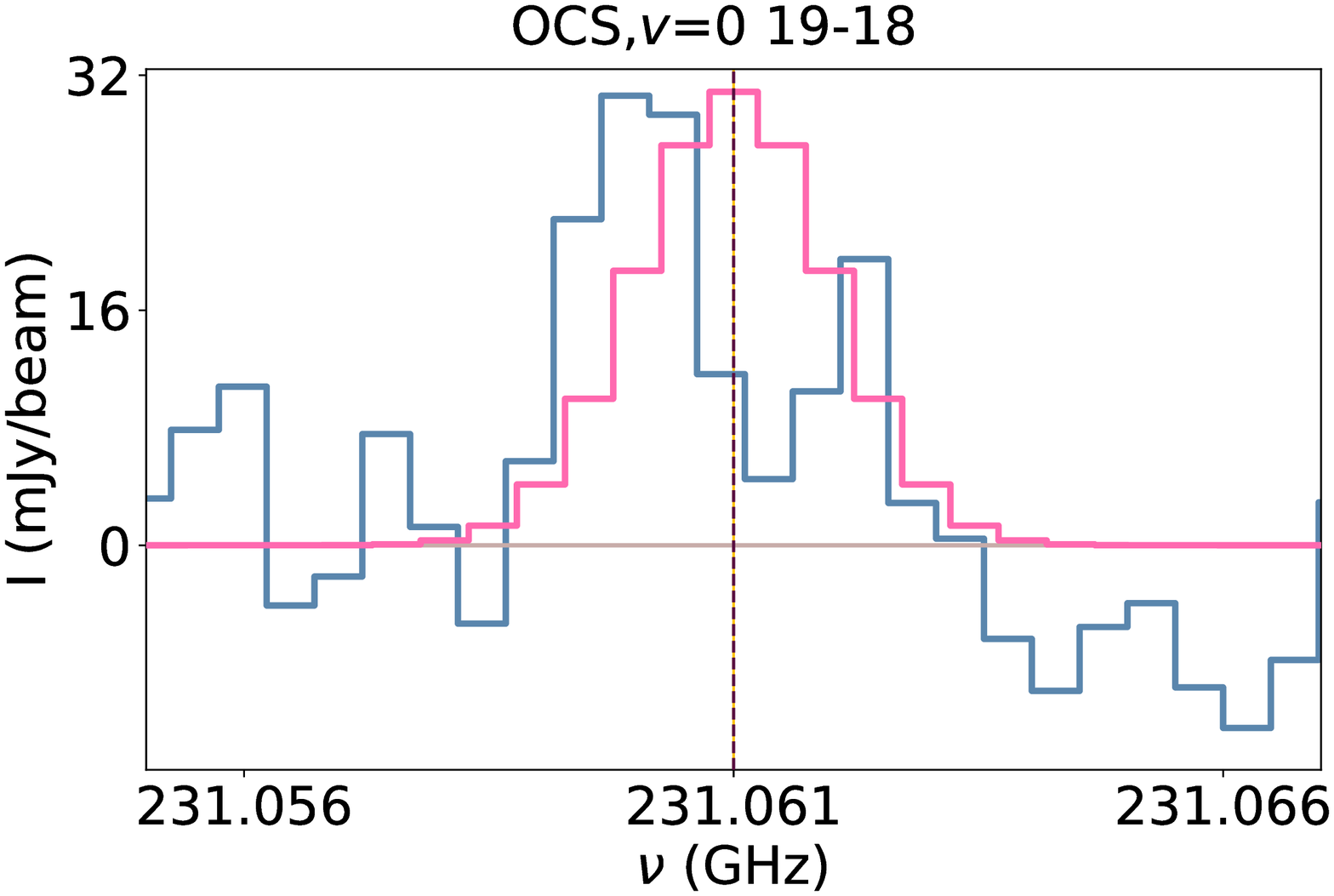}}
    \caption*{$T_{\text{ex}} = 150$ K}
\end{minipage}
\hfill
\begin{minipage}{3.1in}
    \subfigure{\label{200K_H2S}\includegraphics[width=1.5in]{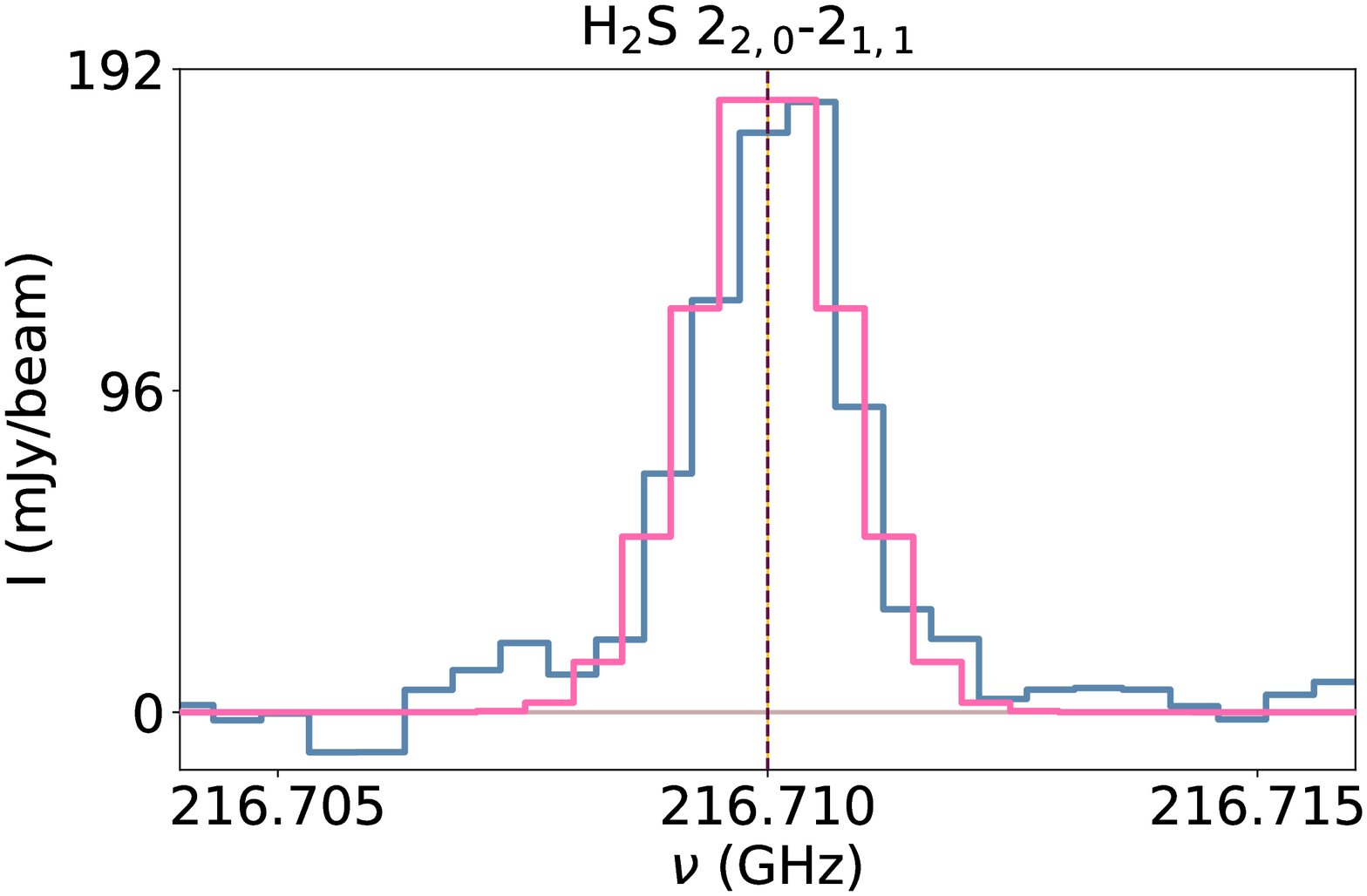}}
  \subfigure{\label{200K_OCS}\includegraphics[width=1.5in]{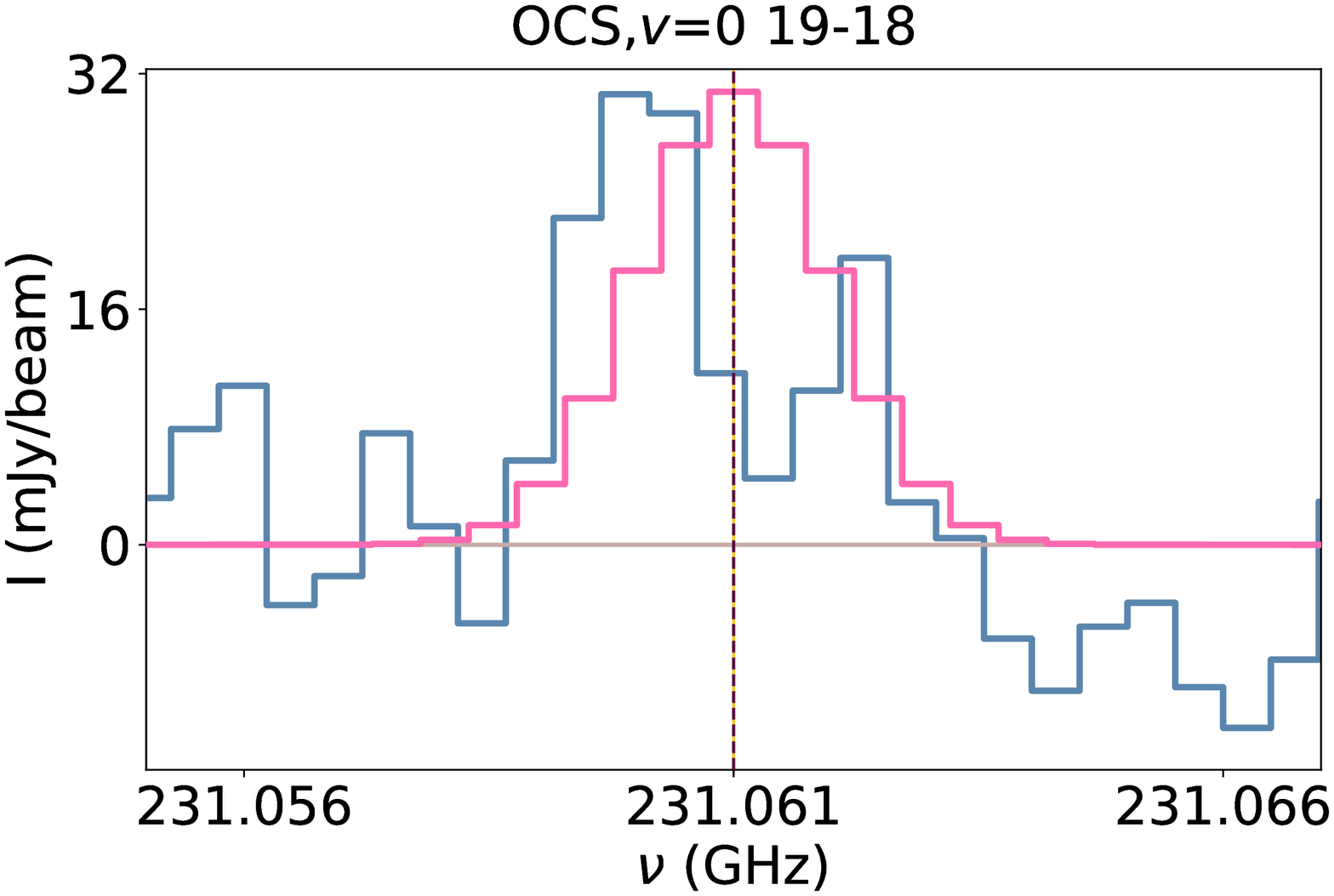}}
    \caption*{$T_{\text{ex}} = 200$ K}
\end{minipage}
\hfill
\begin{minipage}{3.1in}
    \subfigure{\label{250K_H2S}\includegraphics[width=1.5in]{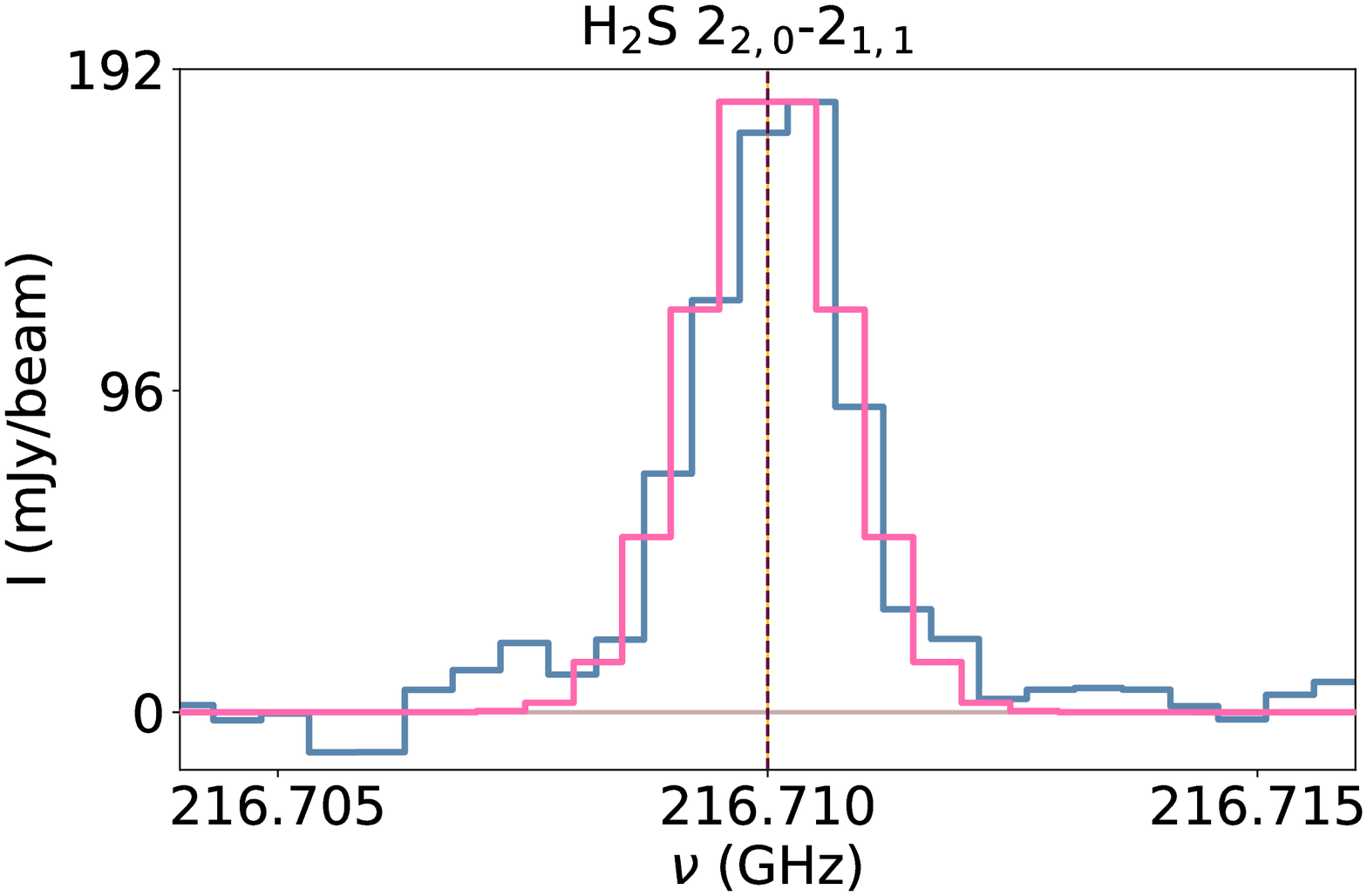}}
  \subfigure{\label{250K_OCS}\includegraphics[width=1.5in]{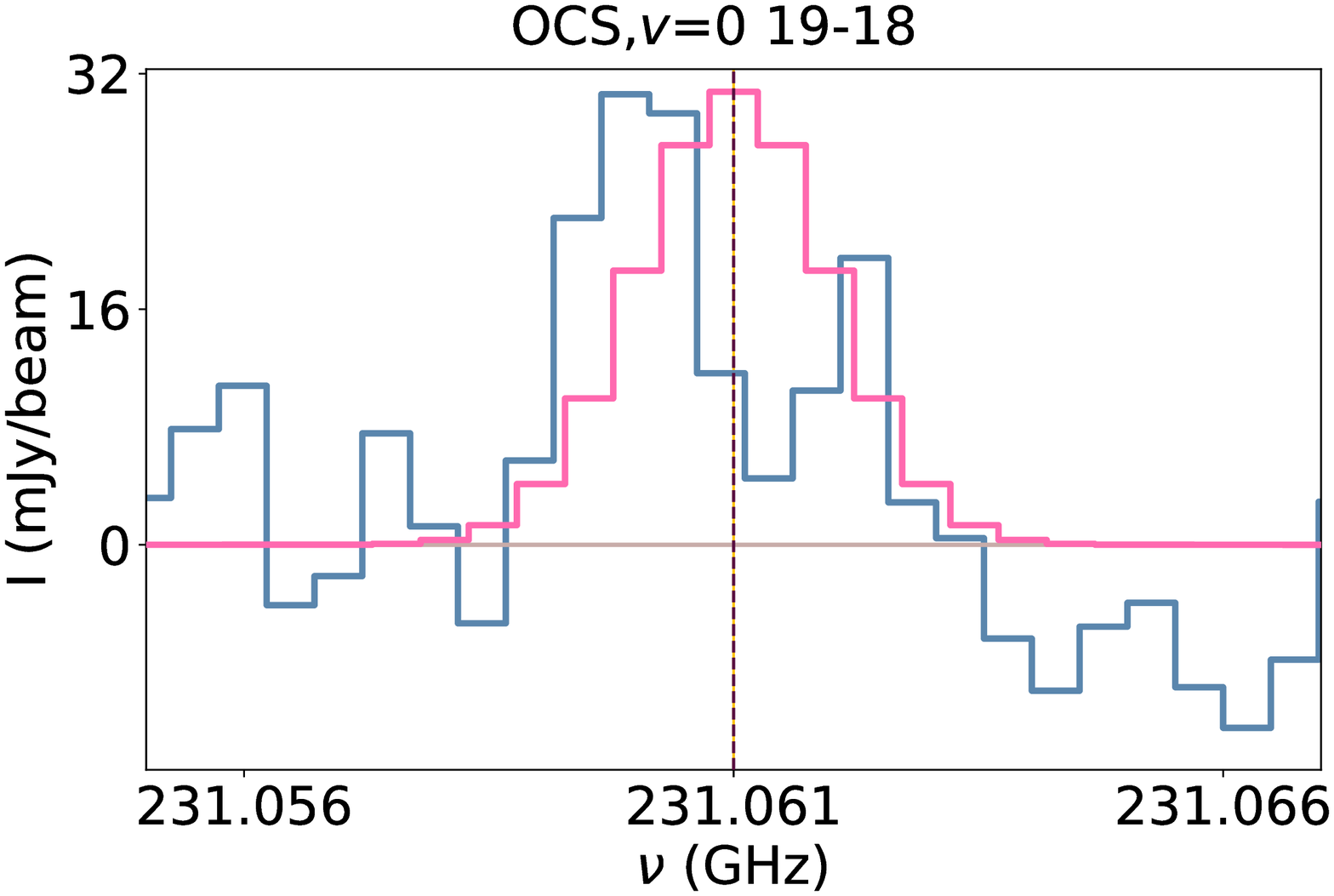}}
    \caption*{$T_{\text{ex}} = 250$ K}
\end{minipage}
\hfill
\caption{Observed spectra (in blue), rest frequency of the detected line (brown dashed line), spectroscopic uncertainty on the rest frequency of the detected line (yellow shaded region), and fitted synthetic spectra (in pink) plotted for the sulfur-bearing species detected towards Ser-SMM3 for a range of excitation temperature between 100 and 250 K.}
\label{detected_SMM3}
\end{figure}

\subsection{Undetected lines in Ser-SMM3}

\begin{figure}[H]
\centering
\begin{minipage}[b]{6.2in}%
    \subfigure{\includegraphics[width=2.0in]{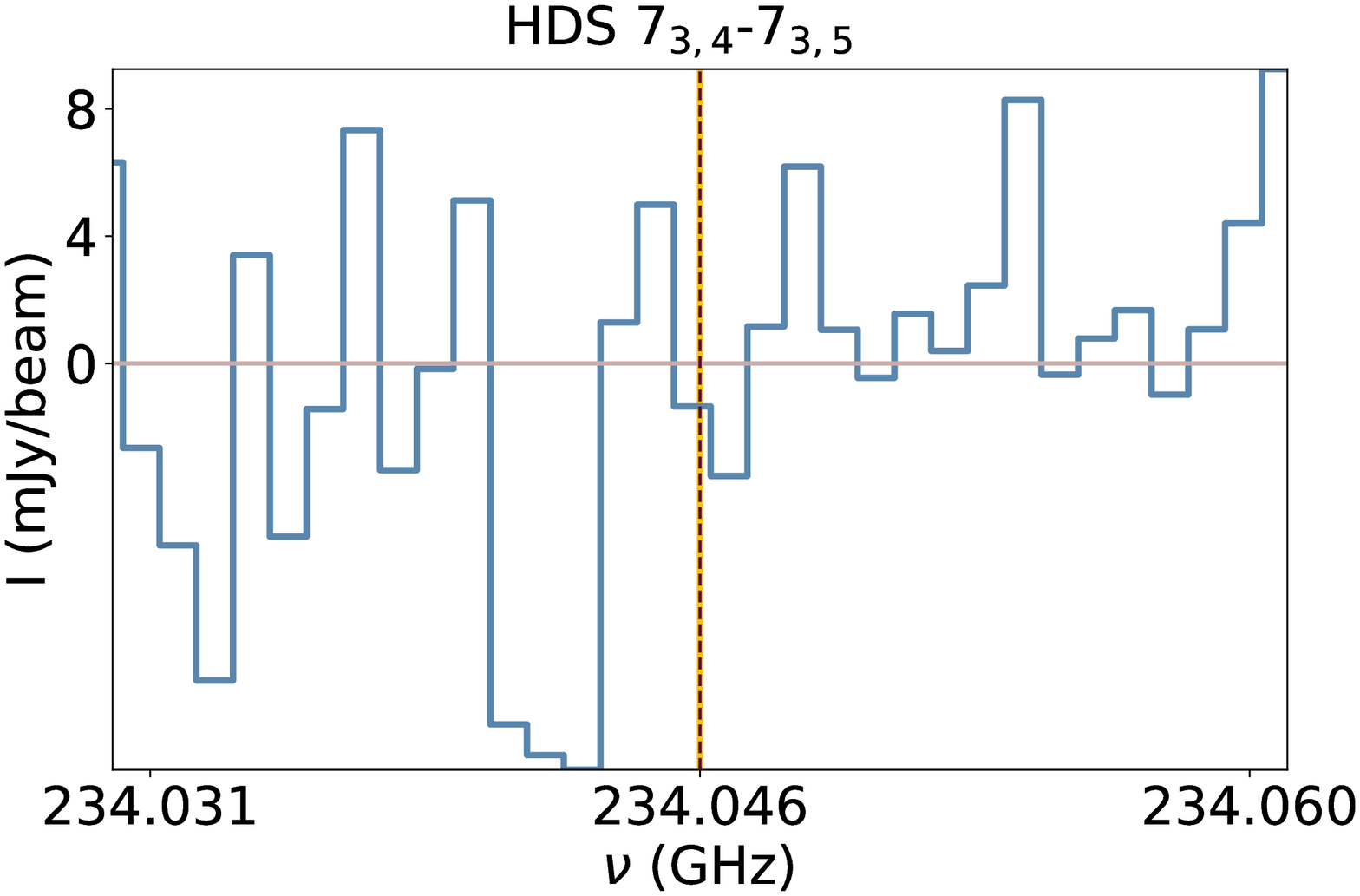}}
    \subfigure{\includegraphics[width=2.0in]{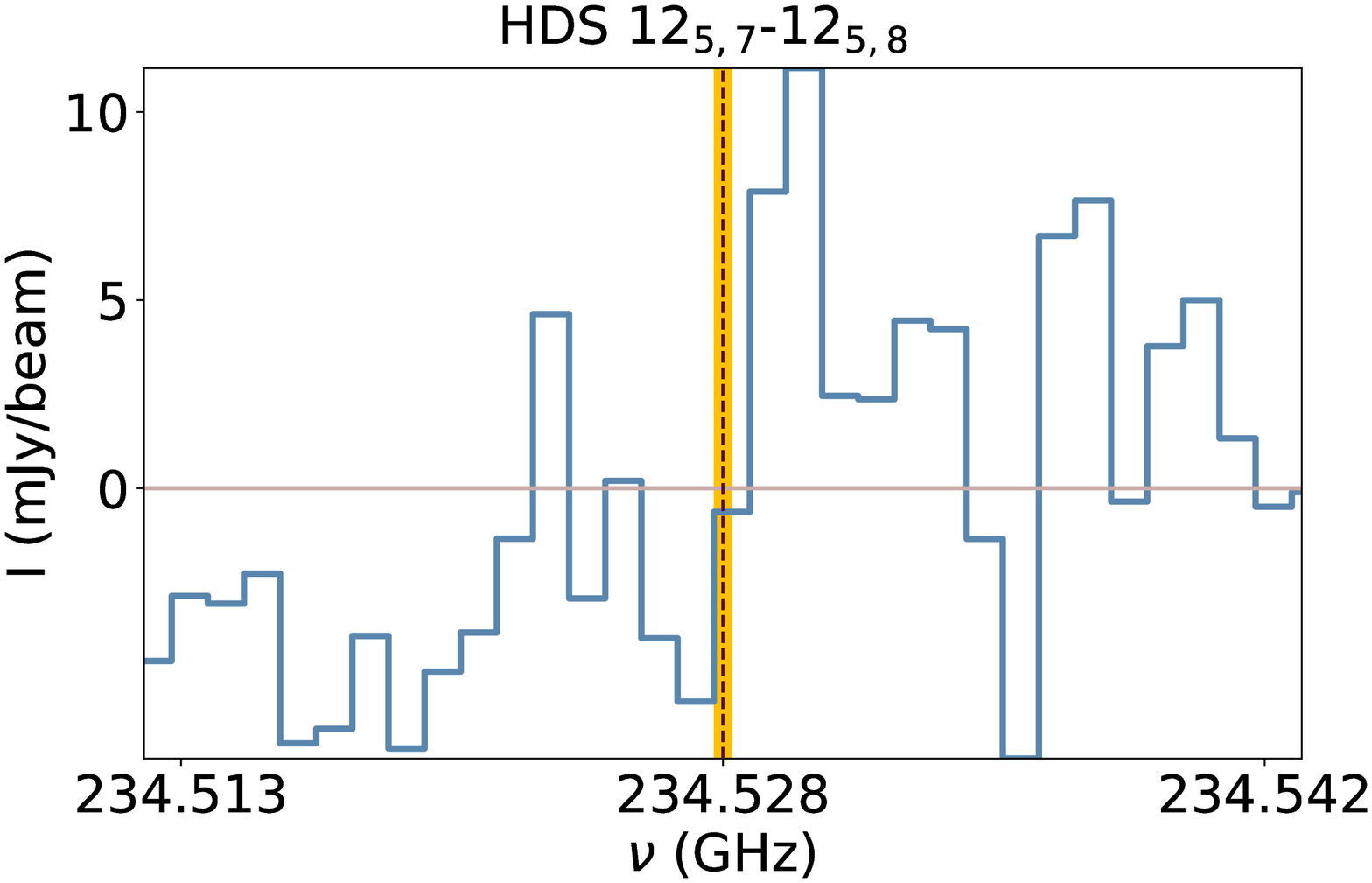}}
    \subfigure{\includegraphics[width=2.0in]{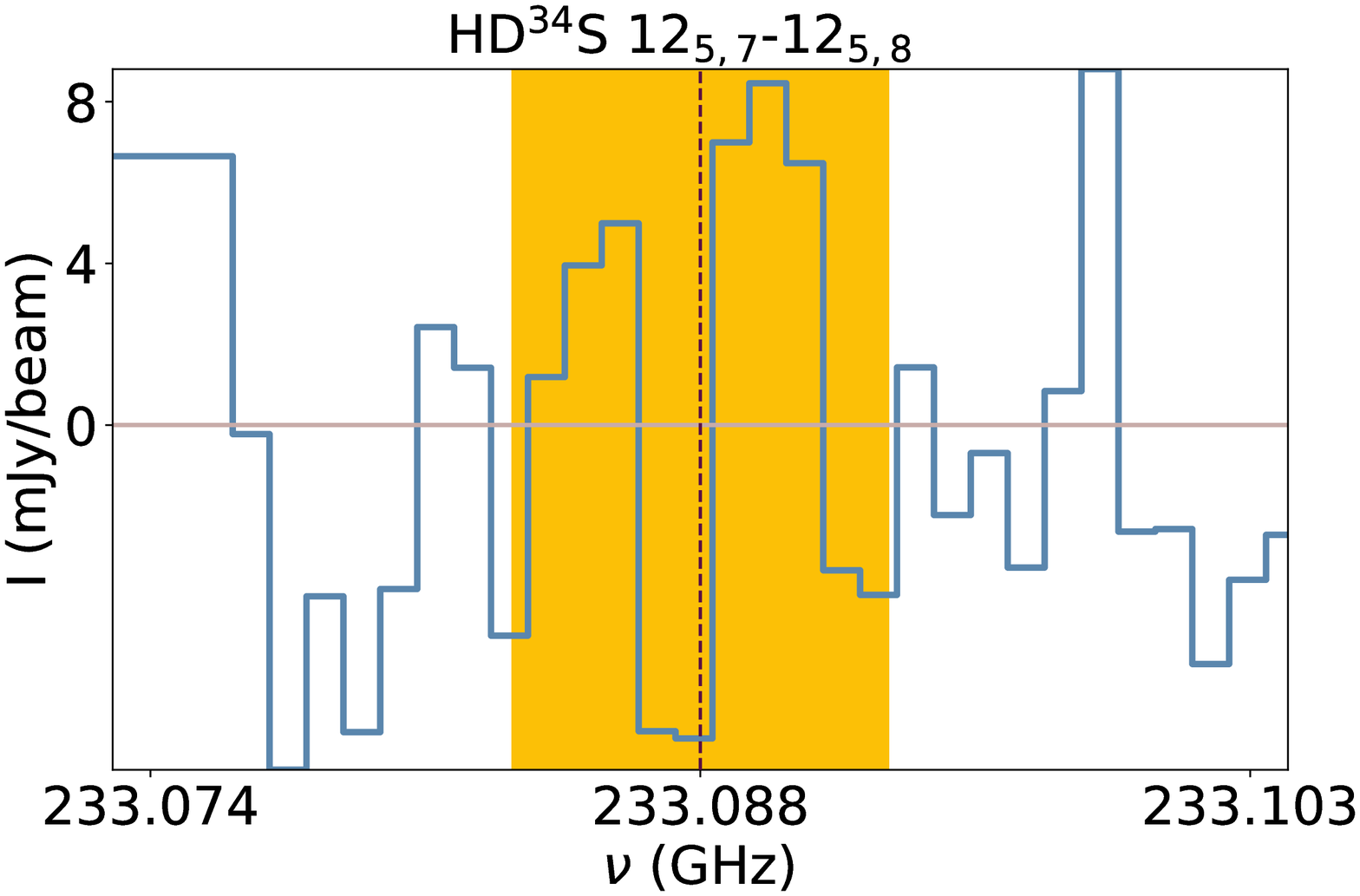}}\\
    \subfigure{\includegraphics[width=2.0in]{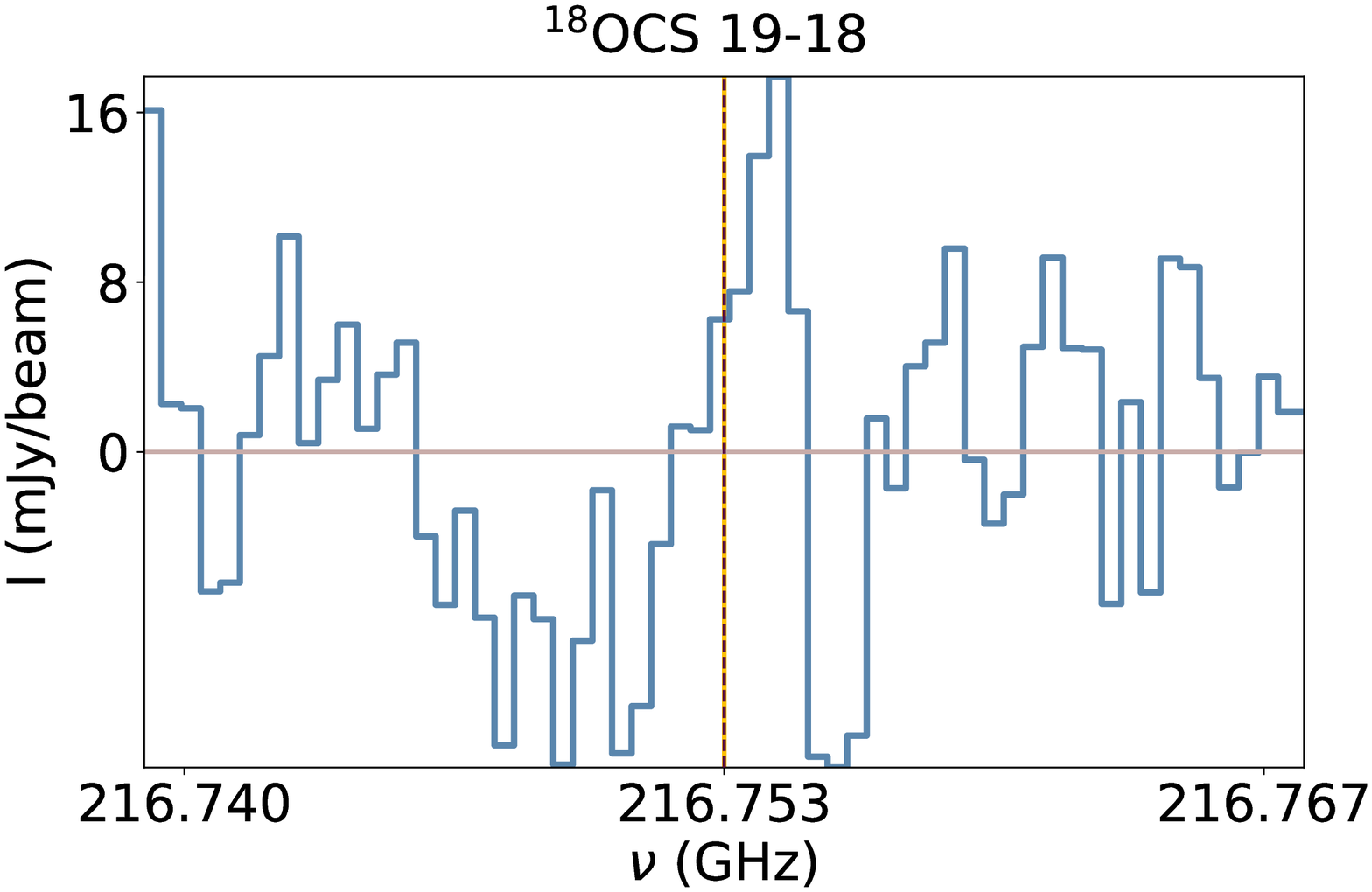}}
    \subfigure{\includegraphics[width=2.0in]{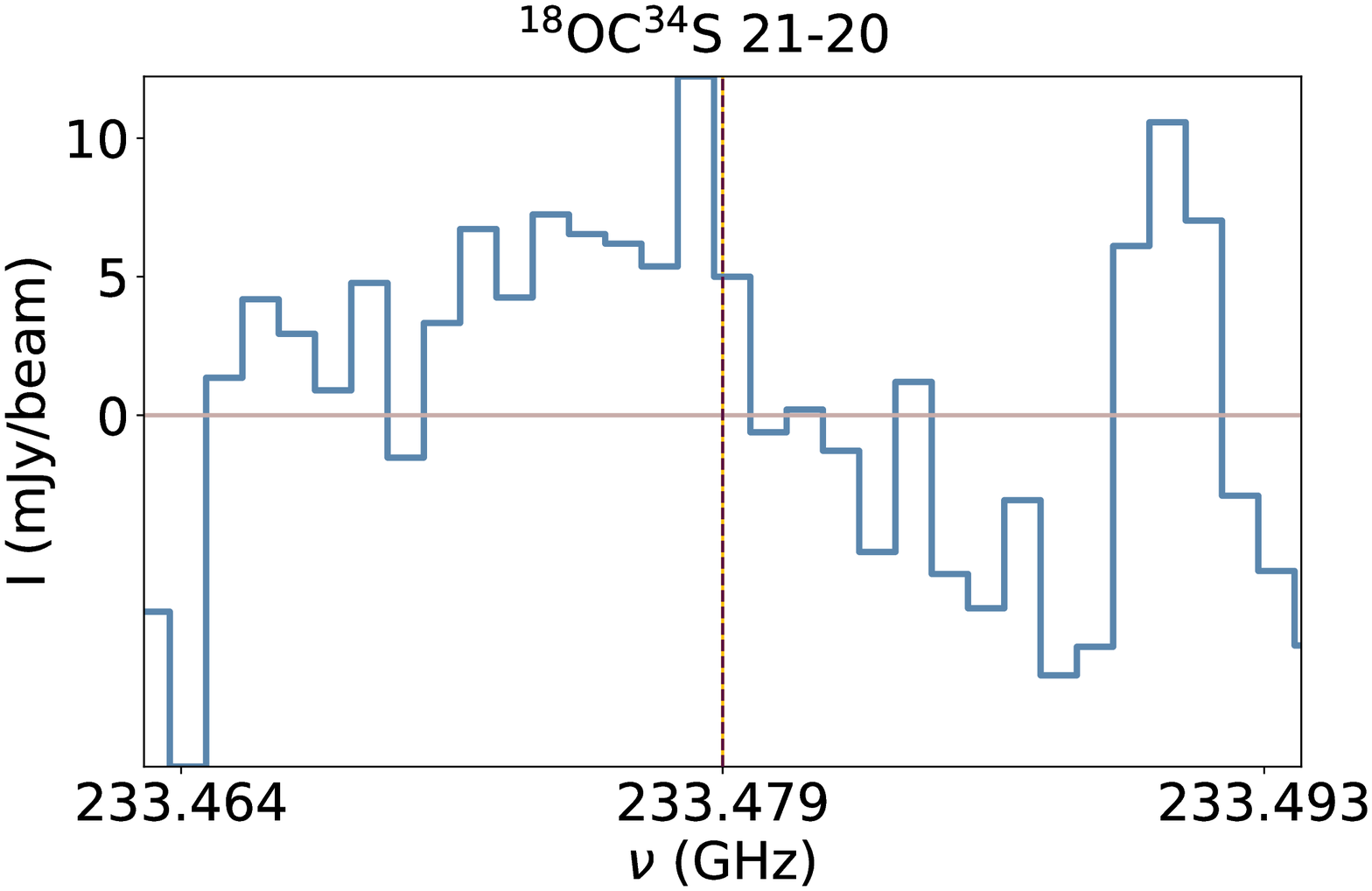}}
    \subfigure{\includegraphics[width=2.0in]{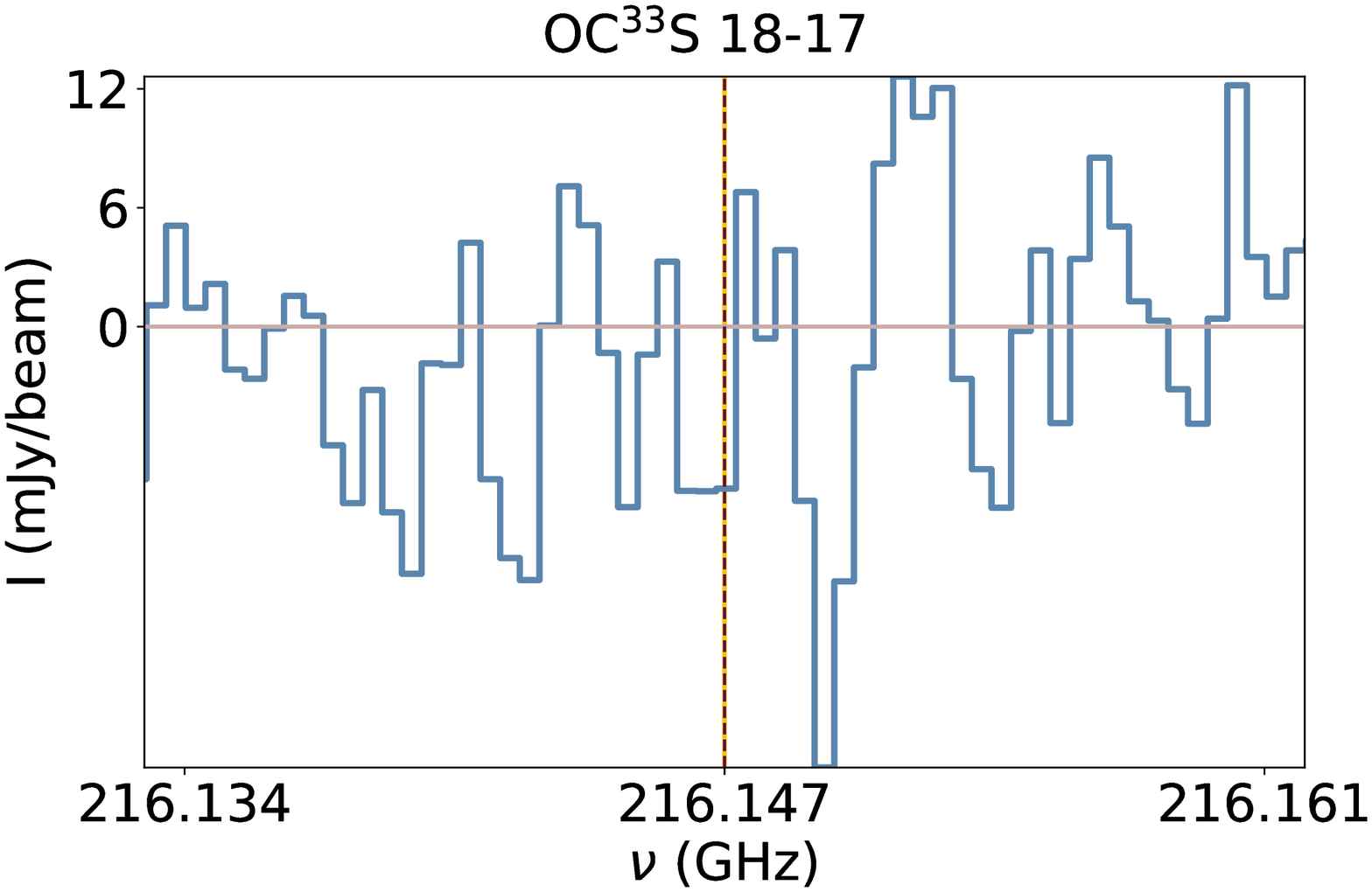}}\\
    \subfigure{\includegraphics[width=2.0in]{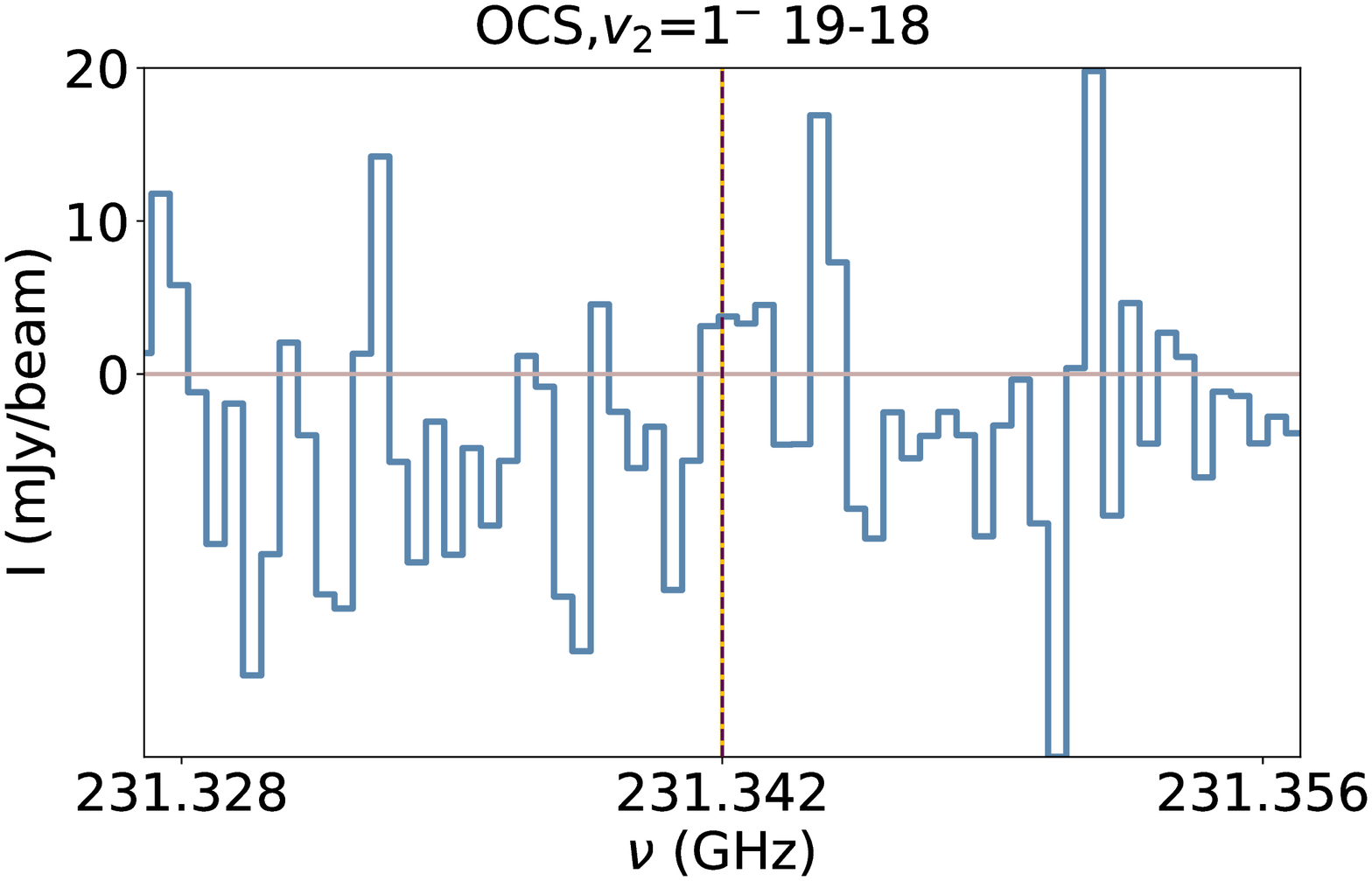}}
    \subfigure{\includegraphics[width=2.0in]{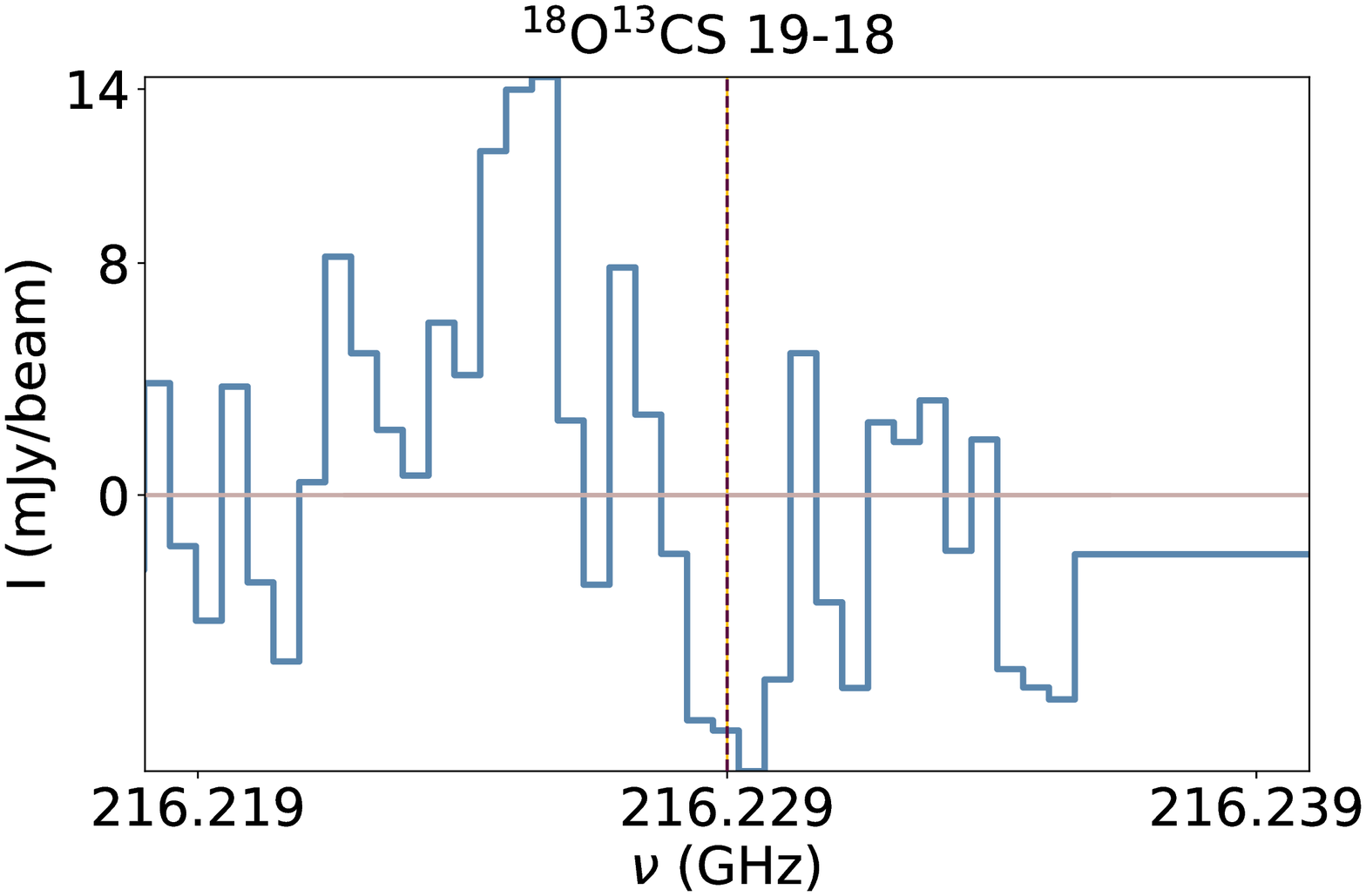}}    
\end{minipage}
\caption{Observed spectra (in blue), rest frequency of the undetected line (brown dashed line), and spectroscopic uncertainty on the rest frequency of the undetected line (yellow shaded region) plotted for the sulfur-bearing species undetected towards Ser-SMM3.}
\label{undetected_SMM3}
\end{figure}

\section{TMC1} \label{A_TMC1}

\begin{figure}[H]
\centering
\begin{minipage}[b]{6.2in}%
    \subfigure{\includegraphics[width=2.0in]{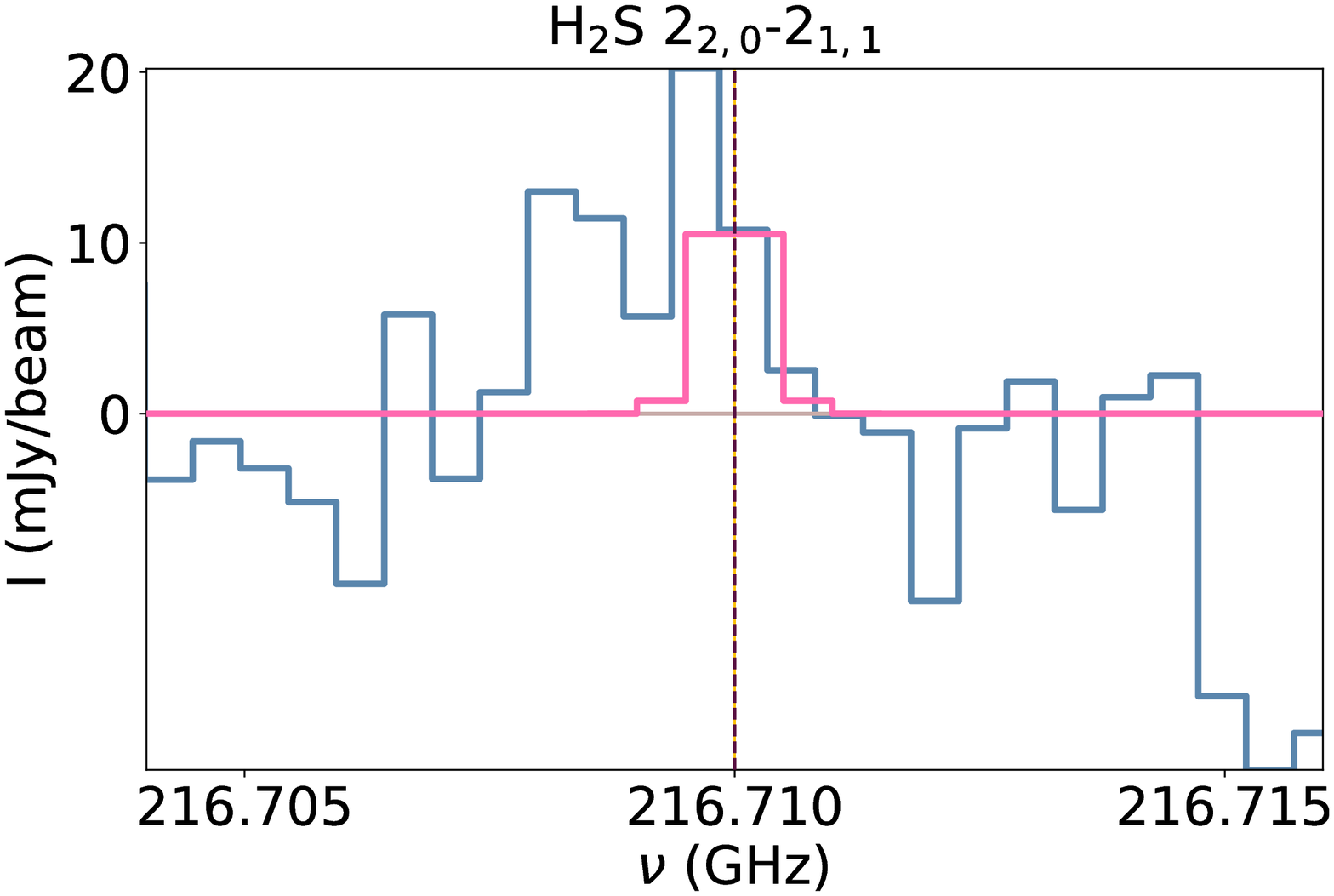}}
    \subfigure{\includegraphics[width=2.0in]{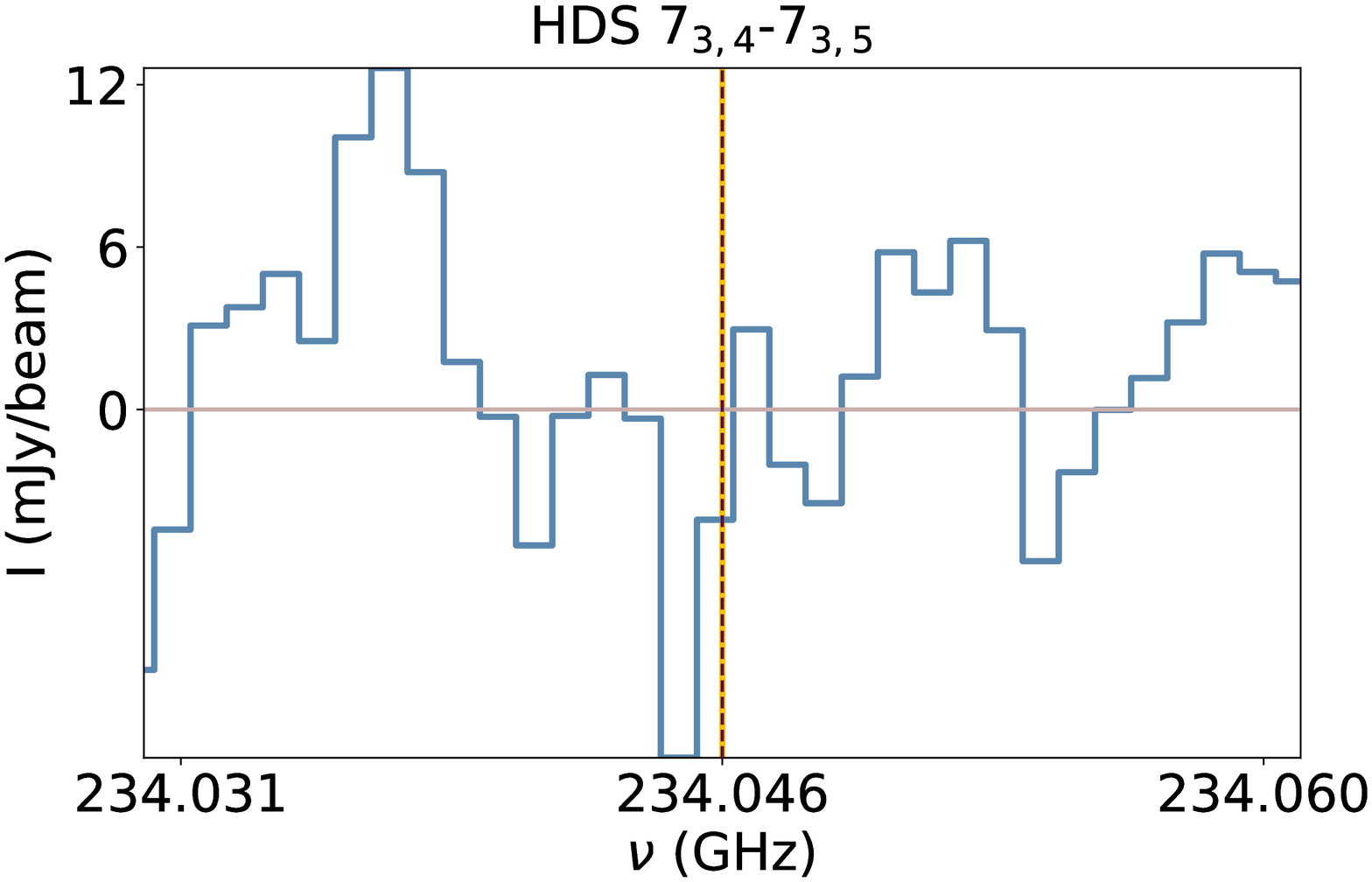}}
    \subfigure{\includegraphics[width=2.0in]{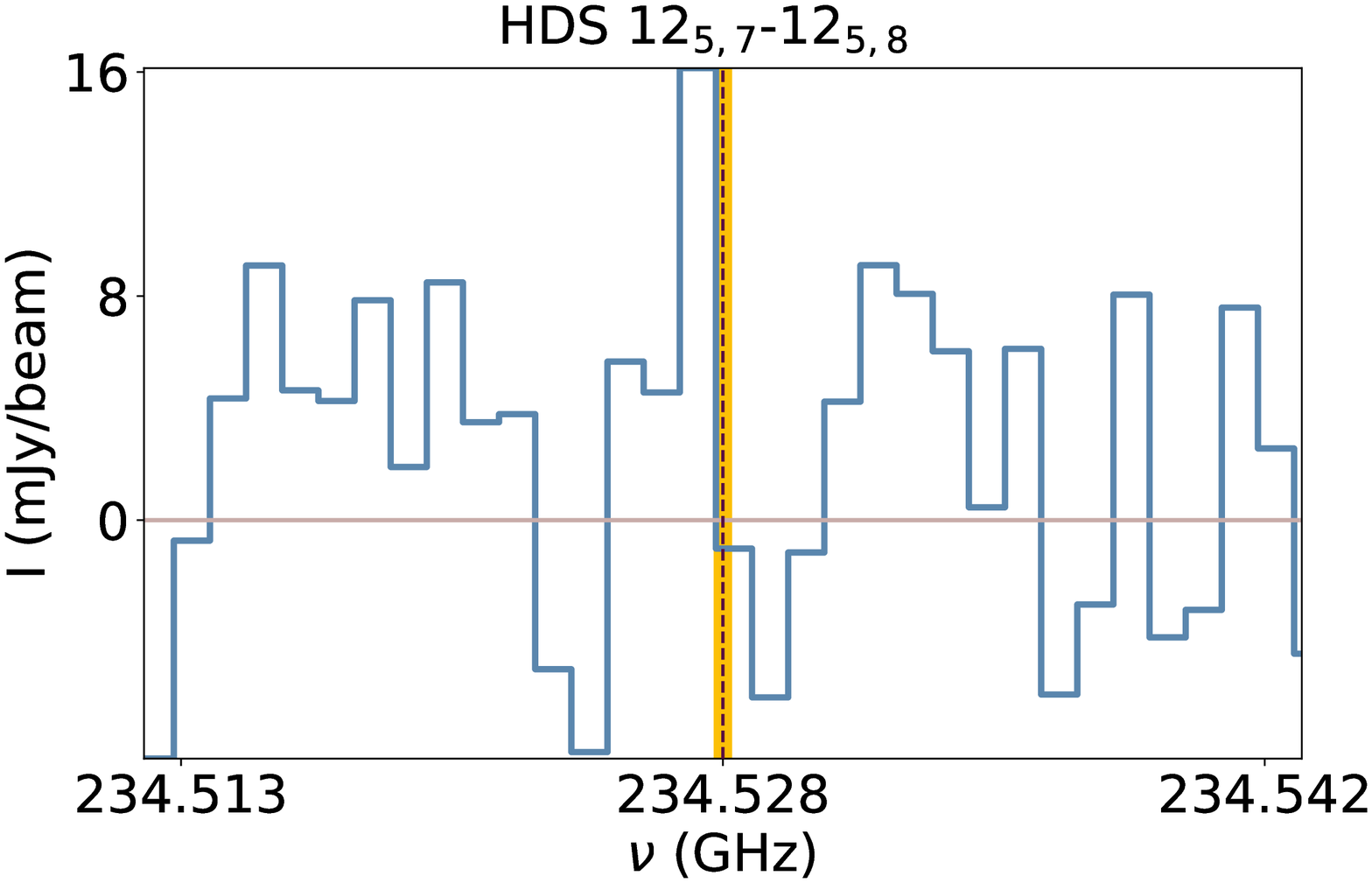}}\\
    \subfigure{\includegraphics[width=2.0in]{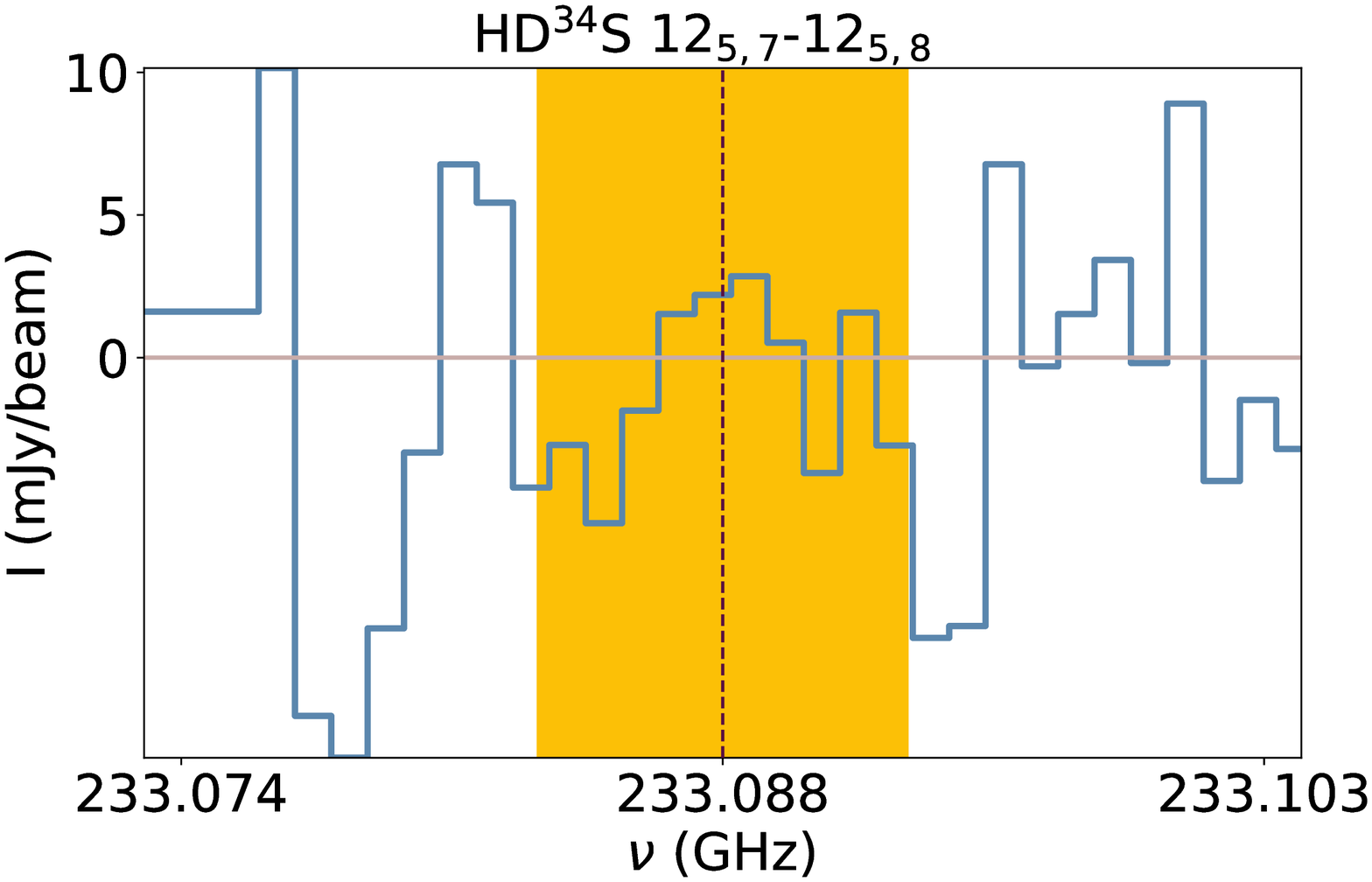}}
    \subfigure{\includegraphics[width=2.0in]{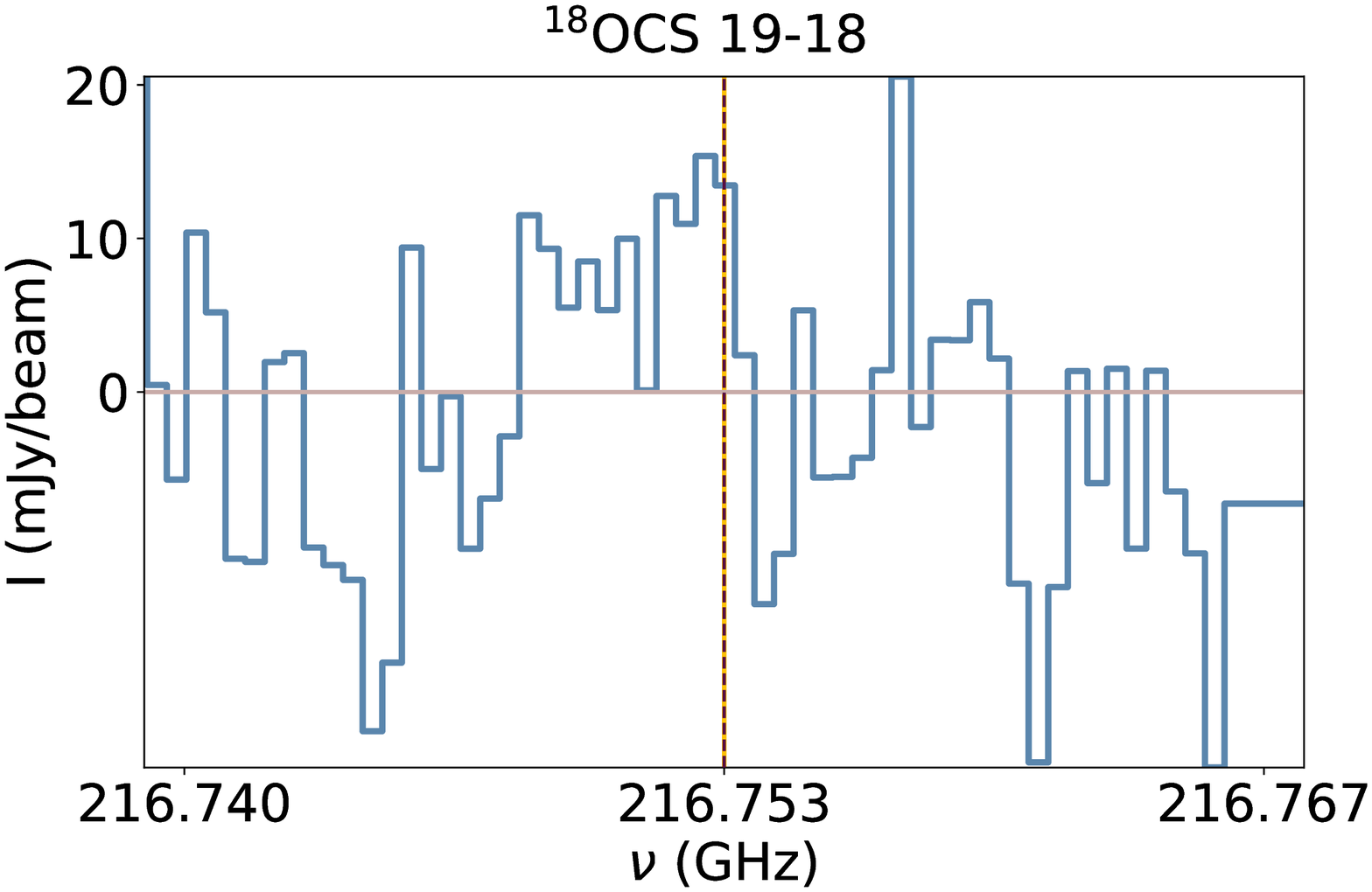}}
    \subfigure{\includegraphics[width=2.0in]{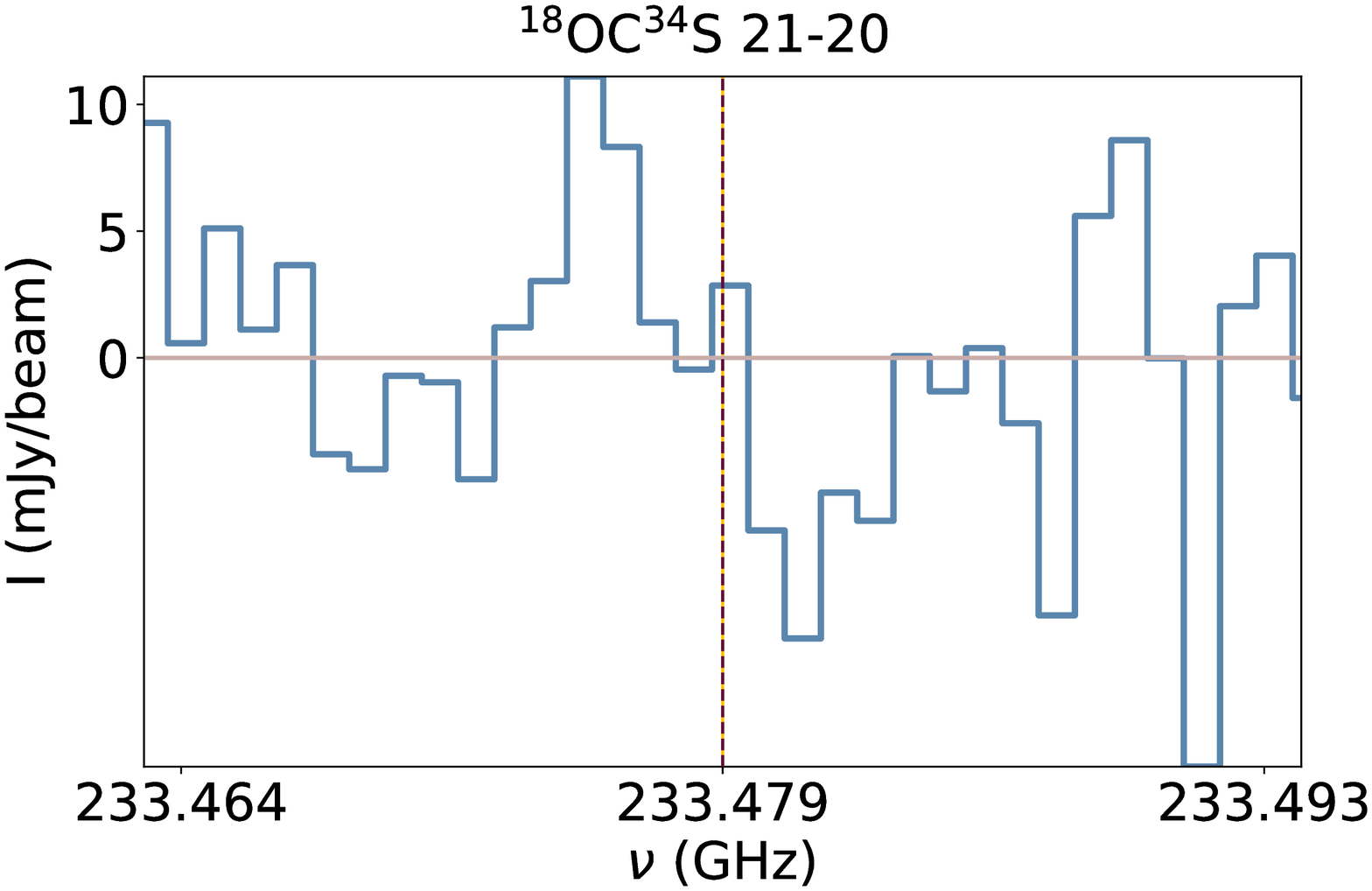}}\\
    \subfigure{\includegraphics[width=2.0in]{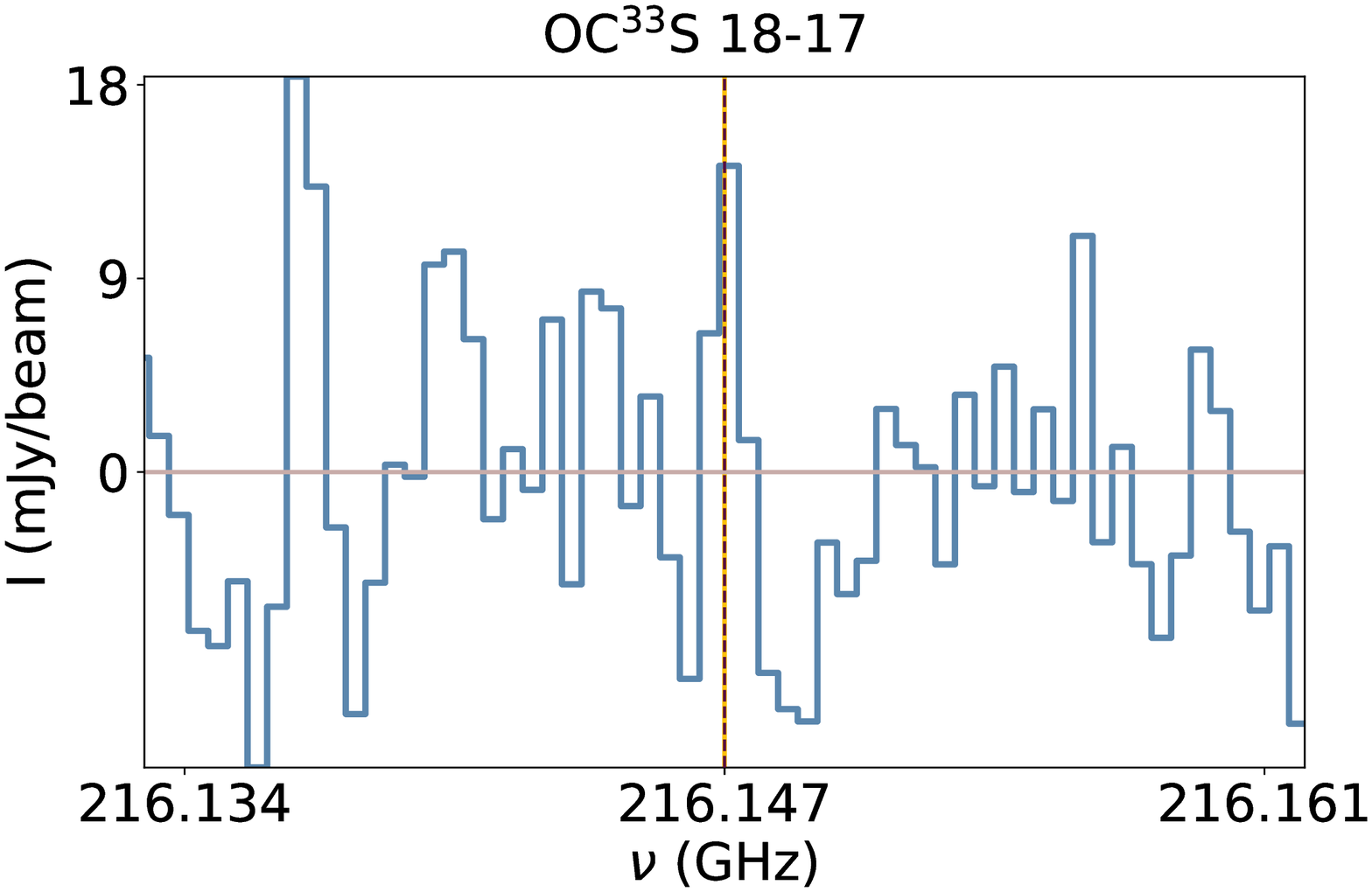}}
    \subfigure{\includegraphics[width=2.0in]{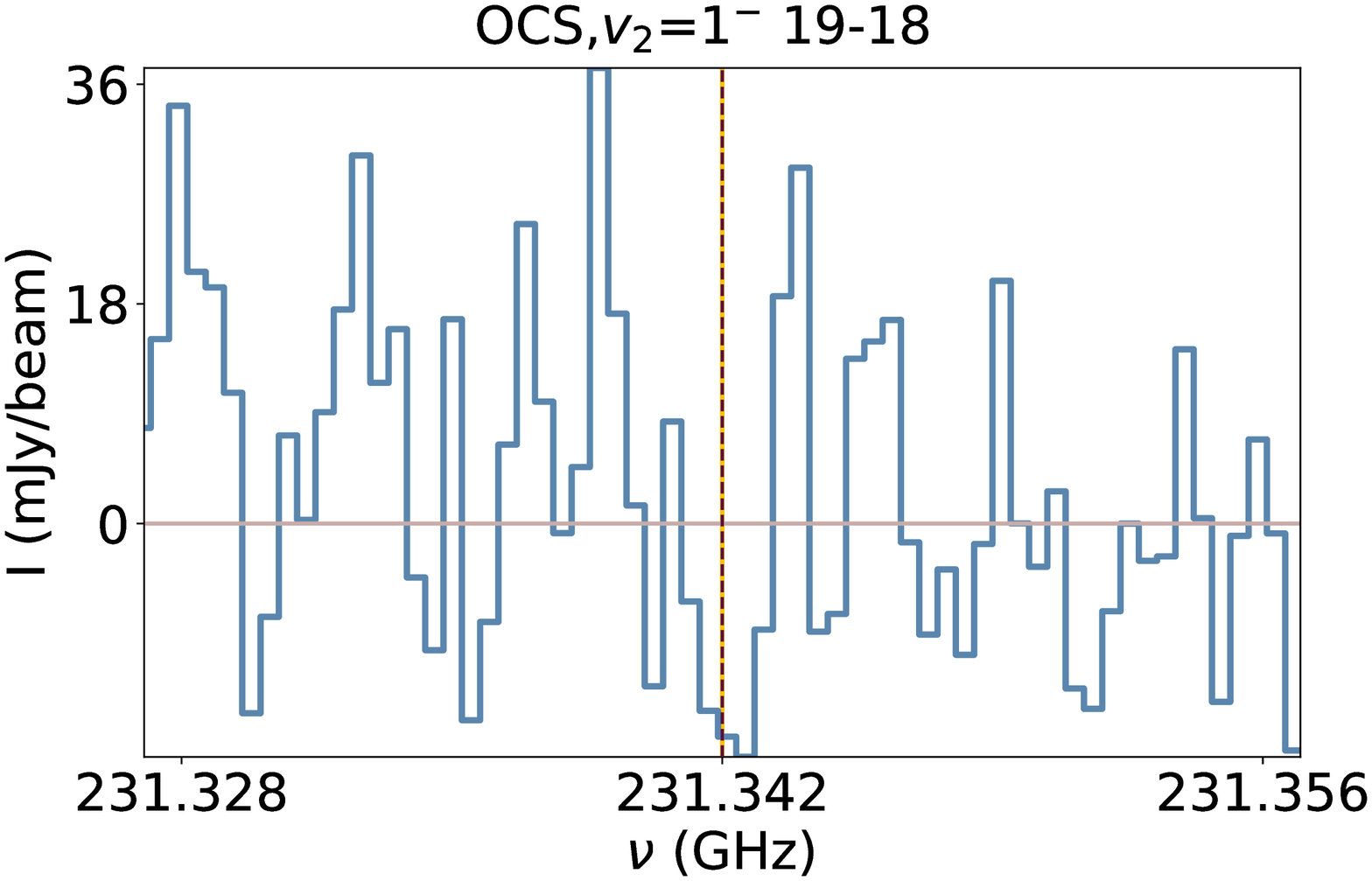}}
    \subfigure{\includegraphics[width=2.0in]{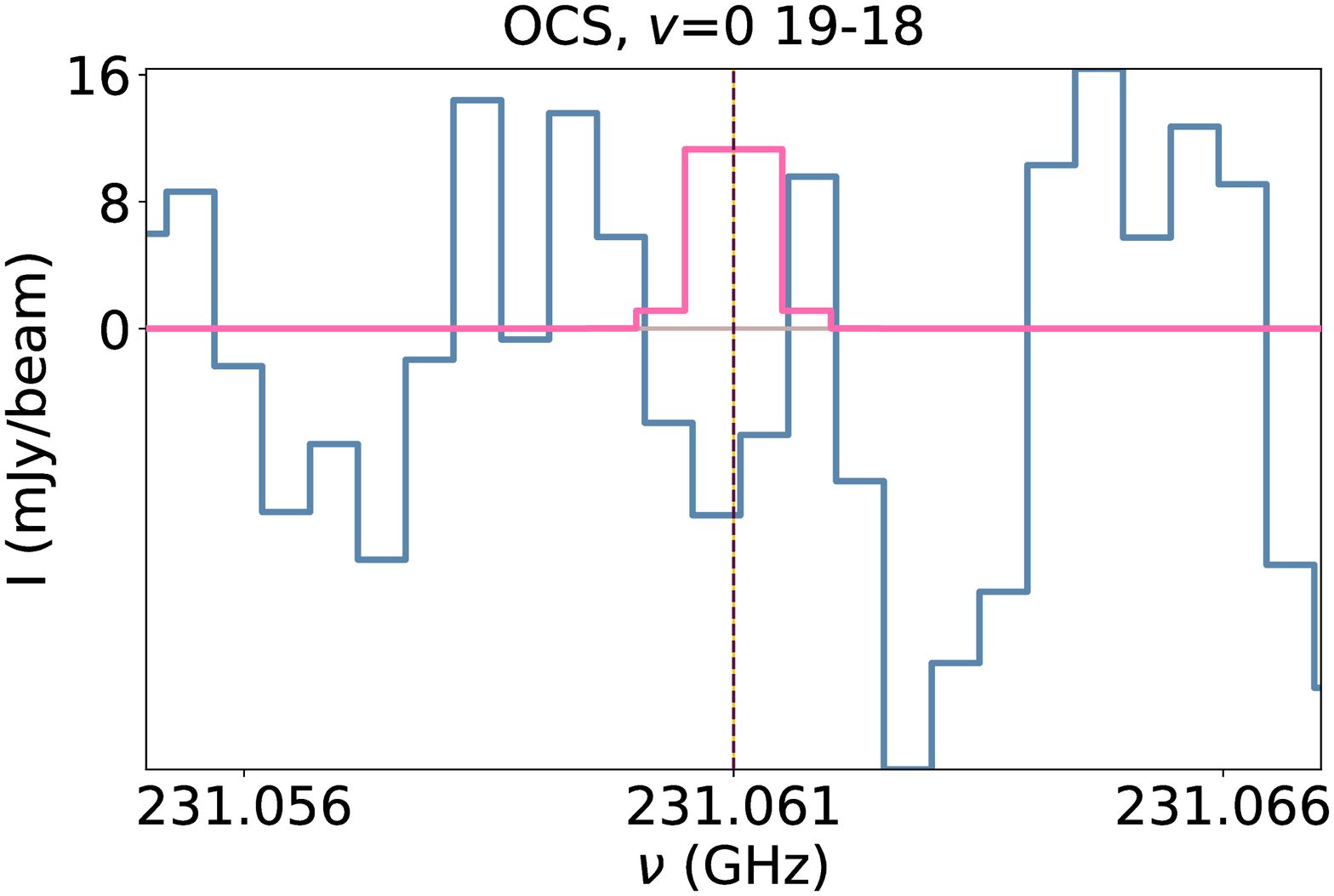}}\\
    \subfigure{\includegraphics[width=2.0in]{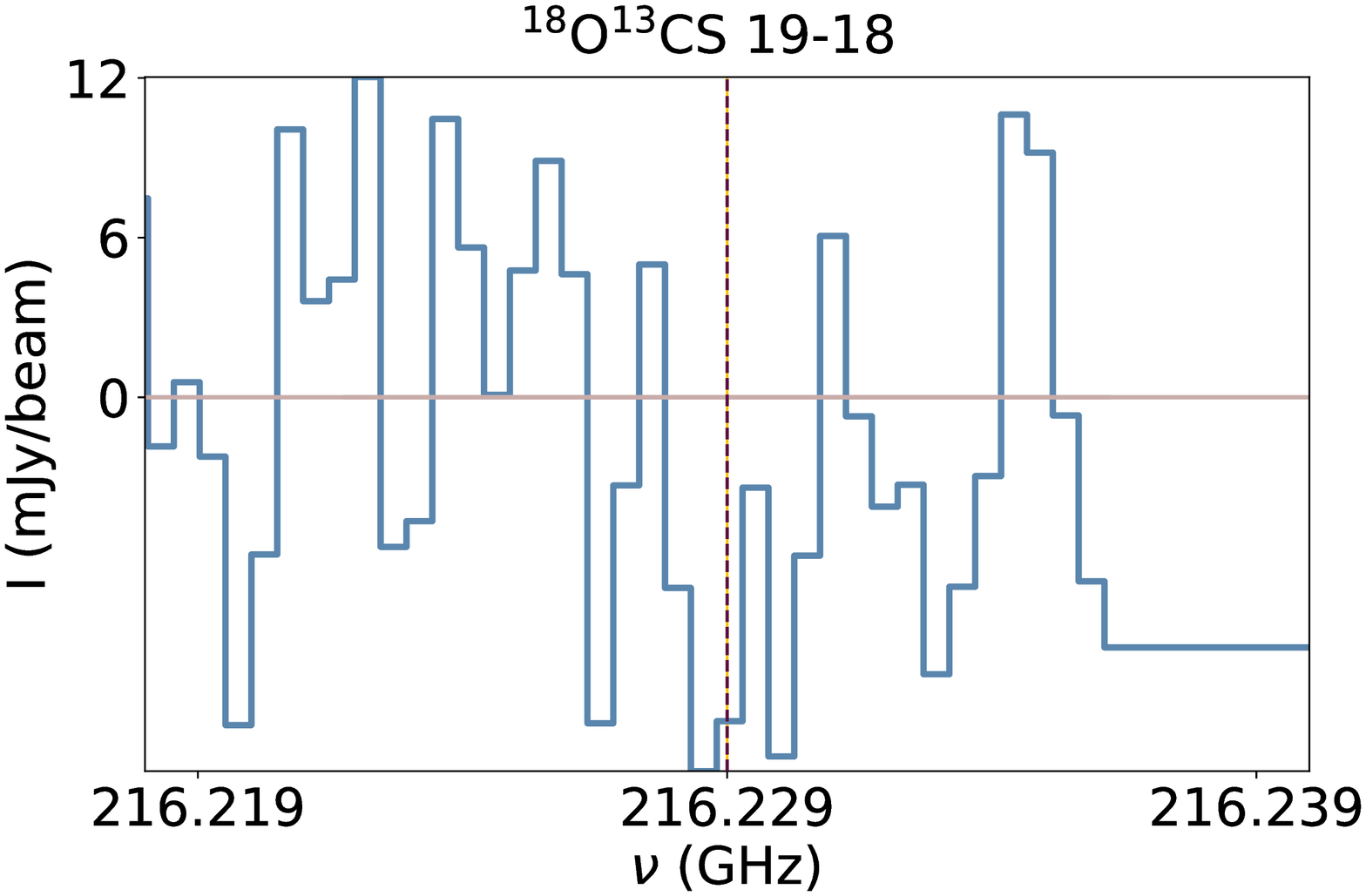}}
\end{minipage}
\caption{Observed spectra (in blue), rest frequency of the undetected line (brown dashed line), and spectroscopic uncertainty on the rest frequency of the undetected line (yellow shaded region) plotted for the sulfur-bearing species undetected towards TMC1. Synthetic spectra (in pink) fitted to the H$_2$S, and the OCS, $v$=0 line with the 1-$\sigma$ upper limit on its column density.}
\label{undetected_TMC1}
\end{figure}

No S-bearing species are detected towards TMC1. Thus, 1-$\sigma$ upper limit on the column density of H$_2$S, 2$_{2,0}$-1$_{1,1}$ and OCS, $v$=0, $J=19-18$ are computed with a FWHM of 1 km s$^{-1}$, for an excitation temperature of 40 K, and a source size of 1.5$''$ to include both the components of TMC1 \citep{Harsono2014}. The upper limits on the column density of H$_2$S and OCS are given in the \autoref{results_TMC1}. The undetected lines are shown in \autoref{undetected_TMC1}.

\begin{table}[H]
    \centering
    \caption{Synthetic fitting of the undetected main S-bearing species towards TMC1 for an excitation temperature of 40 K, a FWHM of 1 km s$^{-1}$, and a source size of 1.5$''$.}
    \label{results_TMC1}
    \begin{tabular}{r r c r c c  c r}
    \hline
    \hline
    Species & Transition & Frequency & $E_{\text{up}}$ & $A_{ij}$ &
    Beam size & $N$ \\
    & & (GHz) & (K) & (s$^{-1}$) & ($''$) & (cm$^{-2}$)  \\
\hline
& & & & &  &  &  \\
H$_2$S & 2$_{2,0}$-2$_{1,1}$ & 216.710 & 84 & 4.9$\times$10$^{-5}$  & 6.4 &  $\leq$1.5$\times$10$^{13}$ \\
OCS, $v$=0 & 19-18 & 231.061 & 111  & 3.6$\times$10$^{-5}$  & 5.9 & $\leq$2.6$\times$10$^{13}$\\
\hline
\hline
    \end{tabular}
\end{table}

\subsection{Undetected lines in TMC1}

\end{appendix}
\end{document}